\documentclass[onecolumn,useamsfonts,mfastrosym]{pasj01}

\Received{$\langle$reception date$\rangle$}
\Accepted{$\langle$acception date$\rangle$}
\Published{$\langle$publication date$\rangle$}

 \usepackage{color,times}
 \SetRunningHead{Astronomical Society of Japan}{Usage of \texttt{pasj00.cls}}
\definecolor{LinkBlue}{RGB}{6,69,173}
\definecolor{DarkBlue}{RGB}{11,0,128}
\definecolor{red}{rgb}{1,0.,0.}
\usepackage{cuted}
\usepackage{numcompress}

\def\be{\begin{equation}}
\def\ee{\end{equation}}
\def\bea{\begin{eqnarray}}
\def\eea{\end{eqnarray}}

\newcommand{\Mie}{\mathcal{M}}
\newcommand{\Sie}{\mathcal{S}}
\newcommand{\il}{~}

  \newcommand{\Qa}{\mathcal{Q}}
  \newcommand{\Sa}{\mathcal{S}}
\newcounter{num}

\begin{document}

\title{{Jet collision with  accreting tori around SMBHs}\\
{GRHD and   light surfaces constraints  in   aggregates of  misaligned tori}}
\author{D. Pugliese and Z. Stuchl{\'{\i}}k}%
\altaffiltext{}{Research Centre for  Theoretical Physics and Astrophysics\\
Institute of Physics,
  Silesian University in Opava,\\
 Bezru\v{c}ovo n\'{a}m\v{e}st\'{i} 13, CZ-74601 Opava, Czech Republic }


\maketitle

\begin{abstract}
We explore the possibility  of jet collisions   with  accreting tori  orbiting around \textbf{SMBHs}. The analysis provides constraints  on formation and  the observational evidences of  the     host configurations.
We  use a  \textbf{GRHD} model, investigating the   light surfaces contraints  in   aggregates of  misaligned tori orbiting a central static Schwarzschild black hole. Each  (toroidal) configuration  of the agglomeration is   a geometrically thick, pressure supported,  perfect fluid torus. Aggregates  include  proto-jets,  the open cusped solutions  associated to the geometrically thick   tori. Collision  emergence and  the stability properties of the  aggregates  are considered at  different inclination angles relative to a fixed distant observer.  We relate the constraints to the relevant frequencies of the configurations and fluid specific angular momentum, separating  the constraints related to the  fluids hydrodynamics  and to the geometric backgrounds. We  analyze existence of accreting tori supporting  jet-emission.
We discuss  the existence of orbit-replicas that could host shadowing  effects in replicas  of the emissions  in two  regions,  close and far from the \textbf{BH} (horizon replicas in  jet shells).
The investigation clarifies  the role of the pressure gradients of the   orbiting matter and  the essential  role of the  radial gradient of the  pressure in the determination of the disk  verticality.
Finally we analyze  the possibility that a toroidal magnetic field could be related to the collimation of proto-jets.

\end{abstract}

\section{Introduction}
We  investigate  the possibility  of jet collisions   with  misaligned accreting tori  orbiting around super-massive black holes  (\textbf{SMBHs}).
 There are many observational evidences  concerning  different periods of accretion of \textbf{SMBHs} hosted in
active galactic nuclei  (\textbf{\textbf{AGNs}}), which are characterized by multi-accreting periods leaving   counterrotating and even misaligned orbiting structures  around the \textbf{SMBHs},   producing sequences of orbiting toroidal structures  with strongly   different features including  different rotation orientations \citep{Dyda:2014pia,Aligetal(2013),Carmona-Loaiza:2015fqa,Blanchard:2017zfe,Gafton:2015jja,natures,Nixon:2013qfa,Dogan:2015ida,Bonnerot:2015ara,Bonnell,WA,Aly:2015vqa}.

Misaligned  tori \textbf{AGNs} can be  then located   at small or
 relatively large distances from the central \textbf{SMBH} \citep{2006MNRAS.368.1196L,Herrnstein1996,Greenhill2003}.
Warped inner accretion
can explain
the relation between radio jets in \textbf{AGN} and the galaxy disk.
Evidences of misalignment    and  of  tilted disks and jets  are discussed  for example in \citet{Miller-Jones:2019zla},  focused on
  relativistic jets in a  stellar-mass
black hole,
 launched  and ``redirected" from the accretion
 and subjected to the  frame dragging effects,
or  on  the   images of
accreting black holes in presence of the
disk and jet misalignment \citep{King:2018mgw,Chatterjee:2020eqc,2011ApJ...730...36D,Fragile:2008sv}.  For the   relation between the   flow structure and  the inner region of the
tilted-disk see for example
\citep{franchini,LLL,Teixeira:2014una}.
There are several  indications of  a jet emission-accretion disk correlation, where
the inclusion of  tilted disks can affect  a very large number of aspects of the attractor characteristics,  such as the mass accretion rates of \textbf{SMBHs} at high red-shift, and  the spin--down or spin--up processes which can be associated  with the   rotational  energy extraction from the central \textbf{BH} due to the  interaction with the surrounding matter.

Jet emissions, significant for active galactic nuclei, are related to outflows of matter along the attractor symmetry axis--see for a general discussion \citet{KJA78,AJS78,Sadowski:2015jaa,Lasota:2015bii,Lyutikov(2009),Madau(1988),Sikora(1981)}.
Generally, the presence of jet emission is intrinsically related to \textbf{BH} and more specifically to \textbf{BH} accretion, although several aspects  of this process are  still under investigation.
Jets are also linked to the extraction energy having a role in \textbf{BH} magnetosphere  \citep{Uz,TNM,Uzdensky:2004qu,CCpuol,Mahlmann:2018ukr,Universe}.

In this work we consider a central static Schwarzschild \textbf{BH},  however, in a wider scenario in which a central Kerr attractor is involved it  is possible to relate directly    jet emission with energy extraction from the  spinning  \textbf{BH}-- due for example  to the
Blandford-- Znajek process. In this scenario there is  the magnetic field lines  torque and the Lense--Thirring precession regulated by  light surfaces, the Killing horizons and   the outer ergosurface of the geometry.
Relevant aspects of jets emission   still remain to be clarified, such as the origin of the jet, the jet launching, the  influence of the spin of the central attractor, role of the magnetic fields, dependence on the accretion matter, dependence on  the accretion  mechanism and disk model  (relation with   the accreting disk  inner edge), the jets  collimation, the different components of the jet (possibly characterized by  an inner articulated  structure) and the  velocity components.

In this analysis we make use of a  \textbf{GRHD} model, providing  constraints for the occurrence of tori-jets collision, framed in an agglomeration of misaligned toroidal structures   known as  Ringed Accretion Disks (\textbf{RAD})  orbiting a central static Schwarzschild black hole\citep{pugtot,dsystem,ringed,Multy,mnras2,cqg2020}.  The \textbf{RAD} model of tori aggregates was first developed as \textbf{eRAD}, featuring tori sharing same symmetry plane which is also the equatorial plane of the central super-massive \textbf{\textbf{BH}}. Each  (toroidal) configuration  of the agglomeration is   a geometrically thick, pressure supported,  perfect fluid torus. Aggregates  include  proto-jets,  the open cusped solutions  associated to the geometrically thick   tori.
 Collision  emergence and  the stability properties of the  aggregates  are considered at  different inclination angles relative to a fixed distant observer.

Tori considered in this analysis as jet source are
{opaque}  (with large optical depth) and
Super-Eddington (with high  matter accretion rates) disk, characterized by
an
ad hoc distributions of constant angular momentum. This model and its derivations are widely studied in the literature with both numerical and analytical methods, we refer  for  an extensive bibliography to \citet{abrafra}.
Gravitational force in these disks constitutes the basic ingredient of the accretion mechanism   independently of any dissipative effects that are   strategically   important   for the accretion process in the thin models \citep{[SS73],[S73],BHawley}.
Geometrically thick accretion disks  are associated to very  compact attractors, origins of strong  gravitational fields, for example \textbf{SMBHs}, and  they characterize the physics of most energetic  astrophysical objects as \textbf{AGN} or gamma-ray bursts.  As these tori are often located in   regions very close to the central \textbf{BH} attractor,  a full general relativistic treatment of the model is often required.
Although restricted by the typical assumptions of these simplified models, thick (stationary) disks provide a striking good approximation of  several  aspects of accretion instabilities in  different and  more refined dynamical models, for example  providing an estimation of the   tori  elongation on their symmetry plane, the  inner edge of quiescent and accreting disks, the tori   thickness, the  maximum height, and the critical pressure points.

The torus shape is defined by the constant Boyer potential  $W$:
closed equipotential surfaces define stationary equilibrium configurations, the fluid
can fill any closed surface.
The tori are associated to
 open  surfaces defining
dynamical situations as, for example, the formation of matter jets \citep{Boy:1965:PCPS:}.
There are also critical, self-crossing  (cusped) and closed equipressure surfaces.
Cusped closed tori govern the accretion onto the \textbf{BH} due to Paczy\'nski
mechanism, where
violation of the hydrostatic equilibrium  leads to accretion onto the central \textbf{BH}.
The relativistic
Roche lobe overflow at the cusp of  the equipotential surfaces is also  the
 stabilizing mechanism against the thermal and viscous instabilities locally,
and against the so called Papaloizou and Pringle instability globally \citep{Blaes1987}.

In this frame the presence of accretion disk can be relevant for the launching, collimation and  replenishment of jet materials.
More specifically, a    related aspect of the jets emission is
role of frame dragging for the spinning  attractor, the   Lense--Thirring effect induced by the  central attractor which can engine also
  the Bardeen--Petterson effect  (a process resulting in the  tearing up  of the orbiting disk).
A second relevant  aspect  in the outflow of orbiting matter is the  role of magnetic fields. We  face here this aspect  by considering the contribution of a toroidal magnetic field to the possibility of jets collimation in the aggregate of  misaligned tori.  The role of the disk  inner edge and, more generally, the accretion mechanism, is  also important for the proto-jets emission,   related to the jet-accretion correlation.  These issues are  correlated to the discussion of the complex morphology and location of the jet and, eventually, the possibility that jet emission of matter can have its own  complex inner structure, constituted by multi-layers and funnels with different velocity components.

In the analysis of  \citet{ringed,dsystem} we considered mainly agglomerated of tori, in the \textbf{RAD} and  \textbf{eRAD} models.
In \citet{Multy} the   Kerr \textbf{SMBHs} in \textbf{AGNs} are related to  \textbf{RADs} configurations, binding the fluid and \textbf{BH} characteristics, and providing indications on the situations where to search for \textbf{RADs} observational evidences. Whereas,  proto-jet configurations in \textbf{eRADs} orbiting a Kerr \textbf{SMBH}  are considered in \citet{proto-jet}.

\textbf{RADs} were  introduced in \citet{pugtot},  and   detailed as a fully general relativistic model of (equatorial) tori, \textbf{eRAD},  in  \citet{ringed}. The possibility of  instabilities, including open configurations related to jets, was discussed in \citet{open,long}. Constraints  on double accreting configurations were considered in \citet{dsystem}. Tthe  energetics of couple of tori corotating and counterrotating and \textbf{eRADs}  tori collisions around a Kerr central super-massive \textbf{BH}  were addressed in    \citet{Letter}.
  The effects of a toroidal magnetic field were included in    \citet{mnrasB},  and analysed in the formation of several magnetized accretion tori aggregated as \textbf{eRAD} orbiting around one central Kerr \textbf{SMBH} in \textbf{AGNs}.
  Charged  fluid tori were considered around magnetized black holes  (or neutron stars) in \citet{Kovar:2010ty,Slany:2013rml,Kovar:2014tla,Trova:2016ton,Kovar:2016kqh,Trova:2018bsf,Schroven:2018agz,Kovar:2011uh,Stuchlik:2004wk,Universe}, influence  of the dark energy was studied in \citet{1983BAICz..34..129S,1999PhRvD..60d4006S,2000A&A...363..425S,2005MPLA...20..561S,2016PhRvD..94j3513S,2008IJMPD..17.2089S,2009CQGra..26u5013S}.
  Special effects  around Kerr or Kerr-Newman  naked singularities  where discussed in \citep{1980BAICz..31..129S,2011CQGra..28o5017S,Blaschke:2016uyo,2017PhRvD..96j4050S}.

  Finally,  in  \citet{mnras2,cqg2020}  \textbf{RADs} with tilted disks were considered with the  limiting effects in clusters of misaligned toroids orbiting static \textbf{SMBHs}. We explored the  globulis  hypothesis, consisting in  the formation of  an embedding of orbiting multipole structures covering the central \textbf{BH} horizon. Together with this effect,  the  possibility that the twin peak high-frequency quasi-periodic oscillations (HF-QPOs) could be related to the agglomerate inner  structure was explored,   considering several oscillation geodesic models associated to the toroids composing the aggregates   \citep{mnras2}.

Here, in the context of aggregates of multi tori with  different matter flows, we included proto-jets, explicitly  exploring
 orbiting toroidal jets shell, the  possible location of  launching point  (here associated to the cusp), the jets  interaction with surrounding matter, considering  tori-proto-jets collisions. Presence of surrounding matter is also relevant for the  possible mechanism of jets  replenishment,  enhancing eventually  chaotic processes due to the impact of accreting material on the inner configuration \citep{Multy}. Constraints on  jets impacting tori  are provided in terms of misalignment  angles and   tori parameters.
 The analysis also  discusses  the observational evidences of     host configurations, investigating the   light surfaces constraints.  We relate the constraints to the relevant frequencies of the configurations and fluid specific angular momenta, showing the constraints related to the hydrodynamics  and to the backgrounds. We point out the  existence of orbit-replicas that could host shadowing  effects in replicas  of the emissions  in regions close and far from the \textbf{BH} (horizon replicas in  jet shells); these orbits have equal limiting photon orbital frequency.  These structures are related to recently introduced Killing metric bundles \citep{Pugliese:2020azr,Pugliese:2020ivz,Pugliese:2019rfz,Pugliese:2019efv,remnants}.
The investigation clarifies   also the role of the pressure gradients of the   orbiting matter and  the essential  role of the  radial gradient of the  pressure in the determination of the disk  verticality.

Finally, we explore the possibility that a toroidal magnetic field could be related to the collimation of  proto-jets:
including
a strong toroidal magnetic field, developed in \citet{Komissarov:2006nz,Montero:2007tc}, we address  the specific question of jets  collimation.
We prove  that the analysis of the    radial pressure gradient  inside each configuration of the aggregate  is sufficient to determine the disk verticality (poloidal projection of the pressure gradients). The use of a static attractor allows us to isolate, in  first approximation, in the determination of the constraints for the jets emission,  the role of the hydrodynamics of the system  and  the geometric causal structure and the magnetic field  in the onset proposed in Sec.\il(\ref{Sec:Komi}), excluding  the dragging effects that has to be present around a Kerr attractor.
The choice of toroidal symmetry in the motion of the fluid indicates actually  progress of a limiting  situation even for cusped disks generally associated to emerging accretion and therefore also the proto-jets interpretation.
The analysis is based on the assumption of the  preponderance of fluid hydrodynamics and the relativistic effects induced by the strong gravity   of the attractor in defining the constraints  of the toroidal  configurations: the limits are essentially
  \textbf{GRHD} induced  and constructed from the  light surfaces which are part of the constraints of the causal structure.

\medskip

This  article is structured as follows:
 In Sec.\il(\ref{Sec:app}) we introduce the model  discussing the  \textbf{GRHD} tori construction, and the main notations and definitions used in this analysis.
We  briefly discuss in Sec.\il(\ref{Sec:protp}) the  proto-jets emission hypothesis and  possibility of the  shells of jets in the \textbf{RAD} frame.
Relevant  frequencies of the  \textbf{GRHD} thick    tori   for the proto-jets emission considered here  are introduced   in Sec.\il(\ref{Sec:freqs}).
Constraints on these  \textbf{GRHD} systems are addressed in Sec.\il(\ref{Sec:constr}).
 In Sec.\il(\ref{Sec:Asymptotic}) we discuss the role of  the   ``asymptotic radius"  $r_{\infty}$ from  the normalization conditions on the fluids four-velocity of each component of the aggregate.
Limiting conditions on frequency and fluid specific angular momentum are  analyzed in Sec.\il(\ref{Sec:fre-mome-supri}).
In Sec.\il(\ref{Sec:hearLS}) we introduce the stationary observers and light--surfaces relevant for the proto-jets.
Consequently    replicas are derived in Sec.\il(\ref{Sec:replics}). These objects are significant for the possible observational evidences of  the jets shells.
{Impacting conditions} are  therefore discussed in Sec.\il(\ref{Sec:glod-powe}).
{Tori characteristics, limiting conditions and pressure gradients in the tori} are the focus of Sec.\il(\ref{Sec:polar-gradient}).
In  Section (\ref{Sec:cross-interaction}) we explore the intersections between the toroidal  surfaces in different topologies.
{Proto-jets collimation in presence of a  toroidal magnetic field} is analyzed  in  section (\ref{Sec:Komi}).
Concluding remarks are in Sec.\il(\ref{Sec:conclusions}).
\section{GRHD tori construction}\label{Sec:app}
  We start  by considering  the Schwarzschild metric
\bea\label{11metrica}
&&
ds^2=-e^{\nu(r)}dt^2+e^{-\nu(r)}dr^2
+r^2\left(d\theta^2+\sigma d\phi^2\right),\quad\mbox{where}\quad
  e^{\nu(r)}\equiv\left(1-2/r\right),
\eea
 written in standard spherical coordinates
$(t,r,\theta,\phi)$, the outer horizon is $r_+=2M$, where $M$ is the \textbf{BH} mass. (In the following where more convenient we often use dimensionless units where $M=1$) and $\sigma\equiv (\sin\theta)^2$.
 It is convenient to list the    quantities
$
{E} \equiv -g_{ab}\xi_{t}^{a} p^{b},$ and $ L \equiv
g_{ab}\xi_{\phi}^{a}p^{b} 
$,
  constants of motion for test particle geodesics with four-momentum $p^a$,   related to the   spacetime  Killing vectors $\xi_t$ and $\xi_\phi$.
From these
 quantities we can define the function  effective potential $V_{eff}(r,\sigma;\ell)$ for the fluids,  together with the  relativistic angular frequency  $\Omega$ and the fluid specific angular momentum $\ell$:
\bea\label{Eq:poll-delh}
&&U^r =\sqrt{E^2-V_{eff}^2},
\quad\Omega \equiv\frac{U^{\phi} }{U^t }=-\frac{g_{tt} L}{{E} g_{\phi \phi}}= -\frac{g_{tt} \ell}{g_{\phi\phi}},
\quad\mbox{and}\\ &&\nonumber
\ell\equiv\frac{L}{{E}}=-\frac{g_{\phi\phi} \Omega }{g_{tt} }=-\frac{g_{\phi\phi}}{g_{tt}}\frac{U^{\phi} }{U^t }.
\eea
Assuming $U^r=0$, where $U^{a}$ is the fluid four-velocity,  there is   $V_{eff}=E$, and we obtain explicitly
\bea&&
U_t=V_{eff}=\sqrt{\frac{-g_{tt} g_{\phi \phi}}{g_{\phi \phi} +\ell^2g_{tt}}}=
\sqrt{-\frac{g_{tt} ^2(U^t)^2}{g_{tt} (U^t) ^2 +g_{\phi\phi}(U^\phi)^2}}=\sqrt{-\frac{{E}^2 g_{tt} g_{\phi\phi}}{{E}^2 g_{\phi\phi}+g_{tt}  L(\ell)^2}}.
\eea
 It is then convenient  to define the following angular momenta
\bea&&\label{Eq:wEo}
 \ell(r)\equiv\sqrt{\frac{\sigma r^3}{ (r-2)^2}}; \quad L(r)=\pm\sqrt{\frac{\sigma   r^2}{r-3}},\quad  L(\ell)=\sqrt{\frac{r^2\sigma }{e^{-\nu}r^2 \sigma^2\ell^{-2}-1}}\quad (\ell\neq0),
\eea
where
 $L(r)$  is the distribution of conserved  angular momentum   for test particle (geodesic) circular  motion.  $L(\ell)$ expresses  $L$ as function of the  fluid specific angular  momentum $\ell$ (asymptotically,  for  large radius  $r$, there is $L\approx \ell$). Function $\ell(r)$  is the
 distribution of fluid specific  angular momentum $\ell$   for the agglomerate of orbiting tori (\textbf{RAD}).
 %
%
%

The {\textbf{GRHD}}  system of multi-tori  we consider here   describes  clusters of  tori orbiting around one center \textbf{BH} attractor, and  composed by  perfect (simple) fluid   governed by the energy momentum tensor
\bea
\label{E:II}&&
T_{ab}=(p+\rho)U_a U_b - p g_{ab},
\eea
$p$ is the HD pressure and $\rho$ is the fluid density, as measured by an observer moving with the fluid whose four-velocity is  $U^{a}$.  The fluid is regulated by   barotropic equation of state $p=p(\rho)$.
The  fluid dynamics  is described by the {continuity  equation} and the {Euler equation} respectively:
\bea\nonumber
&&
U^a\nabla_a\rho+(p+\rho)\nabla^a U_a=0,\quad
(p+\rho)U^a \nabla_a U^c+ \ h^{bc}\nabla_b p=0,
\eea
 $h_{ab}=g_{ab}+ U_a U_b$  is  the projection tensor and $\nabla_a g_{bc}=0$.
The system symmetries  (stationarity and axial-symmetry), imply that the orbiting configurations are  regulated  by the Euler equation only:
 we  assume $\partial_t \mathbf{Q}=0$ and
$\partial_{\phi} \mathbf{Q}=0$, for any quantity  $\mathbf{Q}$.
Within our assumptions $(U^r=0, U^{\theta}=0)$,   we obtain from
the Euler equation
\bea&&
	\frac{\nabla_a p}{p+\rho}=-\nabla_a\ln(V_{eff})+\frac{\Omega \nabla_a \ell}{1-\Omega\ell},
\eea
 expression for the \emph{radial
pressure gradient},  regulated by the radial gradient of the effective potential  and the polar pressure gradient regulated by the polar gradient of the effective potential.
 The  specific angular momentum  $\ell$ is   here   assumed  constant  and conserved \emph{per} configuration ($V_{eff}(\ell)$ is the {torus effective potential} function).
Tori are determined by the equipotential surfaces   $V_{eff}(r;a)=K=$constant.

The \textbf{RAD} model  describes  an  orbiting  macrostructure formed by an agglomeration of tori orbiting  one central attractor  \citep{pugtot,ringed,dsystem,open,Letter}.  The specific angular momentum $\ell$,   assumed  constant  and conserved for each torus,   is a  variable as $\ell(r)$  in the \textbf{\textbf{RAD}} distribution.
We  distinguish a \textbf{RAD} where tori have all possible relative inclination angles (\citet{cqg2020,mnras2}), considering tilted or misaligned disks,
 and  the \textbf{eRAD} where all the tori are on the equatorial plane of the central (axially-symmetric or spherically-symmetric) attractor,  coincident with the symmetry plane of each toroid of the aggregate.
The model  constrains  the  formation of the centers of maximum pressure and density in an agglomeration of orbital extended matter, emergence of \textbf{RADs} instabilities in the
phases of accretion onto the central attractor and tori collision emergence \citep{dsystem,Multy,Letter}.
The instability points in the configurations are the  minima of  density and pressure in the disk related to the extreme of the effective potential as function of the radius $r$.
Introduction of the \textbf{RAD} rotational law $\ell(r)$ (or eventually keeping the explicit dependence on the poloidal angle $\ell(r; \sigma)$) is a key element of the  construction of the model. This function provides  the distribution of maximum and minimum points of density and pressure in the orbiting  extended matter configuration. The minima  of the tori effective potentials, as functions of $r$, are seeds for the tori  formation, coincident with the maxima  of pressure and density, the tori centers $r_{cent}$.  Closer to the central attractor are   unstable points (the maxima of the effective potentials) associated to the tori cusps  $r_{\times}$ (associated to the accreting phase) or proto-jet cusps  $r_{j}$ (the open cusped configurations). Unstable  points closer to the central attractor have  greater  centrifugal component,  leading eventually to proto-jets  open configurations.  The  tori located (center of maximum pressure location)  very far from the attractor have an extremely large centrifugal component  and  no unstable HD phase  (not cusped tori are  also known as  quiescent tori).
(For  the  cusped and   closed equipotential surfaces,   the accretion onto the central black hole can occur through the
cusp $r_{\times}$ of the equipotential surface:  torus  surface exceeds
the critical equipotential surface (having a cusp),  for a mechanical
non-equilibrium process due to a violation of the hydrostatic equilibrium known as Paczy\'nski mechanism \citep{abrafra}).

\begin{figure}
\centering
  \includegraphics[width=6cm,angle=-90]{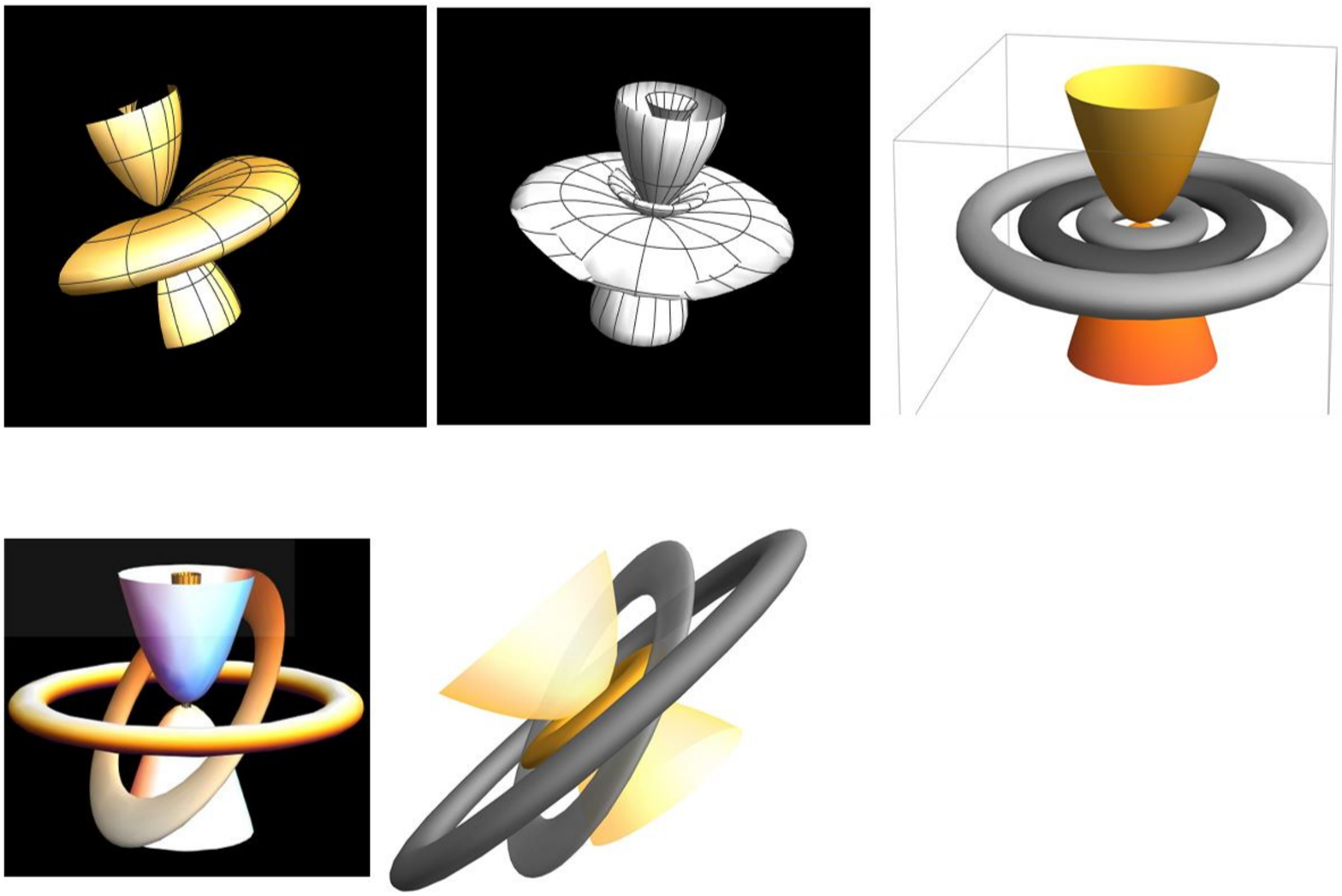}
  \caption{ Pictorial representation of a ringed accretion disk (\textbf{RAD}) with  shells of open
 surfaces with tilted accretion disks.}\label{Fig:Cat1}
\end{figure}

The tori  of the agglomerations  are geometrically thick disks, nevertheless the \textbf{eRAD}   models  are usually  geometrically thin  disks with the  internal ringed structured composition,    with differential rotating inter disks shells of jets--Figs\il(\ref{Fig:Cat1}), and distinctive set of internal activities as tori collision and different  unstable processes as runway instability, runaway--runaway  instability, presence  of obscuring tori\citep{Multy,long,open,proto-jet}.
Around an (almost) static attractor, a \textbf{RAD}  can  evolve in  an embedded  \textbf{SMBH}, i.e.  configurations of  orbiting multi-poles  structure  screening  the central  \textbf{BH} horizon   to an observer at infinity. Eventually this situation   may give rise to the collapse of the orbiting innermost shells,     resulting in   an extremely violent outburst\citep{cqg2020,mnras2}.

\begin{figure}
\includegraphics[width=5cm]{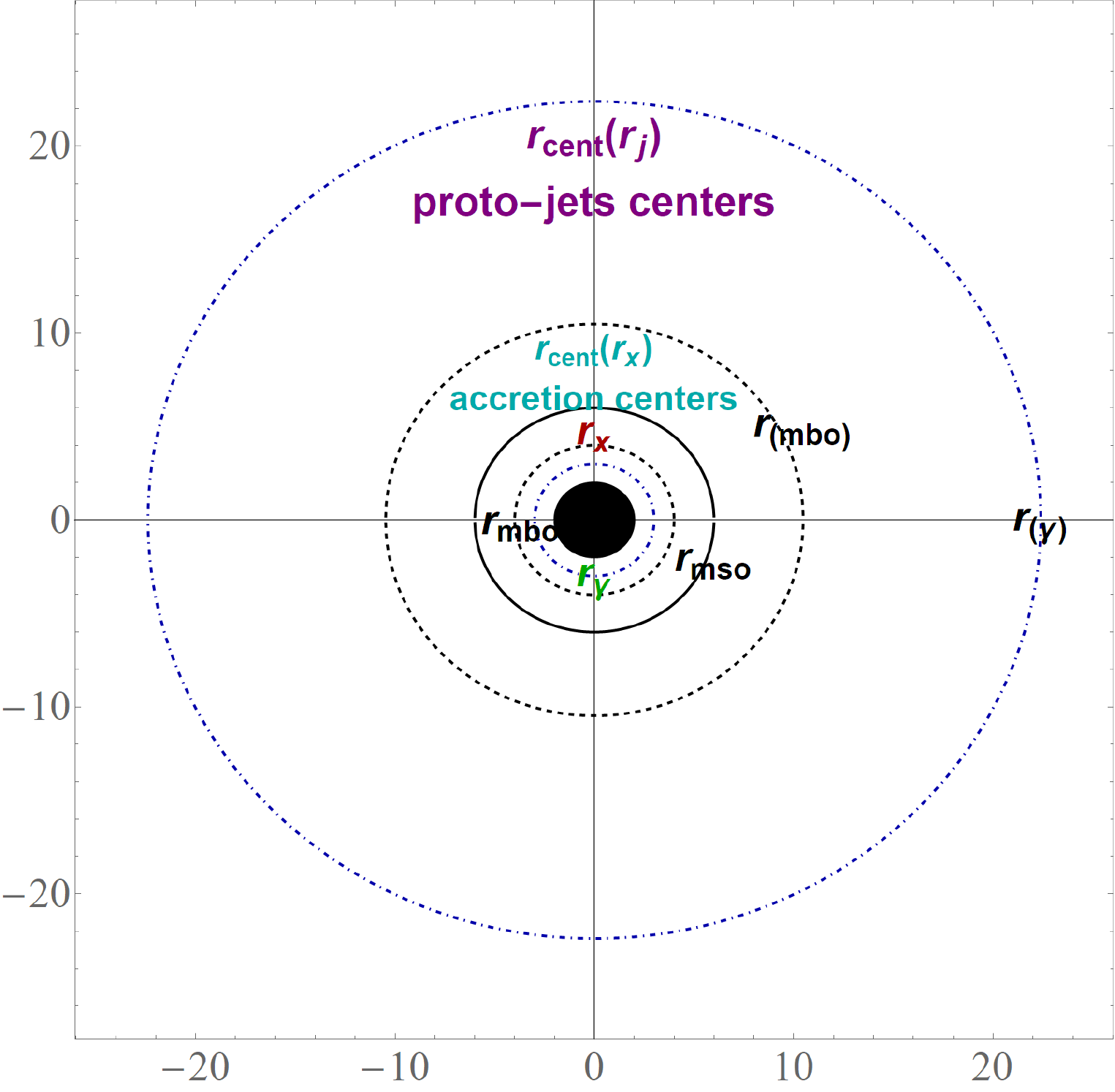}          \includegraphics[width=5cm]{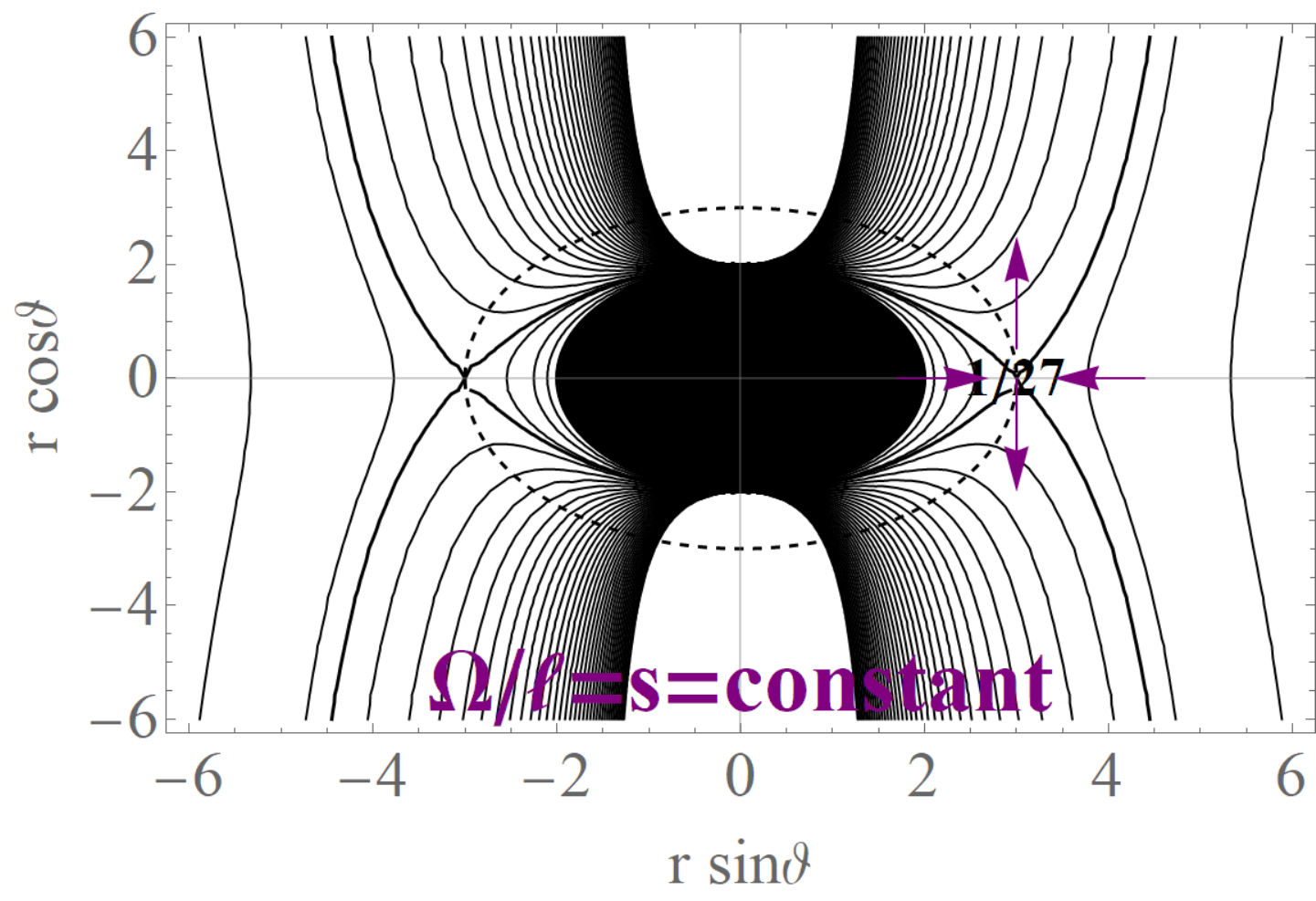}
          \includegraphics[width=5cm]{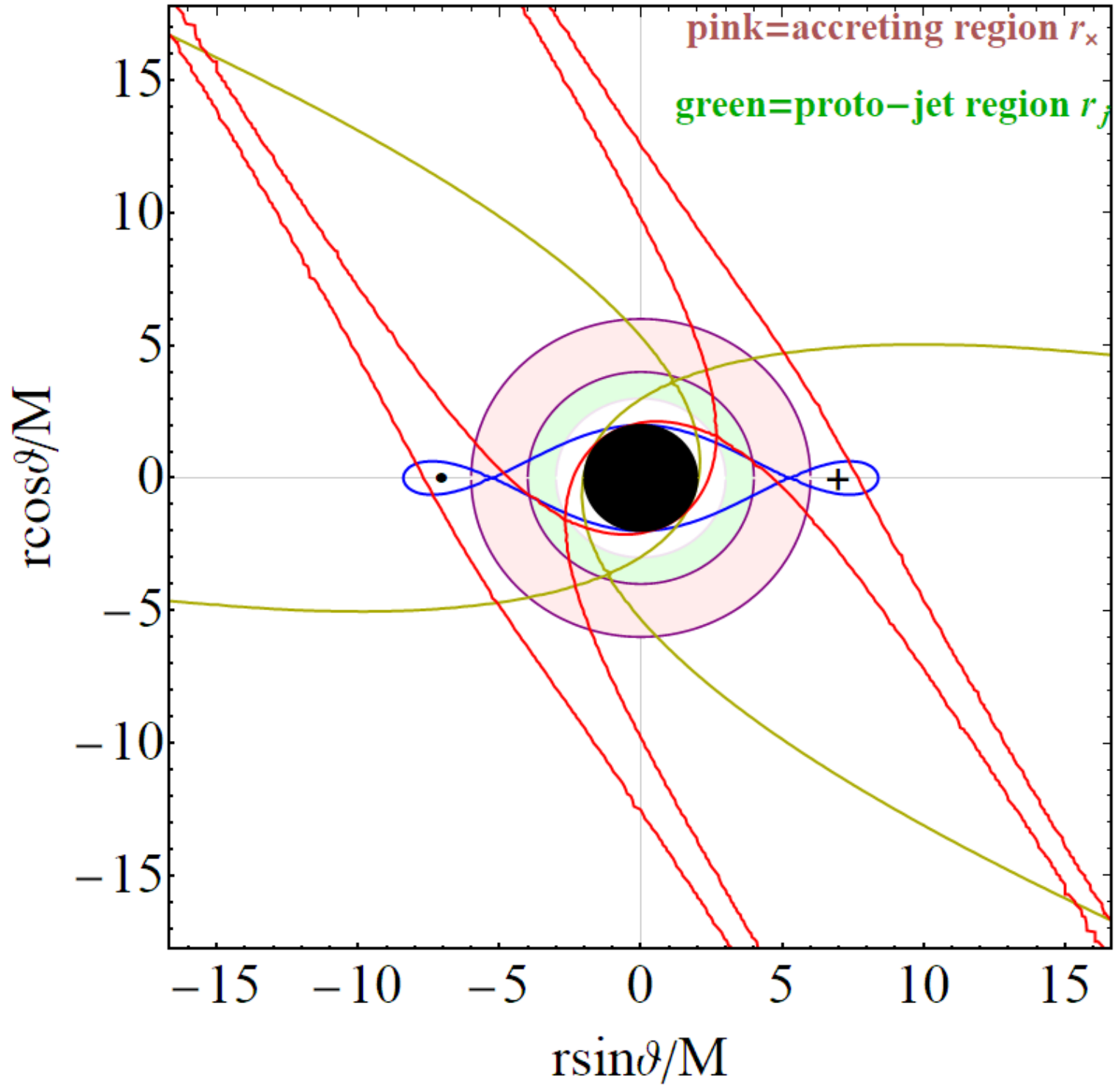}
\caption{Black region is the central \textbf{BH}. Left panel: Cusps of accreting tori are in $r_{\times}\in]r_{mbo},r_{mso}[=]4M,6M[$, the center in $r_{cent}(r_{\times})\in[r_{mso},r_{(mbo)}[$.
Cusps of open cusped proto-jets configurations are in $r_j\in]r_{\gamma},r_{mbo}[=]3M,4M[$, the center in $r_{cent}(r_j)\in[r_{(mbo)},r_{(\gamma)}[$. Configurations in $r>r_{(\gamma)}$ are quiescent. Center panel: $\Omega$ is the fluid relativistic angular velocity, $\ell$ is the fluid specific angular momentum, $\Omega/\ell$ represent the von Zeipel surfaces, black central region in the figures is the central Schwarzschild \textbf{BH}. Here $\sigma=\sin \theta$.  Right panel: The proto-jet configurations. Black region is $r<r_+$, where $r_+$ is the \textbf{BH} horizon. Regions where cusps, $r_{\times}$, in the cusped closed surfaces (accreting toroids) and $r_j$, in the cusped open surfaces (proto-jets) are shown.}\label{Fig:DopoDra}
\end{figure}

More specifically, each (closed) toroidal  component is a thick, opaque (high optical depth) and super-Eddington,  radiation pressure supported  accretion disk,  cooled by advection with low viscosity.  In these toroidal disks, the  pressure gradients are crucial although we shall prove here that, for each disk,  in its own adapted frame,  the radial pressure  gradient is relevant for the  large part of the stability analysis  of the tori and agglomerations and to even fix the disk verticality (essentially defined by the polar pressure gradient).
In this model the entropy is constant along the flow.
Toroidal surfaces   are therefore  provided by the  radial gradient  projection $(U^r)$ of the Euler equation (in the adapted frame), or radial gradient of the effective potential, reducing the analysis  of each accretion disk or \textbf{eRAD} models to  a 1-dimensional problem.
In Sec.\il(\ref{Sec:polar-gradient}) we shall determine the verticality of the disks through the analysis of the radial gradients or equivalently the rotational law.
There exists, in general accretion disks,
an   extended region where the fluids angular momentum in the torus  is larger or equal  (in magnitude) than to the  Keplerian (test particle) angular momentum.
 Equation  $\partial_r V_{eff}$  can be solved for the specific angular momentum of the fluid  $\ell(r)$ defining the  critical points  of the hydrostatic pressure in the torus. We use also  the function
$K(r,\ell)\equiv V_{eff}(r,\ell(r))$, locating  the tori centers and providing  information on torus elongation and  density.
Constant pressure  surfaces are  essentially based  on the application  of the von Zeipel theorem
 (the surfaces of constant angular velocity $\Omega$ and of constant specific angular momentum $\ell$ coincide and  the rotation law $\ell=\ell(\Omega)$ is independent of the equation of state \citep{Lei:2008ui}
        \footnote{
Essentially, the application the  von Zeipel results  reduces to an integrability condition on the Euler equations. In the case of a barotropic fluid, the right hand  side of  the differential equation  is the gradient of a scalar,  which is possible if  and only if
$\ell=\ell(\Omega)$. The exact form of the rotational law  is  linked to scale-times of the main physical processes involved in the disks for transporting angular momentum in the disk, as in the  MRI process.  In the geometrically thick disks analyzed here, the functional form of the angular
momentum and entropy distribution during the evolution of dynamical processes, depends on the initial conditions of the system and not on
the details of the dissipative processes.}).
More specifically this implies  that  if $\Sigma_{\mathbf{Q}}$ is the  surface $\mathbf{Q}=$constant, for any quantity or set of quantities $\mathbf{Q}$,  there is  $\Sigma_{i}=$constant for \(i\in(p,\rho, \ell, \Omega) \),  where the angular frequency  is indeed $\Omega=\Omega(\ell)$ and it holds that  $\Sigma_i=\Sigma_{j}$ for \({i, j}\in(p,\rho, \ell, \Omega) \).
Therefore we consider here again  $\ell $, and  $ \Omega $,  obtaining
\bea&&\label{Eq:it-impo-ger1}
\Omega(\ell(r))=\frac{1}{(r-2) r^{3/2} \sin\theta },\quad
\partial_\ell\Omega(\ell(r))=\frac{\Omega(\ell(r))}{\ell(r)}=\frac{r-2}{r^3 (\sin\theta ) ^2}=s
\eea
$\Omega (\ell (r)) $ is the curve of the  relativistic velocity evaluated on the  \textbf{RAD} rotation curve, $ s $ defines the surface of von Zeipel  which we consider in  Figs\il\ref{Fig:DopoDra}.

Constraints  on the ranges of values of the fluid specific angular momentum $\ell$  follow from the properties of the  geometric background  and are essentially regulated by  the  {marginally stable circular orbit}, at $r_{mso}=6M$, the {marginally bounded circular orbit}, at $r_{mbo}=4M$, and  the {marginal circular orbit} (photon orbit), at $r_{\gamma}=3M$.

 Radius
$r_{\mathcal{M}}$  is instead a  solution  of  the condition $\partial_r^2\ell(r)=0$.
Alongside the geodesic structure, we  introduce also the   radii $r_{(mbo)}$ and $r_{(\gamma)}$, including also $r_{(\mathcal{M})}$,
 relevant to the location of the disk center and outer edge  and radius,
where
\bea&&\nonumber
{r}_{(mbo)}:\;\ell(r_{(mbo)})=
 \ell({r}_{{mbo}})\equiv {\ell_{mbo}},
\quad
  r_{(\gamma)}: \ell(r_{{\gamma}})=
  \ell(r_{(\gamma)})\equiv {\ell_{{\gamma}}}, \quad
r_{(\mathcal{M})}: \ell(r_{(\mathcal{M})})= \ell_{\mathcal{M}}
\\&&\label{Eq:conveng-defini}\mbox{and  there is}\quad
r_{\gamma}<r_{mbo}<r_{mso}<
 {r}_{(mbo)}<
 r_{(\gamma)},
 \eea
 here and in the following we adopt notation $\Qa_{\bullet}\equiv \Qa(r_{\bullet})$ for any quantity $\Qa$ evaluated on a radius $r_{\bullet}$.
 Below we  summarize the constraints on the ranges of fluids specific angular momentum:
\begin{itemize}
\item
$ \mathbf{L_1}\equiv[\ell_{mso},\ell_{mbo}[$. There are quiescent  ($K_{mso}<K_{cent}<K<K_{\max}<1$) and cusped tori $K=K_{\max}\equiv K_{\times}<1$. Here $K_{\max}$ is  $K$ at the maximum point of the effective potential, in $\mathbf{L_1}$, the value at the cusp of a toroid.
Accretion (cusped) point  is located in   $r_{\times}\in]r_{mbo},r_{mso}]$ and the center of  maximum pressure/density   is  at $r_{cent}\in]r_{mso},r_{(mbo)}]$.
\\
\item $ \mathbf{L_2}\equiv[ \ell_{mbo},\ell_{\gamma}[ $. There are
quiescent  tori ($K_{mso}<K_{cent}<K<K_{\max}>1$)  and proto-jets  ($K=K_{\max}>1$) .
 Unstable  cusped point of proto-jets are located at  $r_{j}\in]r_{\gamma},r_{mbo}]$  and   the center of  maximum density/pressure is at $r_{cent}\in]r_{(mbo)},r_{(\gamma)}]$;
\\
\item $\mathbf{L_3}\equiv(\ell>\ell_{\gamma})$. There are   quiescent  tori  with center  at $r_{cent}>r_{(\gamma)}$.
 \end{itemize}
 In this context, the radii $r_{(\mathcal{M})}$ in the unstable region and  $r_{\mathcal{M}}>r_{mso}$ in the stability region,  are interpreted  as extreme  in the distribution of seeds of tori (the stability points) and instability points (corresponding to the left range $r<r_{mso}$). These are interpreted as the  maximum  of distributions of  points of the aggregation seeds  (these radii are related to the derivatives of certain frequencies of oscillations typical of  thick toroidal structures) \citep{mnras2}. Concluding, it is possible to define a further limit where only inner  Roche lobe of matter ``encircling" the \textbf{BH} horizon or  open surfaces can form, namely $\mathbf{L_0}: \ell<\ell_{mso}$ and $\mathbf{K_0}\equiv [K_{mso},1]$ see \citet{open,proto-jet}

\textbf{{Toroidal surfaces as equipotential surfaces}}

From  the Euler equation featuring the radial gradient of the pressure, it is immediate to find  the following form of the  toroidal surfaces on each plane $\theta$:
\bea&&\label{Eq:condizioni}
\forall \theta:\quad\left[\frac{2 \left(\mathcal{B}^2+K^2 Q\right)}{K^2 \left(Q-\mathcal{B}^2\right)+\mathcal{B}^2}\right]^2-\mathcal{B}^2-\mathcal{Z}^2=0,\\&&\nonumber  \Sa_{eff}(\sigma,\ell)\equiv\sqrt{\left[\frac{2 \left(\mathcal{B}^2+K^2 Q\right)}{K^2 \left(Q-\mathcal{B}^2\right)+\mathcal{B}^2}\right]^2-\mathcal{B}^2}
\eea
as equipotential surfaces,
where $Q\equiv \ell^2$, $\mathcal{B}(x,y)$ and $\mathcal{Z}(x,y)$ are functions of Cartesian coordinates $x,y$. We set  $\mathcal{B}=x \cos\theta+y \sin\theta$ and $\mathcal{Z}=y \cos\theta-x \sin\theta$ .
 Function $\Sa_{eff}(\sigma,\ell)$, will also be useful projecting the equipotential surfaces on an adapted plane.
\subsection{On proto-jets emission hypothesis and  shells of jets}\label{Sec:protp}
   \textbf{RAD} models including  proto-jets contain  open cusped solutions  associated to geometrically thick tori,   associated to jet emission empowered by initial unstable fluid  centrifugal component (and eventually the dragging of the spacetime for models with Kerr attractors). The exact significance of these configurations is still under debate \citep{open,proto-jet,Lasota:2015bii}.
Here we consider the appearance of the open cusped surfaces with cusp  $r_{j}$ (correspondent to a  minimum of pressure) in the \textbf{RAD} frame,  where  proto-jets  are possible in the misaligned  tori scenario.  Proto-jets will have different spin orientations (related to the direction orthogonal with respect  to the configuration equatorial plane), creating possibly an intriguing  complex structure  with clear impact on the associated  stability and phenomenology of the accretion disks-jets systems.
    Proto-jets  are  not  ``geometrically  correlated" directly with accretion,  i.e.,  accretion cusp occurs in the range  $r_{\times}\in[r_{mbo}, r_{mso}]$, the  fluid having initial specific angular momentum $\ell\in[\ell_{mso},\ell_{mbo}]$, the fluid supporting proto-jet has instead a  cusp (launching point) in $r_j\in[r_{\gamma}, r_{mbo}]$ (therefore closer to the \textbf{BH} than the accretion  point) and  center  in $[r_{(mbo)},r_{(\gamma)}]$, therefore the proto-jet structure is a shell englobing the accreting configuration distinguishing  these  configurations from other open structures--see Fig.\il(\ref{Fig:Cat1}) and Fig\il(\ref{Fig:compePlottion}) .
  (Details on possible collimation around a Kerr central \textbf{BH}, the collimation angles, and differences between corotating and counter-rotating proto-jet can be found in \citet{proto-jet}.)
In these structures  we should  note how the centrifugal  component  should be not ``too large", i.e. if   $\ell>\ell_{\gamma}$ then the disk is stabilized against the formation of the cusp.
The existence of such configurations, as noted in \citet{proto-jet}, can be interpreted as limiting surfaces for accretion and jet emission, i.e.,  not as actual  matter surfaces but as significant limiting matter funnels for several types of emission.
 (It would also be noted that   boundary conditions on the  Euler equations  have to be  re-interpreted   leading to the case of proto-jets.)
Reconsidering the origin of proto-jets, in fact,  these emerge as the unstable configurations for the closed tori with  the specific  conditions on the angular momentum and $K$ parameter. Therefore, the kind of instability leading to the  formation of cusp in such conditions is not yet  completely understood. It is however possible that tori eventually formed within  the condition $\ell\in[\ell_{mso},\ell_{\gamma}]$, as conjectured in \citet{dsystem}, could more or less rapidly undergo a phase of angular momentum decreasing  bringing the  torus  to the condition for accretion.
The presence of proto-jet cusp is regulated also by the $K$ parameter, $K_{j}$ corresponding to the value of the fluid effective potential at the maximum point, where  $K_{j}>1$ is located very close to the central \textbf{BH}, closer than the accretion point. These values of $K$ correspond to very large tori, supporting therefore also in the case of proto-jets some kind of correlation, although not direct with accretion disks (rather than accretion mechanism). We should also note that  there could be the concomitant formation of internal proto-jet associated to an outer  toroid, and related to a disk  between them,  in accretion, replenishing also the cusp of the proto-jet.
Then we have a cusp  close  to the central \textbf{BH} jet $r_j$, followed by the cusp $r_\times$ of accretion, the center of maximum pressure and density of the accretion disk, followed by the center of maximum pressure and density relative to the proto-jet cusp. The fluid of the inner shell has a higher specific  momentum of the intermediate shell where there is an accretion point and maximum pressure of the accretion disk, which then replenishes the jet with a  fluid with initial lower momentum-- in   the \textbf{eRAD} case with  the same direction of rotation, or with any direction in the \textbf{RAD} case around a Schwarzschild central \textbf{BH}.
The exact proto-jet  shell structure  between the internal cusp $r_j$ and its center is not in fact well known, this shell   eventually incorporates the accretion torus. (We note that in the case of Kerr \textbf{BH} there can be also  an external shell ``breaking" the internal accretion disk. This limiting occurrence, regulated by the background geometry and precisely by the  Kerr \textbf{BH}  dimensionless spin, distinguishes also  the torus direction of  rotation \citep{open,proto-jet,long}.)

The analysis  in \citet{proto-jet}, focused on  proto-jets has ultimately singled out  the role of a broader set of the  limiting surfaces directly connected to proto-jets: \textbf{{(i)}} the  {$\gamma$-surfaces} and {\textbf{(ii)}}{h$\gamma$-surfaces}. It was proved that   there is a strict correlation  between   different $\gamma$-surfaces, which are defined as property of   the spacetime structure
and  the  $h\gamma$-surfaces  emerging from the matter models,  limiting  the  fluid  toroidal configurations as well as the proto-jets.

\textbf{{ (i)}} The {  {$\gamma$-surfaces}} are   related to the geometric properties of the Kerr or Schwarzschild  spacetimes, associated to the  specific  angular momentum $\{\ell_{mso},\ell_{mbo},\ell_{\gamma}\}$.

\textbf{{(ii)}} The limiting hydrostatic surfaces, {{$h\gamma$-surfaces}}, are associated with  each surface of constants $\ell$  whose topology  changes with values of $\ell$ and the \textbf{BH} spin dimensionless spin $a/M$.
For fixed Kerr \textbf{BH}, the $\gamma$-surfaces turn to the   limits  of the $h\gamma$-surfaces, approached by varying $\ell$.

We resume this issue here to characterize the role of the proto-jets in defining the main characteristics of the clusters.
Incidentally we note that these surfaces can be clearly connected with the light-surfaces (LS) in  the  Grad-Shafranov (GS) approach to magnetosphere  in a  Kerr spacetime,  and therefore also in the limiting Schwarzschild case, especially in the  presence of the thick accretion disks \citep{Uz,TNM,Uzdensky:2004qu,CCpuol,Mahlmann:2018ukr}.
There are clear  differences  between the model set up expressed here and the scenario of the  magnetosphere problem, the  divergence  consists  mainly in  the presence of the Kerr central \textbf{BH} and obviously  the presence of  an external magnetic field. There are  however connections between the open surfaces in the clusters  considered in the present analysis  and  the force free magnetosphere GS equation. More precisely, one point  consists in the GS  limiting surfaces, light surfaces and secondly  the boundary conditions considered for the integrations  of the GS problem. The second  relevant aspect consists in the case in  an  \textbf{eRAD} in the presence of an inner obscuring torus covering the \textbf{BH}  horizon, altering  the well known and widely used boundary conditions at the horizon used to integrate the \textbf{GS} equation. For all these reasons the analysis pursued here on the misaligned clusters of \textbf{GRHD} tori and proto-jets is seen as a preliminary analysis towards a more complex set of situations implied by the new accretion paradigm determined by the \textbf{RAD}.

It is   convenient  to
define the stationary observers and corresponding light surfaces (LS).
Stationary observes are observers with
 a tangent vector which is  a spacetime  Killing vector.
 Their    four-velocity  $U^\alpha$ is thus   a
linear combination of the two Killing vectors $\xi_{\phi}$ and $\xi_{t}$:
$
U^\alpha=\mathcal{L}^{\alpha}=  (\xi_t^\alpha+\omega \xi_\phi^\alpha$) and
where  $d\phi/{dt}={U^{\phi}}/{U^t}\equiv\omega$, for an analysis of these we refer to \citep{observers,remnants}. Therefore  stationary observers share the same symmetries of the configurations considered here which are also called stationary tori.
The  quantity $\omega$  is the orbital frequency of the stationary observer.
The coordinates  $r$ and $\theta$  of a stationary observer are constants along  its world-line, i. e. for example in the Kerr background   a stationary observer does not see the  Kerr spacetime changing along its trajectory.
Specifically, the causal structure defined by timelike stationary  observers is characterized by a frequency   bounded in the range $\omega\in]\omega_-,\omega_+[$. The limiting frequencies  $\omega_{\pm}$,   are photon orbital frequencies, solutions of the condition $\mathcal{L_{N}}\equiv \mathbf{g}(\mathbf{U,U})=g_{tt} +g_{\phi\phi}\omega^2= 0$, determine
the frequencies $\omega_H$ of the Killing horizons. Obviously, there is
$\mathcal{L_{N}}= 0$ at the horizons. The GS  nucleus in the approaches leading to the light-surfaces in the Schwarzschild case  is provided by $\mathcal{L_{N}}$.
Thus, the fluid   effective potential,  related to the four-velocity component  $U_t=g_{tt} U^{t}$,
is  not well defined on the zeros of the following  $\Pi(\ell)$ quantity
\bea&&\label{Eq:Pi.I}
\Pi(\ell)={g_{\phi \phi}+\ell^2g_{tt}}, \quad \Pi(U^t, U^\phi)=g_{tt} (U^t)^2+g_{\phi\phi}(U^\phi)^2,
\\&&
\mbox{and}\quad \Pi(L, E)= {E}^2 g_{\phi\phi}+g_{tt}  L(\ell)^2,
\eea
where we used relations in Eqs\il(\ref{Eq:poll-delh}).
Solutions  $\Pi=0$ are open configurations. Note that the $\Pi$ quantities are in fact related to $\mathcal{L}_\mathcal{N}$ (and  these to the von Zeipel surfaces) via Eq.\il(\ref{Eq:it-impo-ger1}) where $\Omega\equiv \omega$.
Thus,  $\Pi$ is related to the normalization factor   for the stationary observers,  establishing thereby the (GS) light-surfaces.
At fixed specific angular momentum $\ell$,  the zeros of the $\Pi$ function  define  {limiting surfaces} of  the fluid configurations.
For fluids with specific angular momentum $\ell>\ell_{\gamma}$, the limiting surfaces are  the cylinder-like  surfaces , crossing the equatorial plane on a point, without cusps, which is  increasingly far from the attractor with $\ell$. A second  closed  surface, embracing the \textbf{\textbf{BH}}, appears,  matching  in the limiting case  the outer surface at the cusp $r_{\gamma}$.
The light-surfaces, for  $\ell=\ell_{\gamma}$, can be interpreted as ``limiting surfaces'' of the  open Boyer  surfaces.
The   solutions of $\Pi(\ell)=0$, for  fixed  parameters   $\ell$ and $a/M$ (the dimensionless spin of the central Kerr \textbf{BH} attractor), define  the limiting hydrostatic surfaces, {$h\gamma$-surfaces}.
Concluding, there are three classes  of open matter configurations  bounded  by the limiting hydrostatic  surfaces.
The $\gamma$-surfaces are approached by changing  the specific angular  momentum, while the  $h\gamma$-surfaces are generally approached by an asymptotic limit of $K$, details on these are in\footnote{
There are three possible open configurations associated to thick tori:
\textbf{[I]:}
proto-jets-matter configuration open  (i.e. $r_{out}=\infty$) having features presented  in \citet{open,proto-jet} and cusped with $r_j$.
The closed associated configurations have clearly $K_{\min}(\ell)<K(\ell)<K_{\max}(\ell)$ \citep{open,long,mnras2,cqg2020}.
\textbf{[II]}
Limiting cusped  surfaces with $\ell=\ell_{\gamma}$ and $\ell_{mbo}$.
\textbf{[III]}
 Open  (not cusped) configurations associated to accreting   tori which have
$\ell\in [\ell_{mso},\ell_{mbo}]$, $r_{cent}\in [r_{mso},r_{(mbo)}]$.
The accreting tori, closed cusped configurations,  have cusp $r_{\times}$ or the inner edge $r_{in}$ in the range $[r_{mbo},r_{mso}]$.
These tori in their cusped form are smaller (lower elongation $\lambda$ and height than the pre-proto-jet). In general, the larger is the centrifugal component,  the largest is the configuration.
\textbf{[IV]}
At $\ell<\ell_{mso}$  there are  very slower rotating open matter funnels.
\textbf{[V]}
At $\ell>\ell{\gamma}$ there are  very large rotating open matter funnels associated to quiescent closed configurations, at $r_{cent}>r_{(\gamma)}$ with $K\in ]K_{mso},1[$.}\citep{proto-jet}.
\subsection{Relevant  frequencies  and  jets emission}\label{Sec:freqs}
In the  \textbf{GRHD}-\textbf{\textbf{RAD}}   frame we  include the jet-emission as   proto-jets  constraining  toroidal surfaces.
\textbf{GRHD} thick    tori  have  several characteristic frequencies. We can perform the analysis of the toroidal systems considered here in terms of the fundamental frequencies.
The four velocity of the (stationary) fluid   defined by
$
U^\nu = \gamma(\xi_t + \omega \xi_{\phi}),
$
(stationary observer),
where $\omega  = d\phi/dt= U^{\phi}/U^t$, particularly  there is $U_t = \gamma g_{tt} $ where   $ \gamma^{2} =-1/({g_{tt} +\omega^2g_{\phi\phi}})$  is a conformal factor {(related to the redshift factor)} defined by the normalization conditions  on the four velocity giving  the causal relation
(with signature $(-+++)$):
\bea&&
\label{omega-covariant}\ln V_{eff} = \ln\frac{1}{\sqrt{
 g^{tt}
 + \ell^2\,g^{\phi \phi}}},\quad  \omega = -\frac{\ell g_{tt} }{ g_{\phi\phi}}, \quad\mbox{and}\quad \ell =
-\frac{\omega g_{\phi\phi} }{
g_{tt}},
\quad
 \ell^2\neq L2_u\equiv-\frac{g^{tt}}{g^{\phi\phi}}.
\eea
from the condition of normalization for the relativistic frequency $\gamma$ we obtain
$\omega^2\neq W2_d\equiv-{g_{tt}}/{g_{\phi\phi}}$. (We note that, according to Eq.\il(\ref{omega-covariant}), this corresponds to the condition $\ell \neq1$.).
These   are related to
 the radial $\omega_r$ and vertical $\omega_{\theta}$ epicyclic frequencies, related to the polar and radial gradients of the effective potential.
 The epicyclic frequencies by the comoving observers (with the  fluid of each torus) are
\bea&&\label{Eq:path-omegas-pressure}
 \omega^2_r = \left.
-\frac{1}{g_{rr}}\partial_r^2 {\ln V_{eff}}\right|_\ell,
\quad \omega^2_{\theta} = \left.-\frac{1}{g_{\theta\theta}}\partial_{\theta}^2 \ln V_{eff}\right|_\ell,
\eea
here we assume $\ell$ is a constant parameters
--see \citet{2013A&A...552A..10S,mnras2}.).
\section{Constraints on the  GRHD systems}\label{Sec:constr}
This section  explores  the  general relativistic origins of the constraints of hydrodynamic  proto-jets, investigating  limiting radii bounding the matter funnels. In Sec.\il(\ref{Sec:Asymptotic}) we discuss the role of  the   ``asymptotic radius"  $r_{\infty}$.
{Limiting conditions on frequency and momentum} are analyzed in Sec.\il(\ref{Sec:fre-mome-supri}).
In Sec.\il(\ref{Sec:hearLS}) we introduce the stationary observers and light--surfaces relevant for the proto-jets.
From these concepts    replicas, significant for the observational evidences of  the jets shells, are derived--Sec.\il(\ref{Sec:replics}).

\subsection{The asymptotic radius}\label{Sec:Asymptotic}
\begin{figure*}
\includegraphics[width=4.2cm]{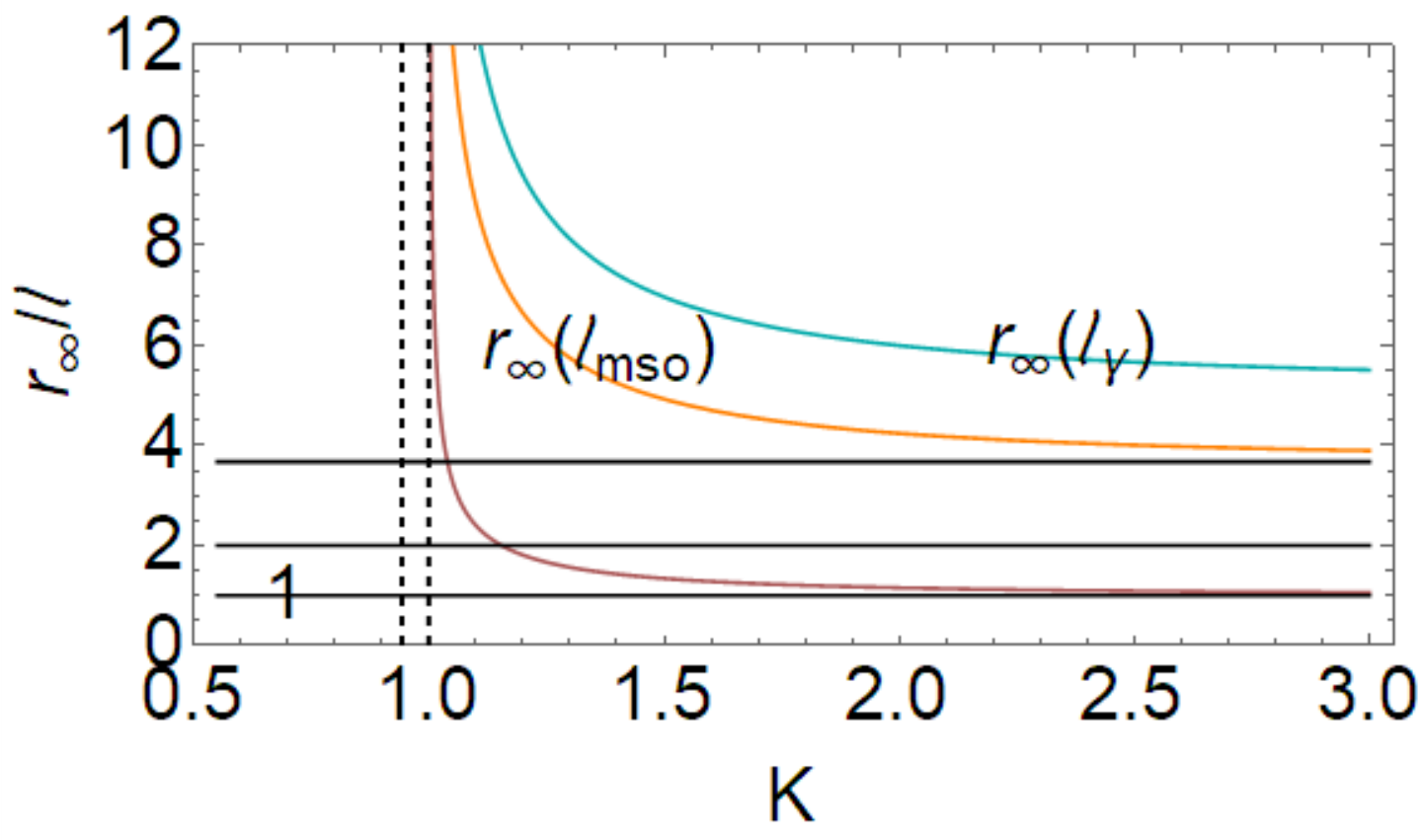}
\includegraphics[width=4.09cm]{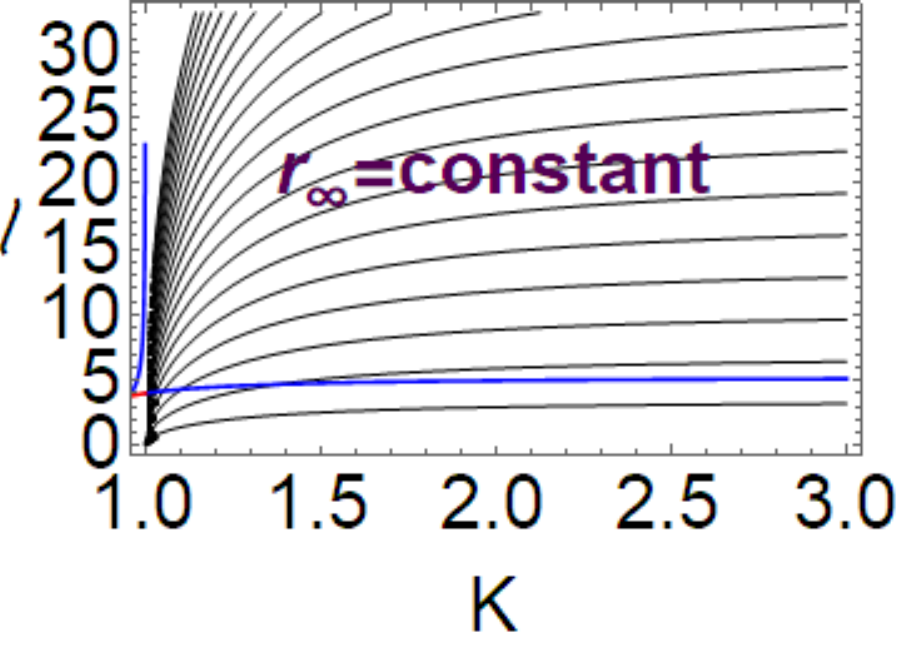}
\includegraphics[width=3.6cm]{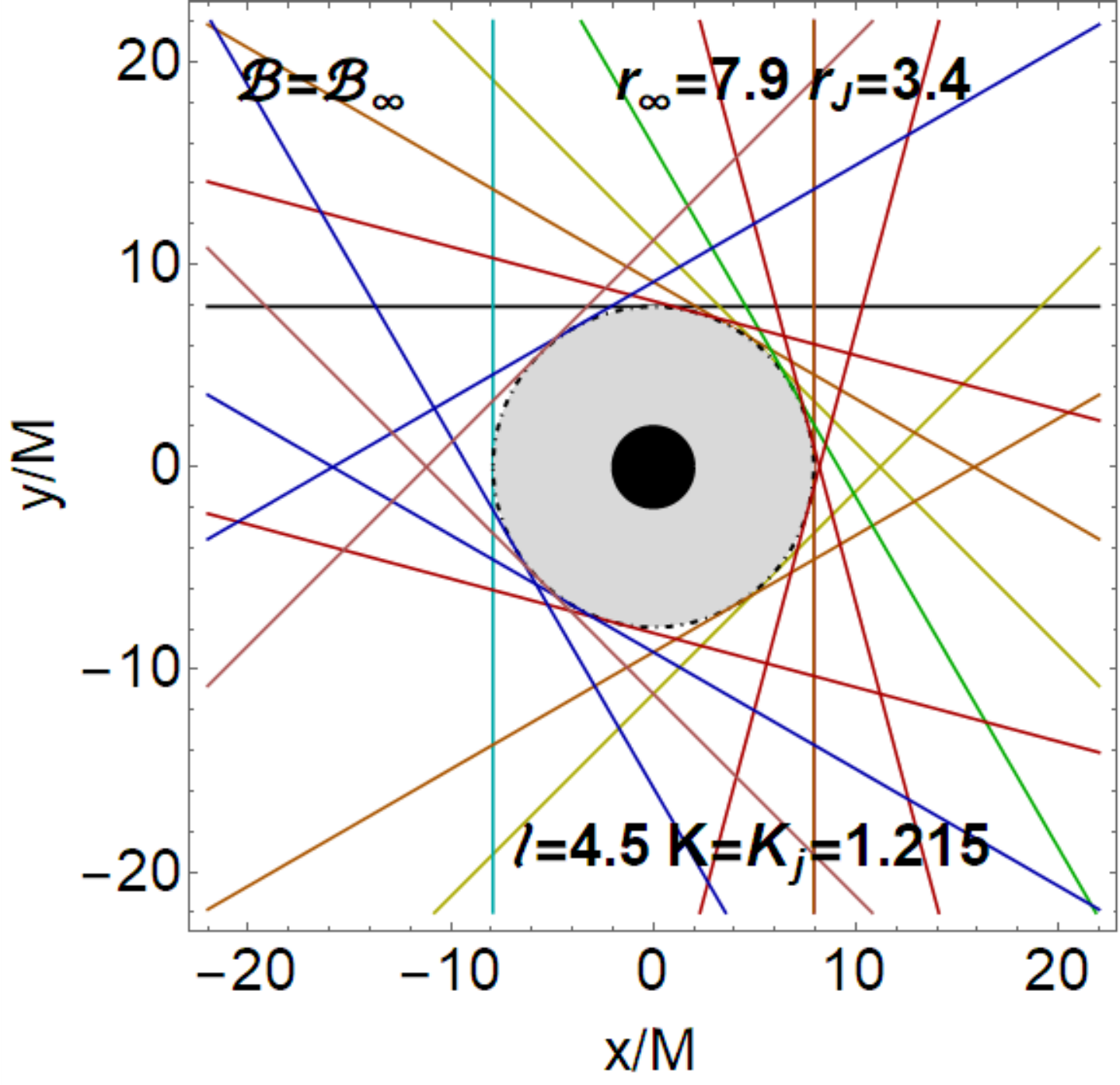}
\includegraphics[width=4.09cm]{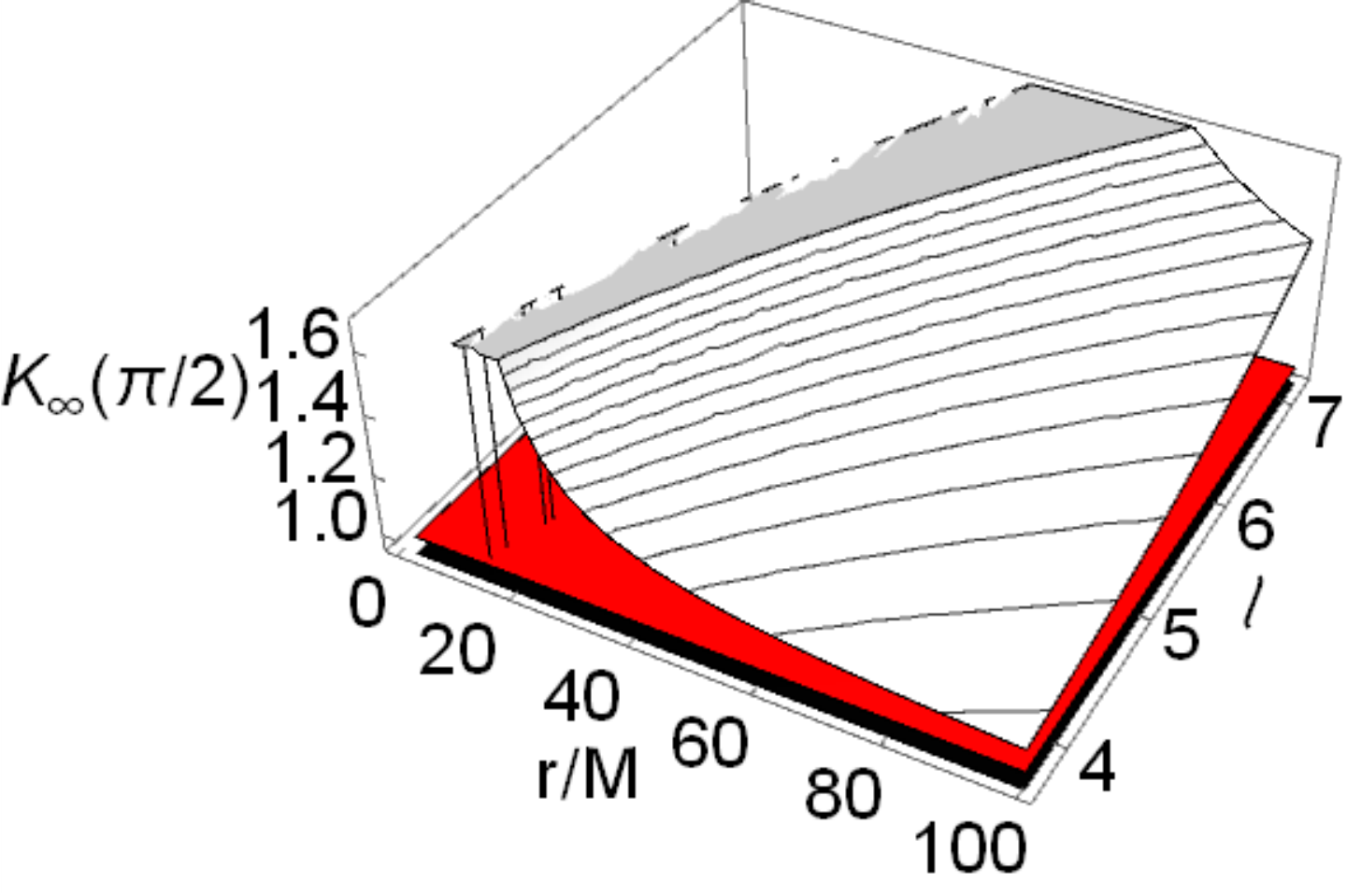}
\caption{
Analysis of the asymptotic radius of  Sec.\il(\ref{Sec:Asymptotic}).  Left plot: asymptotic radius $r_{\infty}/\ell$ in Eq.\il(\ref{Eq:occ-ref-g}) as function of $K$ parameter for fixed values of the angular momentum $\ell$ of the fluid, photon $\ell_{\gamma}$, and  marginally stable orbit $\ell_{mso}$. Second panel: $r_{\infty}=$constant in the plane $\ell-K$.
Third panel: function $\mathcal{B}_{\infty}(r,\theta)$ of Eqs\il(\ref{Eq:binfit-q1})  and (\ref{Eq:occ-ref-g})
which is a condition for the definition of the asymptotic ray of different planes for  values of  $\ell$ and  $K_j$ (K parameter value for the proto-jet) where on fixed equatorial plane  $r_{\infty}$ and the location of proto-jet cusp $r_j$ are known. Lines represent the projection of the asymptotic radii on different planes. Black region is the central \textbf{BH}, gray region  is the corona $r\in [r_+,r_j]$.
Fourth panel: $K_{\infty}$ of Eq.\il(\ref{Eq:K-infty}) on the equatorial plane, as function of $r/M$ and the specific angular momentum  $\ell>\ell_{mbo}$. Black plane is $K_{mso}$, red-plane  is $K=1$.
\label{Fig:ampaerplot} }
\end{figure*}

We introduce the ``asymptotic radius"  $r_{\infty}$  relevant  for the  jet emission  constraints and the HD collimation process.
On each symmetric plane the effective potential can be written as
\bea\label{Eq:condizioneL2S}
V_{eff}^2(r,\theta,\ell)=\frac{(r-2) r^2}{r^3-\mbox{\textbf{L2S}} (r-2)},\quad \mbox{\textbf{L2S}}\equiv {\ell^2}/\sigma,
 \eea
Figs\il(\ref{Fig:eSnatwitda2},\ref{Fig:eSnatwitdpiw}), (here we set $\sigma=\sin^2\theta$),  thus we obtain the condition
 \bea
 \ell^2\neq L2_d=-\frac{g_{\phi \phi}}{g_{tt}},\quad \mathbf{L2S}\neq L2_d/\sigma\quad (\sigma\neq0),
\eea
often we shall consider the limiting condition $\mathbf{L2S}= L2_d/\sigma$.
From  normalization conditions and the definition of the effective potential,
we  obtain, in the Cartesian coordinates $(x,y)$, the limiting conditions
\bea
\sqrt{\mathbf{L2S}}= \sqrt{\frac{L2_d}{\sigma}}=\frac{y^2 \left(x^2+y^2\right)}{x^2+y^2-2 \sqrt{x^2+y^2}}.
\eea
Considering
Eq.\il(\ref{Eq:condizioni}), from the normalization condition on the fluid four velocity we find:
\bea\label{Eq:binfit-q1}
&&
\ell^2=Q\neq Q_{\infty}\equiv \frac{\mathcal{B}^2(K^2-1)}{K^2},\quad\mbox{and}\quad
 K^2\neq K_{\infty}^2\equiv\frac{\mathcal{B}^2}{\mathcal{B}^2-Q},\quad
\mathcal{B}^2\neq\mathcal{B}_{\infty}^2\equiv \frac{K^2 Q}{K^2-1},
\eea
--Figs\il(\ref{Fig:ampaerplot})--
and
\bea\label{Eq:occ-ref-g}&&
\mathcal{B}_{\infty}(r,\theta)\equiv\frac{K \ell}{\sqrt{K^2-1}},\quad \mbox{for}\quad \mathcal{B}=r \quad (\theta=\pi/2),\quad\mbox{and}\quad
r_{\infty}\equiv\frac{K \ell}{\sqrt{K^2-1}},
\eea
where
\bea \lim_{K\rightarrow\infty}r_{\infty}=\ell,
\quad \lim_{K\rightarrow 1}r_{\infty}=\infty
\eea
(without loss of generality we  set  condition  $\mathcal{B}=r$).
We note that these limits are dependent on $K$.
 The equation  $r=r_{\infty}$  is solved for the  parameter $K$
 \bea\label{Eq:K-infty}
 K_{\infty}=\frac{r}{\sqrt{r^2-\ell^2}},\quad \lim_{r\rightarrow\ell}
 K_{\infty}=\infty,\quad \lim_{r\rightarrow\infty}
 K_{\infty}=1
\eea
  which has been used  here directly for the function of energy $K(r)$--
 Fig.\il(\ref{Fig:ampaerplot}).

The radius $r_{\infty}$ is an asymptote for the function $\Sa_{eff}$,
 informing on some relevant aspects of the extended matter configurations.
Firstly,  it is defined only for $K>1$, which means that it has in fact a role for the open configurations only (at $r\rightarrow\infty$ there is clearly
 $K=1$).
  There is a correspondence $r\leftrightarrow \ell$,
and the limiting condition of  Eq.\il(\ref{Eq:occ-ref-g}) hold.
\subsection{Limiting conditions on frequency and momentum}\label{Sec:fre-mome-supri}
Considering
Eq.\il(\ref{Eq:condizioni}),  from the normalization condition on the fluid four velocity  it follows:
\bea
&&
K^2 \left(Q-\mathcal{B}^2\right)+\mathcal{B}^2=0,\quad
\ell^2\neq L2_d=-\frac{g_{\phi \phi}}{g_{tt}}.
\eea
Note that, according to Eqs\il(\ref{omega-covariant}),  constraints on $L2_d$ correspond to the condition  $\omega\neq1$,
this quantity depends on  $\sigma$.
There is then
\bea&&\nonumber
 \omega^2\neq W2_d\equiv-\frac{g_{tt}}{g_{\phi\phi}}=\frac{1}{W2_u}=\frac{\Omega}{\ell},\quad \mbox{and}\quad \ell^2\neq L2_d\equiv\frac{1}{W2_d}=\frac{1}{L2_u}.
\eea
Note that the  leading function $\ell(r,\sigma)=1/s$ is related  to the inverse of the von Zeipel surfaces.
Note also that
$K(r)\equiv V_{eff}(r,\sigma,\ell(r,\sigma))=\sqrt{{(r-2)^2}/{r(r-3)}}$,  is  independent of  $\sigma$.
\subsection{Stationary observers and light--surfaces}\label{Sec:hearLS}
\begin{figure*}
\includegraphics[width=4.cm]{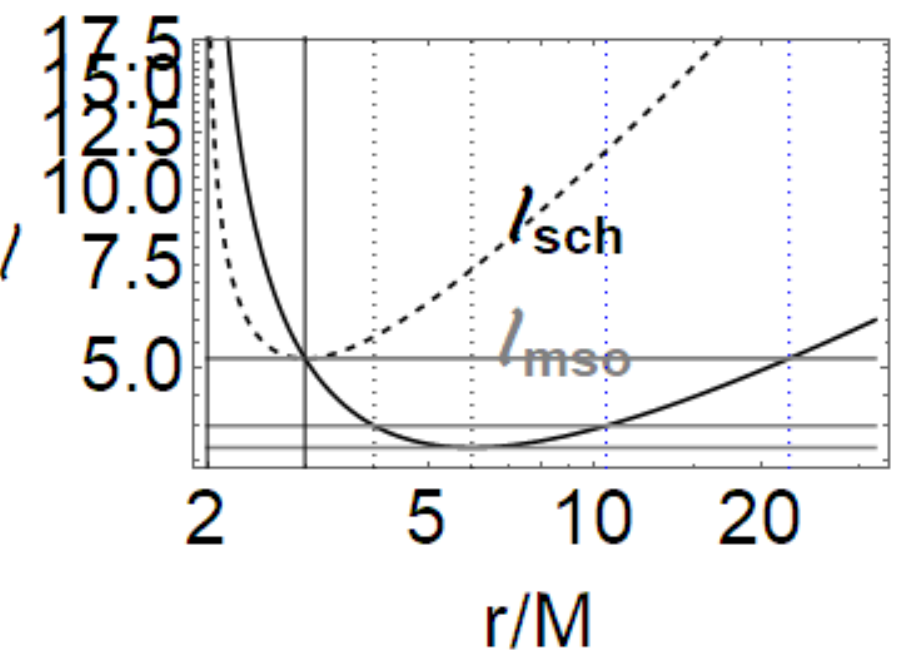}
\includegraphics[width=4.cm]{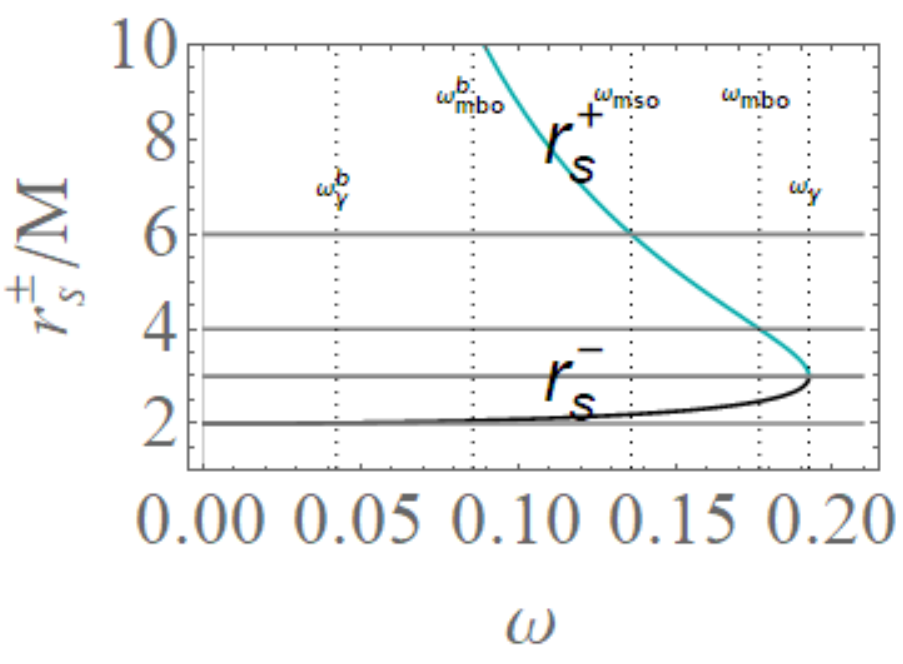}
\includegraphics[width=4.cm]{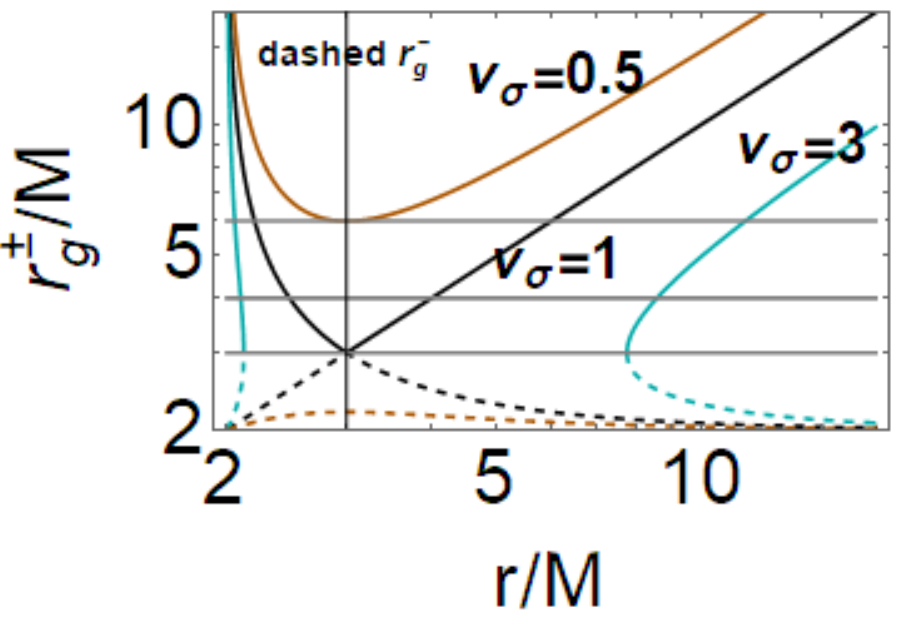}
\includegraphics[width=4.cm]{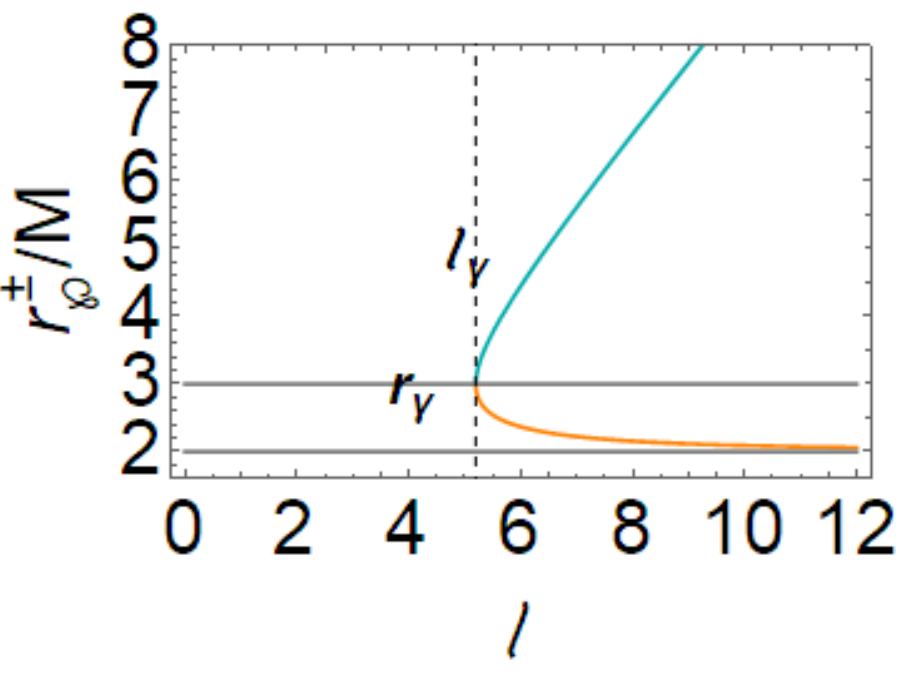}\\
\includegraphics[width=4.cm]{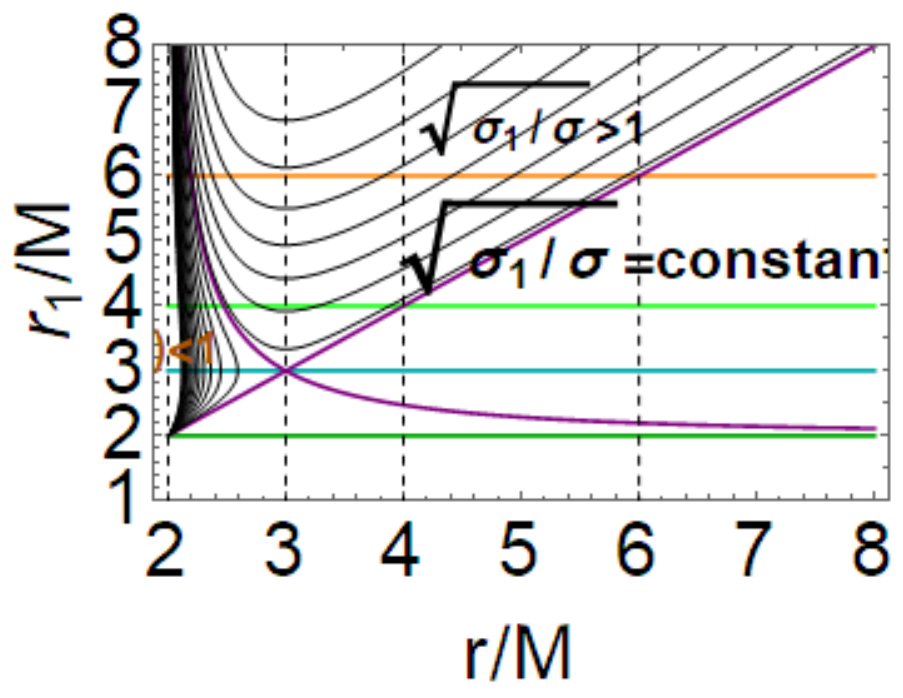}
\includegraphics[width=4.cm]{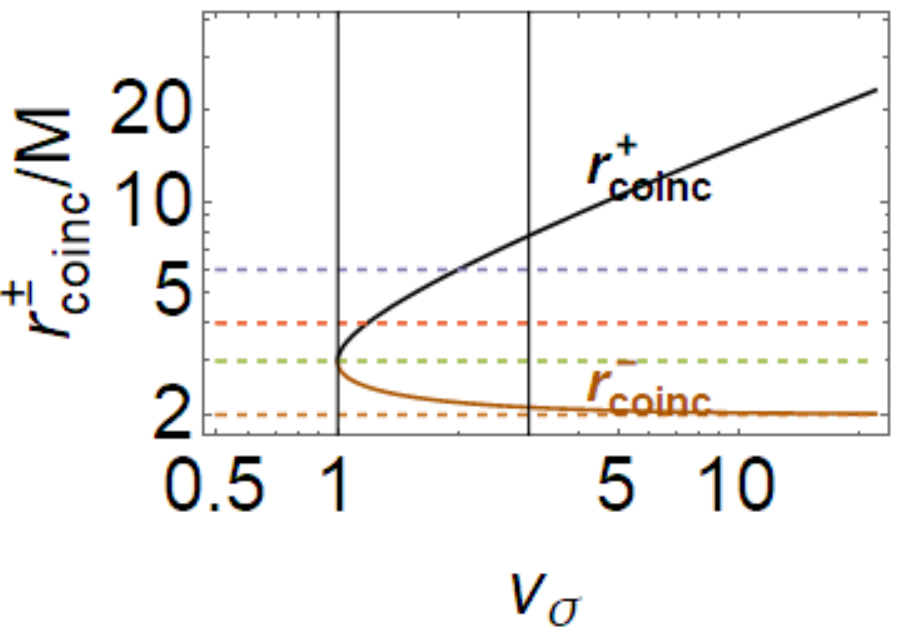}
\includegraphics[width=4.cm]{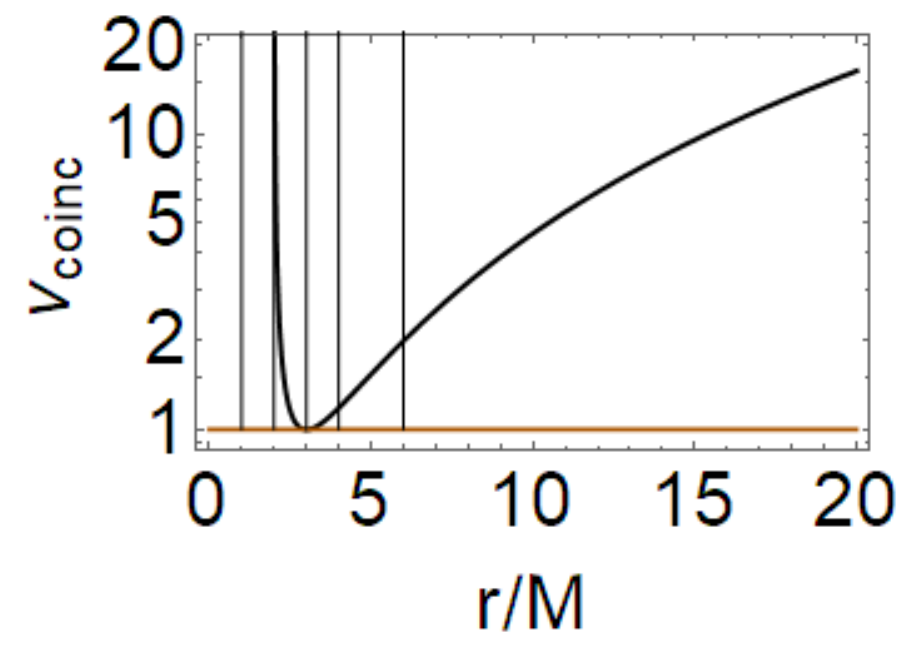}
\includegraphics[width=4.cm]{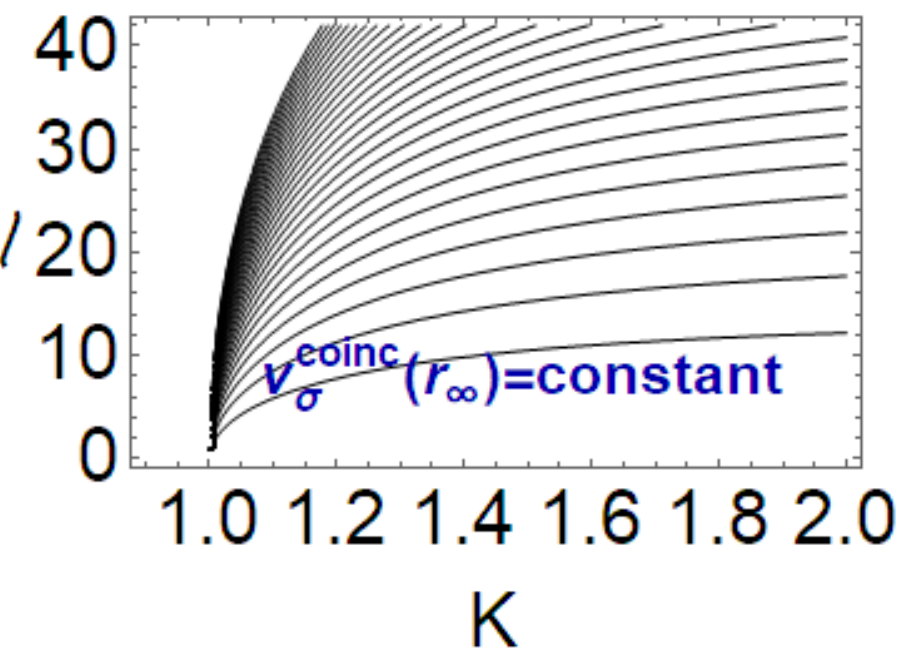}
\caption{Analysis of replicas of Sec.\il(\ref{Sec:replics}) and
 stationary observers and light surfaces of Sec.\il(\ref{Sec:hearLS}). Limiting frequencies of stationary observers $\omega_{Sch}$ are in  Eq.\il(\ref{Eq:omegasc}), $r_{s}^{+}(\omega)$ are the light surfaces  defined in
Eq.\il(\ref{Eq:light-rspone}).
 Relation  $\omega_{Sch}(r,\sigma)=\omega_{Sch}(r_1,\sigma_1)$,  defining the  metric Killing bundles and replicas (solutions $(r_1,r)$),  depends only on the ratio
$\nu_{\sigma}\equiv\sigma_1/\sigma$ where {$\sigma=\sin^2\theta$.}
This relation holds  for
$\nu=\nu_{\sigma}(r)$ in Eq.\il(\ref{Eq:nudisigma}), or   in terms of   radii,  the functions
$r_g^{\pm}(r,\nu)$
of Eq.\il(\ref{Eq:rgpm}).
There is  a solution for the coincidence of the radii
$r_{g}^+=r_{g}^-$  for the angle ratio $\nu_{\sigma}^{coinc}$
in
Eq.\il(\ref{Eq:nu-coinc}) or, in terms of radius, the function
$r_{coinc}^{\pm}$
of Eq.\il(\ref{Eq:rcoinc}).
 Upper panels. First panel:
black curve is the  leading function $\ell(r)$ and the  limiting curve is  $\ell_{Sch}(r)$ of Eq.\il(\ref{Eq:lsch-spaz}). Function $\ell_{Sch}(r)$ is related to definition of photon stationary circular frequency. Second panel:
stationary surfaces $r_s^{\pm}$ of photon surfaces  as functions of photon orbits, notable frequencies  of Eqs\il(\ref{Eq:omegasdefining}) are shown-- it is illustrated  the role of photon circular orbit $r_{\gamma}$.  Orbits at $\omega=$costant define the metric Killing bundles. Third panel:
 radii $r_g^{\pm}(r,\nu)$ (metric bundles)  for fixed angles ratios. Particularly  $\nu=1$  is shown,   we note symmetries  and the emergence of \textbf{BH} horizon in the region $r\geq 2M$.
Fourth  panel:
curves  $r_{\wp}(\ell): \ell=\ell_{Sch}$ solutions of $\Delta (V_{eff})=0$  as functions of   $\ell$.
There is $r_{\wp}(\ell)=r_{s}^{\pm}(\ell=1/\omega)$, having a relevant meaning for the accretion physics. On the equatorial plane there is $\nu_\sigma=1$. Bottom panels. First panel: the ratio $\sqrt{\sigma_1/\sigma}=$constant  in the plane  $(r_1, r)$,
 purple curve   represents $\sqrt{\sigma_1/\sigma}=1$-- equal plane condition. Notable radii, $(r_{\gamma},r_{mbo},r_{\gamma})$ are  represented.
Second panel: $r_{coinc}^{\pm}$ as functions of the $\nu_\sigma$. Notice  the relevance of the ratio $\nu=1$  where   one solution exists. Third panel:
$v_{coinc}$ as function of  $r$, notable radii are shown. For $r=r_{mbo}=4M$ there is the minimum point of the function.
Fourth panel: lines $\nu_{\sigma}^{coinc}=$constant in the plane $(\ell,K)$  evaluated on the asymptotic radius $r_{\infty}$ of Eq. \il(\ref{Eq:occ-ref-g}) providing a limiting condition for the open surfaces.
\label{Fig:Tecnocicoquanu} }
\end{figure*}

Light surfaces play an  essential role in the  constraining the photonic components of  the jet emissions.
It is clear that a major role in the \textbf{RAD} frame is played by the limiting orbital  frequencies on the stationary observers defining  toroidal  light-surfaces. The limiting light-like frequencies and the related light surfaces are respectively
\bea&&\label{Eq:omegasc}
\omega_{Sch}\equiv \sqrt{W2_d}=\sqrt{-\frac{g_{tt}}{g_{\phi\phi}}}=\sqrt{\frac{r-2}{r^3 \sigma }},\quad (\sigma\equiv \sin^2\theta),
\eea
and
\bea
\label{Eq:light-rspone}
&&  r_{s}^{\pm}(w)\equiv\pm
\frac{2 \sqrt{\frac{1}{w ^2}} \cos\epsilon_{\pm} }{\sqrt{3}},\quad \mbox{where}
\\&& \epsilon_+\equiv \frac{\hat{\varepsilon}}{3} ,\quad \epsilon_-\equiv \frac{1}{3} \left(\hat{\varepsilon}+\pi \right),\quad \hat{\varepsilon}\equiv \cos^{-1}\left(-\frac{3 \sqrt{3}}{\sqrt{\frac{1}{w ^2}}}\right),
\eea
(see Figs\il\ref{Fig:Tecnocicoquanu}).
Considering  planes others then the equatorial at $\sigma=1$,  $\omega$  has to be substituted by   $\omega\sqrt{\sigma}$.  Radii $ r_{s}^{\pm}$ are the light-surfaces with light-like orbital frequencies $\omega$.
These frequencies allow  to give immediate limits for jets of material and  the frequencies of photons on limiting  orbits.
It is clear that Eq.\il(\ref{Eq:condizioneL2S}),
being  $r^3-\mbox{\textbf{L2S}} (r-2)\geq0$,  implies $\ell \in [0,\ell_{Sch}]$--see Figs\il\ref{Fig:Tecnocicoquanu}.
In this context jet limiting surfaces are associated with disks.
We obtain the first condition for the existence of these configurations according to
\bea&&\label{Eq:lsch-spaz}
\Delta(V_{eff})=r^3-\mbox{\textbf{L2S}} (r-2)=0,
\quad \frac{\ell_{Sch}}{\sqrt{\sigma}}=\sqrt{\frac{L2_d}{\sigma}}=\frac{r^{3/2}}{\sqrt{r-2}},\quad \ell_{Sch}=\frac{1}{\omega_{Sch}}
\eea
--(see Figs\il\ref{Fig:Tecnocicoquanu}).
The relation between   $\ell_{Sch}$ and the frequency $\omega_{Sch}$  is evident from the definition of the normalization condition.
It is clear then that $\ell_{\gamma}=1/\omega_{\gamma}=\sqrt{27}$, but $\omega_{\hat{(\gamma)}}\equiv \omega_{Sch}(r)$  and clearly this holds only for the light like part  of the geodesic structure of the spacetime.
This relation therefore connects the relativistic frequency to the specific angular momentum $\ell$ and the von Zeipel surfaces--see Figs\il\ref{Fig:DopoDra}.

Then condition $\Delta (V_{eff})=0$, leading to the curve  $\ell_{Sch}$,  provides   the radii
 $r_{\wp}(\ell): \ell=\ell_{Sch}$ solving  also $\Delta (V_{eff})=0$  for a generic  $\ell$.
Surfaces,   $r_{\wp}(\ell)$ and  $r_{s}^{\pm}$ are the same and  this has a relevant meaning for the accretion physics, where  $r_{\wp}(\ell)=r_{s}^{\pm}(\ell=1/\omega)$--see Figs\il\ref{Fig:Tecnocicoquanu}.

\subsection{Light--surfaces and constraints: horizons  replicas in the jet shells }\label{Sec:replics}
\begin{figure}
  \includegraphics[width=3.4cm]{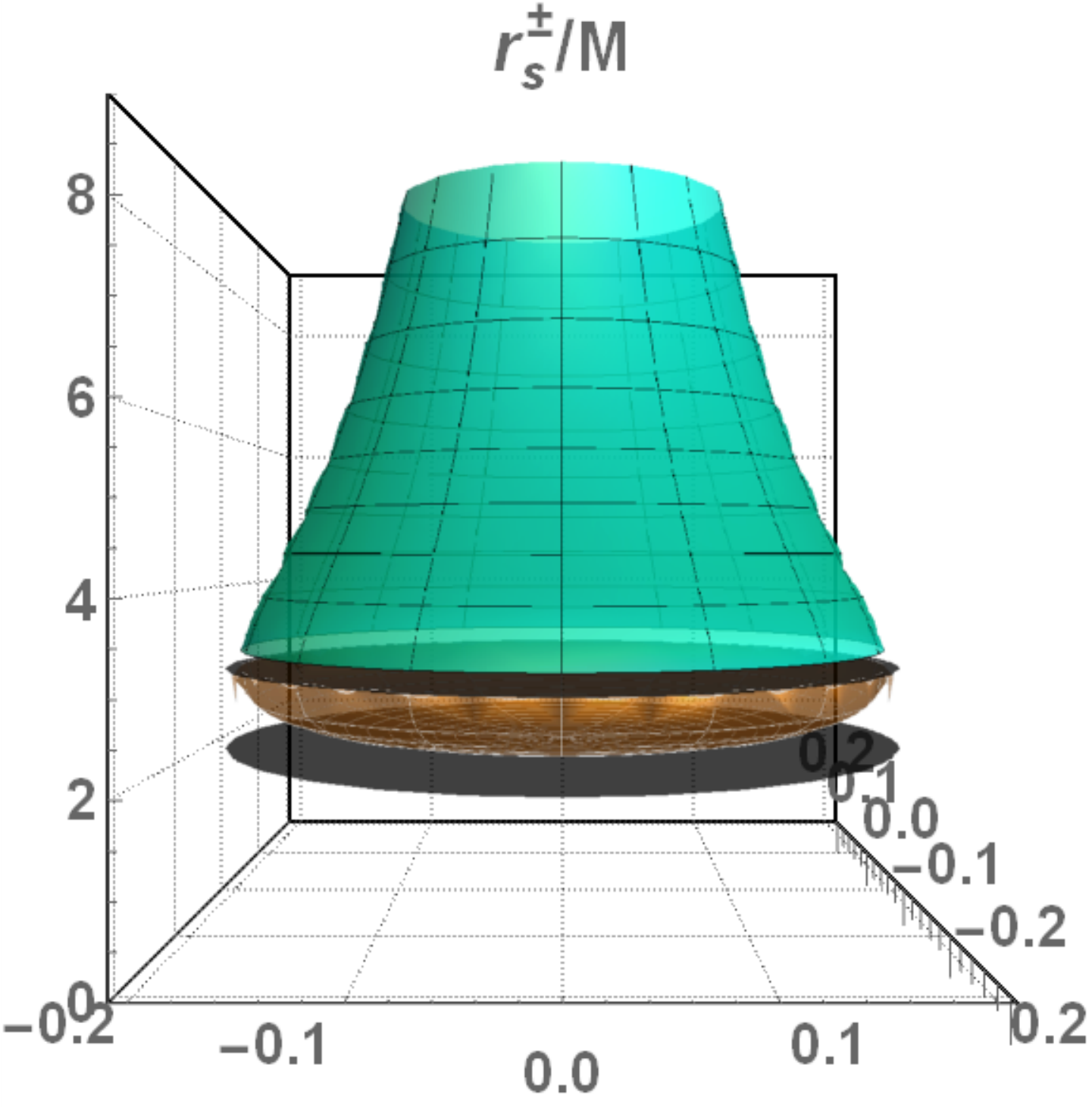}
      \includegraphics[width=4.3cm]{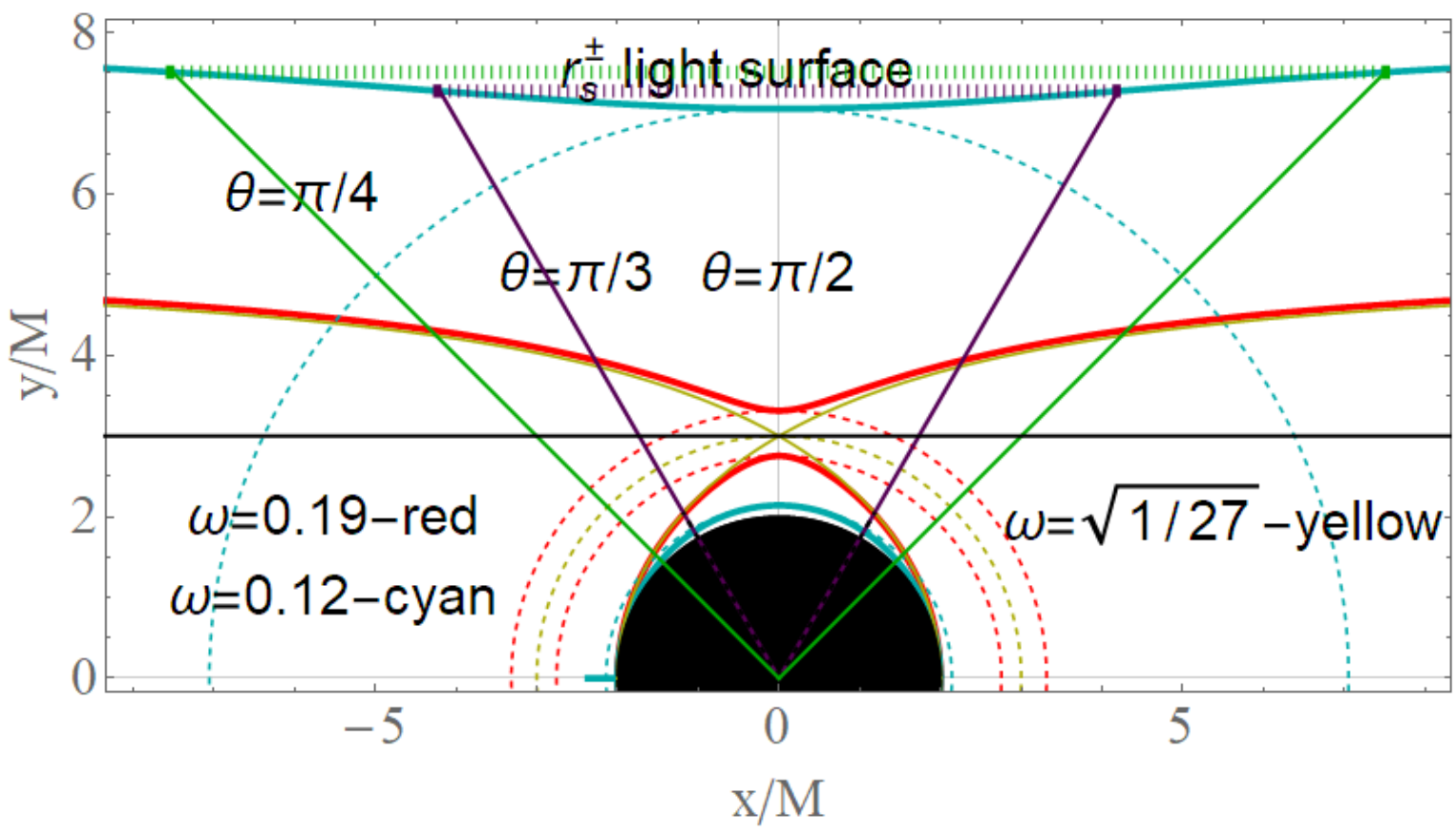}
        \includegraphics[width=3.4cm]{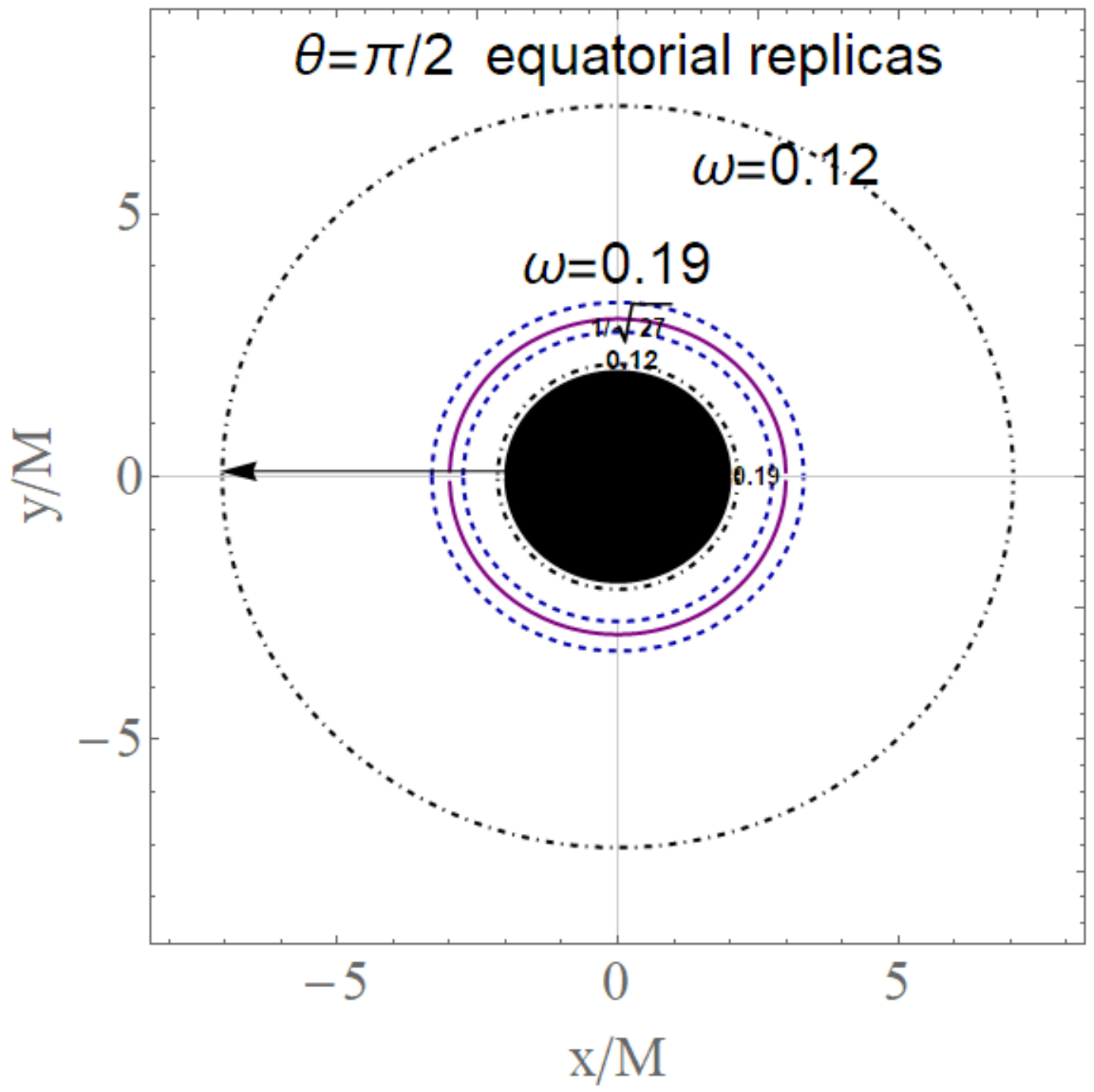}
        \includegraphics[width=4.3cm]{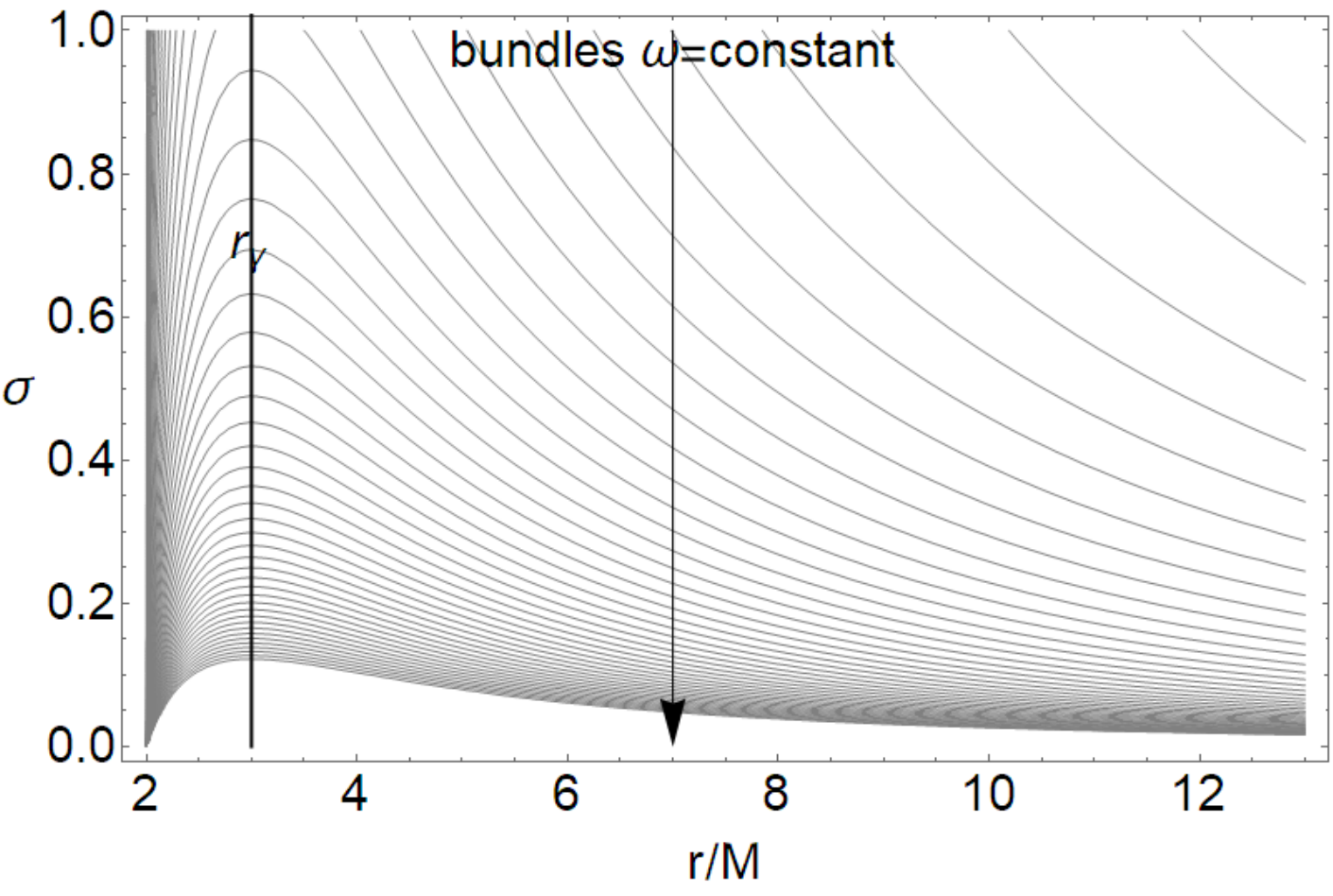}
  \caption{Replicas analysis of  Sec.\il(\ref{Sec:replics}). Left panel:  Plots of the light  surfaces $r_{s}^{\pm}$
 (in units of mass) on the equatorial plane ($\theta=\pi/2$) versus the photon orbital frequency  $\omega$. The surfaces are represented as revolution surfaces with height $r$
 (vertical axes) and radius $\omega$ (horizontal
plane). Surfaces are generated by rotating the two-dimensional curves $r_{s}^{\pm}$  around an axis. Thus, $r =$constant with respect to the frequency $\omega$ is represented by a circle. Second panel: $r_{s}^{\pm}=$constant  in flat Cartesian coordinates $(x,y)$ in mass units on each equatorial plane, therefore are the projection of the surfaces of the left panel on different horizontal  planes. Black region is the  \textbf{BH}, diagonal lines are different planes  as signed by the angle $\theta$.
 Semi-circles are light surfaces (circular orbits at fixed frequencies). There are two orbits at fixed frequency $\omega$ (which is the characteristic frequency of a bundle) signed with equal colors curves, the outer orbit is a replica. Photon  circular orbit $r=3M$ and the frequency $\omega = 1/\sqrt{27}$ are also shown, related to  von Zeipel surfaces and evidencing the replicas role in the QPOs onset  \citep{mnras2}.
Third panel: metric Killing bundles curves in the plane $\sigma-r/M$,  the photon orbit $r_{\gamma}$ is shown, frequencies increase  in the direction of the  arrow.
$\sigma=1$ is the equatorial  plane. Replicas are on the curves, in the same  distance to the axis on the horizontal  lines $\sigma=$constant.
Right panel: equatorial plane replicas. Central  region represents  the \textbf{BH},  replicas curves with equal colors  have same frequencies.
}\label{Fig:QPlot}
\end{figure}

We introduce the concept of replicas for light-surfaces  with equal photon frequencies. We concentrate  on the frequencies  $\omega_{Sch}=\omega_{\pm}$,  null orbits  frequencies, using the  usual notation $Q_{\bullet}\equiv Q(r_{\bullet})$.
The concept of metric Killing bundles introduced in \citet{Pugliese:2020azr,Pugliese:2020ivz,Pugliese:2019rfz,Pugliese:2019efv,remnants}
 defines the replicas
as a couple  orbits   $(r,r_1)$ with  the same values of the  limiting frequency  $\omega_{Sch}$. The problem
$\omega_{Sch}(r,\sigma)=\omega_{Sch}(r_1,\sigma_1)$  is solved for the couple of radii $(r,r_1)$ and planes $(\sigma,\sigma_1)$.
 This relation depends exclusively on the ratio
$\nu_{\sigma}\equiv\sigma_1/\sigma$, getting a relation  $r_1(r)$ parameterized for  $\nu_\sigma$,
and we can solve the problem for $\sigma_1(\sigma)$. We are particularly interested to the conditions $\nu_{\sigma}=$constant and $\nu_{\sigma}=1$, and in  the  spherically symmetric case particularly in  the case  $\sigma=\sigma_1=1$. It is also to be noted that
 $\sigma=\sigma_1=1$ and  $\sigma=\sigma_1$ play an equivalent role in many relations.
Therefore
\bea\label{Eq:nudisigma}
&& \nu_{\sigma}\equiv \frac{\sigma_1}{\sigma}=\sqrt{\frac{r^3 (r_1-2)}{ r_1^3(r-2)}}.\\&&\nonumber
\mbox{For}\quad\nu_\sigma=1 \quad\mbox{there is }\quad r_1\equiv\frac{r \left[2-r+\sqrt{(r-2) (r+6)}\right]}{2 (r-2)}.
\eea
In general,  for a general $\nu_{\sigma}$, the relation  $r_1 (r)$, can be reduced to  the functions $r_{g}^{\pm}$
\bea
\label{Eq:rgpm}
&&
r_g^{-}\equiv-2 \delta_g \cos \left[\frac{1}{3} \left(\cos ^{-1}\left[-\frac{3}{ \delta_g }\right]+\pi \right)\right],\quad
r_g^{+}\equiv2  \delta_g  \cos \left[\frac{1}{3} \cos ^{-1}\left(-\frac{3}{ \delta_g }\right)\right],\\&&\nonumber\mbox{where}\quad \delta_g\equiv\sqrt{\frac{r^3}{3 \nu_{\sigma} (r-2)}}.
\eea
Solution of the problem  $r_{g}=r_{g}^\pm$, for the radius $r_g$ at fixed  $\nu_{\sigma}$, is for
\bea&&\label{Eq:nu-coinc}
\nu_{\sigma}^{coinc}\equiv\frac{r^3}{27 (r-2)}\geq1,\quad\mbox{and}\quad \nu_{\sigma}^{coinc}=1 \quad \mbox{for} \quad
r=r_{\gamma}.
\eea
It is remarkable to consider how, on the same plane, one has only solution for the photon $r_{\gamma}$--see Figs\il\ref{Fig:Tecnocicoquanu}.
Alternately,  condition (\ref{Eq:nu-coinc}) can be reduced to
\bea
\label{Eq:rcoinc}
&&
r_{coinc}^+\equiv6 \sqrt{\nu_{\sigma}} \cos \left[\frac{1}{3} \cos ^{-1}\left(-\frac{1}{\sqrt{\nu_{\sigma}}}\right)\right],\quad  r_{coinc}^-\equiv6 \sqrt{\nu_{\sigma}} \sin \left[\frac{1}{3} \csc ^{-1}\left(\sqrt{\nu_{\sigma}}\right)\right].
\eea
We consider for the notable frequencies evaluated on the  equatorial plane
\bea&&\label{Eq:omegasdefining}
\sigma=1;\quad \omega_{mso}=\frac{1}{3 \sqrt{6}};\quad \omega_{mbo}=\frac{1}{4 \sqrt{2}};\quad \omega_{\gamma}=\frac{1}{\ell_{\gamma}};
 \\
 &&\omega_{(mbo)}=\omega(r_{(mbo)})=\frac{1}{4} \sqrt{\frac{1}{2} \left(\sqrt{5}-2\right)};\quad  \omega_{(\gamma)}=\omega(r_{(\gamma)})=\frac{1}{6} \sqrt{\sqrt{3}-\frac{5}{3}}.
\eea
The investigation of this  special   problem  for the orbit $r=r_{\infty}$ and planes $\nu_{\sigma}=\nu_{\sigma}^{coinc}$  leads to the solution showed in
Figs\il(\ref{Fig:Tecnocicoquanu},\ref{Fig:QPlot}).
 The analysis of the bundles shows the presence   of replicas: a property $\Qa(\omega(r))$ defined on an orbit $r$, function of the frequency $\omega$  is ``replicated"  on an orbit $r_1\neq r$ where there is by definition $\omega(r)=\omega(r_1)$. Eventually this relation includes the polar angle dependence featuring the conditions
    $\omega(r,  \sigma)=\omega(r_1,\sigma_1)$. It can be demonstrated that the maximum number of orbits satisfying this condition is two (defining actually classes depending on the   polar angle dependence)--Figs\il(\ref{Fig:Tecnocicoquanu}) and Figs\il(\ref{Fig:QPlot}).  First orbit  is  very close to the \textbf{BH} horizon and the second orbit is located  far from the central attractor and in general in the stability region for the tori $(r>r_{mso})$. (Clearly we neglect to consider the counterrotating  orbits in the spherically symmetric spacetime.).
  The observation relevance of the metric  bundles  concept relies  in the fact that given a quantity  $\Qa(\omega(r))$ dependent on the frequencies, for example  the constraining functions of the light-surfaces, there are in general  two different orbits such that
 $\Qa(\omega(r))= \Qa(\omega(r_1))$ as there is    $\omega(r)=\omega(r_1)$. The curve defined by the classes of points  $(r,r_1)$ defines the bundles, which can  include eventually the planes dependence from  $\sigma$s such that there is  $\Qa(\omega(r,\sigma))= \Qa(\omega(r_1,\sigma_1))$ as there is  by definition  of replica $\omega(r,\sigma)=\omega(r_1,\sigma_1)$.
 In this sense the regions $(r,\sigma)$ and $(r_1,\sigma_1)$ can be interpreted as  presenting replicas of the property$\Qa$.
 In general, if  $r$ is a circle very close to the attractor, then the second point  $r_1$  is located  far from the attractor.
The  observational  relevance of these structures lies in the fact that it is possible to find replicas  introduced here of effects and quantities dependent on relativistic frequency $\omega$, and since these strictly constrain the jet emission, the presence of frequency replicas should appear in a region  close to the horizon and  in one  far from the \textbf{BH} --Figs\il\ref{Fig:QPlot},\ref{Fig:ampaerplot2k}.
\subsection{Impacting conditions}\label{Sec:glod-powe}
There are two major cases to consider in the investigation  of  the  constraints of  the  toroidal configurations collision.
\textbf{(a)} The  case we mainly consider here is the occurrence of jet collision with accreting configurations.
In this case we consider a one dimensional problem for the accreting disks,  fixing the   angular momentum  parameter $\ell\in \mathbf{L_1}$ for the accretion,  and we consider quantities $\Qa_{\times}$   evaluated in $(r_{\times})$.
For proto-jets  we fix  the parameter $\ell\in \mathbf{L_2}$,   considering quantities $\Qa_{j}$   evaluated at $r_{j}$. Eventually, we can consider a quiescent torus where the specific angular momentum $\ell$ can be in $\mathbf{L_1}, \mathbf{L_2}$ or $\mathbf{L_3}$
\textbf{(b)} It should be noted that  a further possibility is
the collision occurring with an internal  closed configuration (internal Roche lobe ``embracing" the \textbf{BH} horizon) which is considered here in $\mathbf{L_1}$, $\mathbf{L_2}$ or $\mathbf{L_3}$ and  therefore it  comprises different cases.

A further situation to be analyzed comprised the  role of the configurations embracing the \textbf{BHs} at $\mathbf{L_0}$ and $\mathbf{K_0}$.
The cases \textbf{(a)} and \textbf{(b)}  represent very different scenarios.
  The case of quiescent  configurations  is indeed  much more complex than the case of impact between cusped tori and jet-emission, and dependent on the  boundary  conditions.
It is clear that for the impact conditions we have to evaluate  the elongation $\lambda$ on the plane of symmetry of the torus,  the maximum vertical height, that ultimately defines the thickness of the disks increasing  with $K$ and
 $\ell$. However,  for $\ell\in \mathbf{L_3}$, the quiescent configuration is   at large distance from the  central attractor,   and  effects of the torus  self--gravity starts to be relevant. The orbital region of location  for tori  with   $\ell\in \mathbf{L_3}$ extends to infinity. However $|\ell|$ and  $K$  are, for $r>r_{mso}$,  increasing functions  of the radius, implying the presence of closed configurations which can be also  very large. The maximum limit for $K$-parameter of the closed, cusped or quiescent, tori is  $K=1$. For these special tori we are mainly interested    on the location of the inner edge.
Below we enlist the  expression of the equipotential surfaces defining the configurations,  and we introduce a relation  $\ell(K)$ reducing the independent  parameters for cusped  (tori and proto-jets) surfaces.
\medskip

\textbf{The equipotential surfaces}

\begin{figure}
 \begin{center}
  \includegraphics[width=7cm]{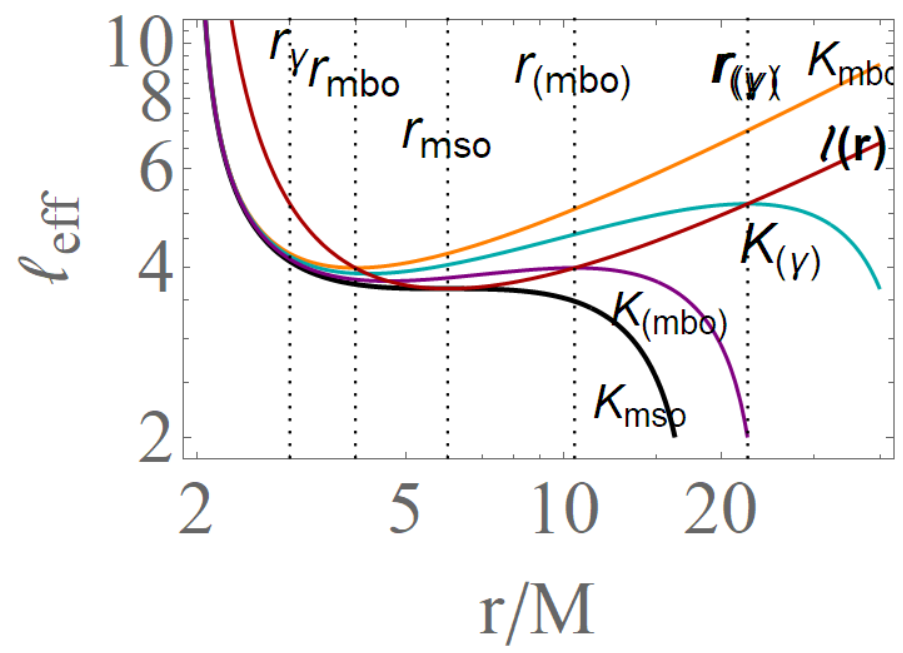}
  \includegraphics[width=7cm]{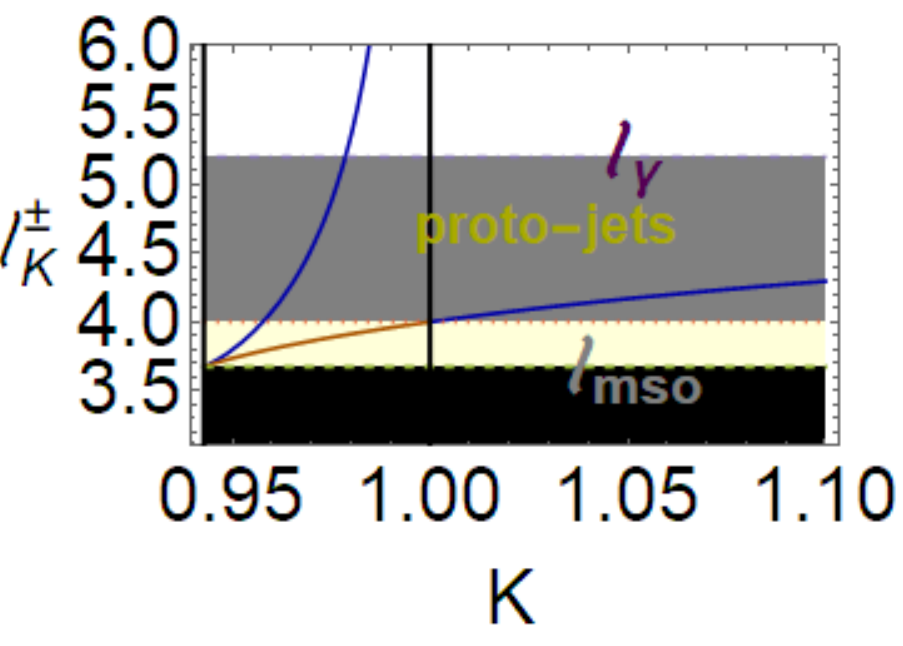}
 \end{center}
  \caption{Left panel: function $\ell_{eff}(r; K): \quad V_{eff}(\ell_{eff},r)=K$ of Eq.\il(\ref{Eq:ell-eff-r-k-wit})  as function of $r$, for different values of $K$ signed  on the curves. The leading function, distribution of specific angular momentum $\ell(r)$ is also plotted (red), and  angular momentum $\ell_{K=1}^{\pm}$ in Eq.\il(\ref{Eq:LKmeasuram}) as for  $K_{(*)}\equiv K(r_{(*)})$. Notable radii are pointed as vertical lines.   Right panel: Angular momentum $\ell_{K}^{\pm}$ in Eq.\il(\ref{Eq:LKmeasuram}) as function of
$K$, and for  $K_{(*)}\equiv K(r_{(*)})$. The solution is obtained by eliminating the radial dependence for functions $\ell(r)$ and $K(r)$ which are the  leading function and the energy function. It relates $\ell$ parameter  with the $K_{min}$ correspondent to the maximum point of pressure and density in the disk, and the $K_{crit}$  corresponding to the  minimum of pressure, which could be related to a proto-jet  $K_j$ or an accreting configuration with $K_{\times}$}\label{Fig:ampaerplot2k}
\end{figure}

We can address the problem directly solving the equipressure-equipotential surfaces  $V_{eff}=K=$constant for a generic $\ell$. Clearly, the constant pressure levels provide the inner edge and,  for closed surfaces, the  outer edges of the configurations.
We therefore introduce the quantity
\bea\label{Eq:ell-eff-r-k-wit}
\ell_{eff}=\frac{r \sqrt{\left(K^2-1\right) r+2}}{K \sqrt{r-2}}: \quad V_{eff}(\ell_{eff},r)=K
\eea
showed in  Figs\il\ref{Fig:ampaerplot2k}.
We obtain in Cartesian  coordinates
{
\bea&&
 \ell_{eff}=\frac{\sqrt{y^2(\bar{r}[\bar{r}(1-K^2)-2])}}{K\sqrt{(2 - \bar{r})\bar{r}}},\quad \bar{r}\equiv x^2+y^2
\eea}
or, alternatively the surfaces
{
\bea\nonumber
        &&x=\sqrt{\frac{4 K^4 \ell^4+2 y^4 \left[(K^2-1) K^2 \ell^2+2\right]+K^2 \ell^2 y^2 \left(8-K^2 \ell^2\right)-(K^2-1)^2 y^6}{\left[K^2 (\ell-y) (\ell+y)+y^2\right]^2}}.
\eea}
The equipotential surfaces $V_{eff}=K$  satisfy both the cancellation of the radial and polar gradient (defining the ``verticality" of the configuration) and therefore the two projections of the Euler equations, fixing also the verticality of the disk, as we shall see  in detail in Sec.\il(\ref{Sec:polar-gradient}).
%
\begin{figure}
  \includegraphics[width=5cm]{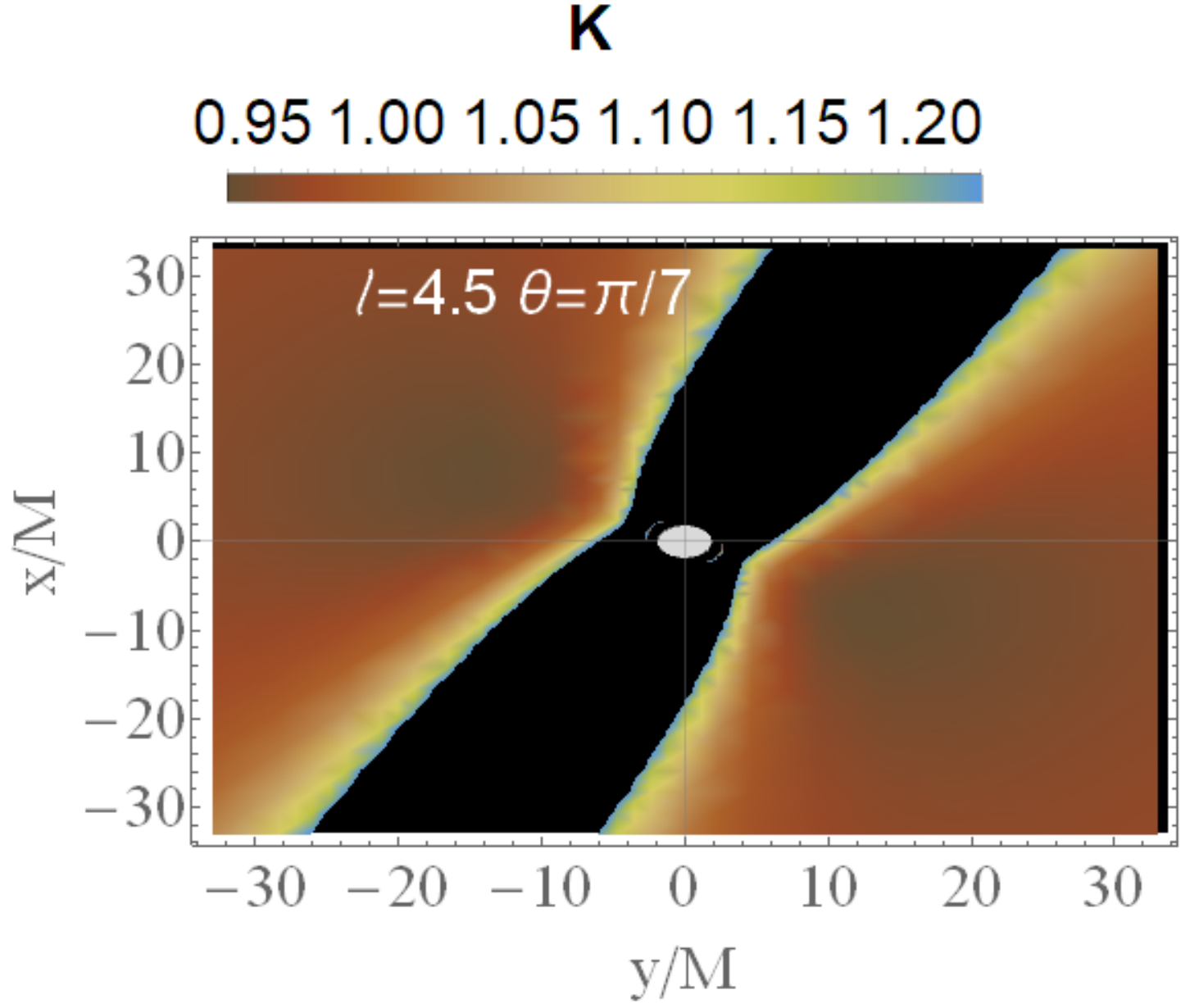}
   \includegraphics[width=6cm]{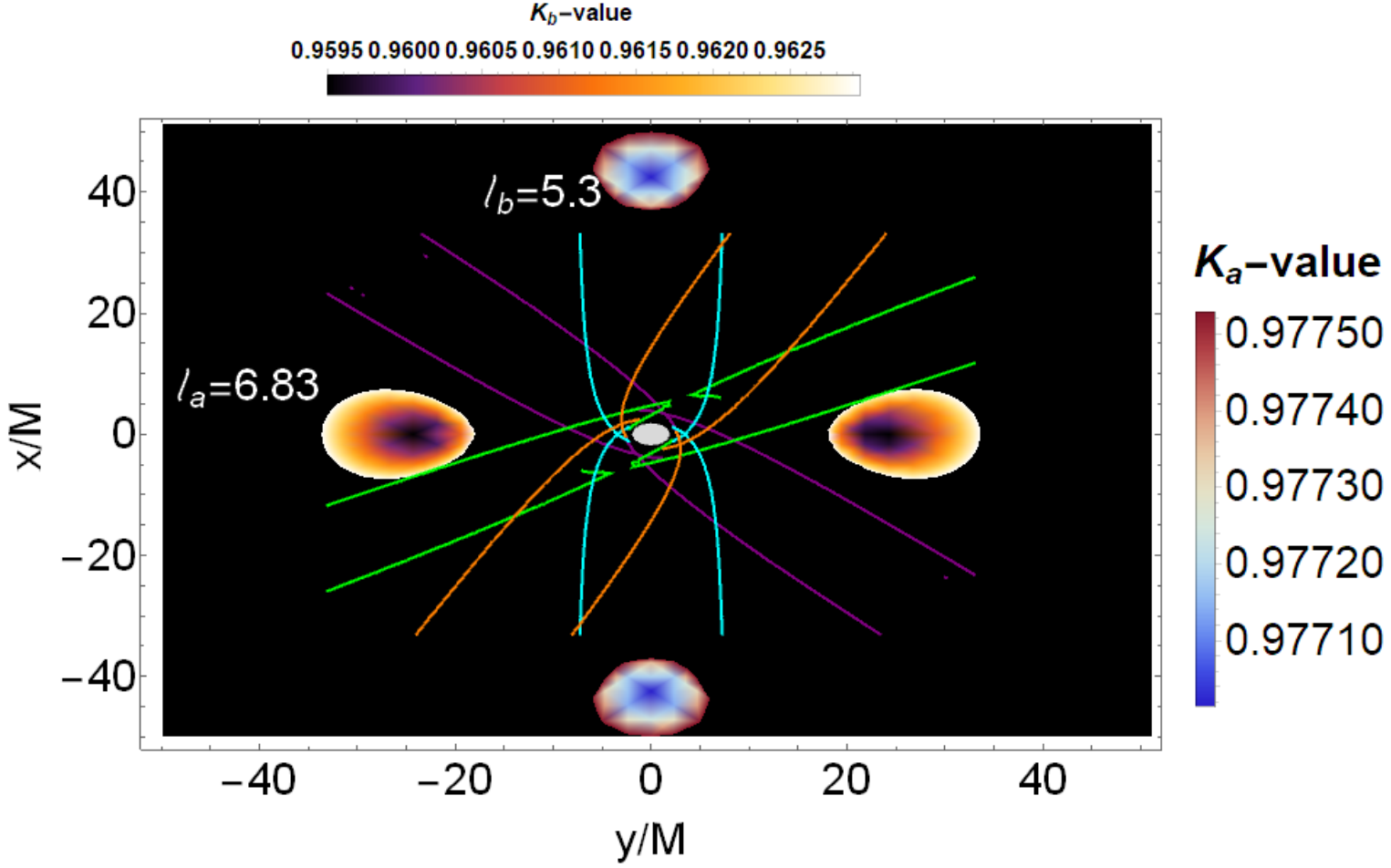}
    \includegraphics[width=5cm]{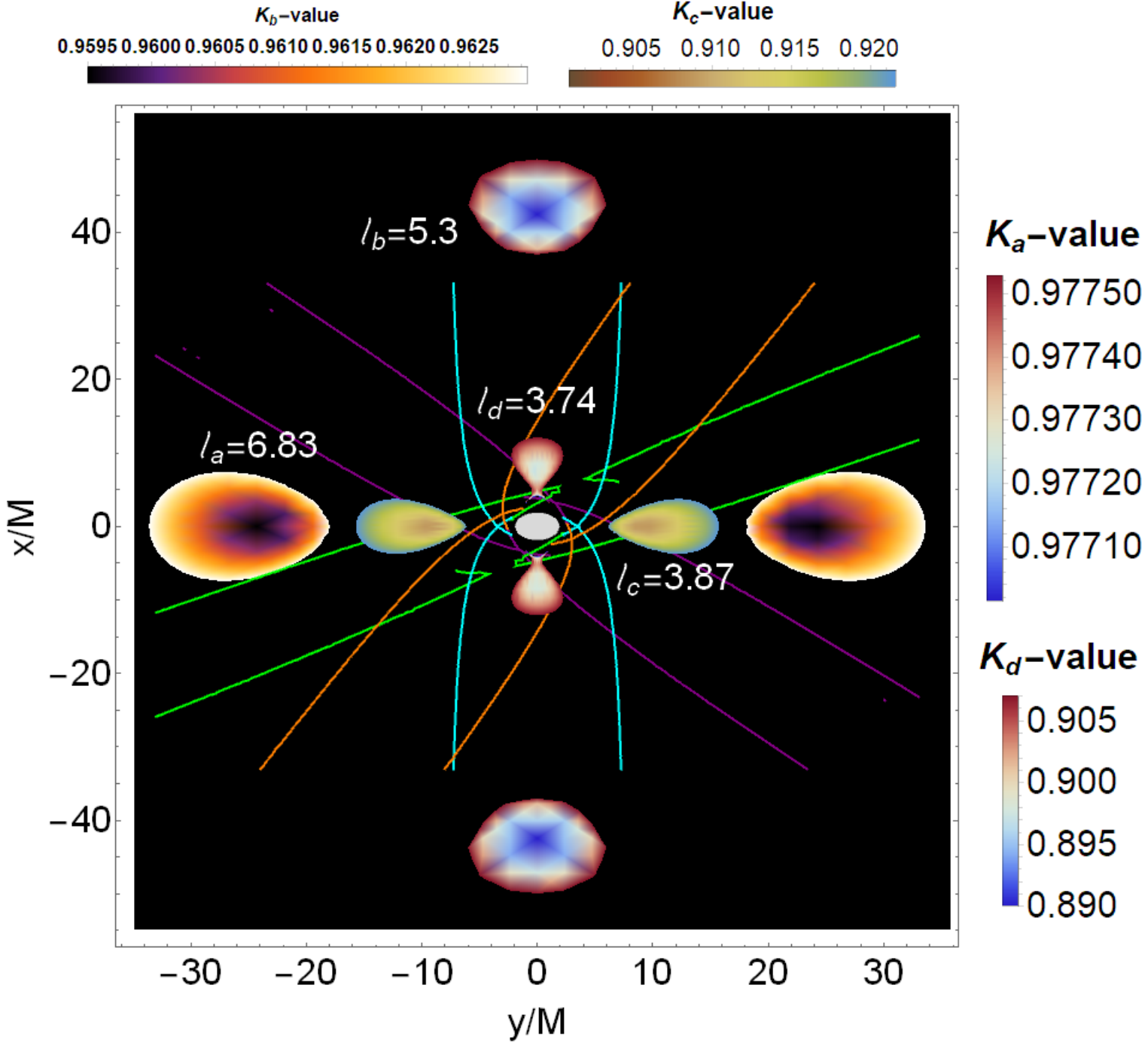}
  \caption{Equipotential (equi-pressure) surfaces $K=$constant,    in flat coordinates, solutions  of the Euler equation  for parameter  values  signed  on the panel. $\ell$ is the fluid specific angular momentum.  White central circle is the central spherical \textbf{BH}.
Left panel shows an open limiting surface. The inclination angle is signed on the panel.
Center and Right  panels show also the limited  configurations at  $\ell=4.5$ with different inclination angles, as the equi-pressure surface at  $K=K_{j}=1.21$ where $r_{j}=3.382M$. Center panel shows  two tori, $(a)$ and $(b)$,  whose momenta are signed in figure. The  right panel shows a \textbf{RAD} of order 4, with tori (a), (b), (c) and (d)-in accretion.}\label{Fig:compePlottion}
\end{figure}
\begin{figure}
    \includegraphics[width=5cm]{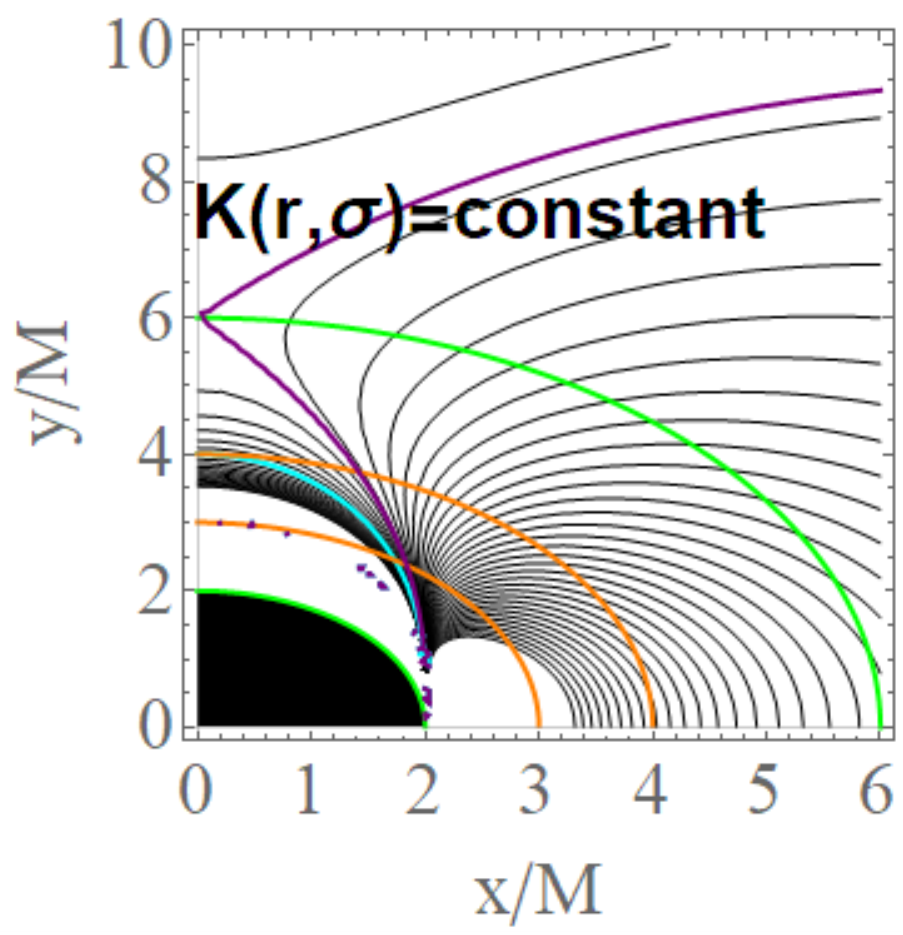}
       \includegraphics[width=5cm]{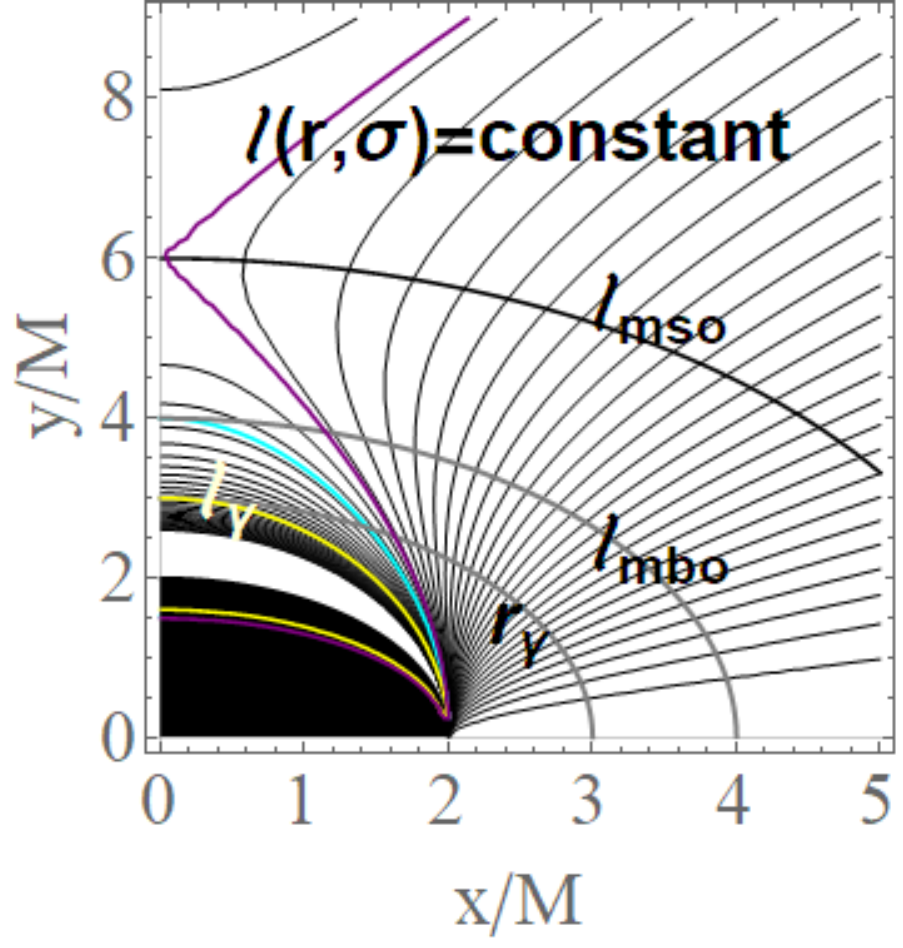}
         \includegraphics[width=5cm]{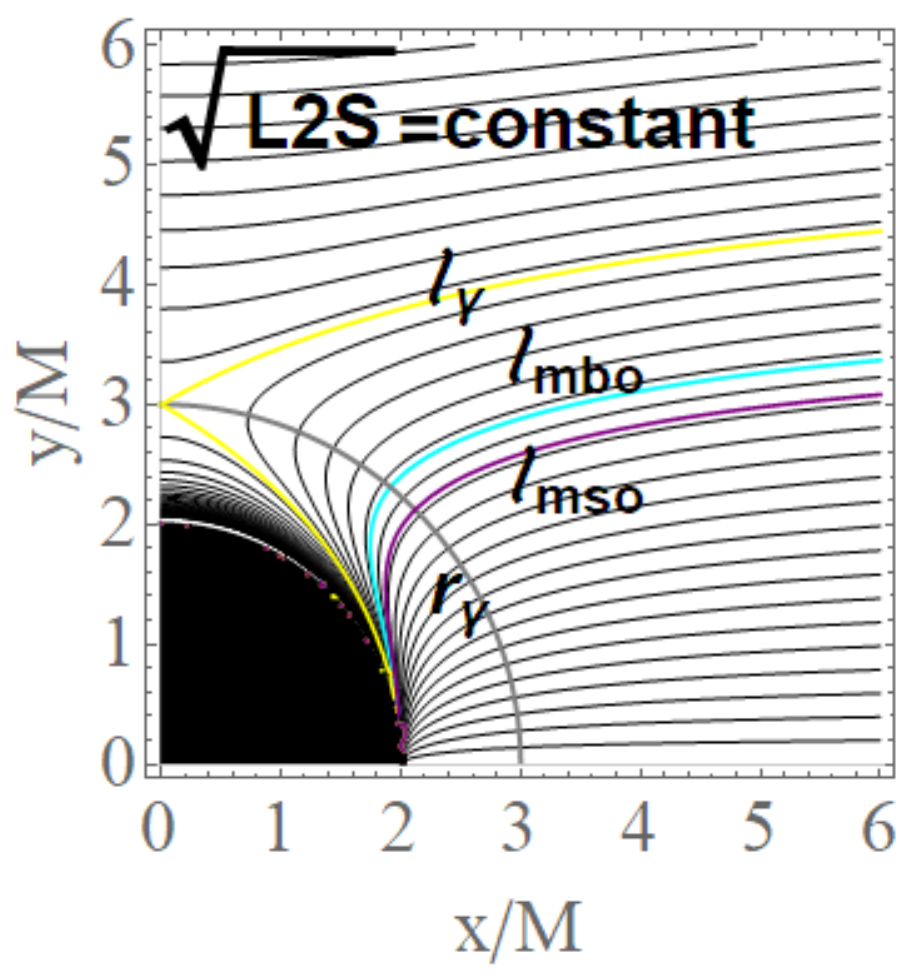}
  \caption{Central black area is the \textbf{BH}. Left panel: surfaces $K(r)$=constant in Cartesian coordinates,  radii
$r_{mso}$ are green, radii $r=r_{mbo}$ and $r_\gamma$  are orange,
$K(r)=1$ is shown as cyan curve,
$K(r)=K_{mso}$ is purple curve. Curves of $K(r,\sigma)$, $\ell(r,\sigma)$ and  $\sqrt{\mathbf{L2S}}$ constant are shown.  Central panel: $\ell(r,\sigma)=$constant,
 cyan curve is $\ell=\ell_{mbo}$,
 purple is $\ell=\ell_{mso}$.
 Right panel: angular momentum
 $\textbf{L2S}$ =constant of Eq.\il(\ref{Eq:condizioneL2S}).
}\label{Fig:eSnatwitda1}
\end{figure}
\begin{figure}
   \begin{center}
  \includegraphics[width=5.cm]{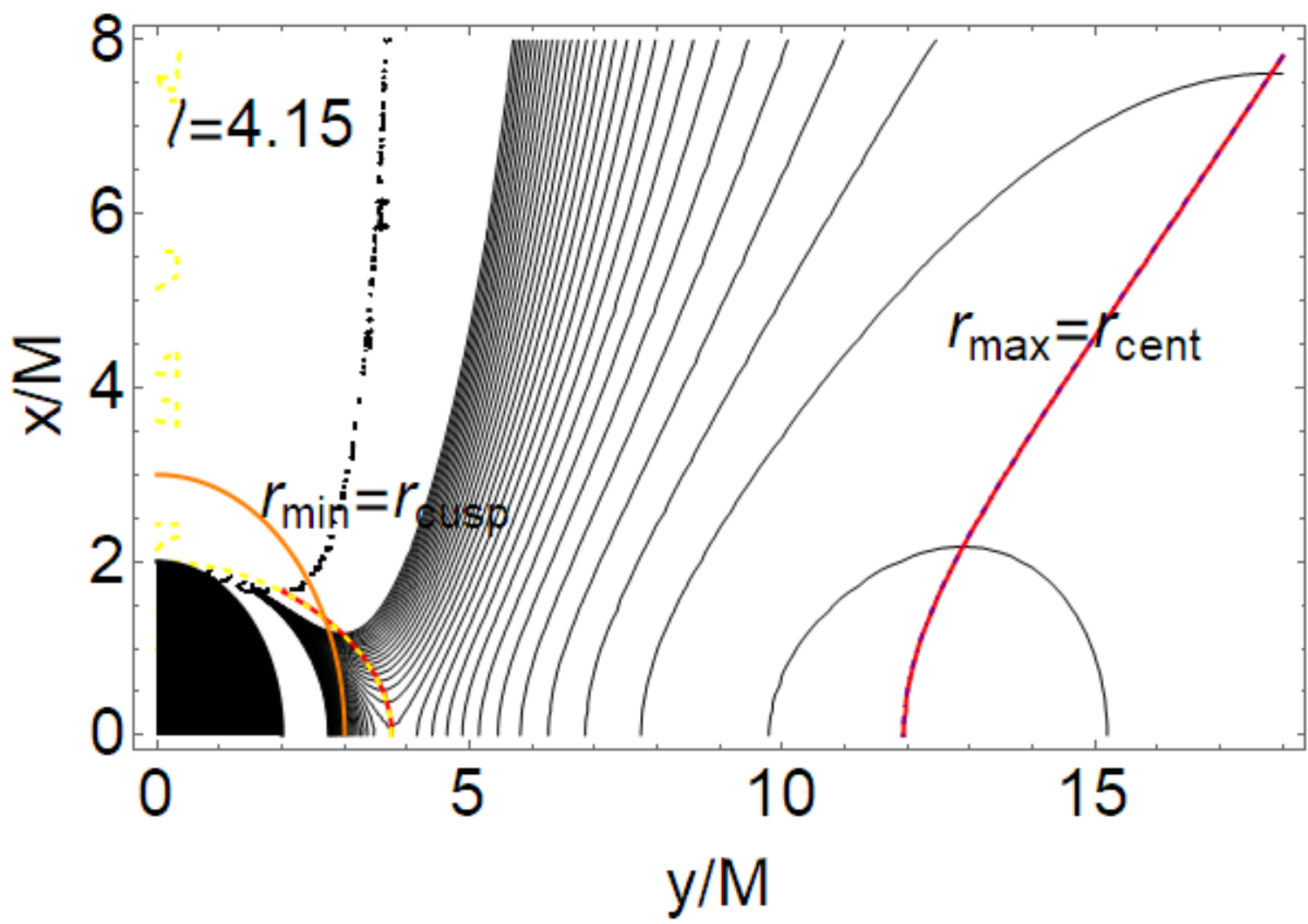}
           \includegraphics[width=5.cm]{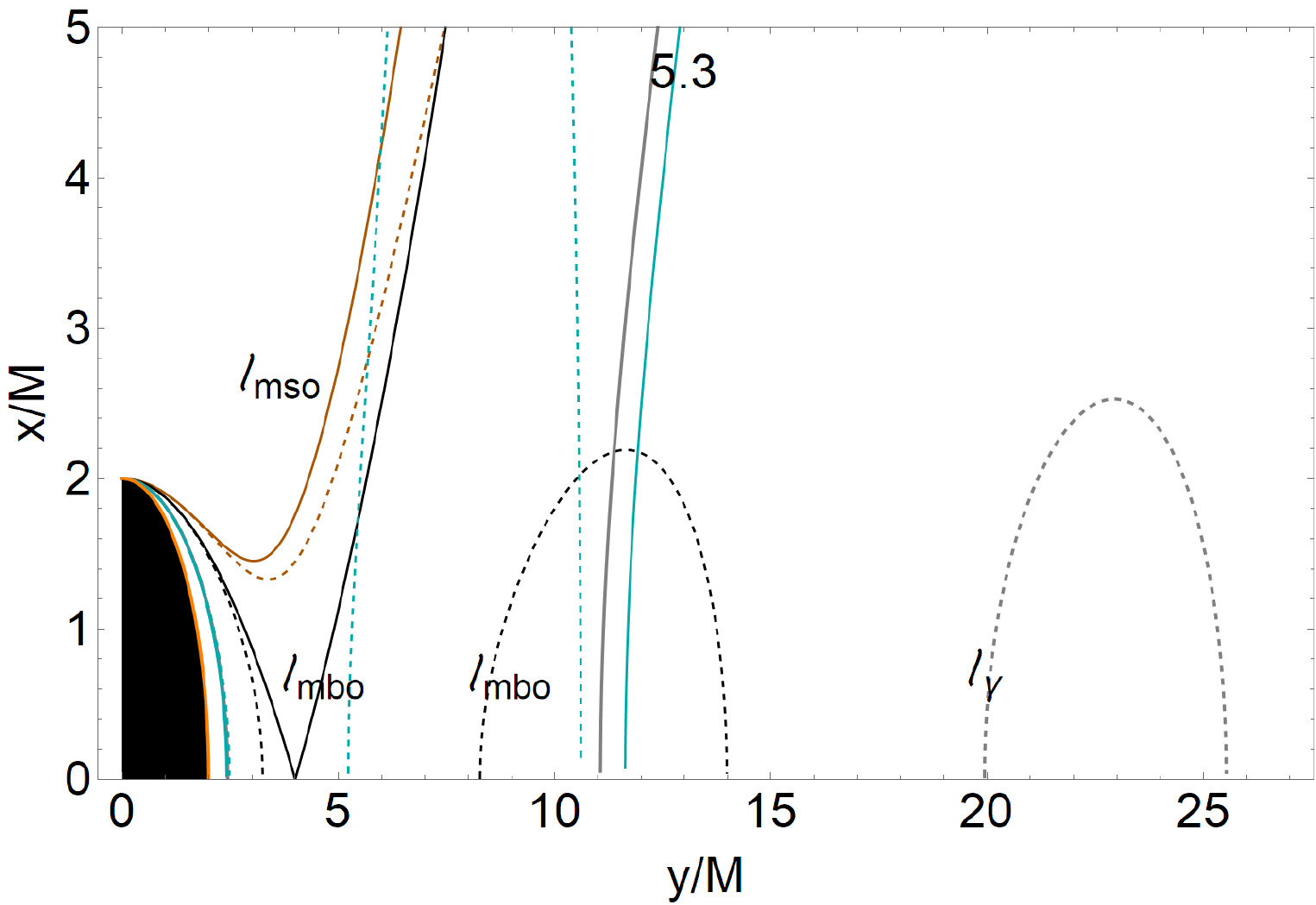}
    \includegraphics[width=5.cm]{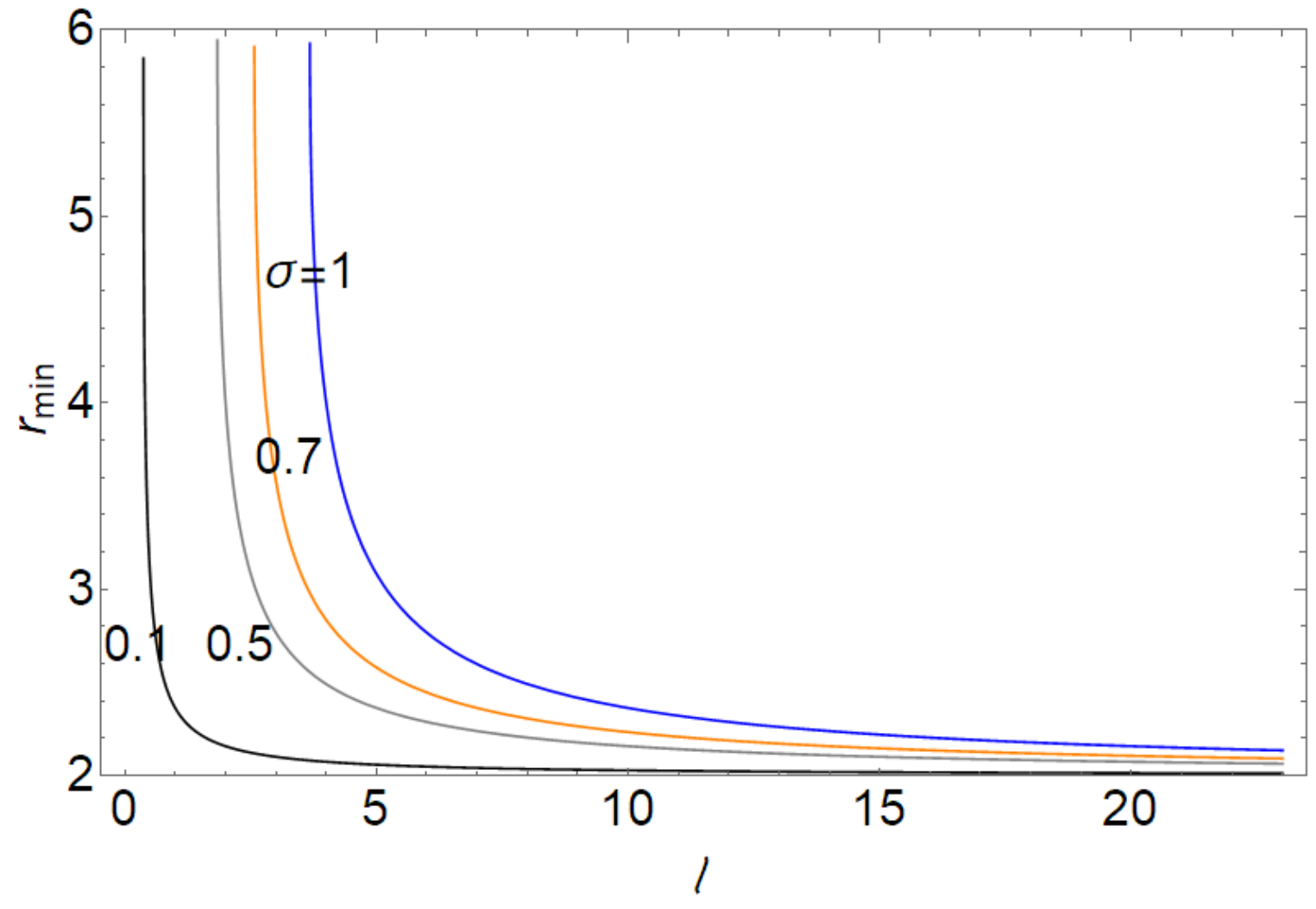}
    \end{center}
  \caption{Left panel: equipotential surfaces for different values of $K$
 for  angular momentum of the  fluid $\ell$ signed on the panel.
 Orange  continuum line is the  photon orbit. The asymptotic open  limiting curves are shown.
  The orange curve is
  $r_{min} =r_{cusp} $
and $r_{max} =r_{cent}$ of  Eq.\il(\ref{Eq:miss-metrc-inenr}) connecting the maximum density point and morphological maximum   point;
 this is also  solution of $\partial_yV_{eff}=0$ that is  curves  $\ell_{extre}=$constant of Eq.\il(\ref{Eq:recurr-exom}).  Right  panel: $r_{min}= r_{inner}(\ell\rightarrow\ell/\sigma)$ for different $\sigma$,
as function of the specific angular momentum of the fluid;  $\sigma=1$ is the equatorial plane.}\label{Fig:eSnatwitda2}
\end{figure}
\begin{figure*}
 \begin{center}
  \includegraphics[width=5cm]{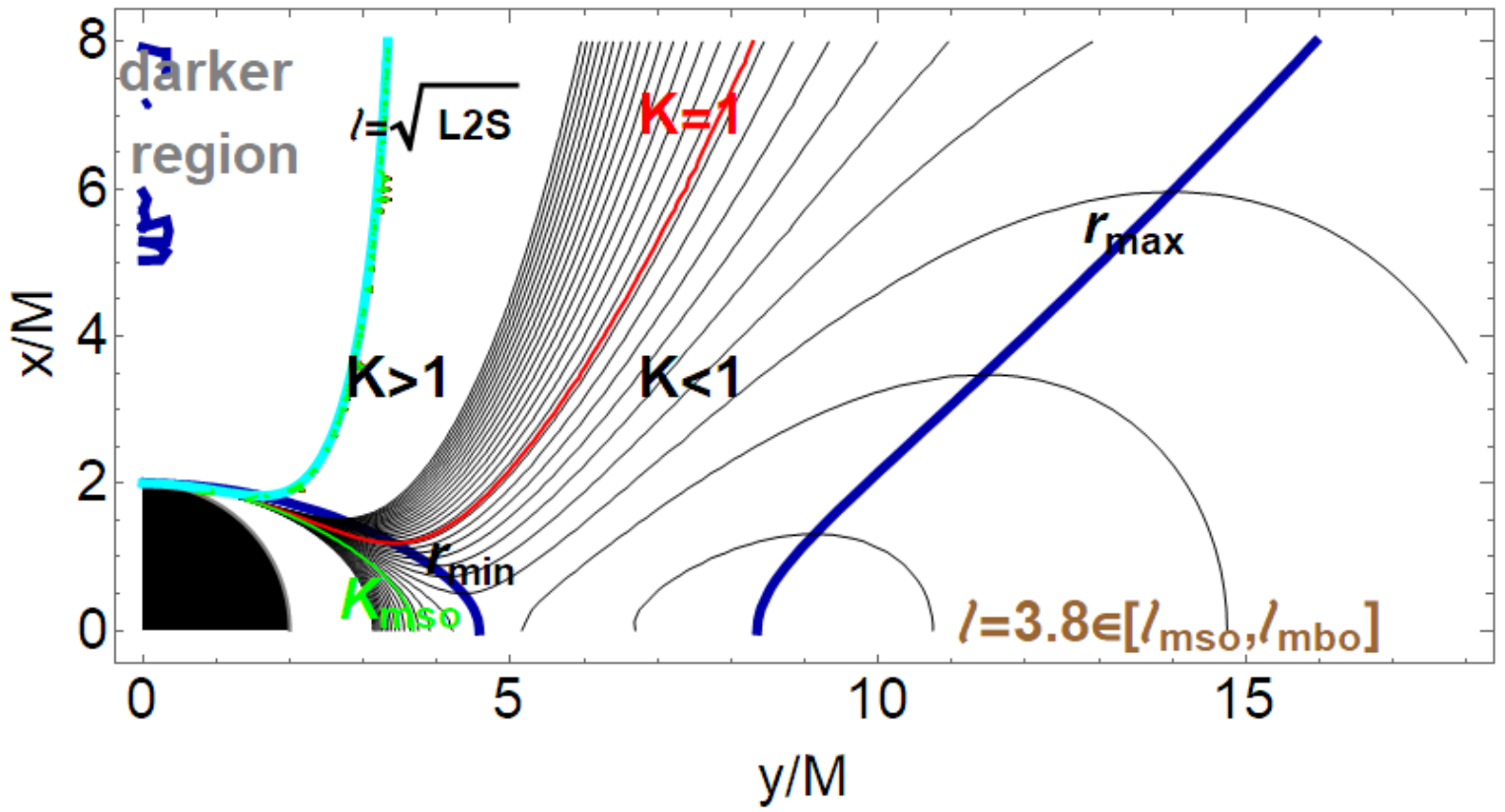}
 \includegraphics[width=5cm]{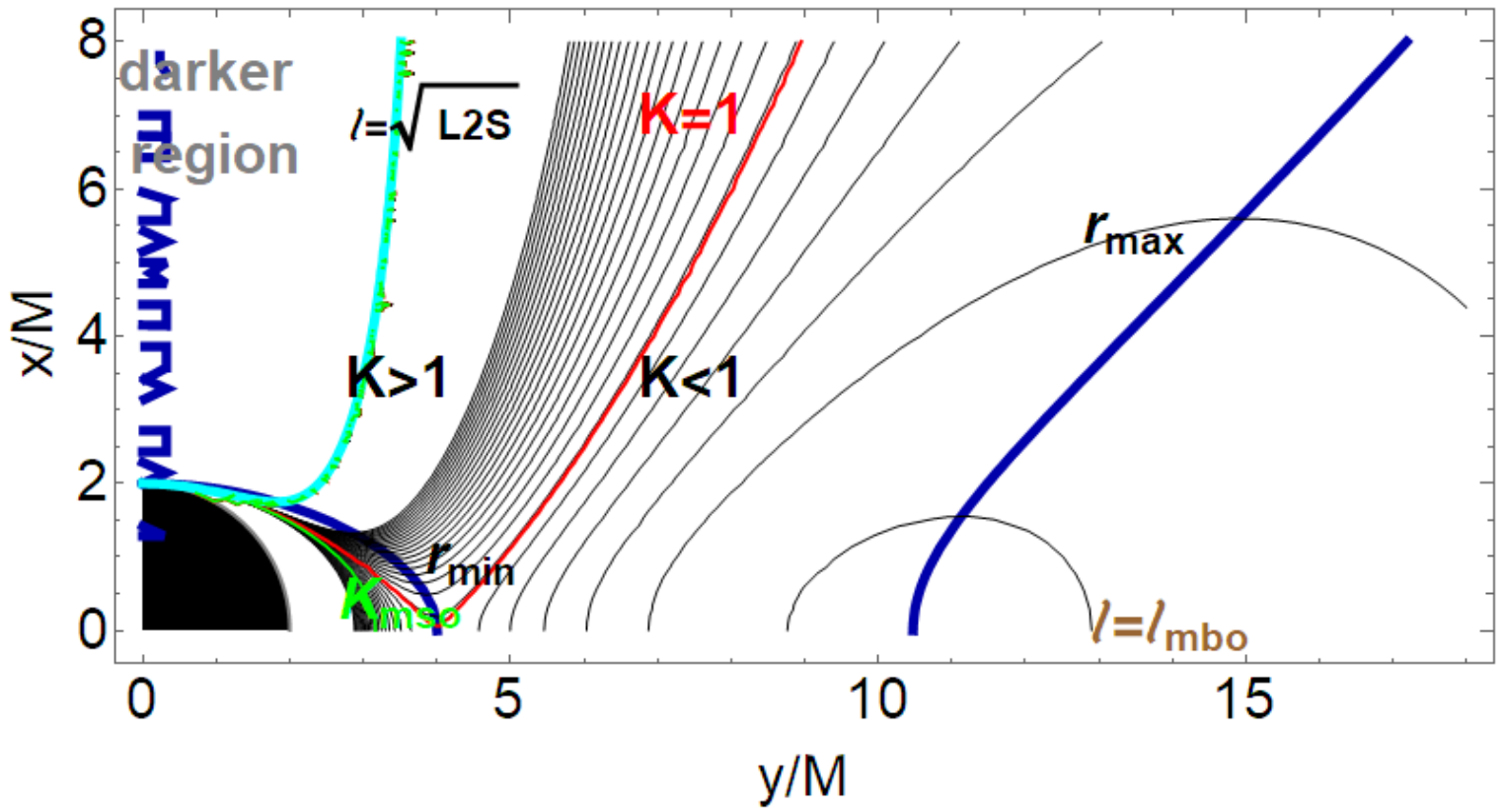}
  \includegraphics[width=6cm]{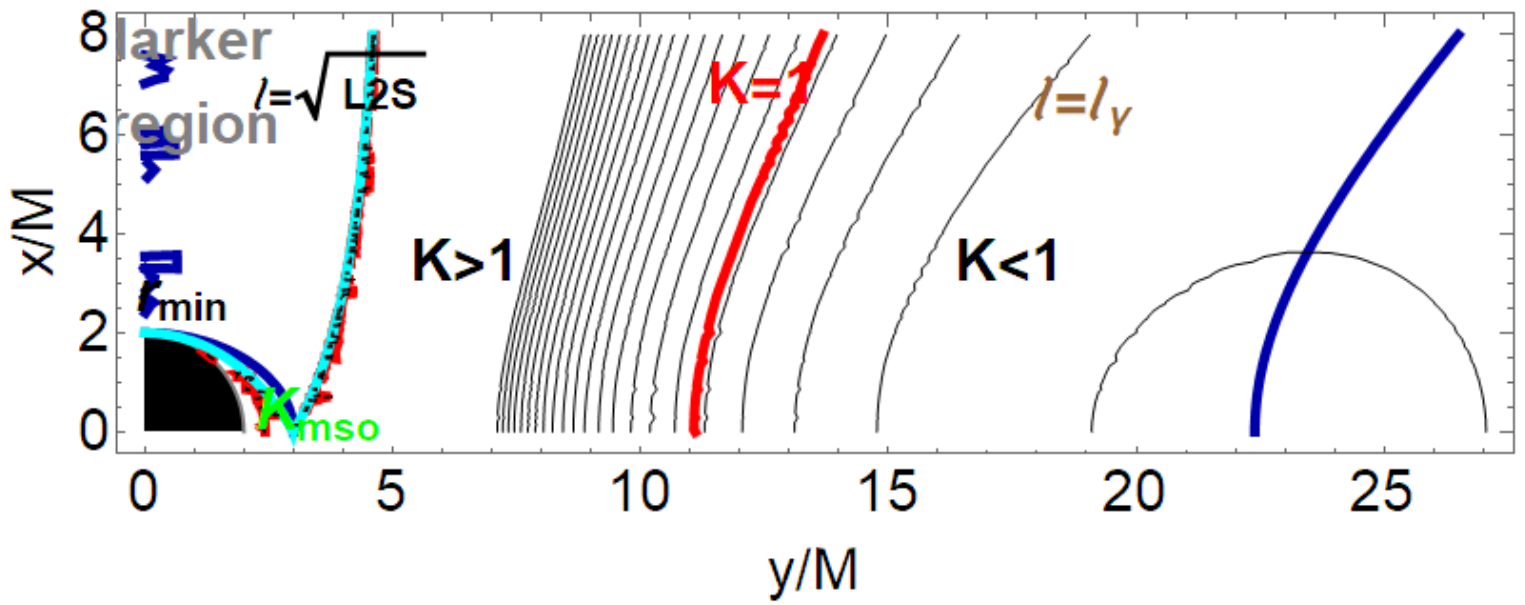}\\
   \includegraphics[width=6cm]{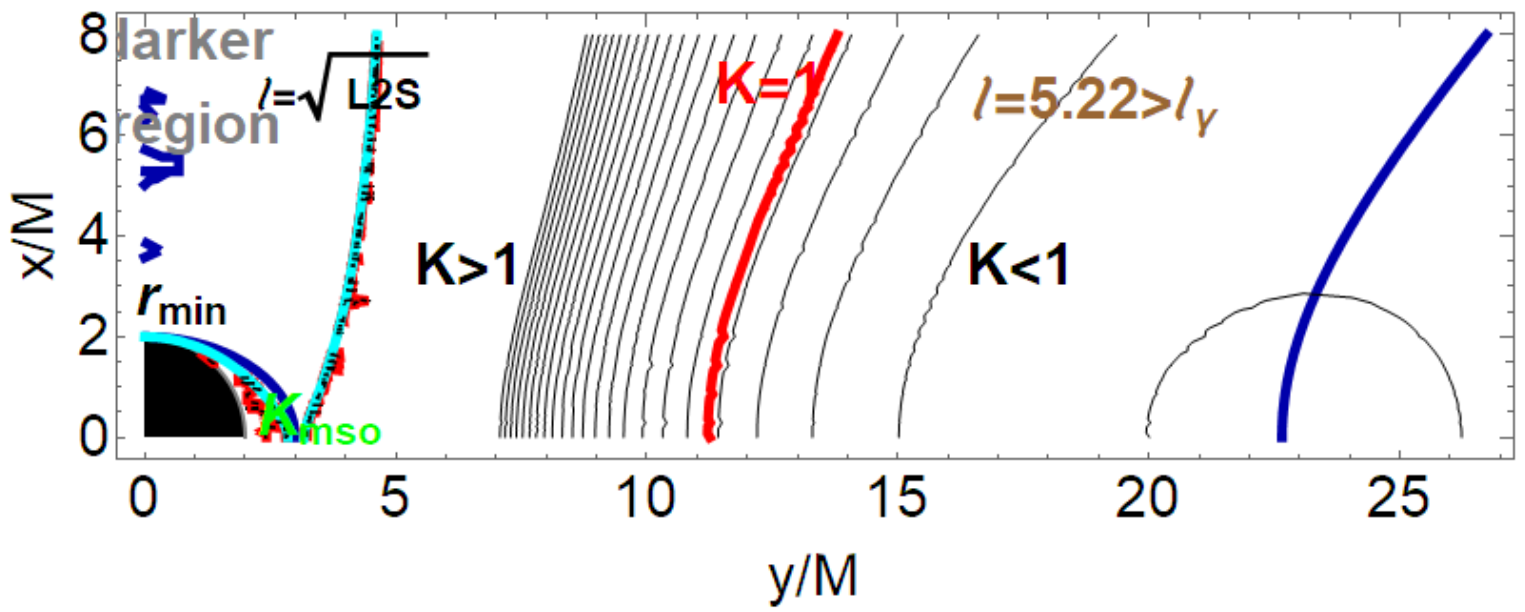}
\includegraphics[width=5cm]{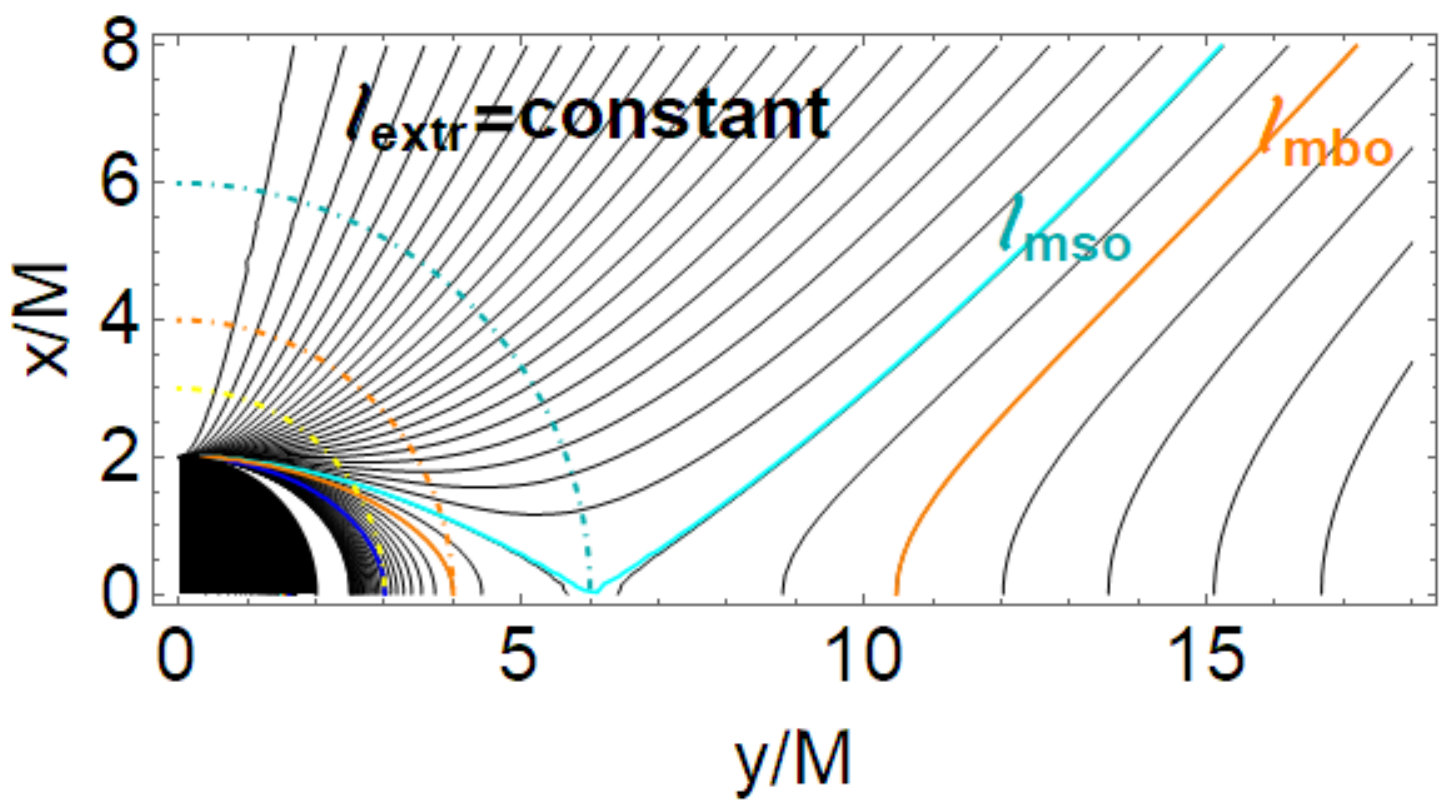}
\includegraphics[width=5cm]{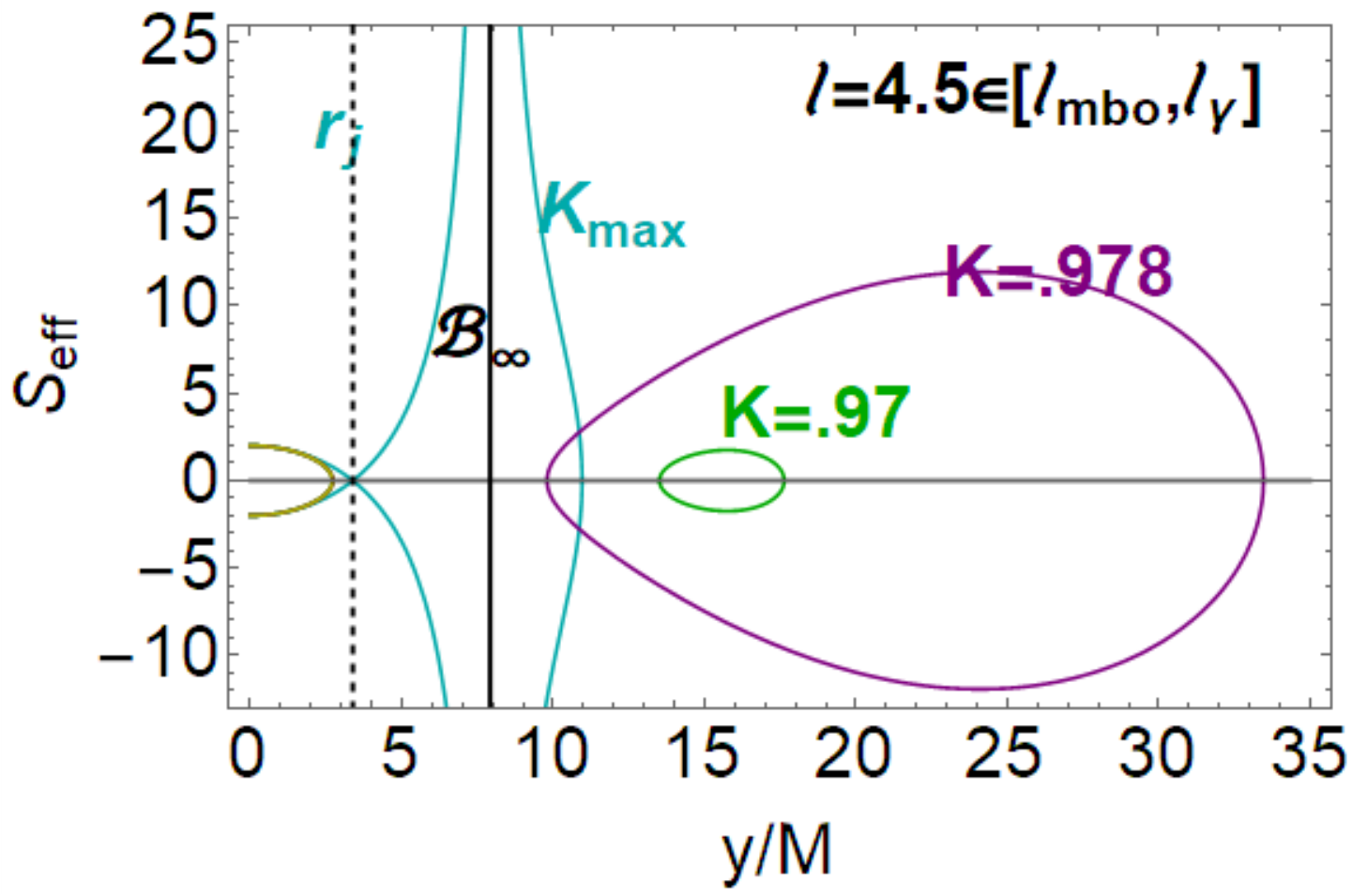}
\end{center}
    \caption{Black regions is the central \textbf{BH}.
Upper panels and left bottom panel:
 equipotential surfaces for different values of $K$ with fixed
  angular momentum of the  fluid $\ell$ signed on the panel.
 The asymptotic open  limiting curves are shown, as well as
  $r_{min} =r_{cusp} $
and $r_{max} =r_{cent}$ of  Eq.\il(\ref{Eq:miss-metrc-inenr}) connecting the maximum density points and the morphological maximum   point,
 also  solution of $\partial_yV_{eff}=0$ that are the   curves  $\ell_{extre}=$constant of Eq.\il(\ref{Eq:recurr-exom}), and angular momentum
$\textbf{L2S}$ of Eq.\il(\ref{Eq:condizioneL2S}). {Notation for  \textbf{L2S} and  $L2_d$ follow the convection of Eq.\il(\ref{Eq:condizioneL2S}). Bottom  Center panel. Curves $\ell_{extre}=$constant of Eq.\il(\ref{Eq:recurr-exom}), special values $\ell_{mso}$ (marginally stable orbits) and $\ell_{mso}$ (marginally bounded orbit) are shown.
Bottom right  panel: for fixed angular momentum, different configurations, closed, open, and cusped open, are shown for  different $K$ values. Surfaces $\mathcal{B}_{\infty}$ are also shown.}}\label{Fig:eSnatwitdpiw}
\end{figure*}

\medskip

\textbf{{Reducing proto-jets parameters $\ell(K)$}:}

We can relate  $\ell(r)$ to  $K(r)$  by eliminating the critical pressure radial  dependence $r$, obtaining the  following function $\ell(K)$:
\bea\label{Eq:LKmeasuram}
\ell_{K}^{\pm}\equiv\frac{\sqrt{\pm\sqrt{\frac{\left(9 K^2-8\right)^3}{K^2 \left(K^2-1\right)^2}}+\frac{27 K^4-36 K^2+8}{K^2 \left(K^2-1\right)}}}{\sqrt{2}},
\eea
represented in  Figs\il(\ref{Fig:ampaerplot2k},\ref{Fig:davmandc}), or alternately the relations $K(\ell)$ defined as
\bea&&\label{Eq:suo-fr1}
K^{\pm}_{crit}(\ell)=\sqrt{\frac{\xi_1^\pm\xi_2^\pm}{3\xi_3^\pm}},\quad \mbox{with}\quad  K^{-}_{crit}(\ell)=K_{cent}(\ell),\quad K^{+}_{crit}(\ell)=\{K_j,K_{\times}\},
\eea
where
\bea\nonumber
 &&\xi_1^+\equiv \ell^2-6-2 \sqrt{\ell^2 \left(\ell^2-12\right)} \cos \left(\frac{\iota +\pi }{3}\right)\quad\mbox{and}\quad \xi_1^-\equiv\ell^2-6+2 \sqrt{\ell^2 \left(\ell^2-12\right)} \cos \left(\frac{\iota }{3}\right),
\\
&&\nonumber \xi_2^+\equiv  \left[\ell^2-2 \sqrt{\ell^2 \left(\ell^2-12\right)} \cos \left(\frac{\pi \iota }{3}\right)\right]^2\quad\mbox{and}\quad   \xi_2^-\equiv \left[\ell^2+2 \sqrt{\ell^2 \left(\ell^2-12\right)} \cos \left(\frac{\iota }{3}\right)\right]^2,
\\
&&\nonumber \xi_3^+\equiv \ell^2 \left[2 \left(12-\ell^2\right) \ell^2 \sin \left[\frac{4 \iota +\pi }{6}\right]+\right.\\
&&\nonumber\left.\qquad \qquad +2 \left(15-2 \ell^2\right) \sqrt{\ell^2 \left(\ell^2-12\right)} \cos \left[\frac{\iota +\pi }{3}\right]+3 \left(\ell^4-13 \ell^2+18\right)\right],
\\
&&\nonumber
 \xi_3^-\equiv \ell^2 \left[2 \left(\ell^2-12\right) \ell^2 \cos \left(\frac{2 \iota }{3}\right)+2 \left(2 \ell^2-15\right) \sqrt{\ell^2 \left(\ell^2-12\right)} \cos \left(\frac{\iota }{3}\right)+3 \left(\ell^4-13 \ell^2+18\right)\right],
\\\label{Eq:suo-fr2}
&&\mbox{and}\quad
\iota \equiv\cos ^{-1}\left[\frac{\ell^2 \left(\ell^4-18 \ell^2+54\right)}{\left[\ell^2 \left(\ell^2-12\right)\right]^{3/2}}\right],
\eea
$K^{\pm}_{crit}(\ell)$ is either $K_{j}\geq1$, for proto-jets, or $K_{\times}\in[K_{mso},1[$  for accreting configurations, which is represented in Figs\il\ref{Fig:davmand} or $K_{cent}$ evaluated in the centers of maximum pressure.
These relations connect   a pair of radii $r$ from the condition   $\ell=$constant,  identifying  a torus and eventually the   associated HD (topological) instability  with the correspondent value of  $K$; accordingly there are two parameters $(K_1,K_2)$ such that  $K_{mso} <K_1<1<K_2$ for  $\ell\in \mathbf{L_1}$ and $K_{mso}<K_1<1\leq K_2$ for  $\ell\in \mathbf{L_2}$ and
  $K_{mso}<K_1<1$ for  $\ell\in \mathbf{L_3}$, where clearly  $K_1$ and $K_2$ are respectively  $K_{\min}=K_{cent}$ and $K_{\max}=K_{\times}$ or $K_j$  for the fixed  $\ell$.
\section{Tori characteristics, limiting conditions and pressure gradients}\label{Sec:polar-gradient}
In this section we discuss the  maximum and minimum density and pressure and the thickness of the disk related to the  radial gradient of the pressure.
We  will explore the disk  verticality by the  analysis of the radial gradient of the pressure.
The analysis developed in the frame of \textbf{RAD} models  is characterized by  the  intensive use of a multi-parametric analysis on important characteristics of the tori.
In this section  we investigate the configuration center, i.e.   the maximum pressure point, focusing in particular on the
 projection of the morphological maxima  on the equatorial plane.
 We discuss the  inner edge, the center and the morphological maximum in dependence on different tori parameters.
The morphological maximum  will be found  from the \textbf{RAD} rotational law, showing   the disk vertical structure determined by  the radial structure through the agglomeration rotation.
The center of the maximum pressure point  of the configurations is  given by
\bea&&\label{Eq:centrol}
r_{centr}(\ell)=\frac{1}{3} \left[\ell^2+\bar{\beta}+\frac{\ell^2 \left(\ell^2-12\right)}{\bar{\beta}}\right],\\&&\mbox{where}\quad \bar{\beta}\equiv\sqrt[3]{\ell^2(\ell^4-18 \ell^2+54+6 \sqrt{81 -6 \ell^4})},
\eea
(for $r_{cent}$ compare with Eq.\il(\ref{Eq:eellH})). The center of the orbiting   torus is the point, on  its equatorial plane, of the maximum pressure and density. The location of this point depends on one torus parameter, $\ell$ or $K$, determined as minimum point of the effective potential function  regulating the force balance in  the  torus.  In Eq.\il(\ref{Eq:centrol}) the center depends on the specific angular momentum of the fluid, and it is obtained by inverting the function $\ell(r)$ in the range $r\geq r_{mso}$.
%
\begin{figure}
 \includegraphics[width=5.36cm]{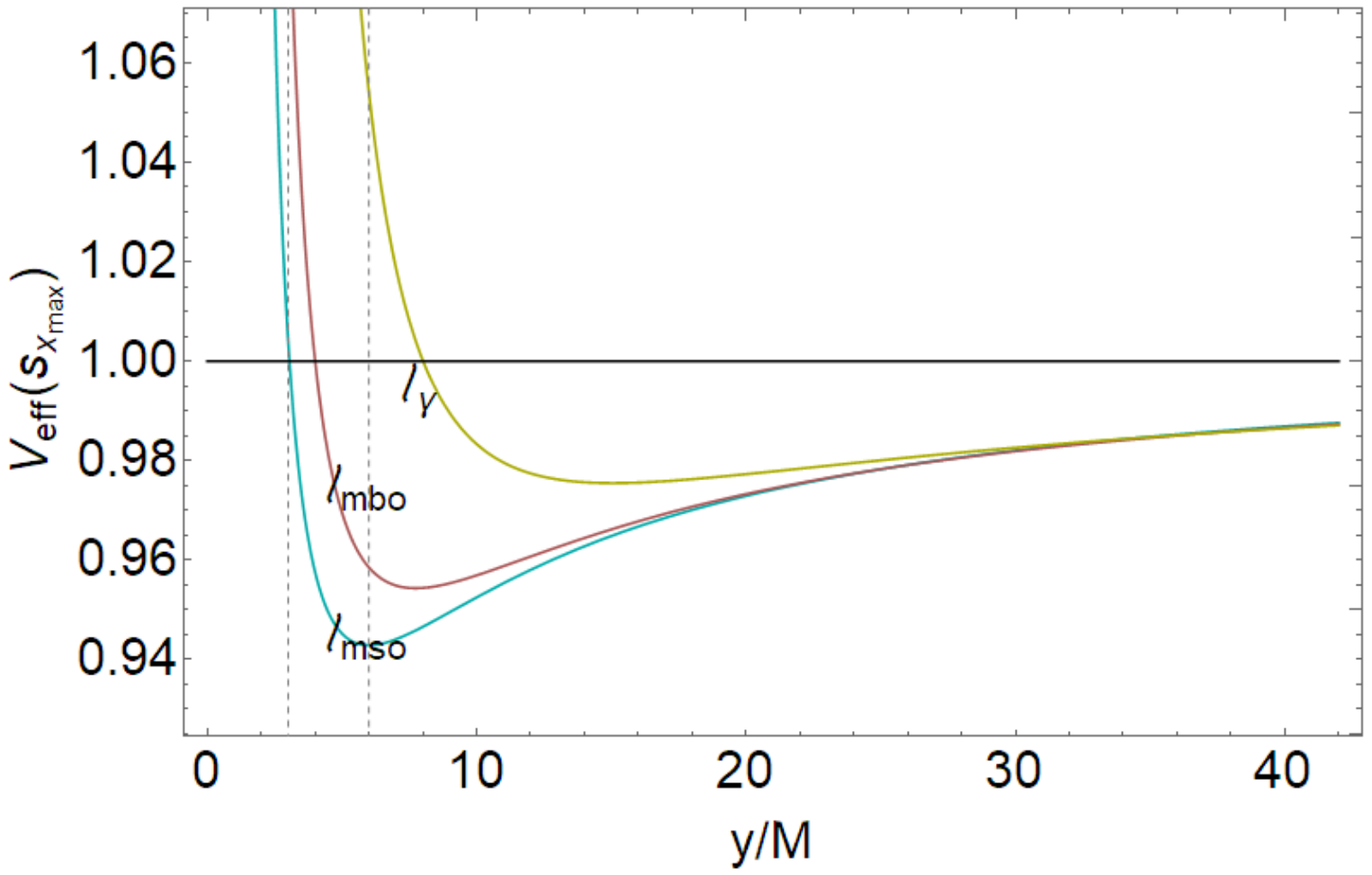}
\includegraphics[width=5.36cm]{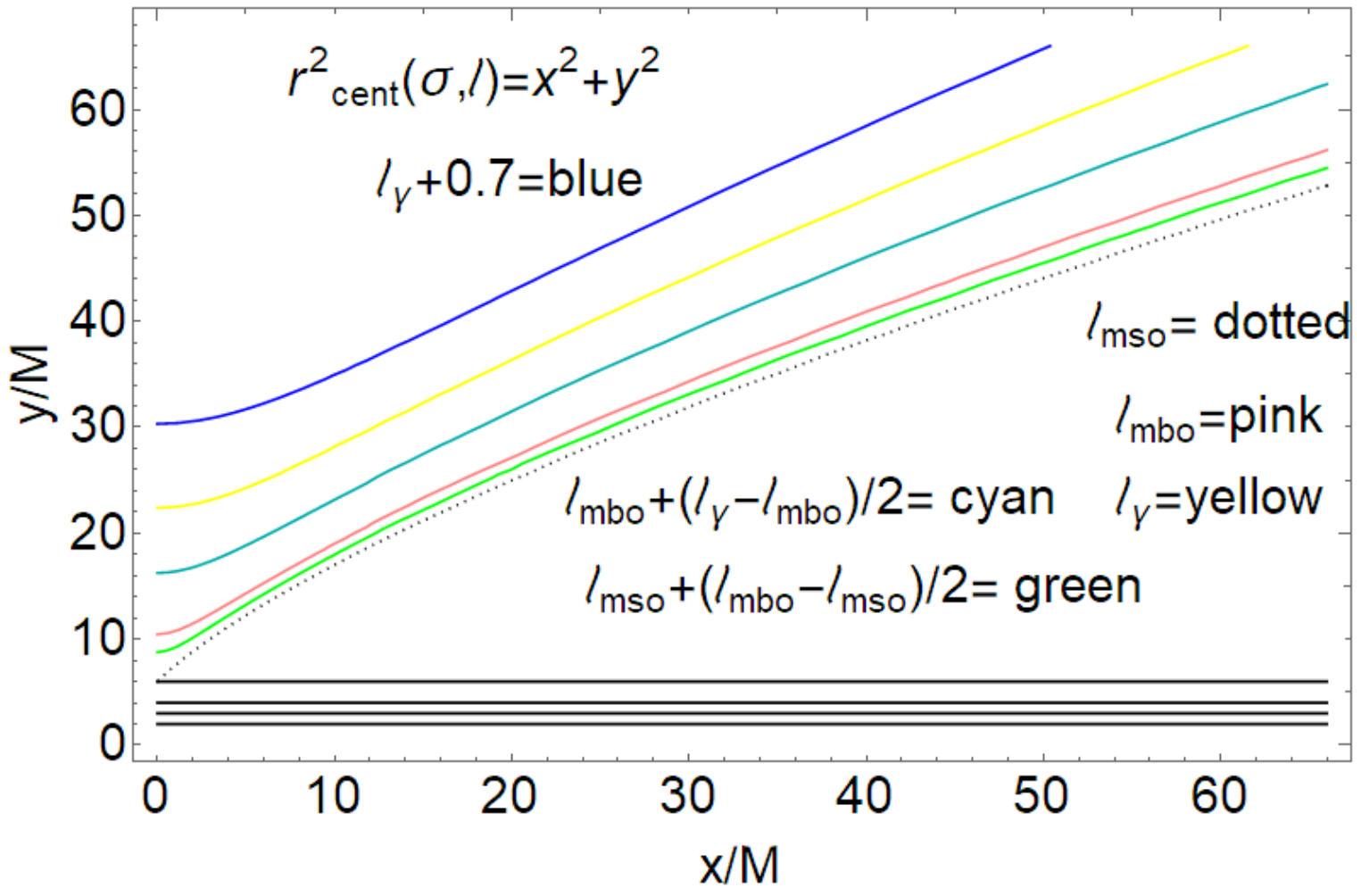}
\includegraphics[width=5.36cm]{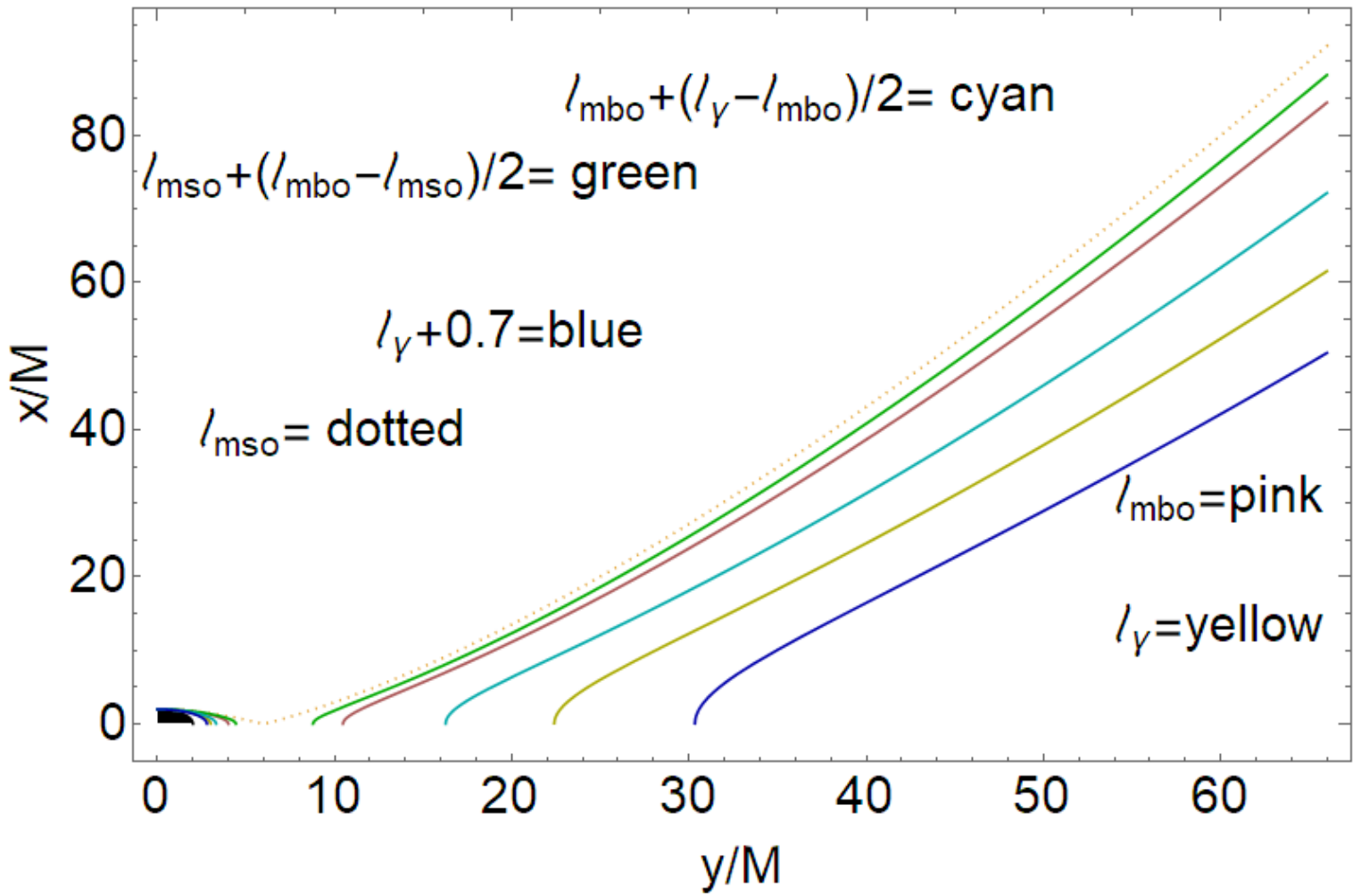}
  \caption{Left panel: $V_ {eff} (x, y)$ for $x =
 x_ {\max} $ evaluated  for the fixed fluid angular momentum
$\ell$ on the marginally stable orbit $mso$ (cyan), marginally
bounded orbit, $mbo$ (pink),  and last circular photon orbit, $\ell_{\gamma} $ (yellow). as function of $y/M$.
Center panel:  surfaces $r_{cent}^2=r^2$, location of maximum density and pressure point in the disk, in the cartesian coordinate for different  specific angular momentum  of the fluid $\ell$.
Right  panel: black region is the central \textbf{BH}, $x_{\max}$ (location of morphological maximum) evaluated on different fluid angular momentum
$\ell$ as function of $y$.}\label{Fig:davmanda}
\end{figure}
\begin{figure*}
 \includegraphics[width=5.35cm]{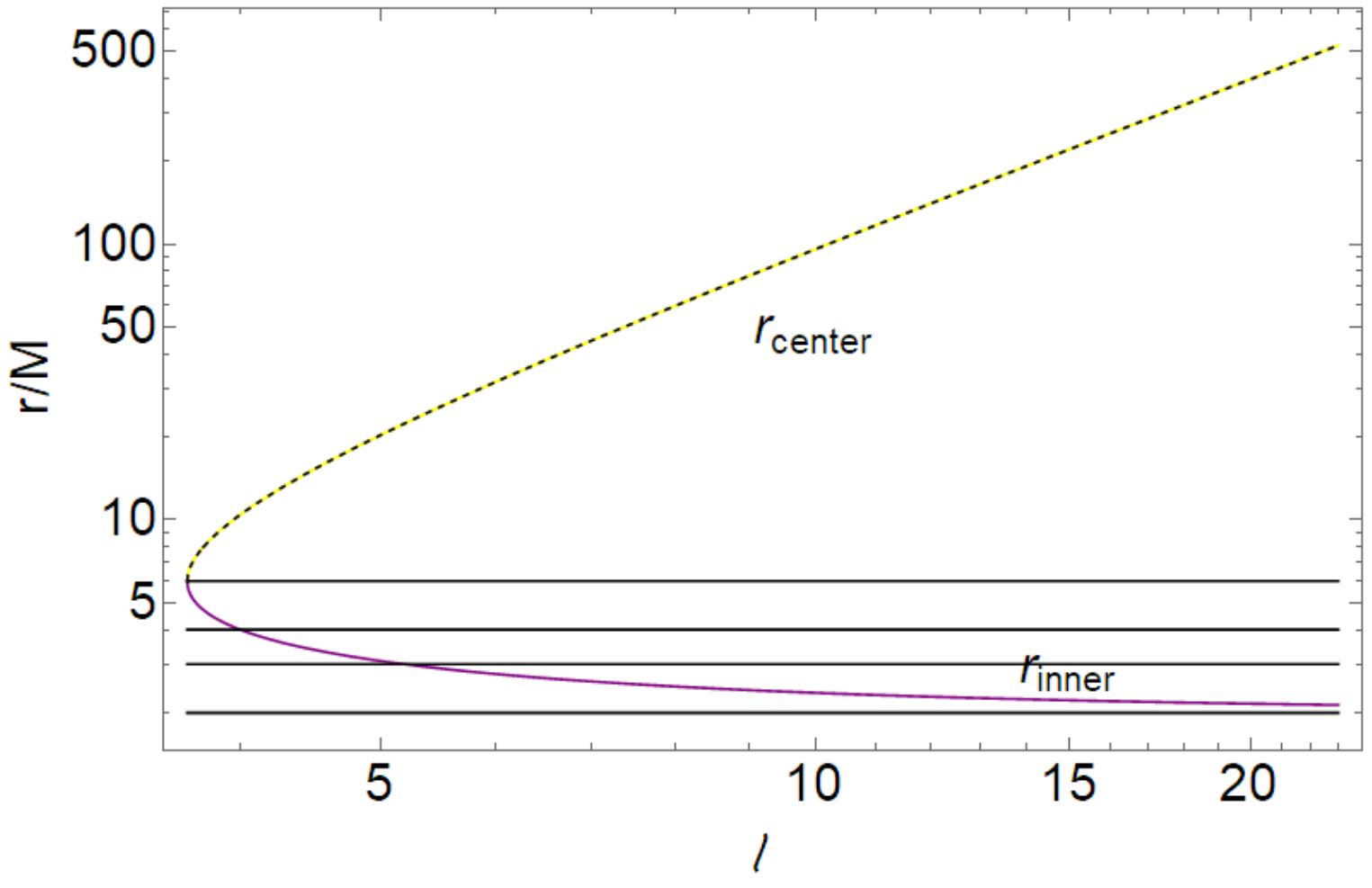}
 \includegraphics[width=5.35cm]{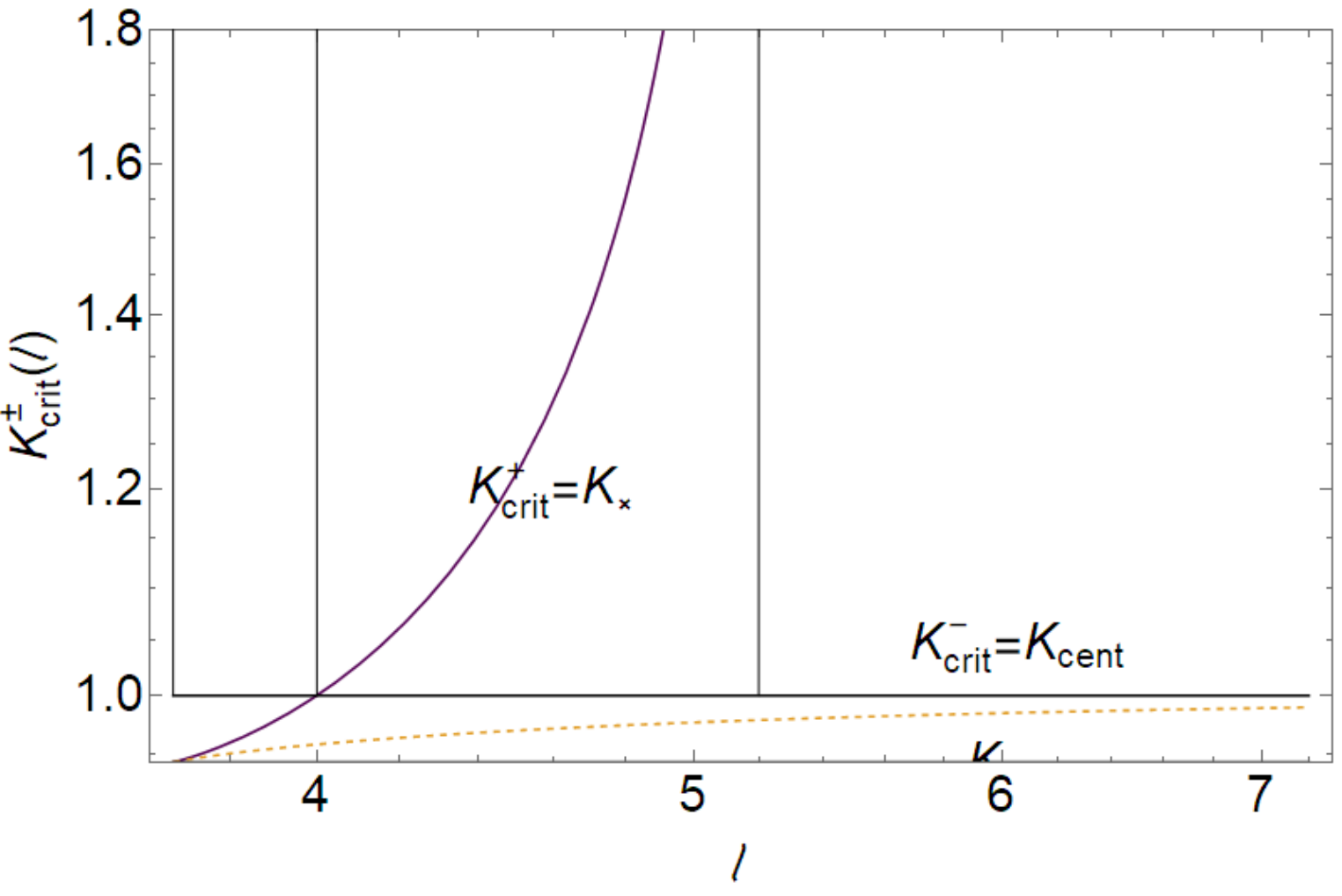}
 \includegraphics[width=5.35cm]{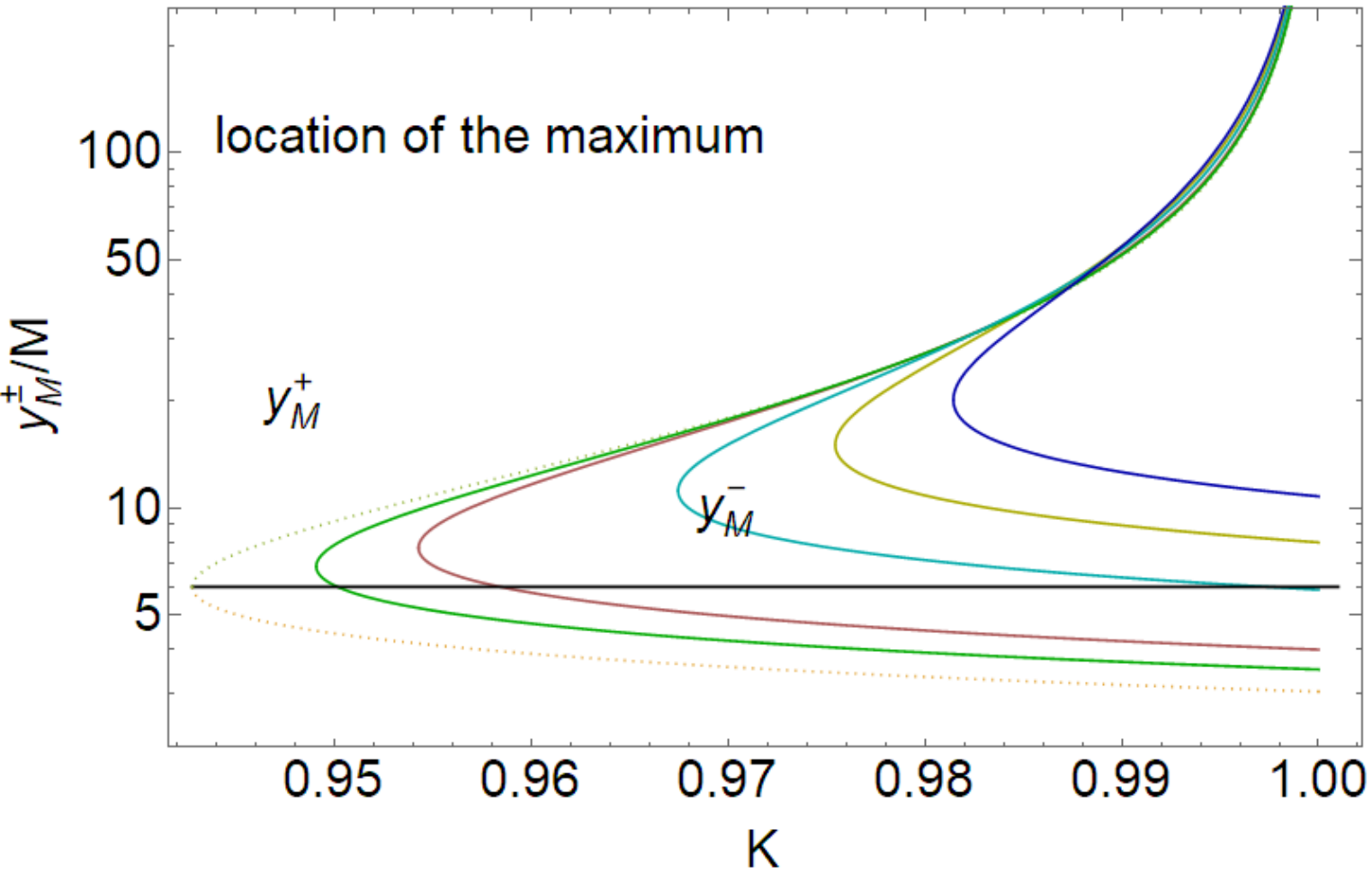}\\
 \includegraphics[width=5.35cm]{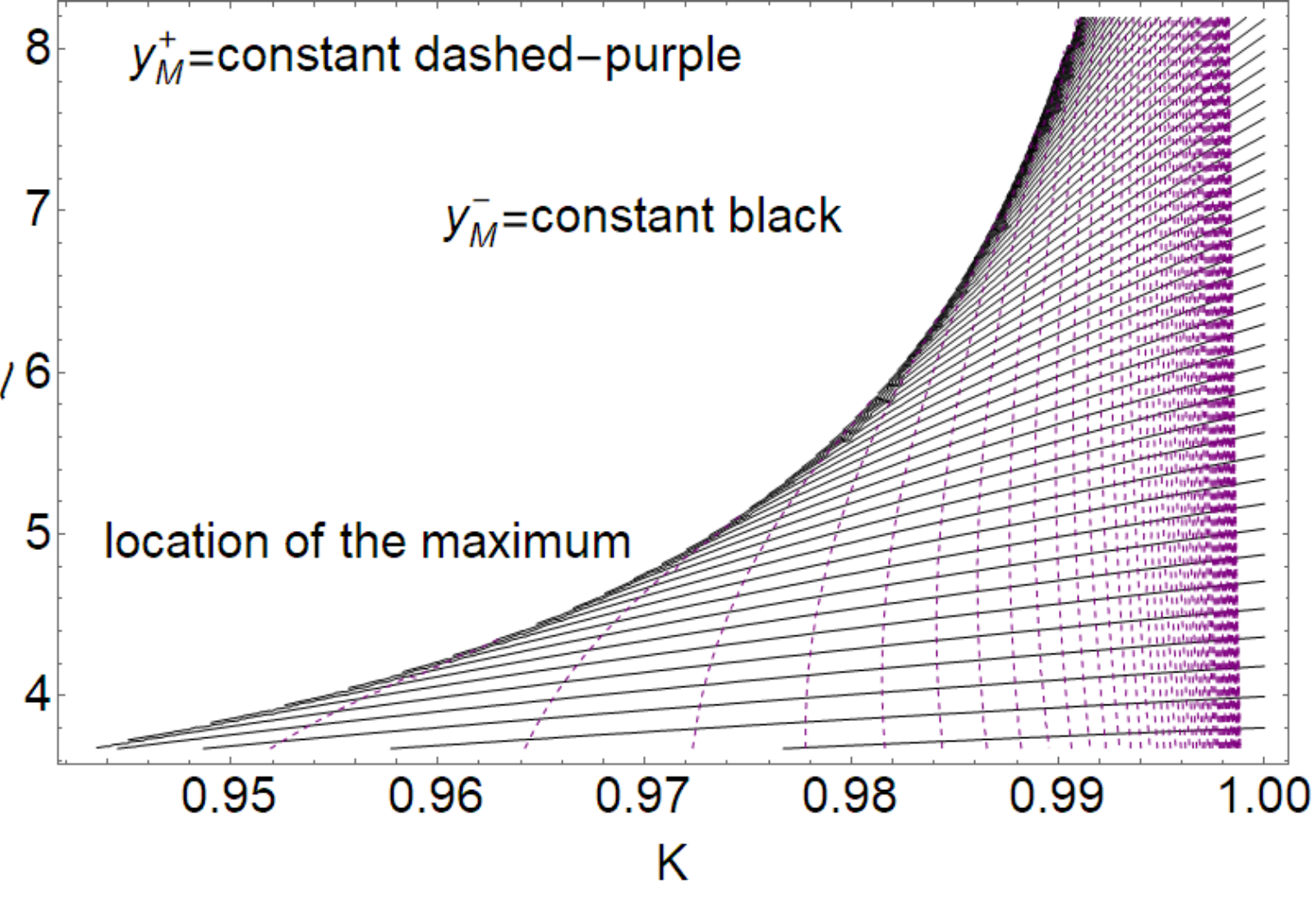}
 \includegraphics[width=5.35cm]{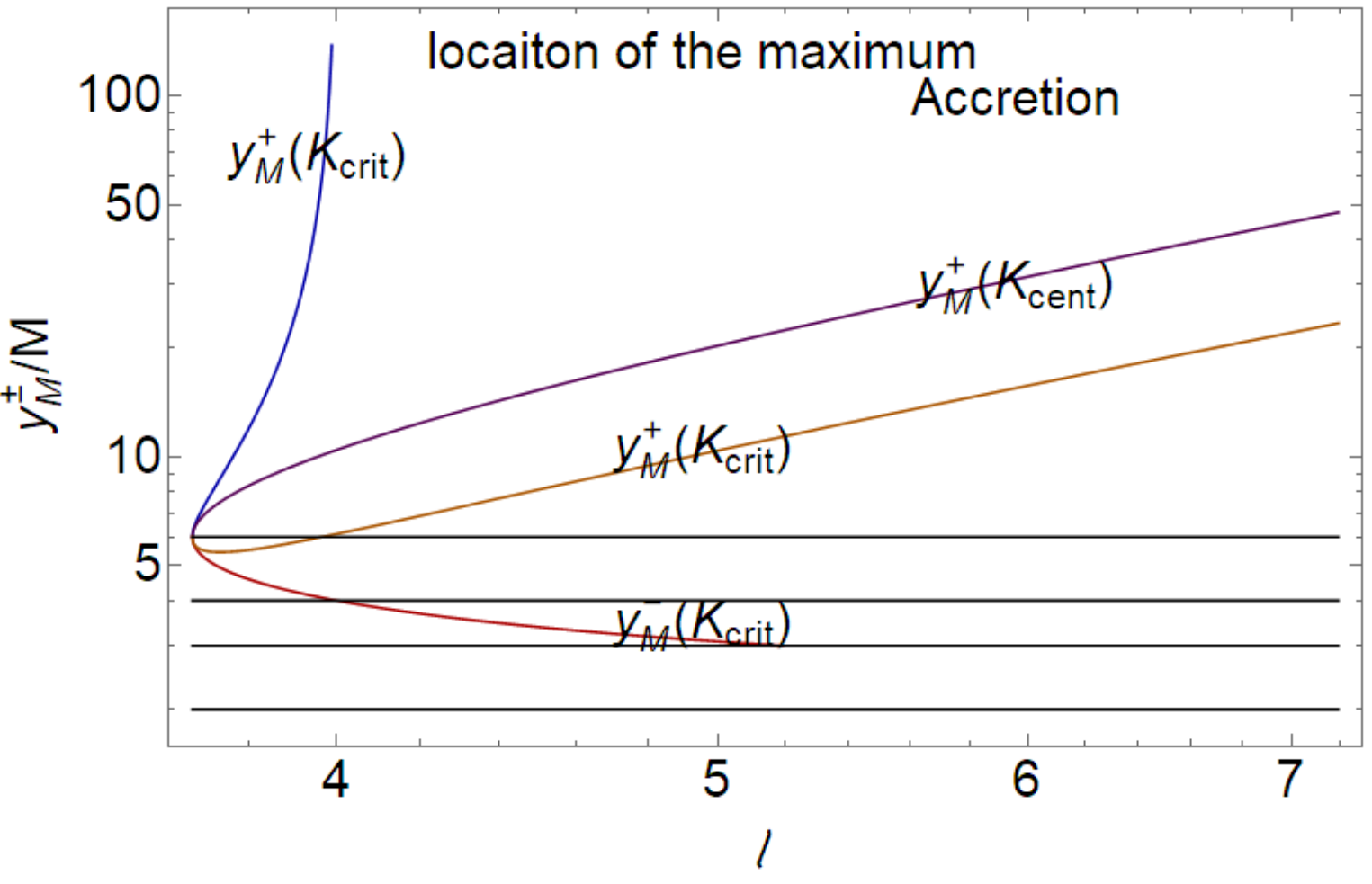}
    \includegraphics[width=5.35cm]{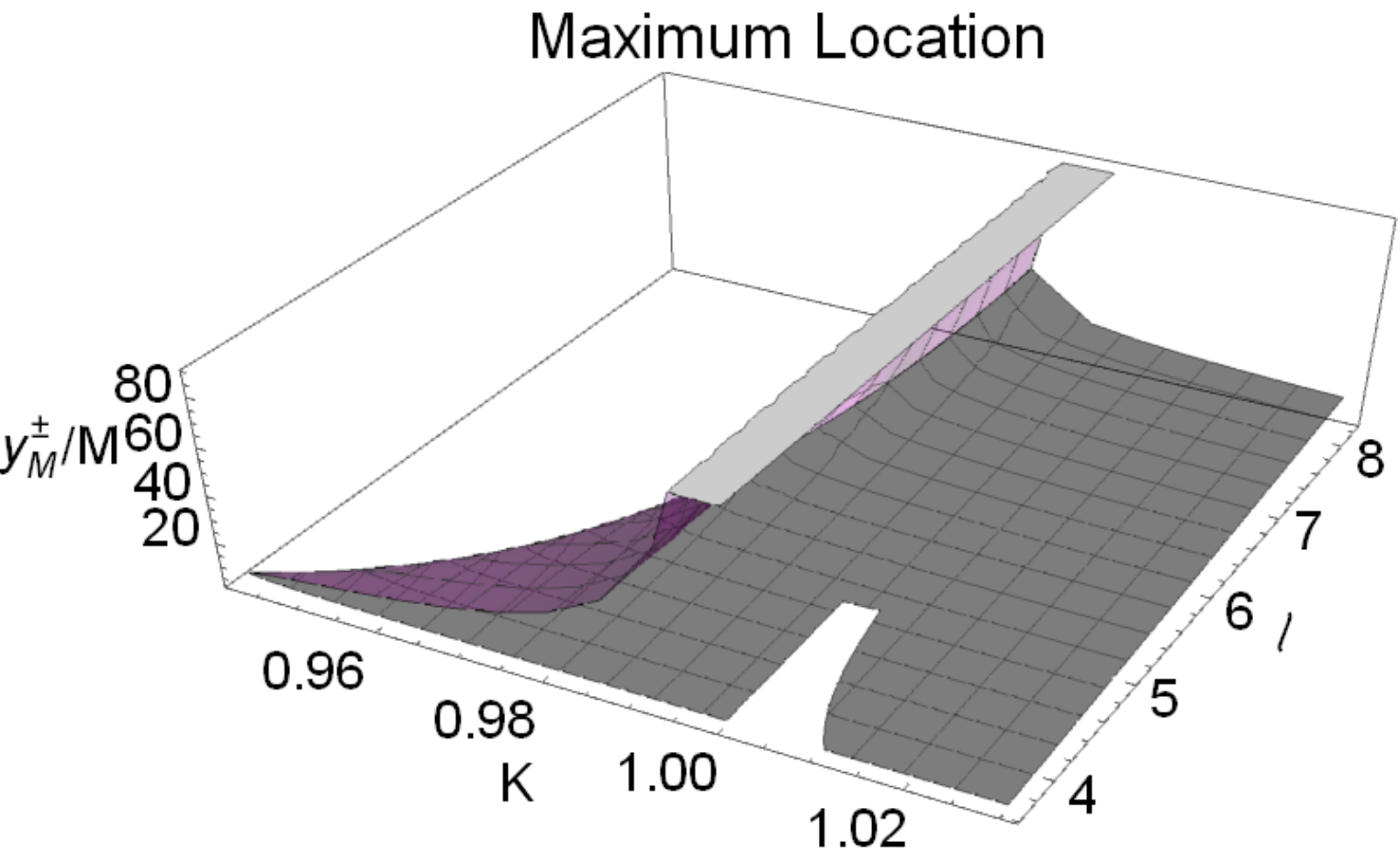}
  \caption{Upper  left-hand panel: the turus center $(r_{center})$,  maximum pressure  point in the disk, and cusp $(r_{inner})$, minimum pressure point  of  the configurations, as functions of the fluid specific angular momentum $\ell$.
  Marginally circular orbit and the  marginally stable orbit  are also shown.
Upper center panel: $K_{crit}^{\pm}$ is shown as function of $\ell$,  providing the  parameter $K $ at the torus center and cusp for critical configurations.
Upper right-hand  panel and Below panels:
  torus geometrical  maximum $y_M^\pm$ (on the  symmetry plane) of the outer and inner  Roche lobes respectively  as function of  $K$  and $\ell$ for critical configurations.  Bottom left-hand panel: curve  $y_M^\pm=$constant  in the plane $(\ell, K)$.
 Center bottom panel: $y_M^\pm$ as functions of $ \ell$ evaluated on $K_{crit}$ and $K_{cent}$ for critical configurations.
Right bottom panel: 3D plot of $y_M^\pm$ as function of $(\ell, K)$.
}\label{Fig:davmand}
\end{figure*}
\begin{figure*}
 \begin{center}
          \includegraphics[width=7cm]{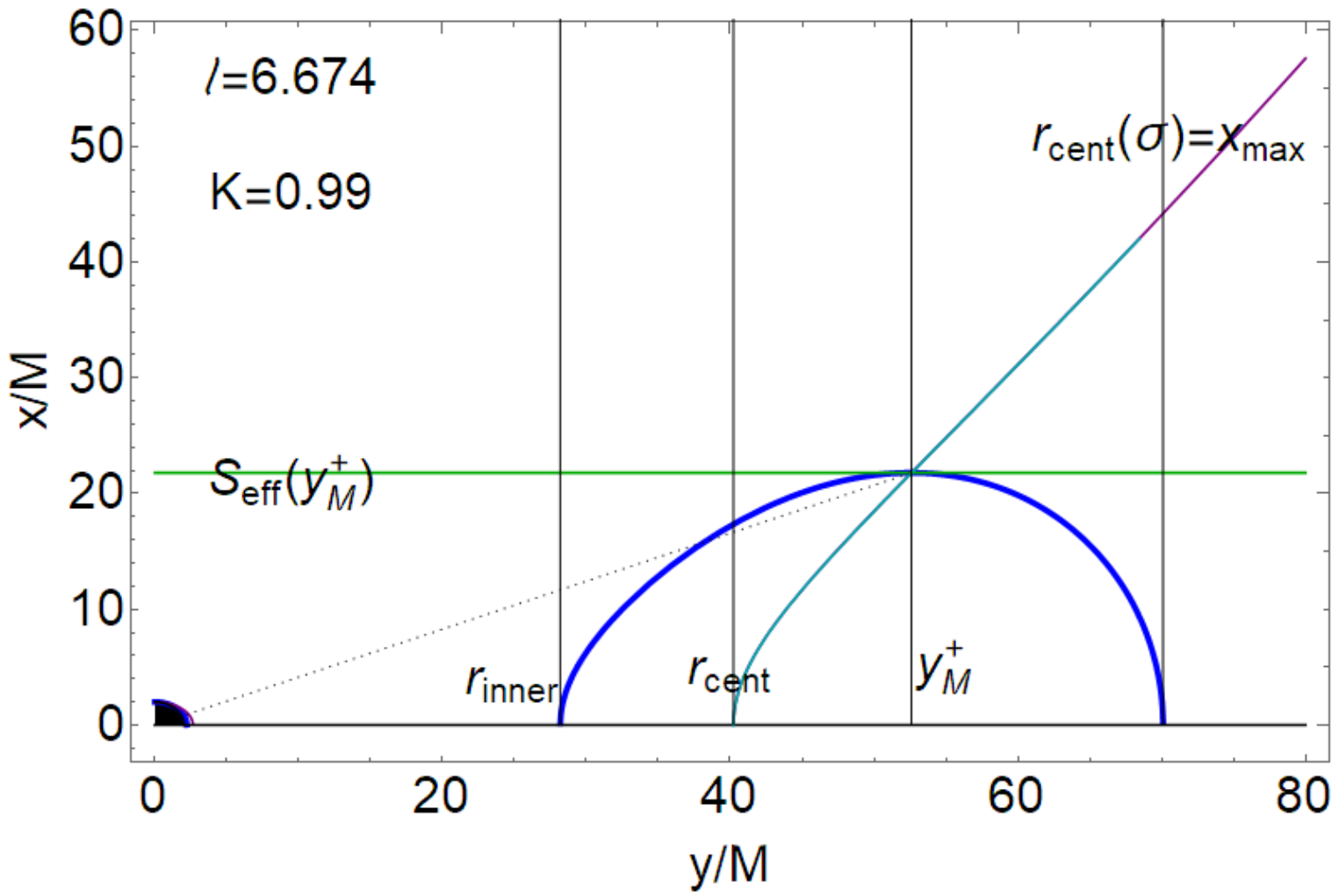}
        \includegraphics[width=7cm]{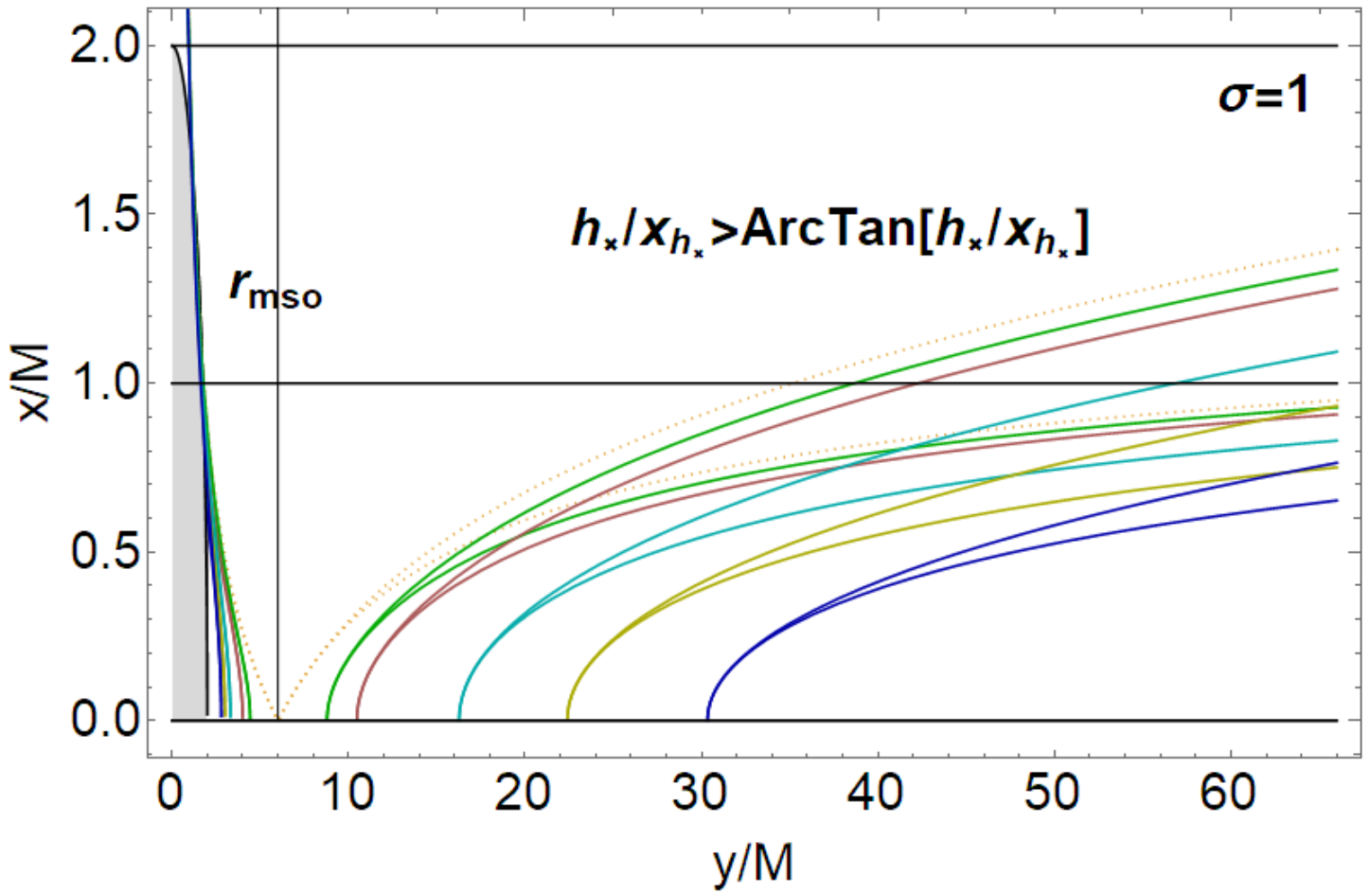}
                  \includegraphics[width=7cm]
                  {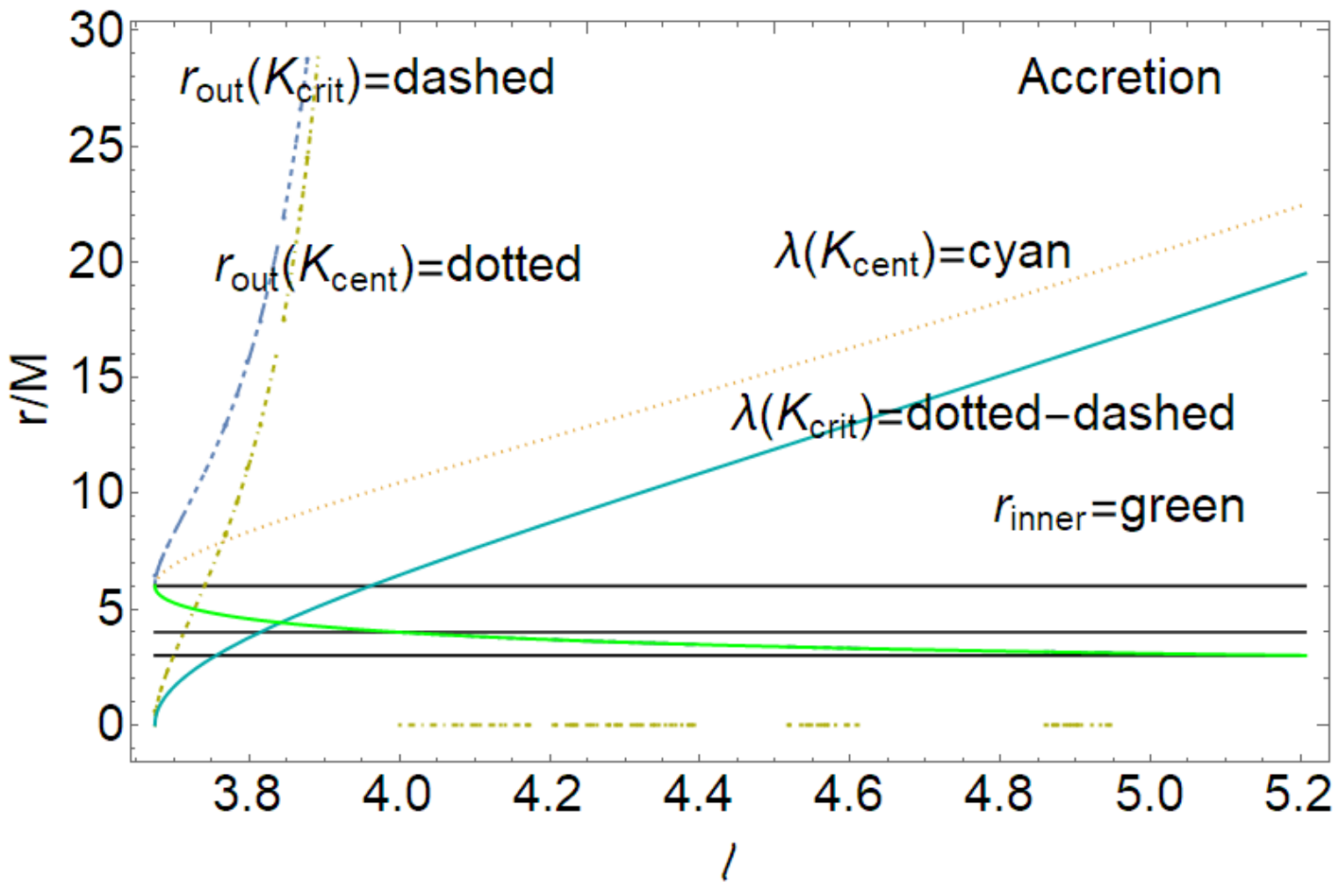}
    \includegraphics[width=7cm]{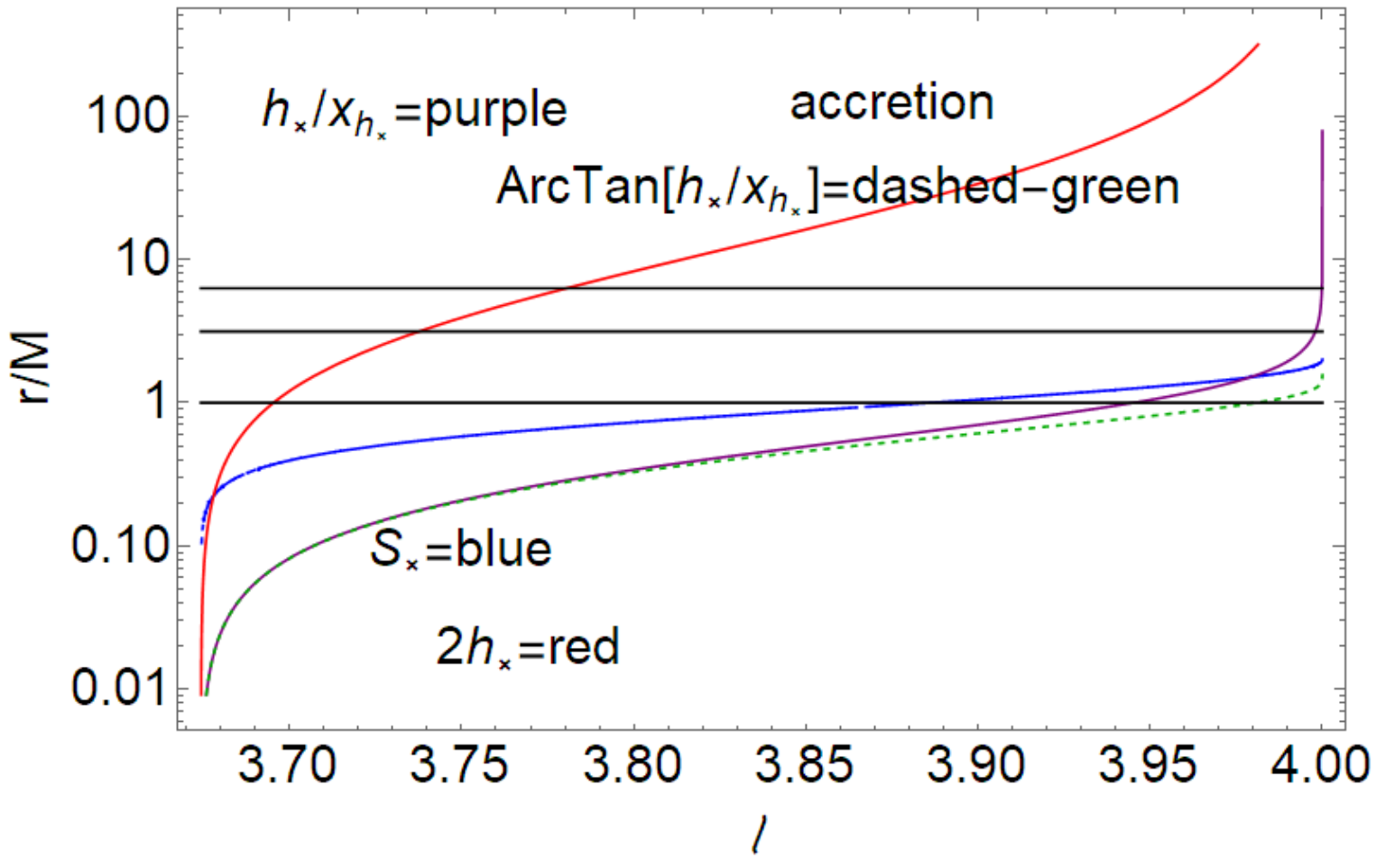}
       \end{center}
  \caption{Upper  left-hand panel: toroidal surface evaluated with the function $\Sa_{eff}$.
We term  the center of maximum pressure and density $r_{cent}$, the inner edge $r_{inner}$,
  and $x_{max}$ the curve $r_{cent}(\sigma)$, connecting the maximum pressure  and density point in the torus with  the torus geometrical    maximum (i.e. for $\sigma=1$ radius  $r_{cent}(\sigma)=r_{cent}$ is the center of maximum pressure, while $r_{cent}(\sigma)$ on the torus surface is  the torus geometrical maximum). We show  the point of geometrical  maximum
 $y_{M}^+$  (outer Roche lobe) and its value on the surface  $\Sa_{eff}(y_M^+)$. $\ell$  is the fluid specific angular momentum. Black  region  is the black hole.
Right  upper panel: $(\times)$ is for accreting (cusped) torus
$h$, is the maximum height of the cusped surface, $x_h$  is the point on the symmetry plane of the maximum height.
The plot is for different specific angular momenta $\ell\in\{\ell_{mso}, \ell_{mbo}, \ell_{\gamma}, (\ell_{mso} + (\ell_{mbo}-\ell_{mso})/2, (\ell_{mbo} + (\ell_{\gamma} - \ell_{mbo})/2), (\ell_{\gamma} +0 .7)\}$ represented as    dotted,
pink, yellow, green, cyan and  blue curve respectively.
Bottom left panel:
outer edge of the disk $r_{out}$  and elongation $\lambda$ evaluated in  $K_{cent}$ and $ K_{crit}$ as function of $\ell$.
Bottom right panel: ratio $h/x $ for cusped surfaces as functions of $\ell$; $S_\times$ is the torus thickness  where
$K^{\pm}_{crit}(\ell)$ is defined in   Eqs\il(\ref{Eq:suo-fr2}).
}\label{Fig:davmandb}
\end{figure*}
\begin{figure}
        \includegraphics[width=5.36cm]{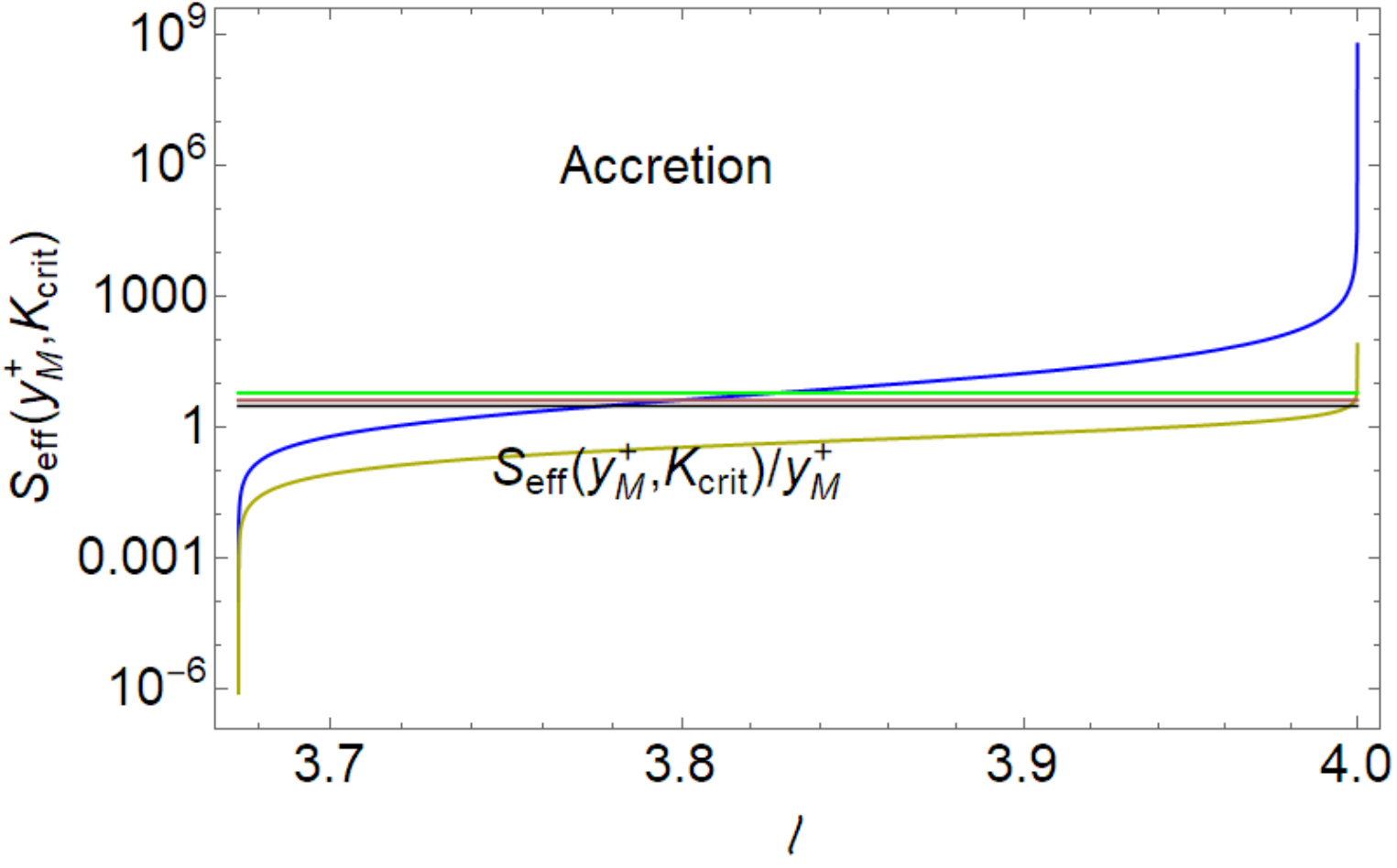}
                 \includegraphics[width=5.36cm]{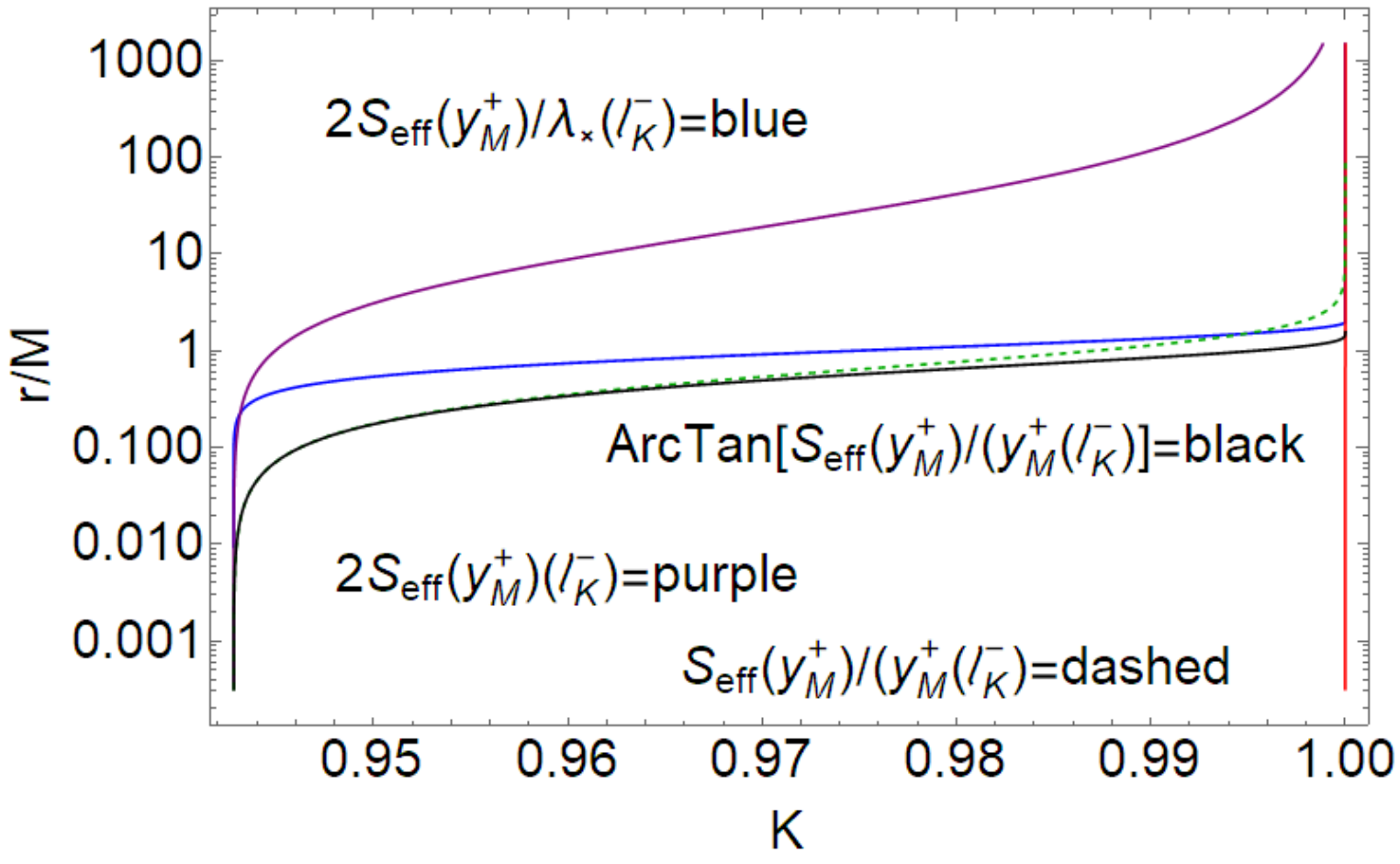}
     \includegraphics[width=5.36cm]{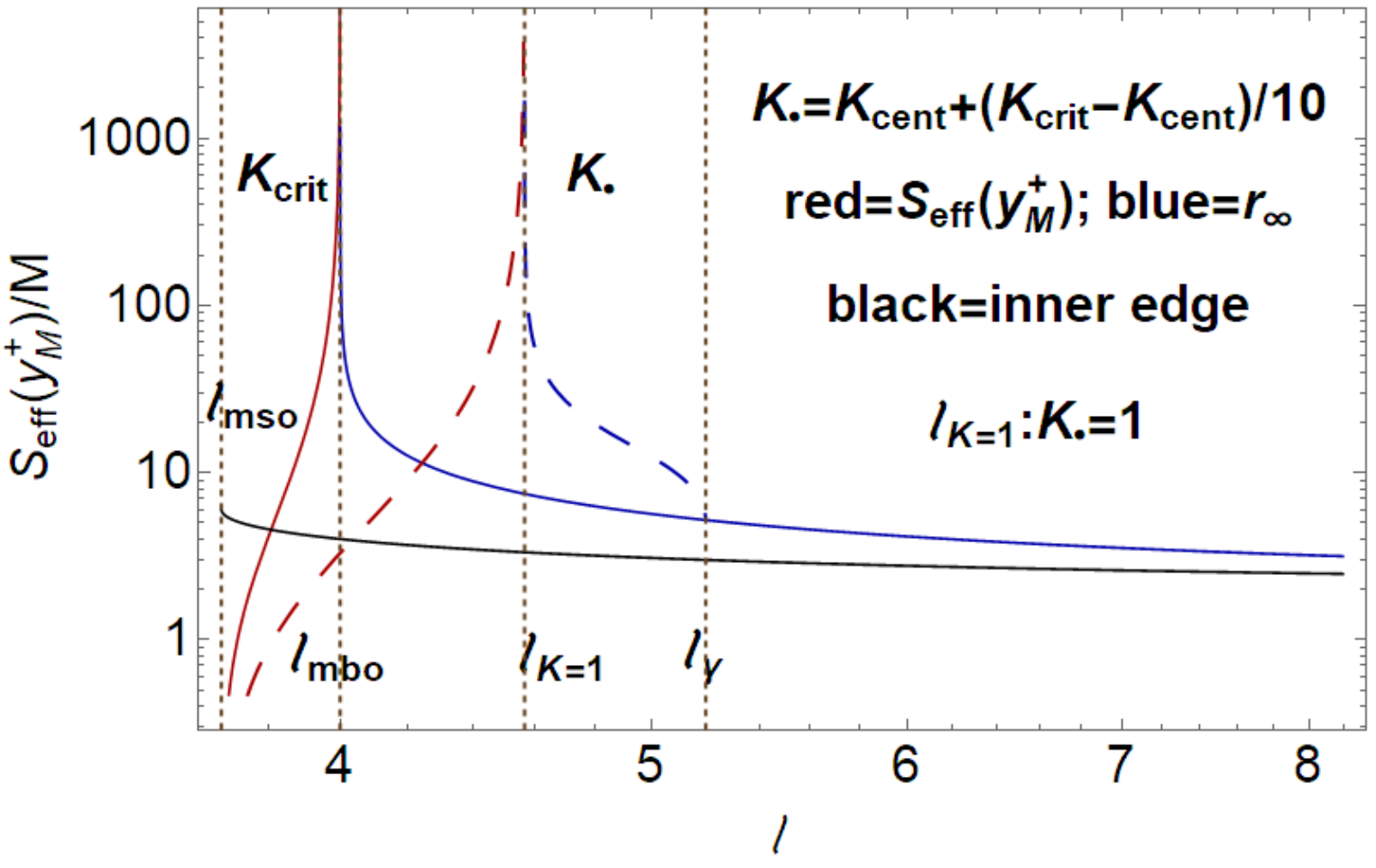}
  \caption{Evaluation of the collision angle
Left: Ratio of maximum of surface
$\Sa_{eff}(y_M^+)/y_M^+$ as function of  $\ell\in[\ell_{mso},\ell_{mbo}]$.
 Central panel:
maximum evaluated on the curve $\ell_K^{\pm}$ as function of  $K$ of Eq.\il(\ref{Eq:LKmeasuram}).
Right panel: $\Sa_{eff}(y_M^+)$ as function of  $\ell$ for different values of $K$. $y_M^+$ is the point of morphological maximum on  the tori symmetry plane. We  show some limiting  values of the fluid angular momentum for marginally stable orbit $\ell_{mso}$, marginally bounded orbit $\ell_{mbo}$,
and the limit $\ell:  K=1$.}\label{Fig:davmandx}
\end{figure}
\begin{figure*}
   \begin{center}
   \includegraphics[width=7cm]{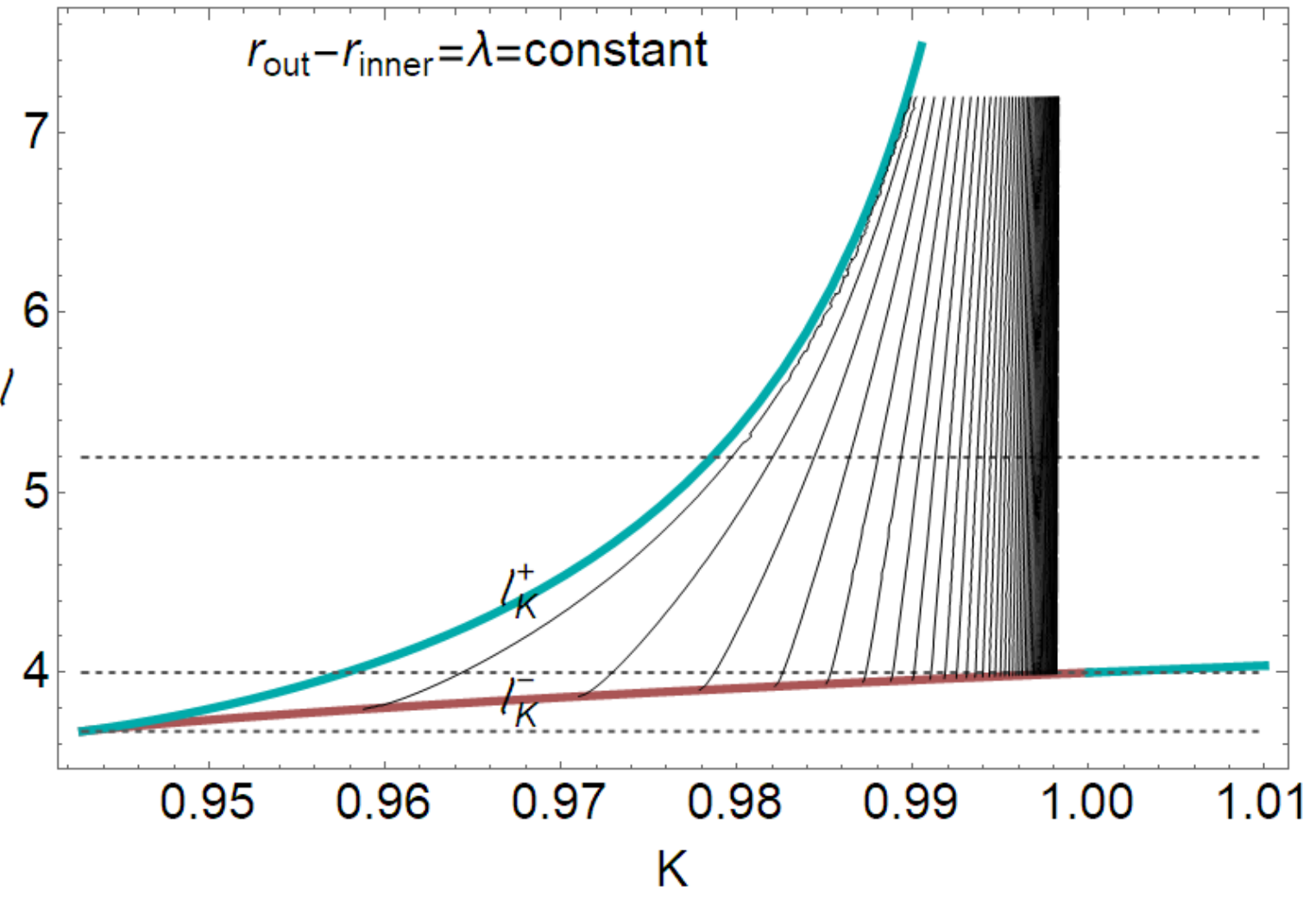}
     \includegraphics[width=7cm]{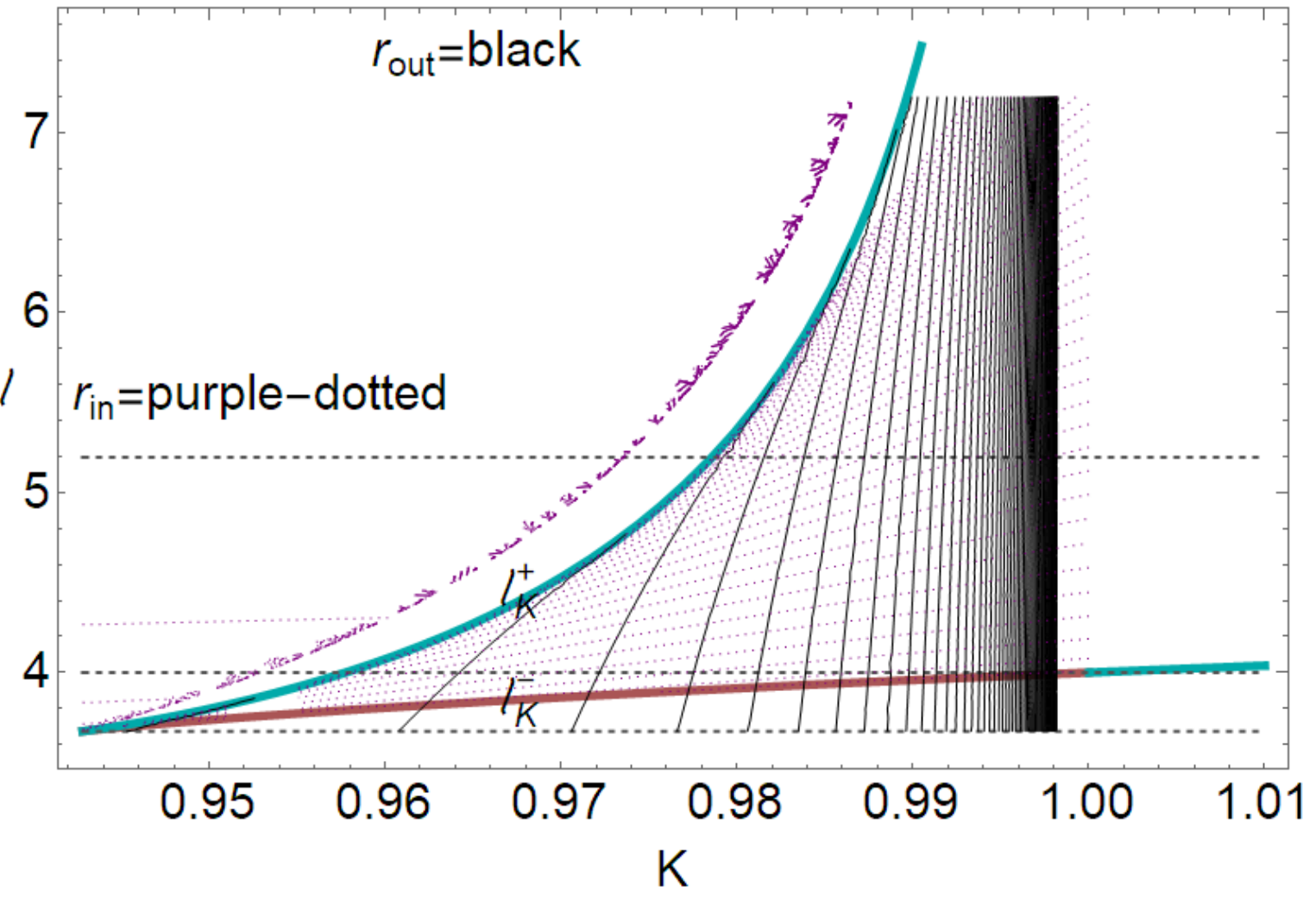}\\
  \includegraphics[width=5.6cm]{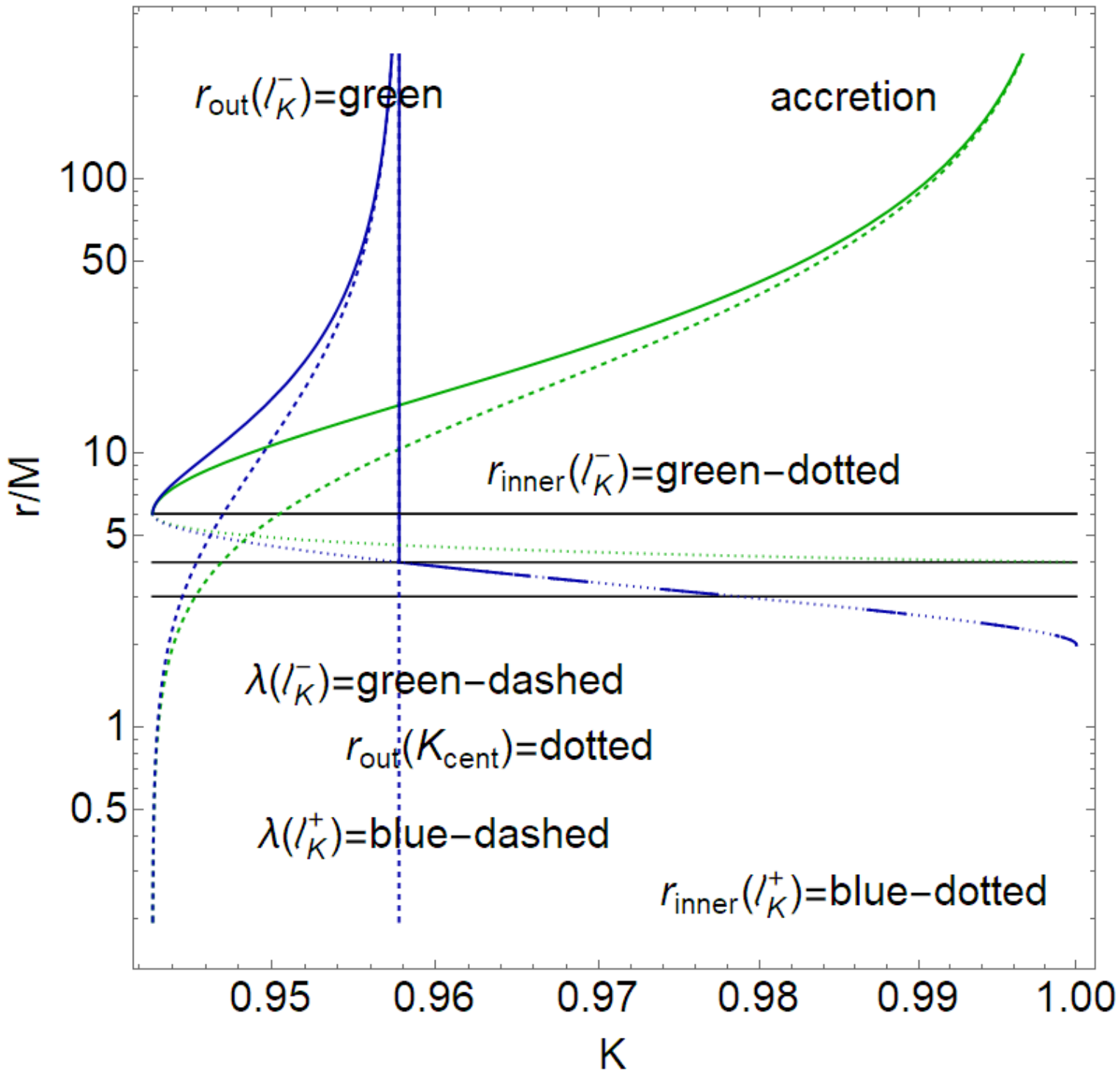}
         \includegraphics[width=5.6cm]{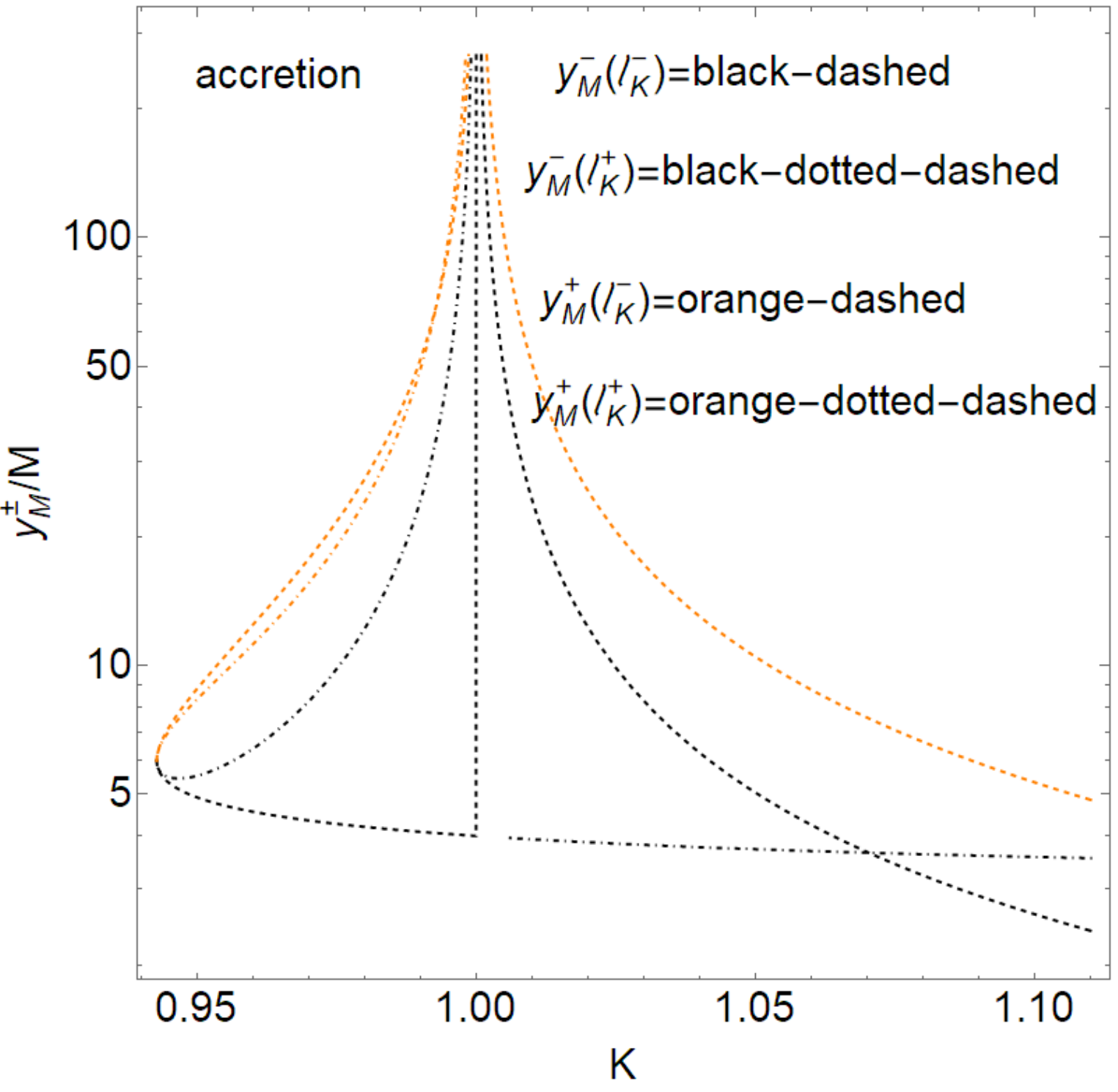}
         \end{center}
  \caption{Left upper panel: constant tori elongation on the  plane $\ell-K $. It is clear that the boundary curves are the critical curves $\ell_K^{\pm}$ in Eq.\il(\ref{Eq:LKmeasuram}).
Right upper panel:  curves of constant outer and inner edges.
Below panels. Left: outer and inner edges  of the disk as functions of  $K$ evaluated on $\ell_K^{\pm}$ and $K_{cent}$, for the  accretion  conditions are also shown.
Right panel: maximum point on the surfaces of the  tori evaluated on $\ell_K^\pm$ of the critical configurations;   the asymptotic $K=1$ is also shown.
}\label{Fig:davmandc}
\end{figure*}
The projections of the morphological maxima  on the equatorial plane are
\bea\label{Eq:spod-ta}
&&
y_M^\pm=\sqrt{\frac{3 K^2 Q\pm4 \sqrt{6} \left(K^2-1\right) \sqrt{-\frac{K^4 Q}{\left(K^2-1\right)^3}} \cos \beta_{\pm}}{3(K^2-1)}}, \eea
where
\bea
&&\nonumber\beta_+\equiv\frac{1}{3} \cos ^{-1}\hat{\beta},\quad\beta_-\equiv\frac{1}{3} \left(\cos ^{-1}\hat{\beta}+\pi \right),\quad
 Q=\ell^2\\&&\nonumber \mbox{and}\quad \hat{\beta}\equiv-\frac{3\sqrt{3}\left(K^2-1\right)^2}{4\sqrt{2}} \sqrt{\frac{K^4 Q}{\left(1-K^2\right)^3}},
\eea
see Fig.(\ref{Fig:davmandb}).
$ y_M^\pm$ is  the point of maximum, on the equatorial plane, of  the  external and internal Roche lobe respectively. This is obtained by using  function $\Sa_{eff}$ of Eq.\il(\ref{Eq:condizioni}) on a fixed equatorial plane (the center lies on the  $y$ axis), therefore
the morphological maximum  points   $s_{\max}$  \emph{on} the toroidal surfaces can be obtained calculating
\bea
\Sa_{eff}\equiv\sqrt{\left(\frac{2 \left(K^2 Q+x^2\right)}{K^2 \left(Q-x^2\right)+x^2}\right)^2-x^2}=y,
 \eea
 where here  there is accordingly  $x= y_{M}^{\pm}$ respectively. For the tori on planes others then the equatorial,  solutions  are rotated according  to Eq.\il(\ref{Eq:condizioni})--see Figs\il(\ref{Fig:eSnatwitda2},\ref{Fig:eSnatwitdpiw},\ref{Fig:davmand},\ref{Fig:davmandb},\ref{Fig:davmandx},\ref{Fig:davmandc}).

 The morphological maximum point for the outer Roche lobe,
 point $y_M^ +$,  the projection of the maximum morphological on the equatorial plane, the  maximum pressure point $ r_{cent}$,  and the  morphological maximum $x_{max} \equiv\Sa_{eff}(y_M^+)$  are shown in  Fig.(\ref{Fig:davmandb}).  (In this work we  focus attention on
the  outer lobus,  especially in the case of quiescent tori. The meaning of the inner configuration at equal $ \ell$ and $K$ embracing the \textbf{BH} needs to be explored in more detail, see for example\citet{pugtot}).

\begin{description}
\item[\textbf{--The inner edge, the center and the maximum}]

It is well known that the point of maximum density  and maximum (HD) pressure in the torus  $r_{cent}=r_{min}$ (the minimum of the effective potential of the fluid  as function of $r$) does not correspond to the morphological maximum point $s_{\max}$ of the torus surface, while this  can happen for the morphological minimum of the surface  $s_{\min}$ and the minimum of the pressure/density (which is $r_{\max}$, maximum point of the effective potential as function of $r$ if it exists).
However the two points, the minimum and maximum point of the toroidal surface and   $r_{cent}$ and $r_{inner}$, i.e.,  the maximum pressure point and center of disk and inner edge of disk (which corresponds to the minimum of pressure) are related. The two points of pressure and morphological  maximum  coincide when projected on the equatorial plane of the torus  in the sense explained  below, and therefore  the maximum $s_{\max}$ is directly given by   the angular distribution calculated on the equatorial planes, as evident from the  Figure\il(\ref{Fig:ampaerplot2k}) and (\ref{Fig:compePlottion}). This analysis fixes also the role of the radial gradient of the pressure in the tori in determining the torus verticality. Explicitly, the following transformations apply
\bea\label{Eq:miss-metrc-inenr}
&&
r_{inner}(\ell\rightarrow\ell/\sigma)=s_{\min},\quad r_{cent}(\ell\rightarrow\ell/\sigma)=s_{\max},\\&&\nonumber\mbox{where}\quad s_{\min}(x=0)=r_{inner}(\sigma=1)
\eea
or also $ r_{\max}=r_{cent}$---see Figs\il\ref{Fig:eSnatwitda2},\ref{Fig:eSnatwitdpiw},\ref{Fig:davmandb}-- where
\bea
&&\nonumber
r_{inner}(\ell)=\frac{1}{3} \left[\ell^2+2 \sqrt{\ell^2 \left(\ell^2-12\right)} \cos \left(\frac{1}{3} \cos ^{-1}\upsilon\upsilon\right)\right],
\\&&
 r_{cent}(\ell)=\frac{1}{3} \left[\ell^2-2 \sqrt{l^2 \left(\ell^2-12\right)} \cos \left(\frac{1}{3} \left(\cos ^{-1}\upsilon\upsilon+\pi \right)\right)\right],
  \\\nonumber
 &&\label{Eq:eellH}\mbox{and}\quad
s_{x_{\max}}= \frac{[(2 \ell^4 \left(8-3 y^2\right)+2^{2/3} \iota\iota\iota+
\frac{16 \sqrt[3]{2} \left(2 \ell^8+3 \ell^6 y^4\right)}{\iota\iota\iota})(\ell^{-4})]^{1/2}}{\sqrt{6}}
\eea
where
\bea\nonumber
&& \upsilon\upsilon\equiv\frac{\ell^2 \left[\ell^4-18 \ell^2+54\right]}{\left[\ell^2 \left(\ell^2-12\right)\right]^{3/2}};\\&&\nonumber
\iota\iota\iota\equiv\left[\ell^8(576 \ell^{2} y^4-128 \ell^{4}+27  y^8)+ 3 \sqrt{3} \sqrt{\ell^{16} y^4 \left(27 y^4-32 \ell^2\right) \left(y^4-16 \ell^2\right)^2}\right]^{1/3}
\eea
for $r_{cent}$ see Eq.\il(\ref{Eq:centrol})--Figs\il\ref{Fig:davmanda}.
$s_{x_{\max}}$ is the component  $x_{\max}$  found from   $r_{cent}(\ell)$  substituting $\ell\rightarrow\ell/\sigma$ where $\sigma ={y^2}/({x^2+y^2})$.
We note that the quantity  $\ell/\sigma$  is related to the frequency.
In fact the center is a point of curve $\ell(r)$ at $r>r_{mso}$,   for any  $K>K_{mso}$. The maximum of the surface  depend on $K$ and it can exist for  $K\leq1$, for  $\ell\in \mathbf{L_1}$ and  $K=K_{max}$.
These are related for the critical cusped configuration in the  implicit relation $\ell(K)$.
To clarify this point we report below the exact form
\bea
&&
s_{\max}=\frac{1}{3} \left[\frac{\ell^2 \left(x^2+y^2\right)^2}{y^4}+\iota\iota\cos \left(\frac{1}{3} \cos ^{-1}\varsigma\right)\right],
\\&&\nonumber
x_{inner}=\frac{1}{3} \left[\frac{\ell^2 \left(x^2+y^2\right)^2}{y^4}-\iota\iota\left(\cos ^{-1}\varsigma+\pi \right)\right],
\eea
where
\bea
&&
\iota\iota\equiv2 \sqrt{\frac{\ell^2 \left(x^2+y^2\right)^2 \left[\frac{\ell^2 \left(x^2+y^2\right)^2}{y^4}-12\right]}{y^4}},\\&&\nonumber
\varsigma \equiv \frac{8\ell^2 \left(x^2+y^2\right)^2 \left[\frac{\ell^4 \left(x^2+y^2\right)^4}{y^8}-\frac{18 \ell^2 \left(x^2+y^2\right)^2}{y^4}+54\right]}{y^4\iota\iota^3}
\eea
where $s_{\max}(\sigma=1)= r_{cent}$, $x_{inner}$ and $s_{\max}$ are solutions of $\ell(r,\sigma)=\ell$,   the morphological maximum is actually connected to the maximum of pressure and density.
Notably  this relation is independent from  $K$ but it holds, for each  $\ell$, for any  $K$, therefore it holds also for non-critical configurations.
Interestingly, this seems to prove that the distribution of specific angular momentum for the fluid has a predominant role  in the determination of the disk structure with respect to the effective potential function (values   $V_{eff}=K\in[K_{\min},K_{\max}]$). Therefore we  bounded the maxima and minima of pressure / density to the maxima and minima of the toroidal surface.
Then, we note that  $s_{\max}$ corresponds to  $y_{\max}$ (on each equatorial plane) and therefore  $x_{\max}$   can be obtained as solution of $s_{\max}(\ell,\sigma)=\sqrt{x^2+y^2}$.  It is clear that the $r_{cent}$ solves the problem $s_{\max}(\ell,\sigma)=\sqrt{x^2+y^2}$ for $y=0$. In  this context  $\sigma$  is related to  $K$.

\textbf{Maximum from the RAD rotational law}
We can prove that the analysis of the morphological maximum leads to the   an angular momentum distribution $\ell(r,\sigma)$ with explicit dependence on  $\sigma$:
\bea&&\nonumber
\ell(r,\sigma)=\frac{r^2 \sigma }{\sqrt{(r-2)^2 r}}=\\&&\nonumber\ell_{extre}(x,y)=\frac{y^2}{\sqrt{x^2 \left(\sqrt{x^2+y^2}-4\right)+y^2 \left(\sqrt{x^2+y^2}-4\right)+4 \sqrt{x^2+y^2}}},
  \\\label{Eq:recurr-exom}
  &&\mbox{where}\quad y=r \sqrt{\sigma },\quad x= r \sqrt{1-\sigma };
\eea
$\ell_{extre}(x,y)$ solves the problem $\partial_yV_{eff}(x,y)=0$ and there is $\ell_{extre}/\sigma = \ell (r)$---see Figs\il\ref{Fig:eSnatwitda2},\ref{Fig:eSnatwitdpiw}.
\begin{figure*}
 \begin{center}
  \includegraphics[width=5.3cm]{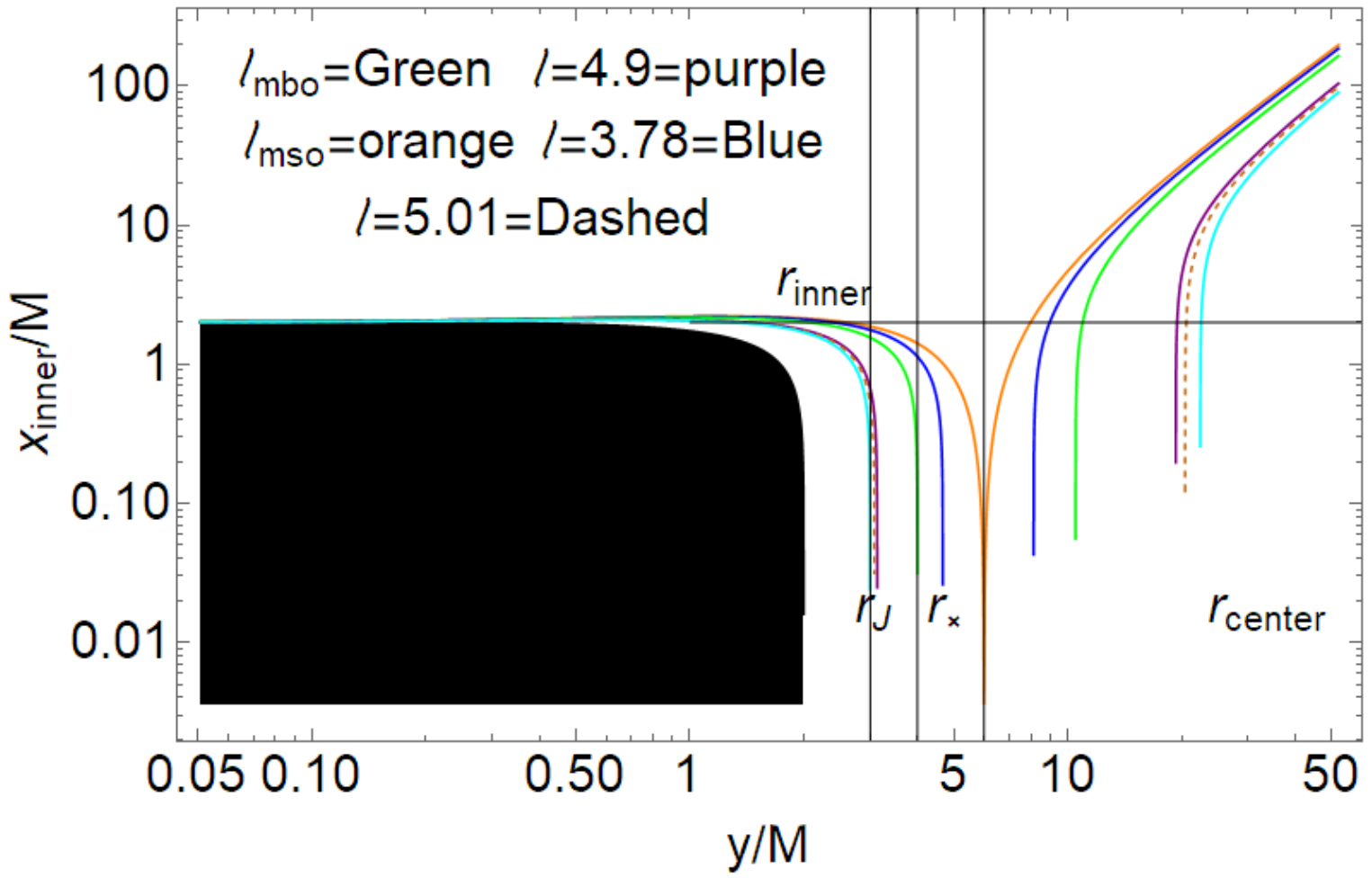}
   \includegraphics[width=5.3cm]{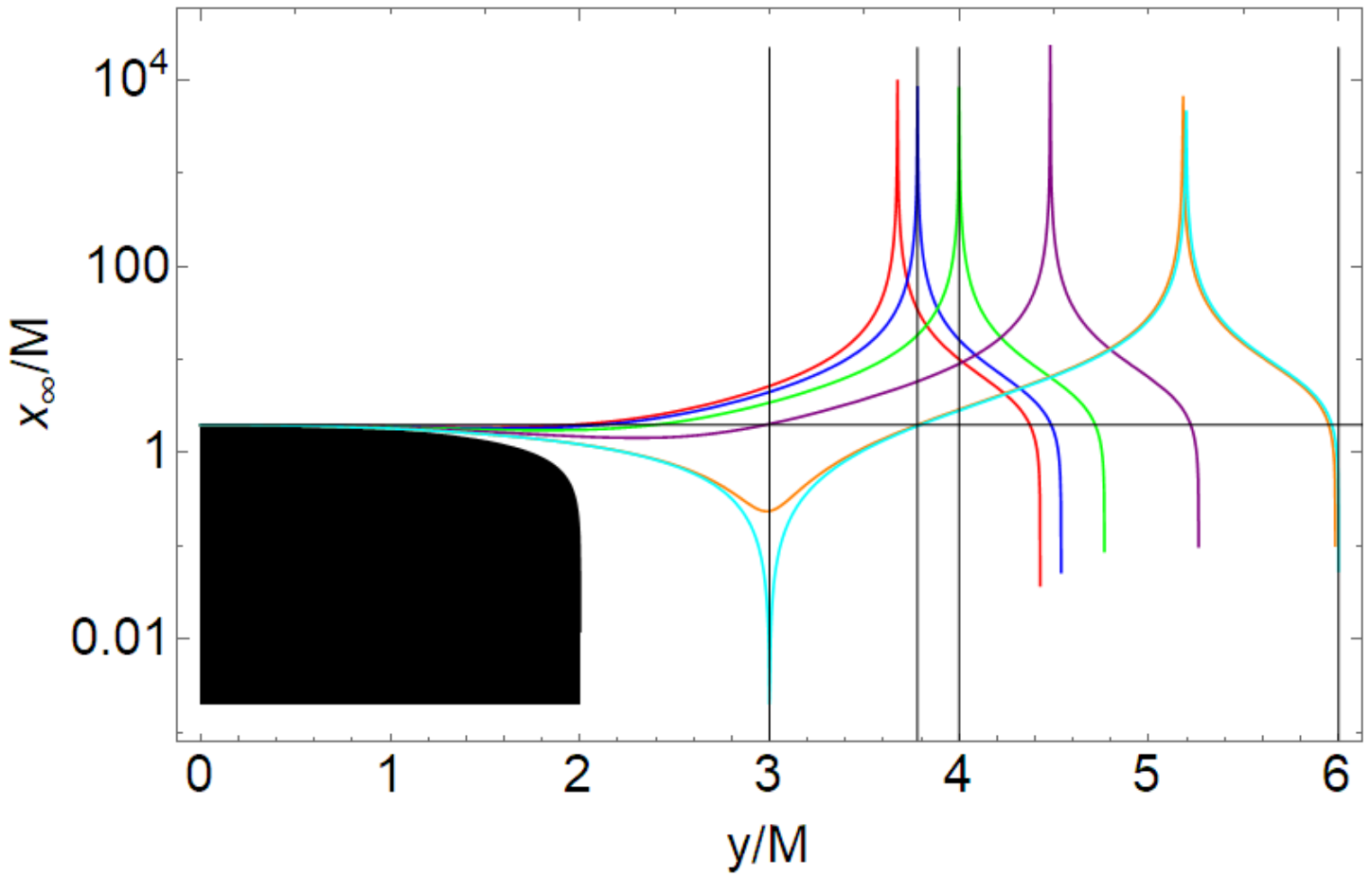}
    \includegraphics[width=5.3cm]{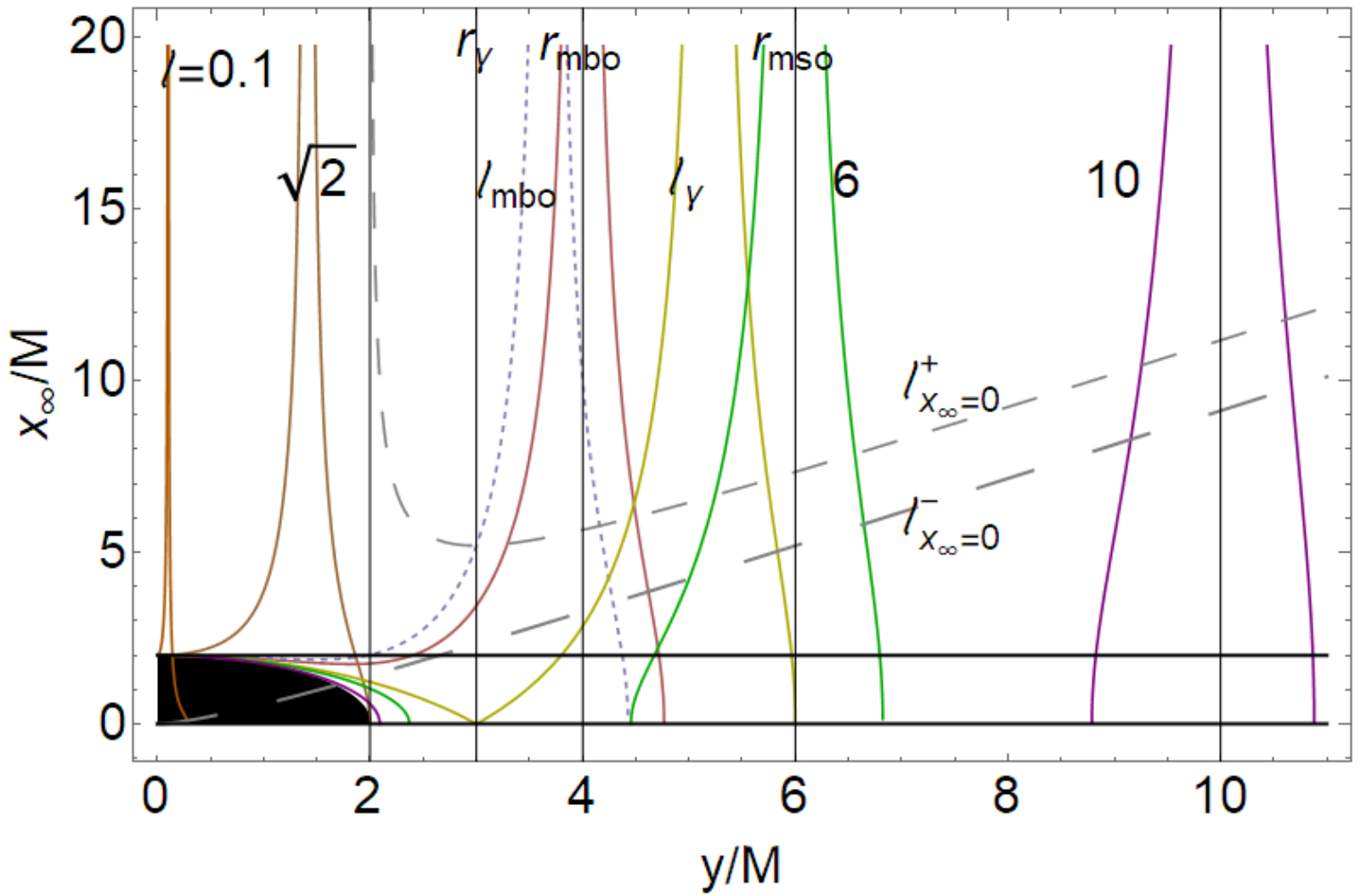}\\
   \includegraphics[width=5.7cm]{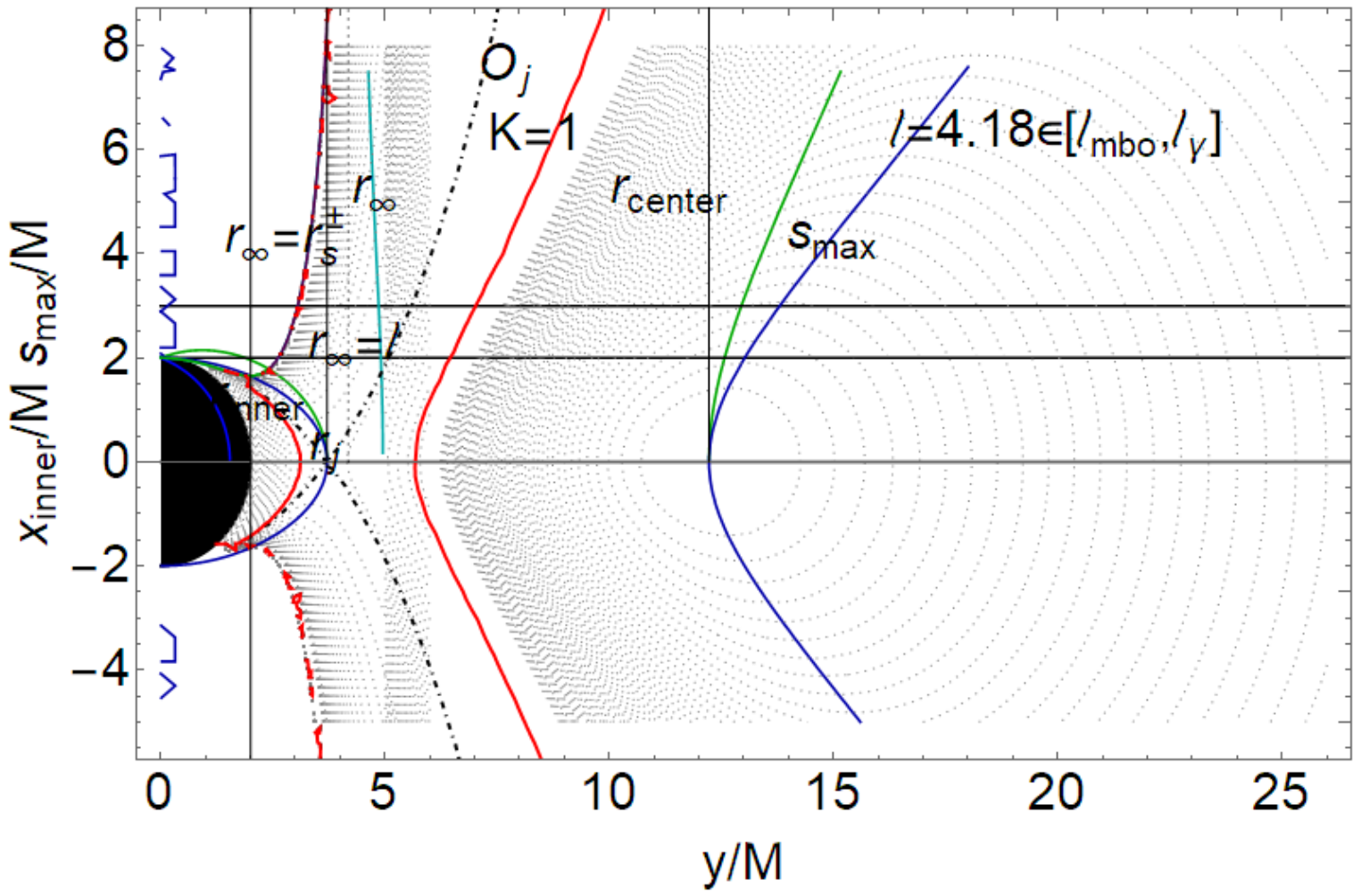}
 \includegraphics[width=5.7cm]{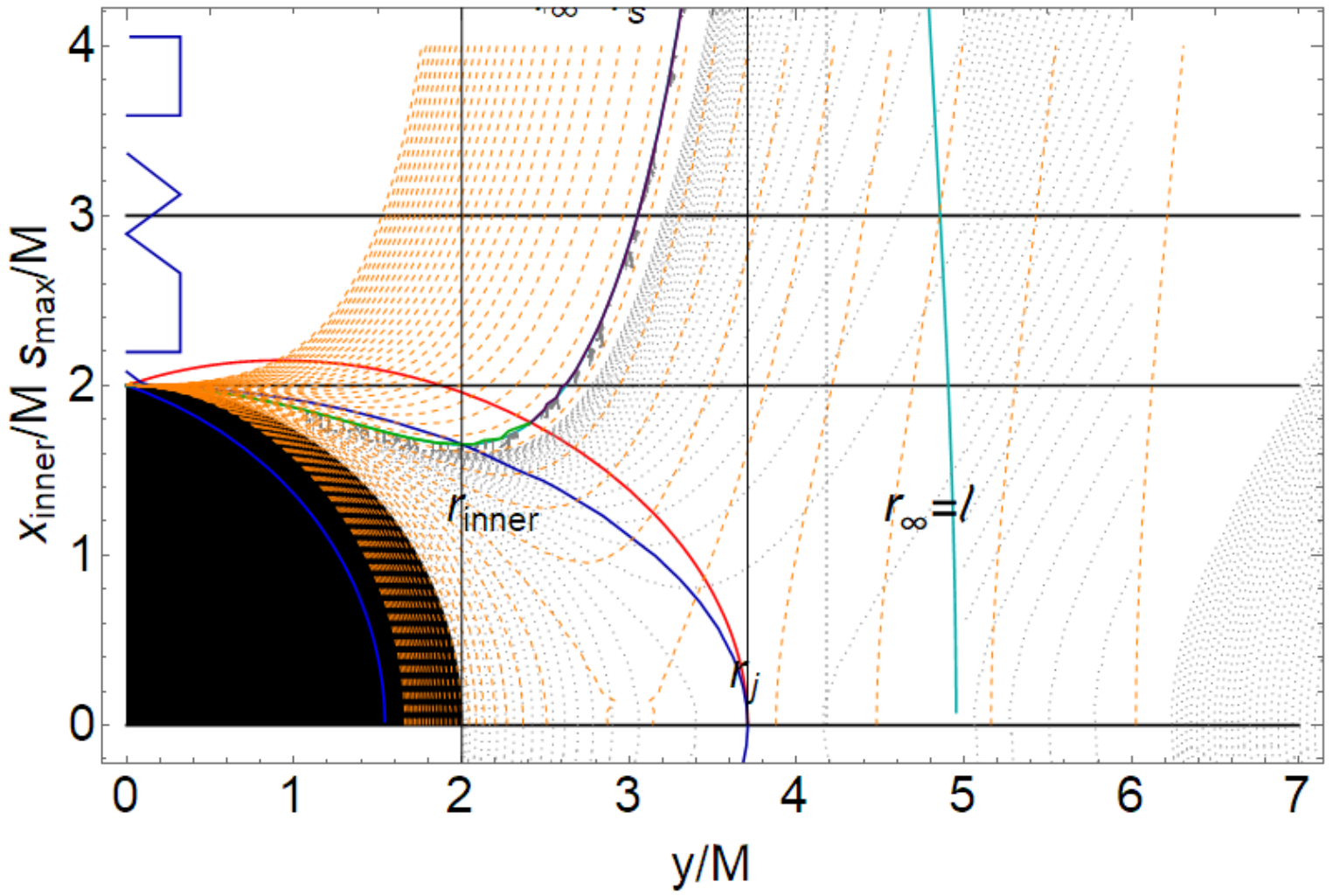}
 \end{center}
  \caption{Black region is the \textbf{BH}. Left upper panel: $x_{inner}$ of Eq.\il(\ref{Eq:xinner}) as function of $y/M$ for different   values of the specific angular momentum $\ell$. Center upper panel: limiting radius
$x_{\infty}$  Eq.\il(\ref{Eq:xansti})  for different values of the specific angular momentum $\ell$, colors are as right panel. Radii $r_{mso},r_{mbo},r_{\gamma}$ (marginally stable orbit, marginally bounded orbit and  photon circular orbit) are shown. Right upper  panel: a zoom. Solutions $\left(\ell _{x_{\infty }=0}\right){}^\pm$ for  $x_{\infty }=0$ are shown. Other solutions are
$\ell=\pm\sqrt{2}$ (with $y=2$). Bottom left panel:
$x_{inner}$ defined in Eq.\il(\ref{Eq:xinner}) providing  the center, and the cusp edge, see also Figs\il\ref{Fig:eSnatwitda1}. Dotted curves are configurations at various $K$. $x_{\infty}$ represent the limiting condition for these configurations. Bottom right  panel: a zoom including the
surfaces $\Omega/\ell$=constant (dashed-orange curves) related to von Zeipel curves.}\label{Fig:eurcorowsuperr}
\end{figure*}
Explicitly,  $r_{cent}$ and  $r_{inner}$,  locating the cusp  can be given as unique solution as follows:
\bea&&\label{Eq:xinner}
x_{inner}=\sqrt{\frac{2 y \left(\bar{\upsilon_{\bullet}}+2 y\right)}{\ell^2}+\frac{y^3 \left(\bar{\upsilon_{\bullet}}+y\right)}{2 \ell^4}-y^2+4},\quad \mbox{where}\quad \bar{\upsilon_{\bullet}}\equiv \sqrt{8 \ell^2+y^2}.
\eea
\\
\item[
\textbf{--Centers  and inner edges as functions of $ K$}]
It  is convenient to express the center and inner edge of critical configurations (the cusps)  explicitly in terms of the $K$ parameter
\bea&&\label{Eq:form-KK-rr-Kcoro}
r_{center}(K)=r^+_K\equiv\frac{3 K-\sqrt{K} \sqrt{9 K-8}-4}{2 (K-1)},\quad\mbox{and}\\&&\nonumber r_{cusp}(K)=r^-_K=\frac{3 K+\sqrt{K} \sqrt{9 K-8}-4}{2 (K-1)},
\eea
where
\bea\nonumber
&&
\mbox{for} \quad K\in[K_{mso},1]:\; r_{cusp}(K)=r_{\times}(K);\;  \mbox{for }  \; K>1:\; r_{cusp}(K)=r_{j}(K)
\eea
solutions of equation $K(r)=K$  which provides $(r_{center},r_{\times},r_j)$ as functions of $K$--see Fig.\il\ref{Fig:PRKRK1}. (The function $V_{eff}(r,\ell,\sigma)$ evaluated on  $\ell(r,\sigma)$ provides
$K(r,\ell(r,\sigma),\sigma)=K(r)$ which is independent from  $\sigma$ and therefore we cannot use  the \textbf{RAD} energy function to directly provide limits on the morphological  maximum of the surface.). Note  we can  use function $\ell(K)$ in Eq.\il(\ref{Eq:LKmeasuram},\ref{Eq:miss-metrc-inenr},\ref{Eq:eellH},\ref{Eq:xinner}).
\begin{figure}
 \begin{center}
  \includegraphics[width=8cm]{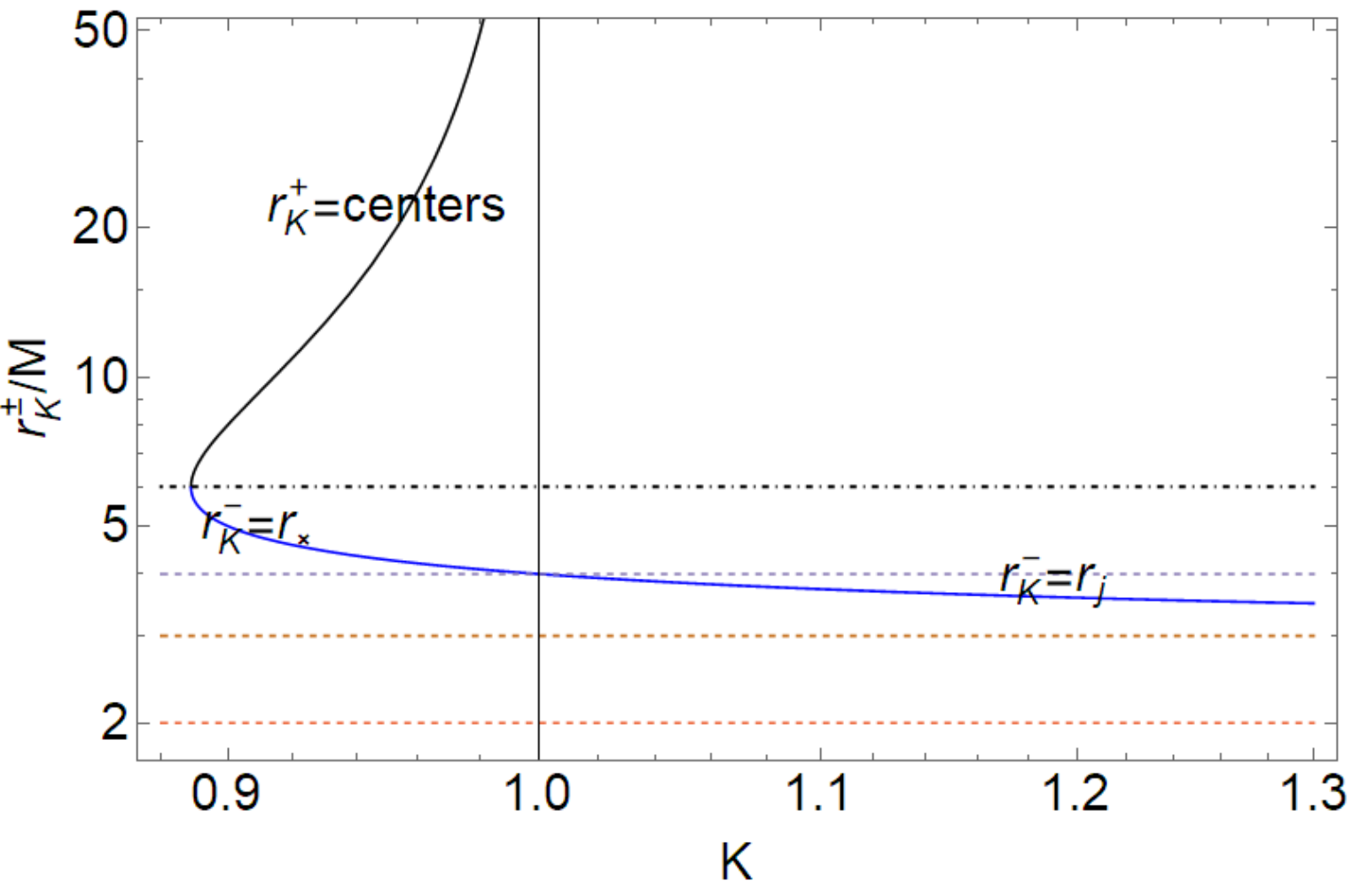}
  \end{center}
  \caption{Points of maximum pressure and density in the disk $r_{center}(K)=r^+_K$ (disk center) and  minnimum points of pressure and density $ r_{cusp}(K)=r^-_K$, ($r_{cusp}(K)=r_{\times}(K) ,r_{cusp}(K)=r_{j}(K)$) of Eq.\il(\ref{Eq:form-KK-rr-Kcoro}). Radius $r_{\times}$ is the cusp of closed tori, $r_j$ is the cusp of open configurations (proto-jets). }\label{Fig:PRKRK1}
\end{figure}
\end{description}
Below we list some limiting surfaces   constraining the proto-jets emission  considered above: the von Zeipel surfaces, the radius $r_{\infty}$, the light-surfaces, the surfaces derived from the normalization conditions and the surfaces at $K=1$.

\textbf{The limiting conditions}

For the limiting conditions, from   Figs\il\ref{Fig:eurcorowsuperr}  it is clear that we have to consider the three regions bounded by $K=1$ and $r_j$ where the surfaces are open,  or  from $r_j$ to $r_{\infty}$ (which is the stationary surface).
The results of this analysis are shown in Figs\il\ref{Fig:eSnatwitda2},\ref{Fig:eSnatwitdpiw},\ref{Fig:eurcorowsuperr},\ref{Fig:davmandb},
\ref{Fig:davmanda},\ref{Fig:davmand}.
These surfaces also include the  von Zeipel surfaces
  role as limiting conditions for jets--
Eq.\il(\ref{Eq:it-impo-ger1}).

\medskip

\begin{figure}
 \begin{center}
  \includegraphics[width=7cm]{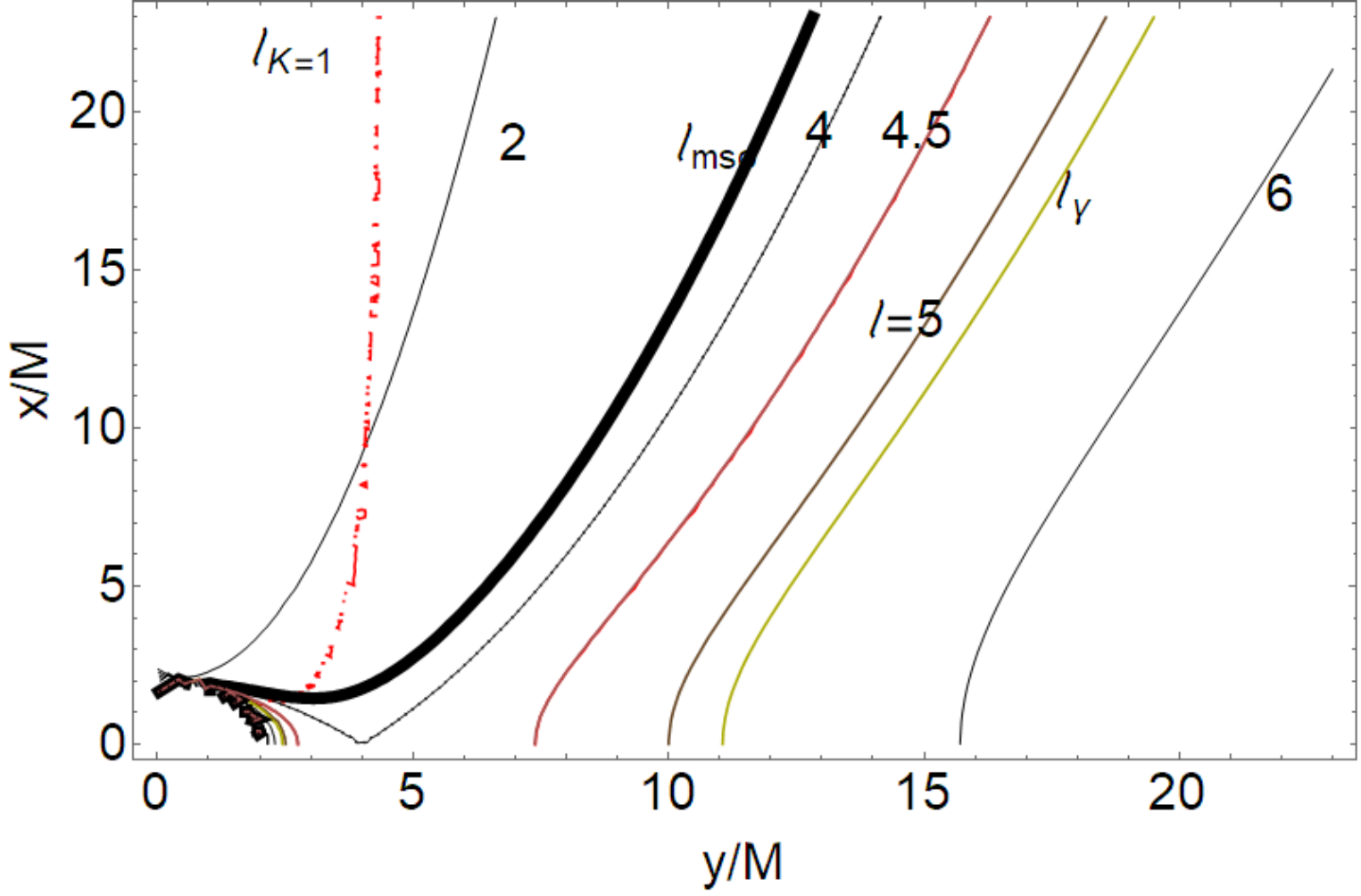}
   \includegraphics[width=6cm]{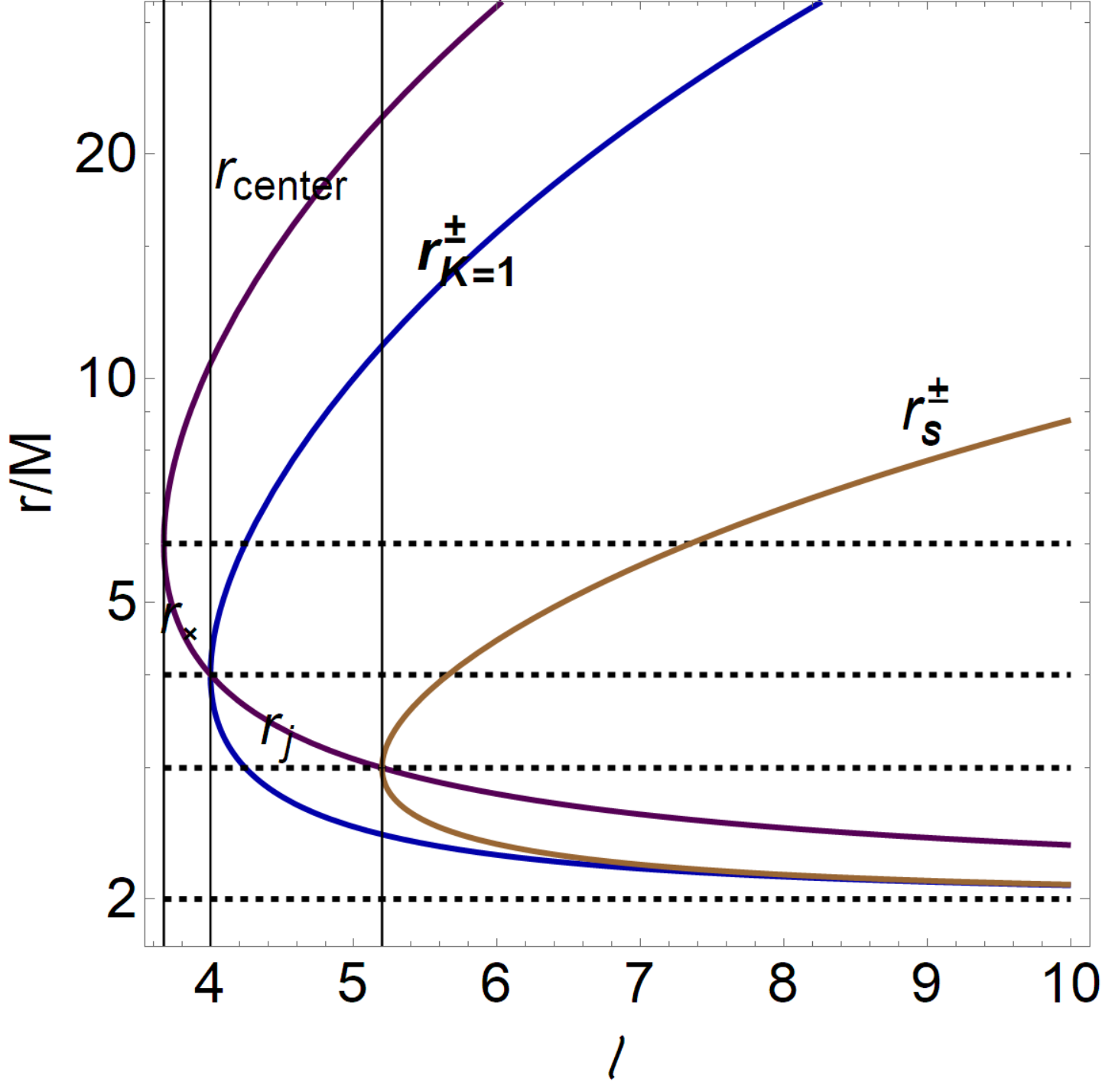}
   \end{center}
  \caption{Left panel: $\ell_{K=1}=constant$ of Eq.\il(\ref{Eq:ga-elect-risp}), in the plane $x--y$, values of the  specific angular momentum of the fluid : $\ell_{K=1}=\ell=$constant are signed on the curves. We note the limiting values of the $\ell_{mbo}$ and $\ell_{mso}$.
Right  panel:  radii $ r_{center}$  of maximum pressure inside the disk as function of the specific momentum, radii
 $r^{\pm}_{K=1}$ of Eq.\il(\ref{Eq:ga-elect-risp}), defined by condition $K=1$, and radii $r_{s}^{\pm}$ limiting radii for stationary observers, which are the light surfaces--here as functions of the angular momentum (instead of photon orbital frequencies $\omega$).  Vertical black and dotted lines mark the ranges  of $(r,\ell)$ where accretion points $r_{\times}$ and proto-jets cusps $r_j$ can be located.}\label{Fig:Plotfincostnti}
\end{figure}

\begin{itemize}
\item\textbf{--The radius $r_{\infty}$}
From the normalization condition on the fluid four velocity:
 using
Eq.\il(\ref{Eq:occ-ref-g})  we have $
\mathcal{B}_{\infty}(r,\theta)\equiv{K \ell}/{\sqrt{K^2-1}}$, ($\mathcal{B}=r \quad (\theta=\pi/2)$)
which can be reduced to
$
r_{\infty}=\mathcal{B}_{\infty}(r,\theta) $,  the limiting $r=\ell$  ($\mathcal{B}=\ell$), and
\bea&&\label{Eq:xansti}
x_{\infty}=\frac{\sqrt{\ell^4 \left(4-y^2\right)+ y^4(2\ell^2-y^2)}}{\sqrt{\left(\ell^2-y^2\right)^2}},\\&&\nonumber  x_{\infty}=0: \quad  y_{x_{\infty}=0}^\pm\equiv\frac{y^{3/2}}{\sqrt{y\mp2}},  \quad \ell=\sqrt{2}, \quad  y=r_+
\eea
shown in Figs\il\ref{Fig:eurcorowsuperr}, we note the limiting value  $x=\ell$.
\\
\item\textbf{--Light-surfaces }
of Eq.\il(\ref{Eq:light-rspone})
\bea&& \nonumber r_{s}^{\pm}(w)\equiv\pm
\frac{2 \sqrt{\frac{1}{w ^2}} \cos\epsilon_{\pm} }{\sqrt{3}},\quad\mbox{where}\\&&\nonumber \epsilon_+\equiv \frac{\hat{\varepsilon}}{3}, \quad \epsilon_-\equiv \frac{1}{3} \left(\hat{\varepsilon}+\pi \right),\quad \mbox{and}\quad \hat{\varepsilon}\equiv \cos^{-1}\left(-\frac{3 \sqrt{3}}{\sqrt{\frac{1}{w ^2}}}\right)
\eea
(we can consider also the substitution  $\omega\rightarrow1/\ell$, expressing  the light surfaces in terms of the specific fluid angular momentum).
\item\textbf{--Normalization conditions}
From the normalization condition on the fluid four velocity (constraining the stream by the causal structure) we obtain the quantity $\textbf{L2S}$. Considering Eq.\il(\ref{Eq:lsch-spaz}) there is
$ {\ell_{Sch}}/{\sqrt{\sigma}}=\sqrt{{L2_d}/{\sigma}}={r^{3/2}}/{\sqrt{r-2}}$, where
$ \ell_{Sch}={1}/{\omega_{Sch}}
$. Therefore the limiting value is
 $\ell_{\gamma}=1/\omega_{\gamma}=\sqrt{27}$, but frequency $ \omega_{Sch}(r)$
  provides  limiting conditions
 $r_{\wp}(\ell): \ell=\ell_{Sch}$, solving  also $\Delta (V_{eff})=0$  for a generic  $\ell$.
 {Surfaces   $r_{\wp}(\ell)$ and  $r_{s}^{\pm}$  are related, as it is  $r_{\wp}(\ell)=r_{s}^{\pm}(\ell=1/\omega)$.}
\end{itemize}
Among these surfaces we also use
condition  $K^2=1$  implying  the orbits and momenta
\bea\label{Eq:ga-elect-risp}
r^{\pm}_{K=1}(\ell)=\frac{\ell^2(1\pm\sqrt{\ell^2-16 \sigma })}{4 \sigma },\quad\mbox{and}\quad
\ell_{K=1}=\frac{ \sqrt{2\sigma }r}{\sqrt{r-2}}
\eea
$\ell_{K=1}$  is shown in Figs\il\ref{Fig:eSnatwitdpiw}--Fig.\il\ref{Fig:Plotfincostnti}-clearly there is a critical point in $ r_{K=1}^{\pm}=4$,  for $\ell=\ell_{mbo}$.
Some of these limiting surfaces are related to the  HD  structures, solutions of the Euler equations for the problem, others are more strictly related to  the  geometrical constraints provided by the causal structure.

As clear from Figs\il\ref{Fig:compePlottion},\ref{Fig:eSnatwitda2},\ref{Fig:eSnatwitdpiw}, we  can identify a  region around the  rotational axis of the toroidal configurations where there are no solutions of the Euler equation, to represent matter funnel constraints, being  therefore   darker regions--Figs\il(\ref{Fig:compePlottion}).  The extension of this region, developed along two boundaries estimated as  $2 r_{\infty}\leq 2 \ell$. The configurations very close (embracing) the horizon are in the range $]r_j, r_1[$ ($r_1$ is defined in Eq.\il(\ref{Eq:nudisigma})). Region to be considered  has extension  by $2 r_{\infty}$, implying the condition
$2h_{\max}\in[r_{j}, r_{\infty}]$ (where $h_{\max}$ is the  torus height  at its  morphological maximum  point),
 where
 both $r_{\infty}$ and $ r_j$, defining a further limit, depend on the  fluid angular momentum.
Boundaries of darker regions are, asymptotically, collimated to  $r_{\infty}$ with   increasing $K$, while the center of maximum pressure for these configurations moves outwardly.
	 In this context we can further reduce the   darker region, for example in Figs\il\ref{Fig:compePlottion},\ref{Fig:eSnatwitda2},\ref{Fig:eSnatwitdpiw},  to
a region bounded by the surfaces
	  $2 s_{\max}< r_{\infty}$.
To evaluate collimation conditions, we fix an open or closed  solution
and proceed to study  the intersection with a second  configuration  under particular conditions.

 \medskip

 \textbf{On the polar gradient}

 \medskip

\begin{figure}
 \begin{center}
   \includegraphics[width=5.3cm]{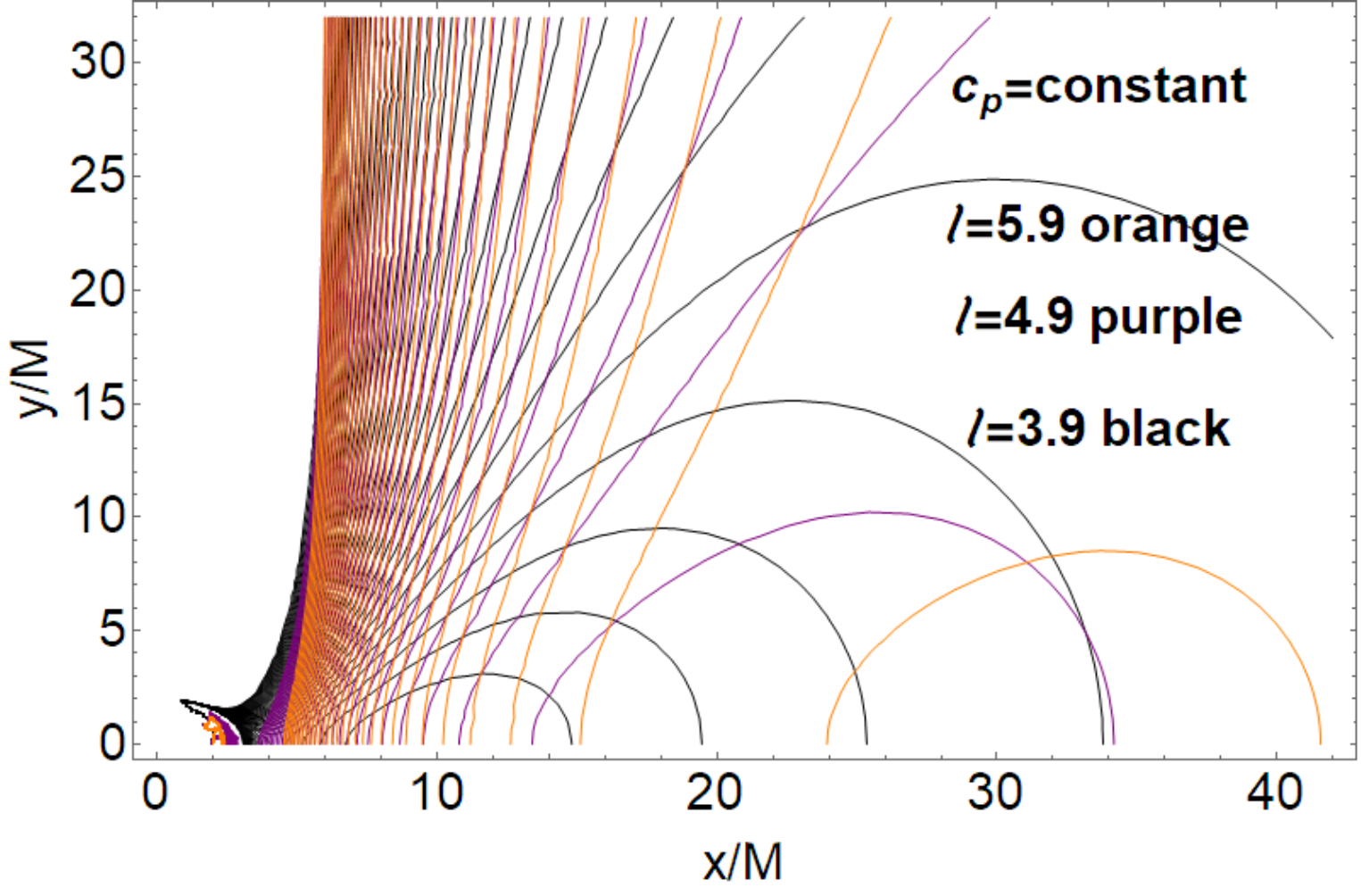}
              \includegraphics[width=5.3cm]{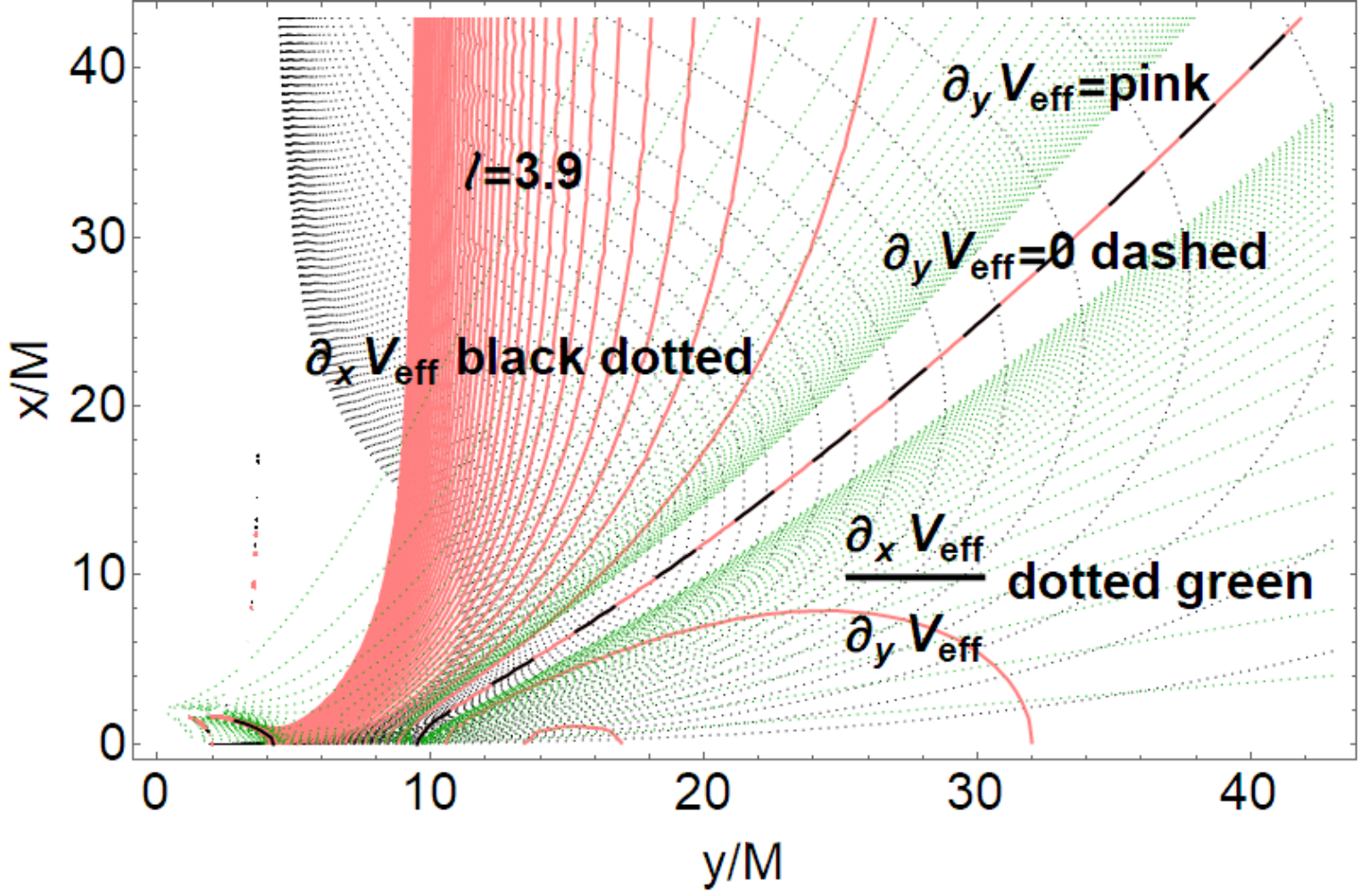}
        \includegraphics[width=5.3cm]{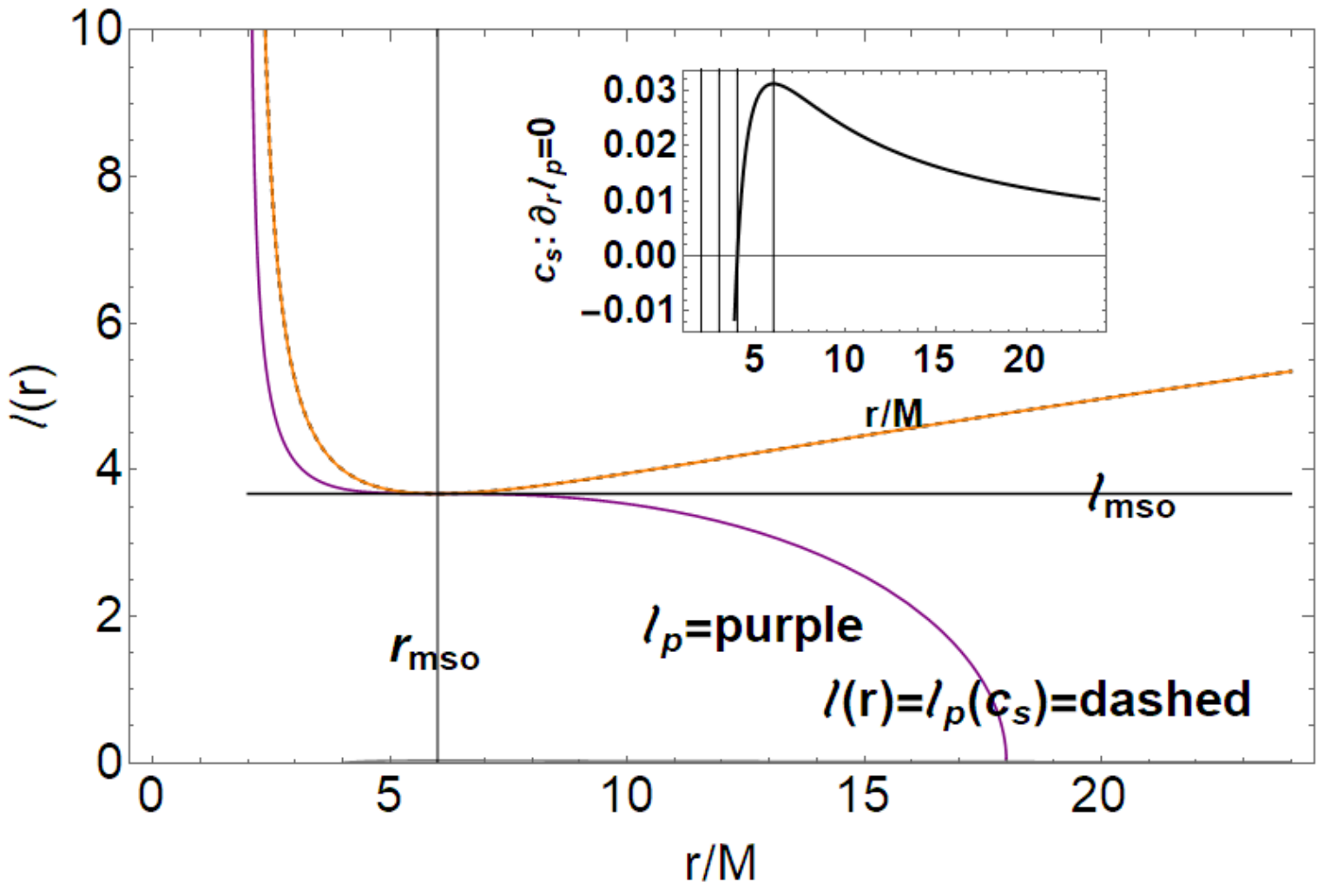}
        \end{center}
                \caption{Analysis  of the pressure polar gradients  and study of the  tori verticality  of Sec.\il(\ref{Sec:polar-gradient}).
Left panel:   solutions  $c_p=$constant of
Eq.\il(\ref{Eq:bod-orton}), for  different  values of fluid specific angular momentum $\ell$. There are closed, cusped and open surfaces.
 (There is $r=\sqrt{x^2+y^2}$, $\sin\theta= {x}/{\sqrt{x^2+y^2}}$)
Center panel: curves of constant gradients of  effective potential  for   $\ell=3.9$. (There is $r=\sqrt{x^2+y^2}$, $\sin\theta= {y}/{\sqrt{x^2+y^2}}$).
Right panel:  specific fluid angular momentum $\ell_p$ and $\ell_p(c_s)$ (the  \textbf{RAD} rotational law) as functions of the radius $r$--see Eqs\il(\ref{Eq:conf-ce1}),(\ref{Eq:conf-ce2}).
Inside panel: function $c_p$ of Eq.\il(\ref{Eq:bod-orton}) providing the   zeros of the radial gradient of the function $\ell_p(r)$. }\label{Fig:Plotcomdensite}
\end{figure}

The polar and radial gradients of the effective potential  $V_{eff}$ are related to the pressure gradients due to the Euler equations. The gradients are  studied in Figs\il\ref{Fig:Plotcomdensite}. The effective potential gradients ratio   coincides with the ratio of the pressure  gradients and related  density gradient  ratios in the disk. We can study the integrals of related differential equations   investigating more closely  the equi-pressures surfaces. This investigation shows the role of  polar and radial  gradients in systems  with toroidal symmetries. Furthermore,  integration of the partial differential equations shows the role  of the  gradients in the determination of the geometric thickness of the torus. The analysis proves also that   the  rotational law  of the \textbf{RAD} can be derived  from the radial gradient and the radial
 gradient determines also   the points of maximum pressure density in the disk (the center)
  connected to  the (morphological) maximum  point of their external Roche lobe.
The solution of the partial differential equation of the first order for the pressure, in the ratio of the gradients,
 provides the function
\bea\label{Eq:bod-orton}&&
p(r,\theta )=c_1(\tilde{f}(r,\theta )),\quad\mbox{where}\quad \tilde{f}(r,\theta )\equiv c_p=\frac{\ell^2 \csc ^2(\theta)(2- r)+2 r^2}{4 (r-2) r^2},
\eea
solutions for the ratio of the pressure gradients. We take   $p(r,\theta )=\tilde{f}(r,\theta)\equiv c=$constant, as in Figs\il\ref{Fig:Plotcomdensite}.
We have already seen how the  disk verticality is  determined by its radial gradient, derived by its rotational law $\ell(r)$.
By using the parametrization introduced in  Sec.\il(\ref{Sec:replics}) on the  metric Killing bundles  we solve the problem for
\bea&&\label{Eq:conf-ce1}
\ell_{sol}\equiv\frac{\ell}{\sin (\theta )}=\sqrt{2} \sqrt{r^2 \left(\frac{1}{r-2}-2 c\right)}.
\eea
However to fix the class of solutions,  for parameter $c$ to get the toroidal surfaces,  the constant  $c$ has to be properly chosen.  For this purpose we  consider the radial gradient of the function $\ell_{sol}$ and solving  the  problem of its zeros we find a function  $c_{sol}$  which provides the correct rotational law $\ell(r)$
\bea&&\label{Eq:conf-ce2}
c_{sol}\equiv \frac{r-4}{4 (r-2)^2}:\quad \partial_r\ell_{sol}=0;\quad  \ell_{sol}(c_{sol})=\ell(r)=\sqrt{\frac{r^3}{(r-2)^2}},\\&&\nonumber  c_{sol}(r_{mso})=\frac{1}{32},
\quad c_{sol}(r_{\gamma})=-\frac{1}{4},
\eea
  we note that this can be also written immediately using  Eqs\il(\ref{Eq:path-omegas-pressure}) in terms of epicyclic frequencies,  interpreted therefore in terms of oscillation frequencies. It is worth noting however,
as also clear from Figs\il\ref{Fig:Plotcomdensite}, that  $r_{mso}$ is a maximum of the function which is actually negative at $r>r_{mbo}$, null at $r_{mbo}$, and the horizon is an asymptote.
\section{Collisions and intersections}\label{Sec:cross-interaction}
In this section we explore the intersections between the toridal  surfaces in different topologies and  with fluid specific angular momentum in  the range $\mathbf{L_1},  \mathbf{L_2}$ or $\mathbf{L_3}$, and the open surfaces.  In particular we consider the open surfaces with cusps and the limiting surfaces  $r_{\infty}$ and  $r_{s}^{\pm}$ (light surfaces).  In general, we assume the open surfaces are on a generic plane $\sigma\in[0,1[$, while we fix the torus on the  equatorial plane $\sigma=1$.
More generally,  the analysis of the collision conditions   with the surrounding matter addresses the problem of jet launching point  location and jet structure (collimation and velocities).
Considering the toroidal surfaces defined by the functions $\Sa_{eff}$, the first immediate way to  obtain the collision conditions, according to the values of different parameters,  is to explore the surfaces crossing. This analysis is mostly  reduced to     an algebraic  multi-parametric condition.
 We take explicitly the four  topological solutions: closed, quiescent, closed cusped surfaces and  the proto-jets, we consider also the limiting surfaces (for example the  three   radii  $r_j$, $r_{\wp}$ and
$ r_K^{\pm}$).
All these surfaces  depend on one  or both parameters $\ell$ and $K$,  according to the morphological conditions assumed for the toroidal surfaces.
In some circumstances, for example in the case of cusped surfaces, we can make use of Eqs\il(\ref{Eq:suo-fr1},\ref{Eq:suo-fr2},\ref{Eq:LKmeasuram})  to fix   $K$ as function of  $\ell$  or   viceversa $\ell$ as functions of $K$.
For a non-cusped  surface there is  $K_{cent}\in K(r)<K <1$ (or  $K_{cent}\in K(r)<K <K_{\max}$). It is clear that in the determination of a  (unique) couple  $(\mathcal{B}, \mathcal{Z})$ for the surfaces collision we obtain a relation  $\theta_1(\theta, P,P_1)$ where $(P,P_1)$ are a couple of reduced parameters,
$P=(\ell,K)$, for the two surfaces respectively, index $1$ is used for quantities related to the configuration $T_1$.
In general we consider a couple constituted by a closed torus conveniently considered on its equatorial plane $\theta=\pi/2$ or $\sigma=1$, reducing then
$\theta_1(P,P_1)$. One condition can be
$\mathcal{B}^2_1=\mathcal{B}^2$, $\mathcal{Z}^2=\mathcal{Z}_1^2$, leaving undetermined the angular relation, we set $x=x_1$ and  $y=y_1$.
Therefore we consider
\be\left(\frac{2 \left(\mathcal{B}^2+K^2 Q\right)}{K^2 \left(Q-\mathcal{B}^2\right)+\mathcal{B}^2}\right)^2-\mathcal{B}^2-\mathcal{Z}^2=0,\ee
where  $\mathcal{B}=y$, $ \mathcal{Z}=-x$, and  $
\mathcal{B}_1=x \cos (\theta_1)+y \sin (\theta_1)$, $\mathcal{Z}_1=y \cos (\theta_1)-x \sin (\theta_1)$ . Using the disks symmetries, we concentrate mainly on the  plane $x>0$ and $y>0$, although clearly a jet on $(x<0,y>0)$ and $(x>0,y<0)$ is in this case not considered.
Very special cases  are    $\sigma_1=\sigma=1$, or
$\sigma_1=0,\;\sigma=1$ i.e. orthogonal configurations. A limiting case for the equations is  therefore the case  $\sigma\neq0$ (note that there is always a phase difference in these relations due the configurations relative orientations).
We consider also the case
$\sigma_1=1/2,\;\sigma=1$.

\begin{figure}
  \includegraphics[width=8cm]{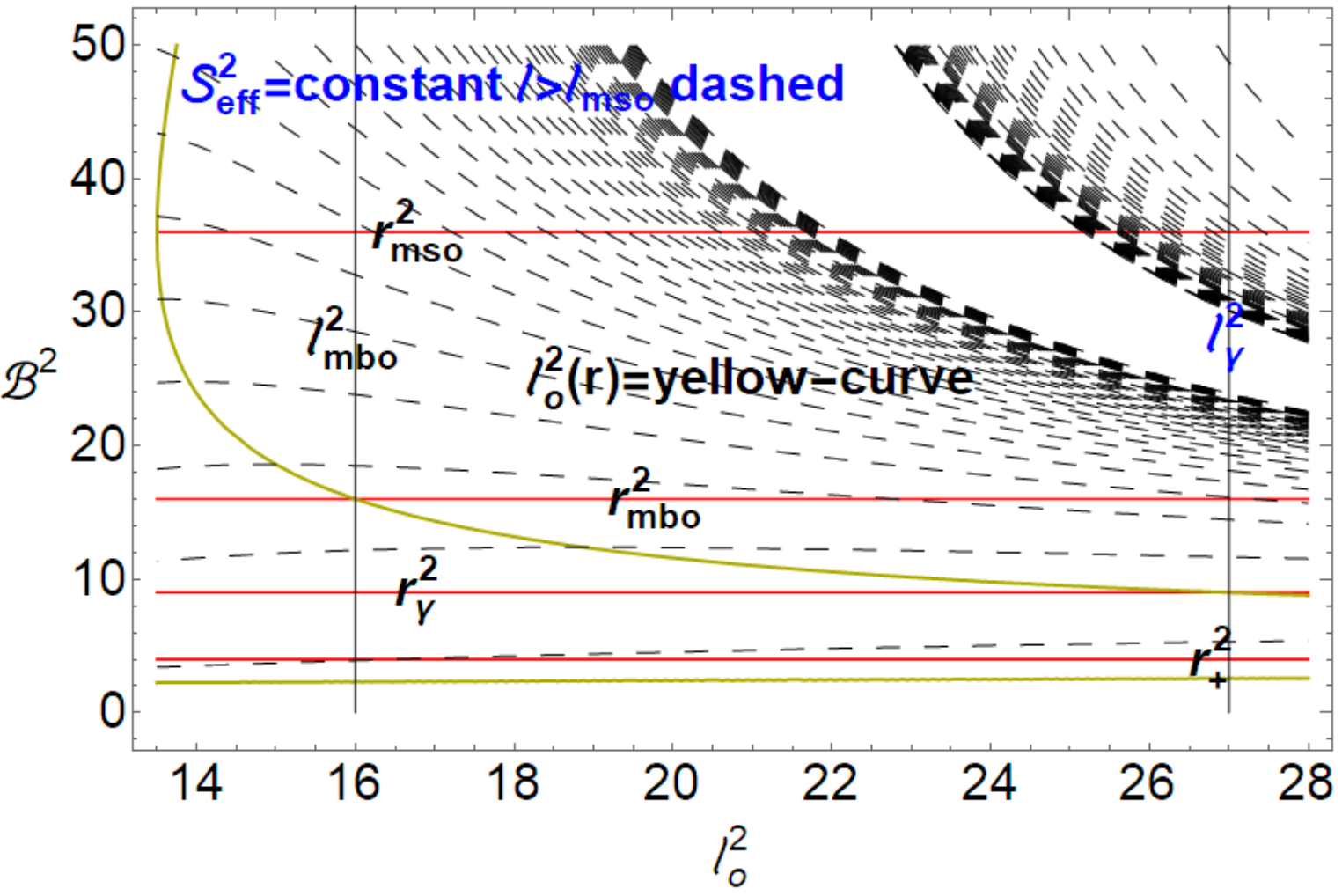}
   \includegraphics[width=8cm]{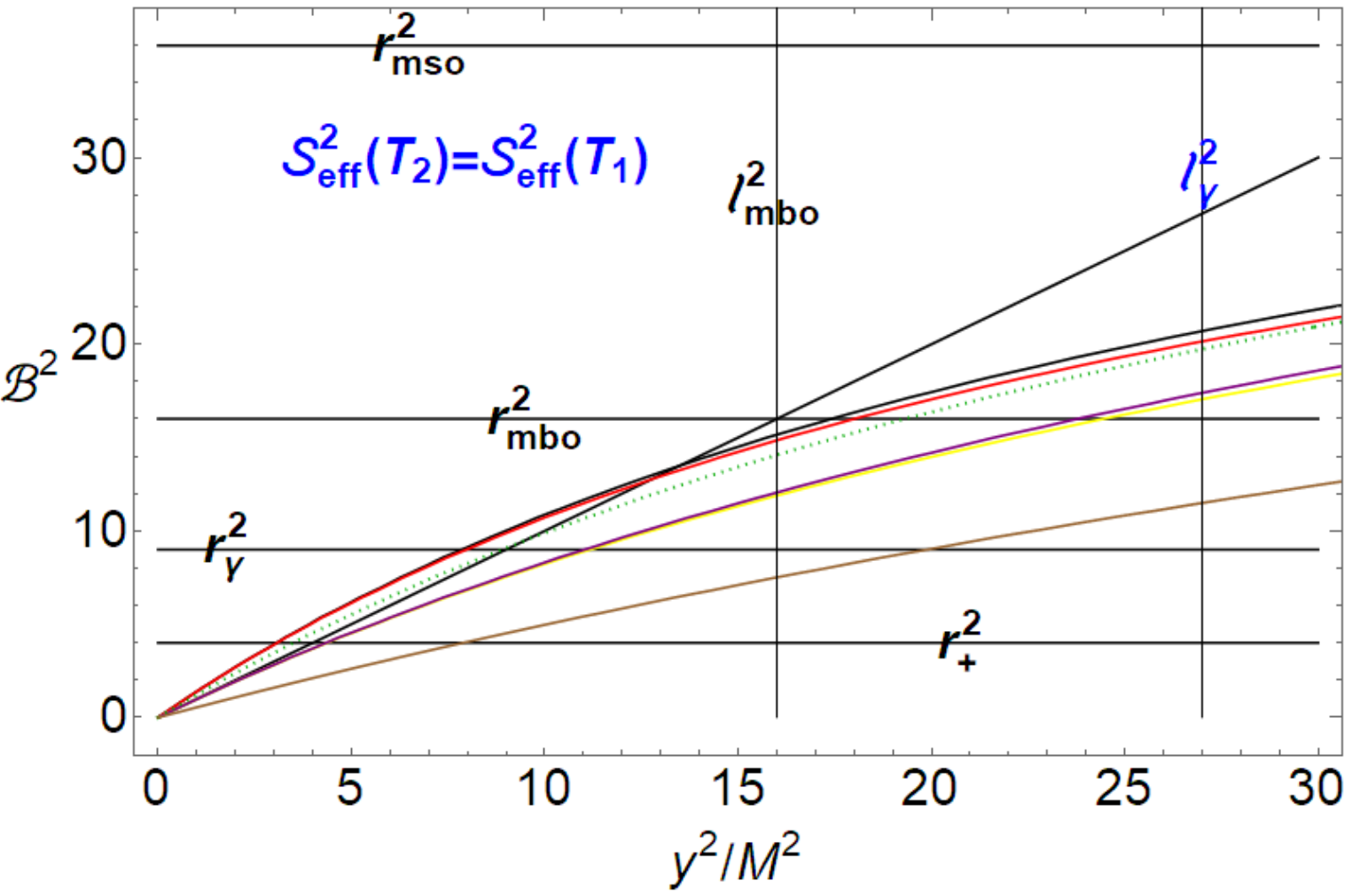}
\caption{
  Left panel:  tori surface  represented by function $\Sa_{eff}(\ell_o,K_o,\sigma)=constant$, in the plane $(\ell^2_o, \mathcal{B})$, where $(\mathcal{B}=x \cos (\theta )+y \sin (\theta ))$, $\sigma=\sin^2\theta$ (units are in \textbf{BHs} mass $M$), $\ell$ is the fluid specific angular momentum, $K$ sets the equipotential level and it is related to tori energetics. Symbols $mso$ is for marginally stable orbit, $mbo$ marginally stable orbit, $\gamma$ is for last circular orbits which is a photon orbit.
  Right panel: solution  $\Sa_{eff}^2 (T_1)=\Sa_{eff}^2(T_2)$
as  in the plan $\mathcal{B}^2$-$y^2/M^2$ for different parameter values:
purple line $K^2= K_{\alpha }^2$, $Q= \ell^2$, $ \ell^2= \ell_{\bullet}^2$, yellow curve: $K^2= K_{\beta }^2$ $\ell\in \mathbf{L_2}$ (quiescent) $\ell^2_{\bullet}$.  Black curve-- for torus $T_1$: $K^2=K_{{crit}}^2$, cusped, $\ell_{\star}^2$. Red curve $K^2= K_{\delta }^2$,  $\ell^2=\ell_{\star}^2$ where $K_{\alpha }^2\equiv\left[({1-K_{cent}})/{2}+K_{cent}\right]^2$; $K_{\beta }^2\equiv\left[({1-K_{cent}})/{10}+K_{cent}\right]^2$; $\ell_{\bullet}^2\equiv\left[({\ell_{\gamma}-\ell_{mbo}})/{2}+\ell_{mbo}\right]^2$;
 $K_{\delta }^2\equiv\left[({K_{crit}-K_{cent}})/{10}+K_{cent}\right]^2$; $\ell_{\star}\equiv ({\ell_{mbo}-\ell_{mso}})/{2}+\ell_{mso}$; Solutions of  $\Sa^2_{eff}(T_2)=\Sa^2_{eff}(T_1)$, $T_2$ with parameters $\Qa_o\equiv \ell_o^2$ and  $K^2_o$, $T_1$ with parameters  $\Qa\equiv \ell^2$ and  $K^2$---see Sec.\il(\ref{Sec:cross-interaction}).}\label{Fig:plotmemo1}
\end{figure}

The relation to be considered to evaluate the crossing conditions  is
$\Sa_{eff}^2(T_1)=\Sa_{eff}^2(T_2)$, where  $T_1$ and $T_2$  are two configurations  under analysis, one configuration will be denoted with parameters $\ell_o$, $K_o$ and plane $\sigma_o$.
We assume a torus $T_1$  (closed configuration)  fixed, without loss of generality, on its equatorial plane $\sigma=1$  and  the second configuration $T_2$ defined by different parameters  on different planes.
Clearly  the points $y^2$ and  $x^2$  in the frame adapted to the  torus $T_1$ are assumed to be the crossing point with the second configuration.
The solution on the equatorial plane is as follows ($Q=\ell^2$):
\bea&&
Q_o= y^2 \left(\frac{1}{K^2}-\frac{1}{K_o^2}\right)+Q,\quad Q_o= \frac{y^2 [K^2 (K_o^2-2) Q+y^2 (K^2+K_o^2-2)]}{K_o^2 (2 K^2 Q-K^2 y^2+2 y^2)}
\eea
{alternatively},  in terms of $K$ parameter
\bea&&
K_o^2= \frac{K^2 y^2}{K^2 Q-K^2 Q_o+y^2},\quad K_o^2= \frac{y^2 (2 K^2 Q-K^2 y^2+2 y^2)}{K^2 y^2 (Q+Q_o)-2 K^2 Q Q_o+y^2 (y^2-2 Q_o)}.
\eea
One can express the collision conditions directly in terms of the contact point $y$
\bea&&
y^2= \frac{K^2 K_o^2 (Q-Q_o)}{K^2-K_o^2},
\\&&\nonumber
y_{\mp}^2= \frac{\mp\sqrt{[K^2 (K_o^2-2) Q+(K^2-2) K_o^2 Q_o]^2+8 K^2 K_o^2 Q Q_o (K^2+K_o^2-2)}}{2 (K^2+K_o^2-2)}\\
&&\qquad \qquad \nonumber +\frac{2 K_o^2 Q_o-K^2 [(K_o^2-2) Q+K_o^2 Q_o]}{2 (K^2+K_o^2-2)}.
\eea
In general  the crossing of a configuration with a $T_2$  surface  on any plane $\sigma$ with the open limiting configurations    is rendered by the following conditions
$\Sa_{eff}(T_2)^2=\Qa_{\lim}$, where   $\Qa_{\lim}=\{r_{\infty}^2,(r_{s}^{\pm})^2\}$. We obtain for  this problem
the solution
\bea&&
\mathcal{B}_\pm^2\equiv \frac{K^2 Q (\Qa_{\lim}\pm2)}{(K^2-1) \Qa_{\lim}\mp2},
\quad
K_\pm^2\equiv \frac{1}{-\frac{Q}{\mathcal{B}^2}\pm\frac{2}{\sqrt{\Qa_{\lim}}\mp2}+1},
 \eea
alternatively
\bea&&
K_{\pm}^2= \frac{\mathcal{B}^2 (\Qa_{\lim}\pm2)}{\mathcal{B}^2 \Qa_{\lim}-Q (\Qa_{\lim}\pm2)},
\quad  Q_{\pm}= \frac{\mathcal{B}^2 [(K^2-1) \Qa_{\lim}\mp2]}{K^2 (\Qa_{\lim}\pm2)}.
\eea
We then proceed to consider the crossing between  a closed torus and an open surface.
  Some points are of particular interest, for example, when the collision point is close to the  edges of the configuration in accretion, or  the cusp  of a proto-jet, showing a geometrical correlation between the two processes \citep{dsystem}.
  For this purpose we consider
 $Y \geq r_{inner}^2$,
or  $Y \leq (r_{out}^{\times})^2 $ (the outer edge of a cusped torus), here and in the following we consider $X\equiv x^2$ and $Y\equiv y^2$. In general the  conditions to  be considered are $x\in (T_1,T_2)$ ($x$ belongs to the Boyer surface associated to $T_1$ and $T_2$ tori) and  $y\in [r_{inner},r_{out}]$  for  cusped or quiescent tori. However,  in the case of cusped tori, the condition can be  $y<r_{\times}$ or, for  the equipotential level  superior of the maximum critical value ($K>K_{\times}$), we can adopt  the condition  $K(r_{s})>K(r_\times)$ where $r_s$ is a fixed point.  Therefore,
we first fix $K^2 =
 K_{\times}^2$  for accretion condition, where  $Q =
 \ell^2\in \mathbf{L_1}$, while the second case we consider is a proto-jet emission. We can find the exact analytical form of these solutions, represented also in Figs\il\ref{Fig:plotmemo1}.
\begin{figure*}
    \includegraphics[width=5.6cm]{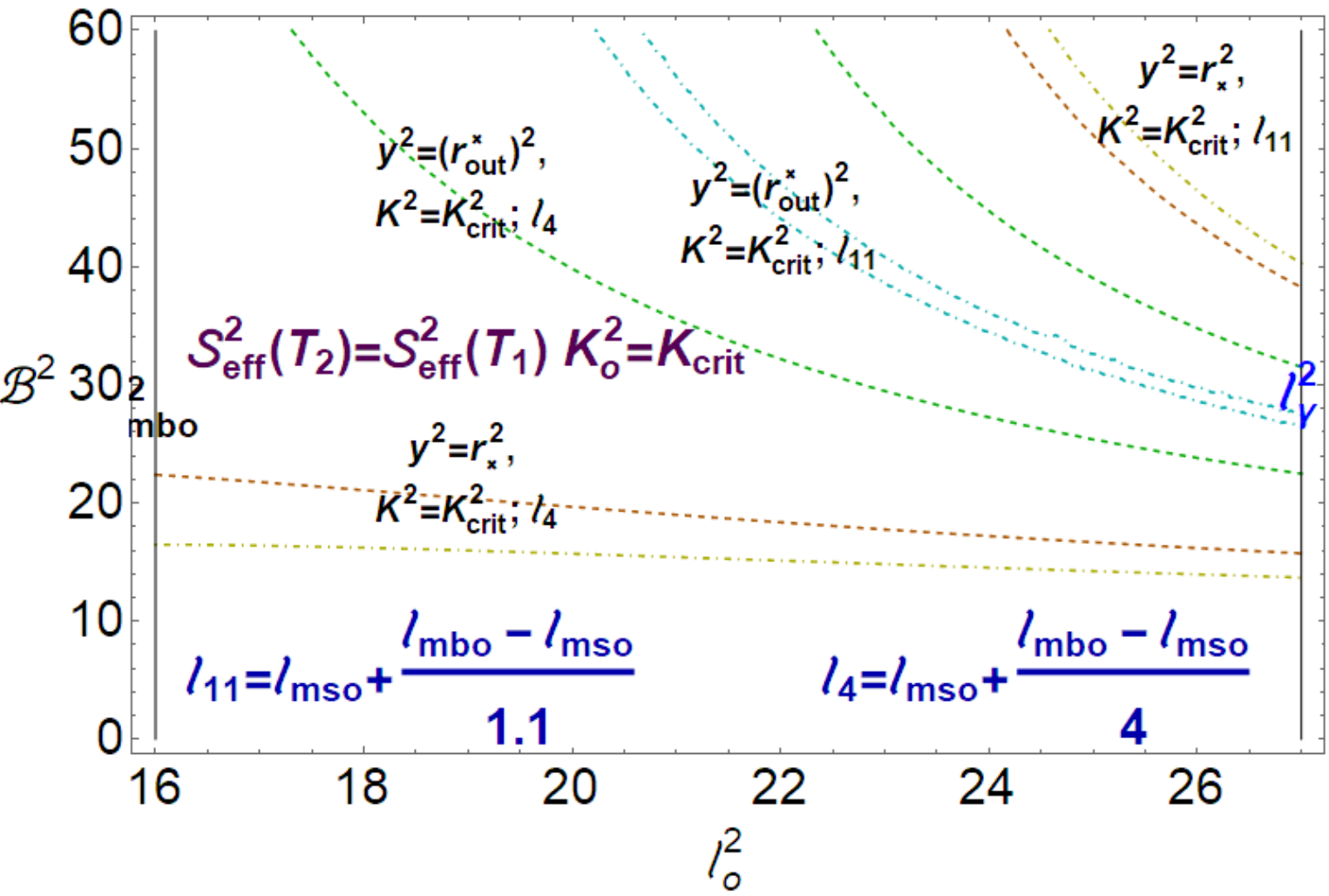}
     \includegraphics[width=5.6cm]{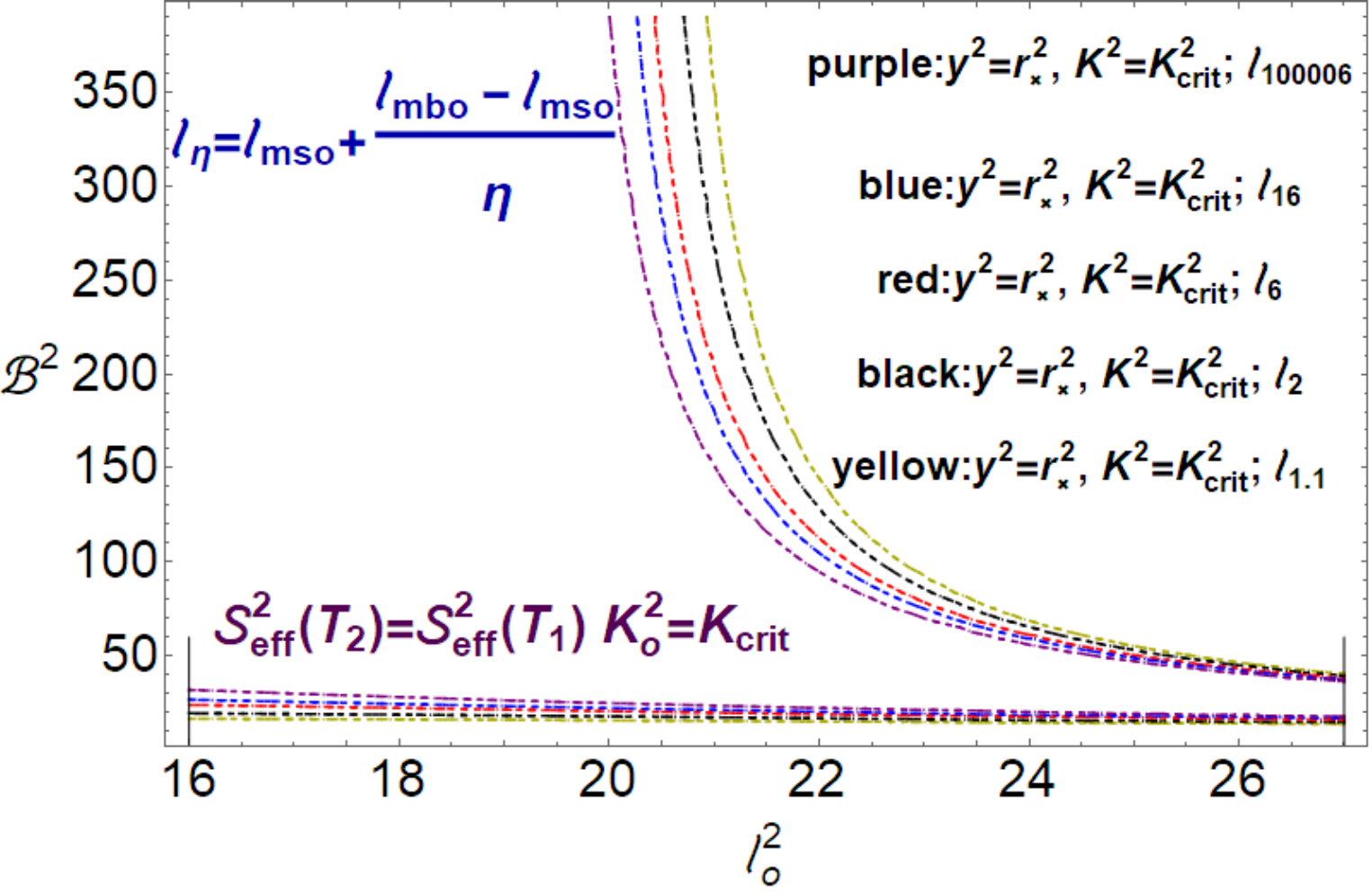}
        \includegraphics[width=5.6cm]{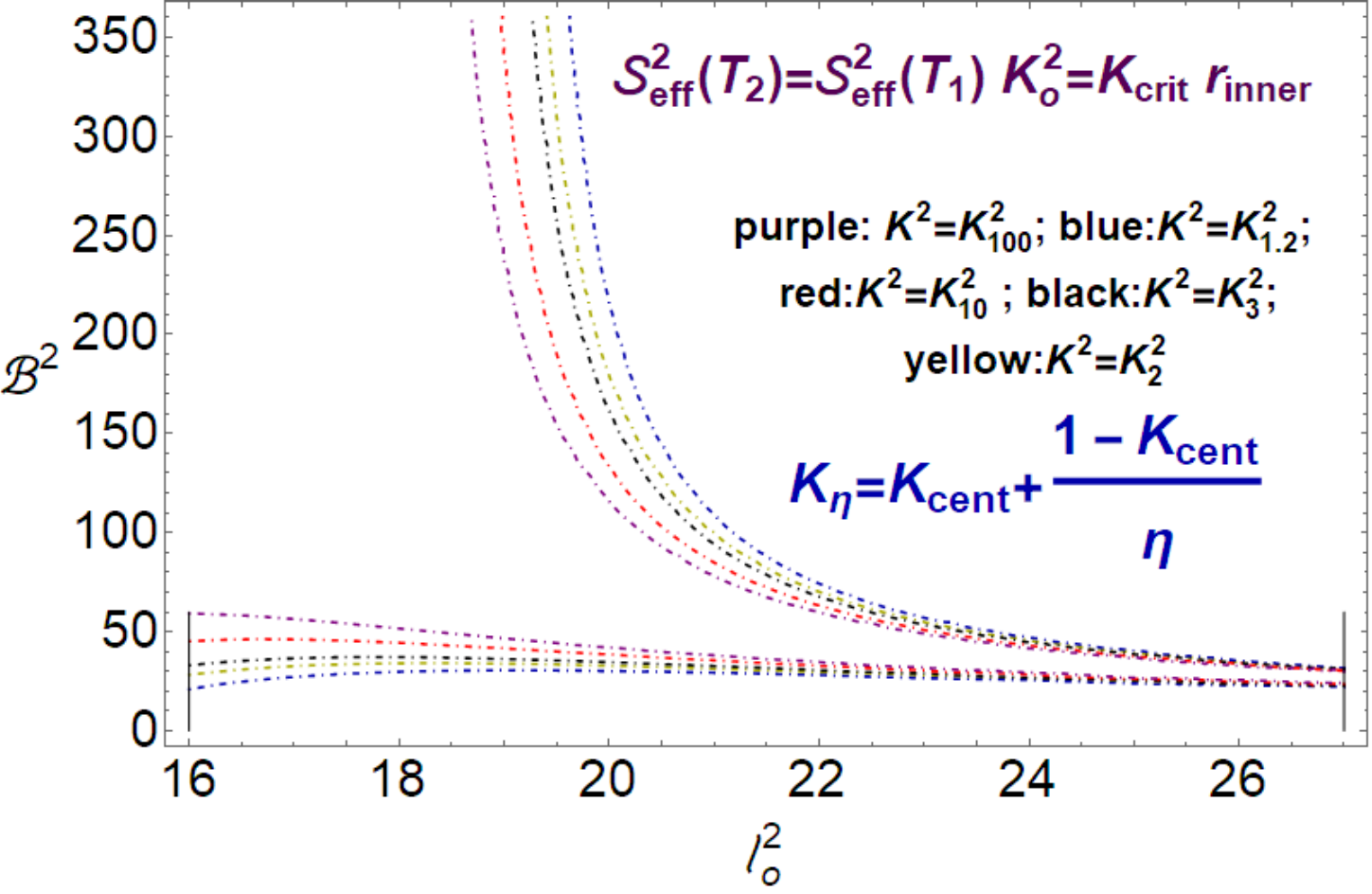}
              \includegraphics[width=5.6cm]{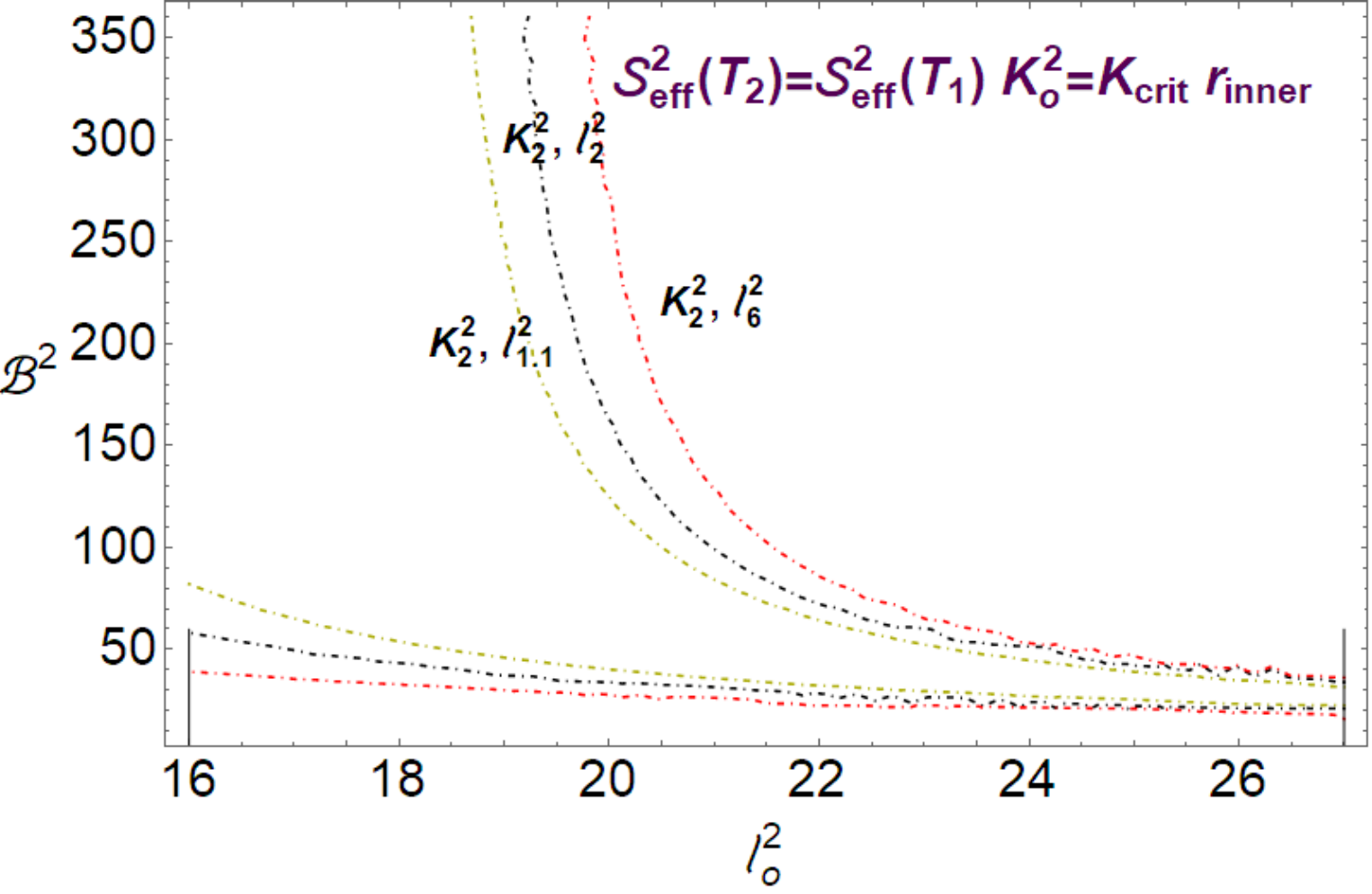}
              \includegraphics[width=5.6cm]{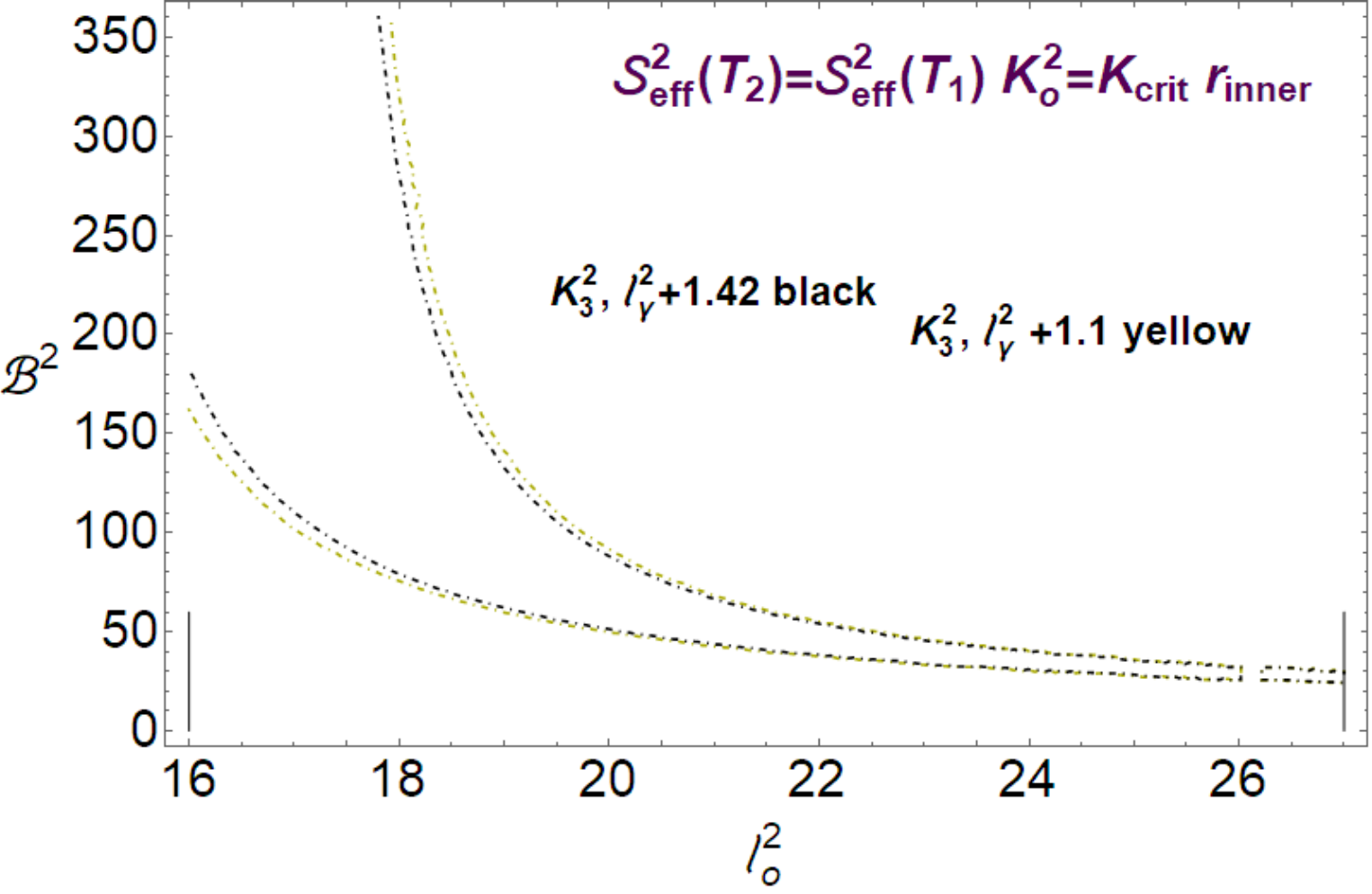}
                   \includegraphics[width=5.6cm]{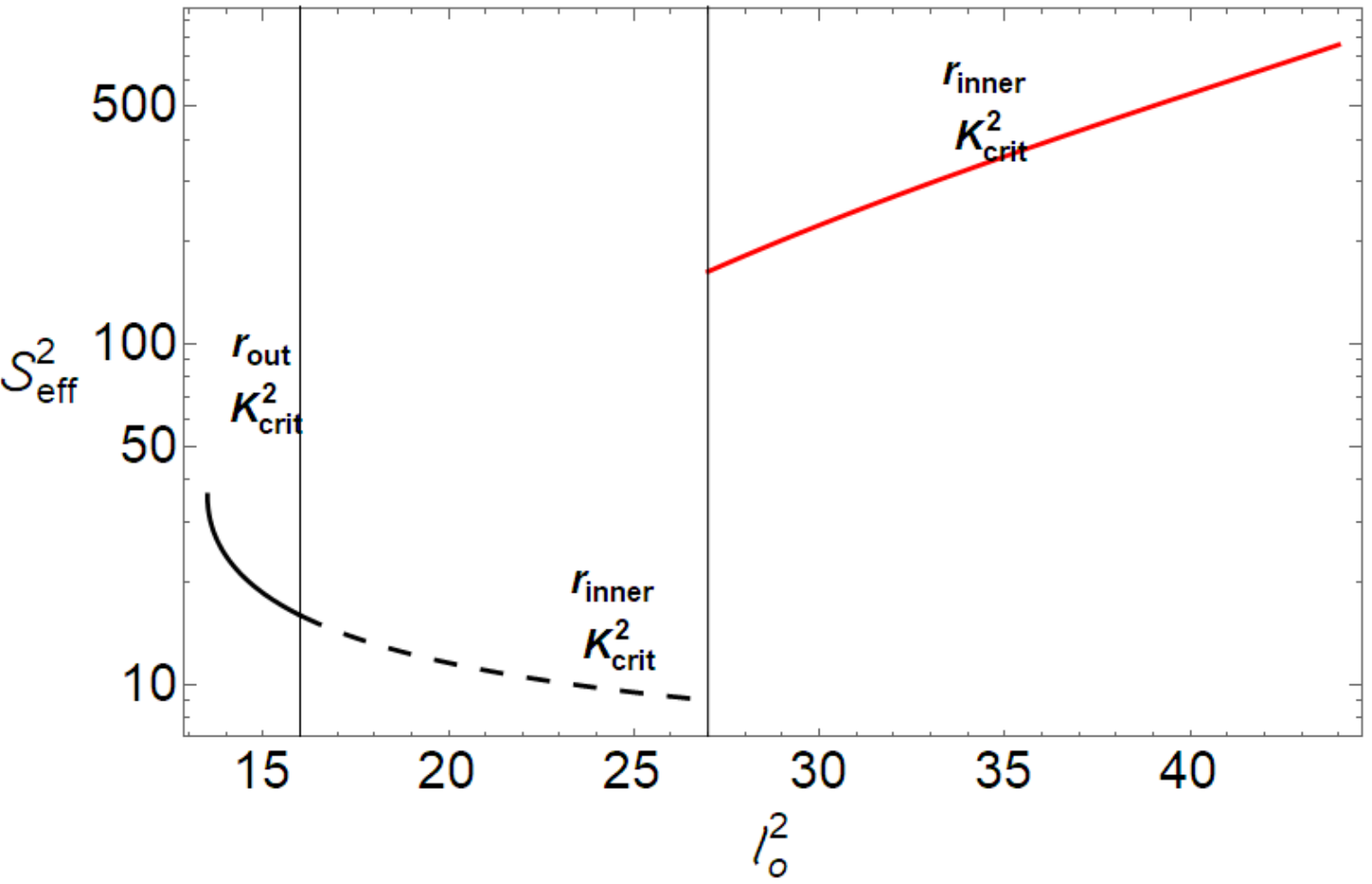}
\caption{Analysis of Sec.\il(\ref{Sec:cross-interaction}): collision jet-torus.
   Solutions  of equations  $\Sa_{eff}(T_1)= \Sa_{eff}(T_2)$,  for two configurations $T_1$ and $T_2$,  with parameters $(\ell_o,K_o,\sigma_o)$ and $(\ell,K,\sigma)$, where   $\Sa_{eff}(\ell_o,K_o,\sigma_o)$, in the plane $(\ell^2_o, \mathcal{B})$, where $(\mathcal{B}=x \cos (\theta )+y \sin (\theta ))$, $\sigma=\sin^2\theta$ (units are in \textbf{BHs} mass $M$), $\ell$ is the fluid specific angular momentum, $K$ sets the equipotential level and it is related to tori energetics. Symbols $mso$ is for marginally stable orbit, $mbo$ marginally stable orbit, $\gamma$ is for last circular orbits which is a photon orbit.
   Parameter sets are defined in figure, we evaluated some conditions for $K=K_{crit}$ on inner edge-cusp of the torus $r_{inner}$. Last panel:   $\Sa_{eff}^2$ as function of $\ell_o^2$  evaluated for $K_{crit}(\ell)$ on $r_{inner}$ for the cusps location and $r_{in}(\ell,K)$ for the inner edge of tori, quiescent or cusped; different regions of existence are shown in the panel.}\label{Fig:plotmemo1a}
\end{figure*}
\begin{figure*}
 \begin{center}
  \includegraphics[width=6cm]{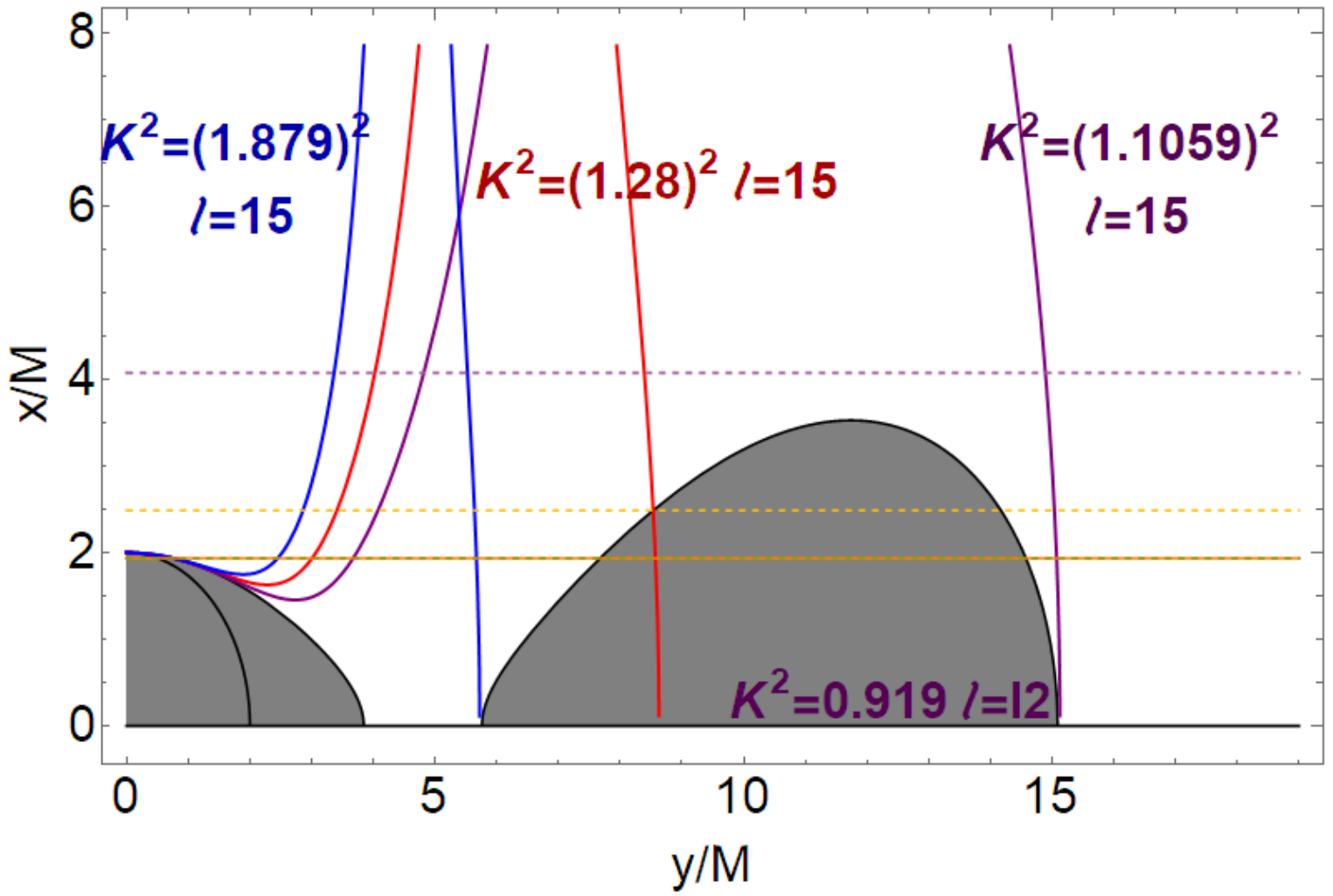}
   \includegraphics[width=6cm]{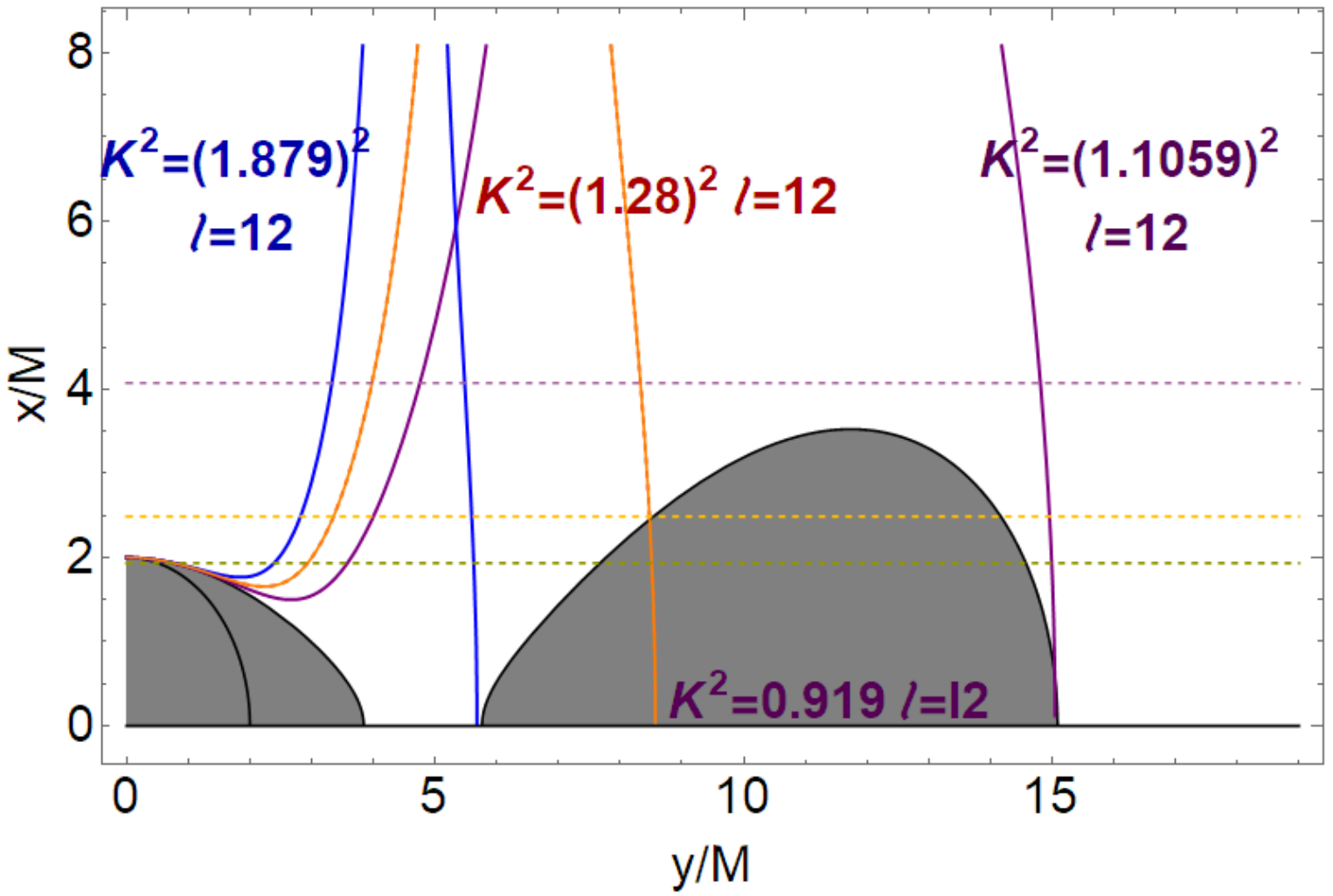}
    \includegraphics[width=6cm]{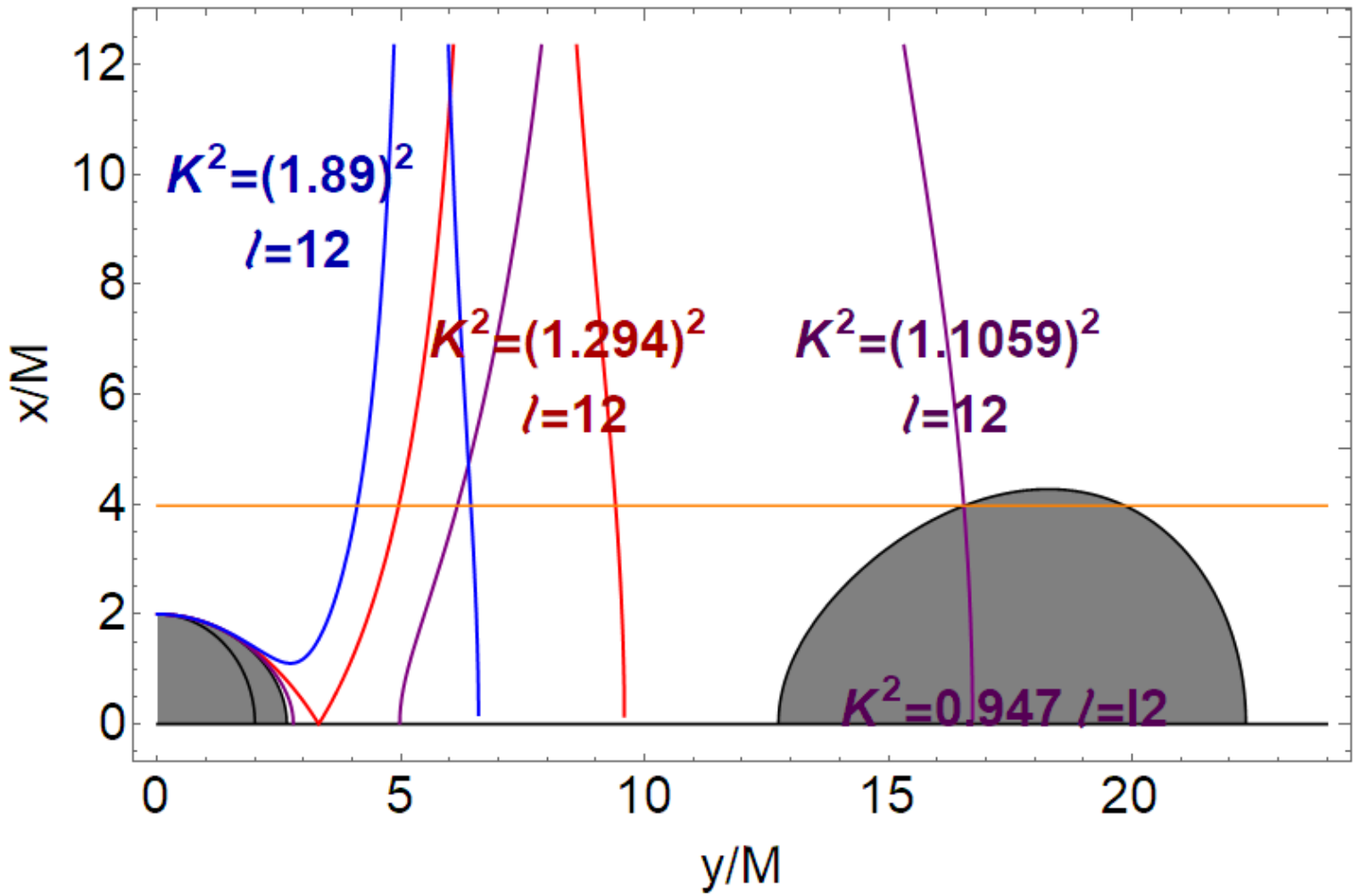}
     \includegraphics[width=5cm]{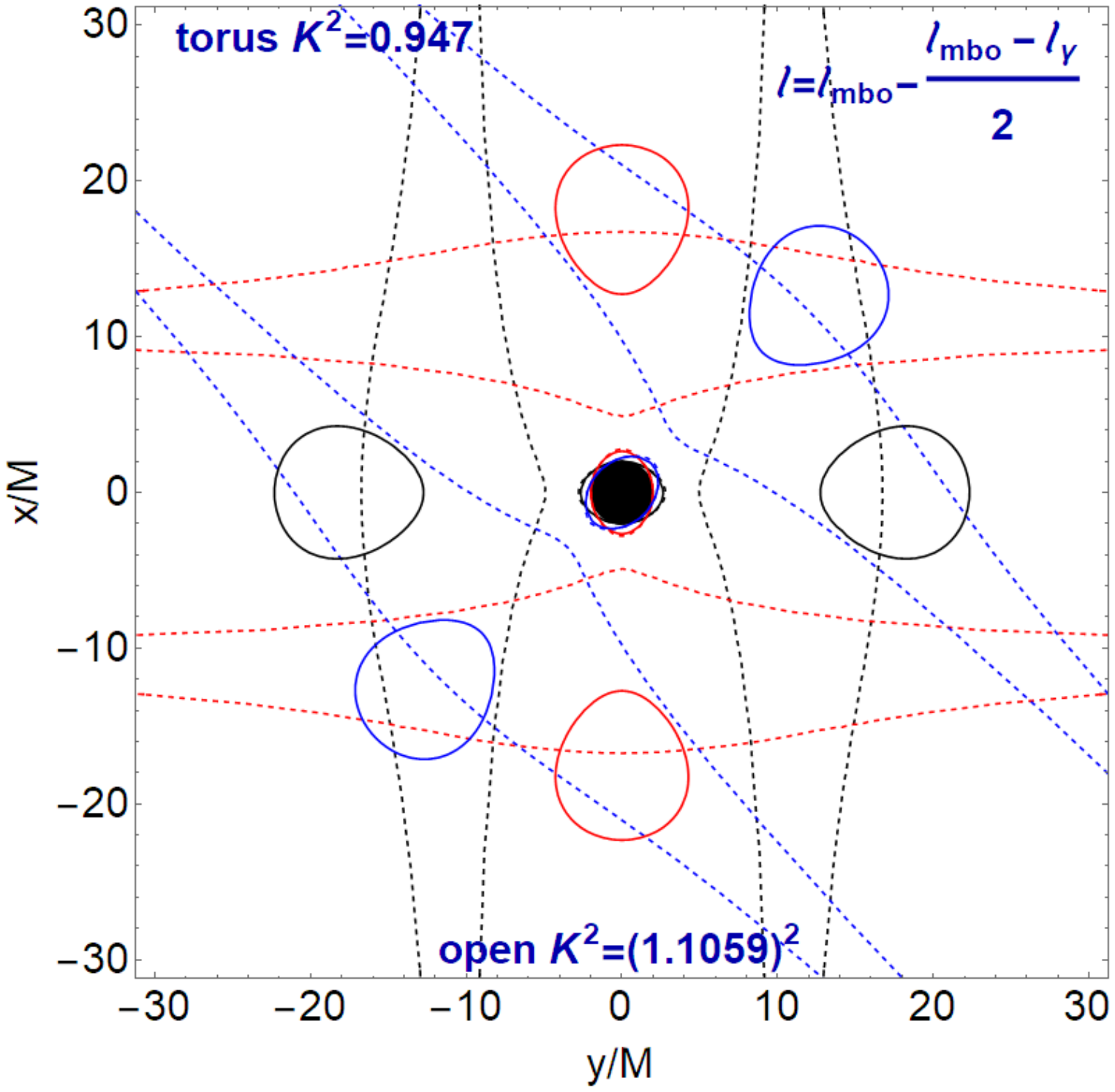}
     \end{center}
\caption{Refers to analysis of Sec.\il(\ref{Sec:cross-interaction}) on   of  jet-torus collisions. Configurations are plotted for different values of the parameters $K$ and the specific fluid angular momentum $\ell$. Open and closed configurations are represented, particularly  intersecting  the  inner edge and outer  edge of the  torus.
Curve   $r=2M$, in Cartesian coordinates, is the outer horizon.
 Radii  $r_{mbo}=4M$ (marginally bounded orbit)  and $r_{\gamma}=3M$ photon orbit are shown. Last panel shows the rotated open and closed configurations and the  cusped open surfaces. Configurations are plotted considering functions $ \Sa_{eff}=x$ on different planes.}\label{Fig:plotmemo1b}
\end{figure*}
Crossing  solutions are
\bea&&
\mathcal{B}^2= \frac{K^2 K^2_o \Qa_o Y}{K^2 K^2_o Q-K^2 Y+K^2_o Y},\quad
\mathcal{B}^2= \frac{K^2_o \Qa_o (2 K^2 Q-K^2 Y+2 Y)}{K^2 (K^2_o-2) Q+Y (K^2+K^2_o-2)},
\\&&\label{Eq:ase-ft}
X_{\pm}= -\frac{\left(\tilde{f}(y)\pm\sqrt{\sigma_1} \sqrt{Y}\right)^2}{\sigma_1-1},
\\&&\nonumber
\sigma^{\mp}_1= \frac{\tilde{f}(y)^2 (Y-X)\mp2 \sqrt{\tilde{f}(y)^2 X Y \left(-\tilde{f}(y)^2+X+Y\right)}+X (X+Y)}{(X+Y)^2},
\eea
we solved for the solutions $x^2\equiv X$ and $y^2\equiv Y$ and
$\mathcal{B}^2= \tilde{f}(y)^2$ for  Eq.\il(\ref{Eq:ase-ft}) (on the equatorial plane there is $X=0$).
More specifically, solving the problem $\Sa_{eff}(T_1)=\Sa_{eff}(T_2)$   on the unique  symmetry plane we find:
%
\bea&&
Y= 0,\quad Y= \frac{K^2 K_o^2 (Q-Q_o)}{K^2-K_o^2},
\\\nonumber
&&\breve{Y}_{\pm}= \frac{\pm\sqrt{[K^2 (K_o^2-2) Q+(K^2-2) K_o^2 Q]^2+8 K^2 K_o^2 Q Q_o (K^2+K_o^2-2)}}{2 (K^2+K_o^2-2)}+\\
\\&&\qquad \qquad \nonumber +\frac{+2 K_o^2 Q_o-K^2 [(K_o^2-2) Q+K_o^2 Q_o]}{2 (K^2+K_o^2-2)}.
\eea
%
%
%
The condition of crossing at equal $K$ parameters implies  $\mathcal{B}_o=Y Q_o/Q$, leading to  the set of solutions:
{
\bea&&\nonumber
Y= 2 \sqrt{\frac{Q^3 Q_o (\sigma_1-1)^2 \sigma_1 X^2}{(Q_o-Q \sigma_1)^4}}-\frac{Q (\sigma_1-1) X (Q \sigma_1+Q_o)}{(Q_o-Q \sigma_1)^2},\quad X= 0;
\quad\mbox{and}\\&&\nonumber Y= \frac{X (Q-Q_o)}{4 Q_o},\quad \sigma_1= \frac{Q_o}{Q},\\&&\nonumber \{Q_o= Q,\sigma_1= 1\},\{Q_o= 0,X= 0,\sigma_1= 0\};
\\\nonumber
&&\breve{\sigma}^\pm_1= \frac{\pm2 \sqrt{Q_oX Y^2 [\bar{\tau}_{\bullet}-Q_oY]}+\bar{\tau}_{\bullet}X+Q_oY(Y-X)}{\bar{\tau}_{\bullet}^2/Q},\quad\mbox{and}
\\&&\nonumber \breve{X}_{\mp}= \mp\frac{2 \sqrt{Q} \sqrt{Q_o (\sigma_1-1)^2 \sigma_1 Y^2}\pm(\sigma_1-1) Y (Q \sigma_1+Q_o)}{Q (\sigma_1-1)^2},
%
\eea}
where $\bar{\tau}_{\bullet}\equiv Q(X+Y)$
(conditions at equal $\ell$ can be easily reduced as well)--(see Figs\il\ref{Fig:plotmemo1},\ref{Fig:plotmemo1a},\ref{Fig:plotmemo1b}).
\section{Proto-jet collimation and toroidal magnetic field}\label{Sec:Komi}
We  address the  role of a toroidal magnetic field in the possible open surface collimation along an axis  crossing the  attractor center. Here we consider again the equatorial plane case, $\sigma=1$, investigating if  the proto-jet  cusp and the open matter funnel are shifted  inwardly  tending to collimate  with respect to the non-magnetized case.
 In \citet{epl}  it has been shown that  the torus geometrical thickness remains basically unaffected by the toroidal magnetic field,  tending however  to increase or decrease slightly  depending  of a combination of many factors such as  the magnetic, gravitational and centrifugal effects.
We used the toroidal ``Komissarov" magnetic field model developed in \citet{Komissarov:2006nz,Montero:2007tc}.

Assuming a barotropic equation of state, we consider the force-free approximation with  an infinitely conductive plasma, $F_{ab}U^a=0$, where $F_{ab}$ is the Faraday tensor and $U^a$ is the fluid four-velocity.
Using the equation $U^a B_a=0$, where $B^a$ is the magnetic field,  we find  the relation  $B^t=\Omega B^{\phi}$, $\Omega$ is the relativistic velocity of the fluid. Moreover, we assume $\partial_{\phi}B^a=0$ and $B^r=B^{\theta}=0$.
 Within the conditions  $B^r=0$ from the Maxwell equations there is,
\(
B^{\theta}\cot\theta=0
\)
 that is satisfied for $B^{\theta}=0$ or $\theta=\pi/2$ ($\ell$=constant).
This  implies that the assumptions $\partial_{\phi}  B^{\phi}=0$ and $B^r=0$ lead to $B^{\theta}=0$.
(The presence of a magnetic field with a predominant toroidal component can be   reduced to the disk differential rotation, which plays the part of  the generating  mechanism  for the  the magnetic field \citep{Komissarov:2006nz,Montero:2007tc,Horak:2009iz,Hamersky:2013cza,Parker:1955zz,Parker:1970xv,Y.I.I2003,R-ReS1999}.)
The magnetic field is therefore \citep{adamek,Universe}
\bea\label{RSC}
B^{\phi }=\sqrt{{2 p_B}({g_{\phi \phi }+\ell ^2g_{tt}})^{-1}},
\eea
where
$
p_B=\mathcal{M} \left(-g_{{tt}}g_{\phi \phi }\right){}^{q-1}\varpi^q$ is the magnetic pressure,
$\varpi$ is the fluid enthalpy,  and $q$  and $\Mie$ are constant.
The introduction of the Komissarov magnetic field maintains the integrability conditions on the  Euler equation which can be exactly integrated, resulting
\bea&&\label{Eq:wgg}
\frac{\partial_\flat p}{\rho+p}=\mathcal{G}_\flat^{(f)}+\mathcal{G}_\flat^{(em)},\quad\mbox{and}\quad
\int\frac{dp}{\rho+p}=-(W^{(f)}+W^{(em)}),\quad\mbox{where}
 \\&&\nonumber
\mathcal{G}_{\flat}^{\natural}=-\frac{\partial}{\partial\flat}W^{\natural}_{\flat};\quad
\flat=\{r,\theta\},\quad \natural=\{(em),(f)\}
\quad
\mbox{and}\\&&\nonumber W_{\flat}^{(f)}\equiv\ln V_{eff},\quad
W_{\flat}^{(em)}\equiv \mathcal{G}_\flat(r,\theta)+g_{\flat}(\theta),
\eea
where $g_{\theta}(r)$ and $g_{r}(\theta)$ are functions to be fixed by the integration.
For $q\neq1$,  there is
\(
\mathcal{G}_r(r,\theta)=\mathcal{G}_{\theta}(r,\theta)=\mathcal{G}(r,\theta).
\)
We consider then %
\(
\mathcal{G}_r(r,\theta)=\mathcal{G}_{\theta}(r,\theta)=\mathcal{G}(r,\theta),
\)
where
\bea\label{Eq:Kerr-case}
\mathcal{G}(r,\theta)&\equiv& \frac{\Mie q \left[(r-2) r \sigma^2\right]^{(q-1)} \varpi^{(q-1)}}{q-1}.
\eea
{Therefore there is $
\partial_{a}\tilde{W}=\partial_{a}\left[\ln V_{eff}+ \mathcal{G}\right]\,
$
and we consider  the equation for the
\(
\tilde{W}=
\)constant.
The effective potential function, modified by the introduction of the magnetic field reads
\bea\label{Eq:goood-da}
&&\widetilde{V}_{eff}^2\equiv V_{eff}^2 e^{2 \Sie \left(\mathcal{A} V_{eff}^2\right){}^{q-1}}=
\frac{\left(-g_{tt} g_{\phi \phi }\right) \exp \left[2 \Sie \left(-g_{tt} g_{\phi \phi }\right)^{q-1}\right]}{\ell ^2 g_{tt}+g_{\phi \phi }}=K^2,
\eea
where $\mathcal{A}\equiv\ell ^2 g_{tt}+g_{\phi \phi }$,  assuming the enthalpy $\varpi$ to be a constant,  where  the ratio  $\Mie/\varpi$  gives the comparison between the magnetic contribution to the fluid dynamics through $ \Mie $, and  the hydrodynamic   contribution   through the specific enthalpy $\varpi$.
Potential $\tilde{V}_{eff}^2$ for $\Sa=0$ reduces to  the effective potential ${V}_{eff}^2$ for the non-magnetized case
$V_{eff}$.
%
%
%
%
%
For $q=0$  the magnetic field does not affect  the Boyer potential and therefore the Boyer surfaces.
(In the limit case $q=0$,
the magnetic field $B$, does not depend on the fluid enthalpy.).

Note that
the magnetic pressure  is regarded  here  as a   perturbation of  the hydrodynamic component, it is assumed that the  Boyer theory of rigid rotating surfaced in GR remains   valid and applicable in this approximation.
Conveniently one can introduce as done in \citet{mnrasB,cqg2020,epl}
 the parameter $\Sie\equiv{\mathcal{M} q\varpi^{q-1}}/(q-1)$.

The modified specific angular momentum distribution $\tilde{\ell}(r)$ of the \textbf{eRAD} and the modified function $\tilde{K}(r)$, with the contribution of the magnetic field are
%
\bea\nonumber
&&
\tilde{\ell}(r)=\pm
\frac{\sqrt{r^3 \left[4 (q-1)^2 (r-1)^2 r \Sie^2 [(r-2) r]^{2 q-1}+2 (q-1) (r-1)^2 r \Sie [(r-2) r]^q+(r-2)^2 r^2\right]}}{2 (q-1) (r-1) \Sie ((r-2) r)^q+(r-2)^2 r}
\\
&&
\tilde{K}^2(r)=\frac{e^{2 \Sie [(r-2) r[^{q-1}} \left[2 (q-1) (r-1) \Sie [(r-2) r]^q+(r-2)^2 r\right]}{(r-3) r^2}
\eea
%
where $q$ and  $\Sie$  in the limiting cases \textbf{(1)} $q=0$ and $ \Sie=0$; \textbf{(2)} $\Sie=0$;  or \textbf{(3)} $q=1$ provide the rotational law $\ell$ in absence of magnetic field. For a deeper discussion on the interpretation and role of the limiting values on the magnetic parameters we refer to  \citet{mnrasB,cqg2020,epl}\footnote{The range of $q$ parameter is  divided into two regions, say  $ 0 <q <1 $ and $ q> 1 $, with a subrange  extreme $ q = 2 $.
 The  case $q=2$ is indeed  interesting because the magnetic field loops  wrap around with toroidal topology along the torus surface \citep{epl}.
For $ q <1$  closed surfaces  (tori) are approximately  those studied in the case $q\approx0$, for $ q> 1 $  there are  closed surfaces in the  limit $\Sie\approx0 $, which is in agreement with the situation when the contribution of the magnetic pressure to the torus dynamics is  regarded as a perturbation  with respect to the HD solution.}. We note however  that for  $q=1$ we obtain  $\tilde{K}^2=e^{2\Sie} (r-2)^2/[(r-3) r]=e^{2 \Sie} K^2$.
In Figs\il\ref{Fig:Plotnom} we show the modified   $\tilde{K}$ and some toroidal surfaces.
\begin{figure*}
 \begin{center}
  \includegraphics[width=5.3cm]{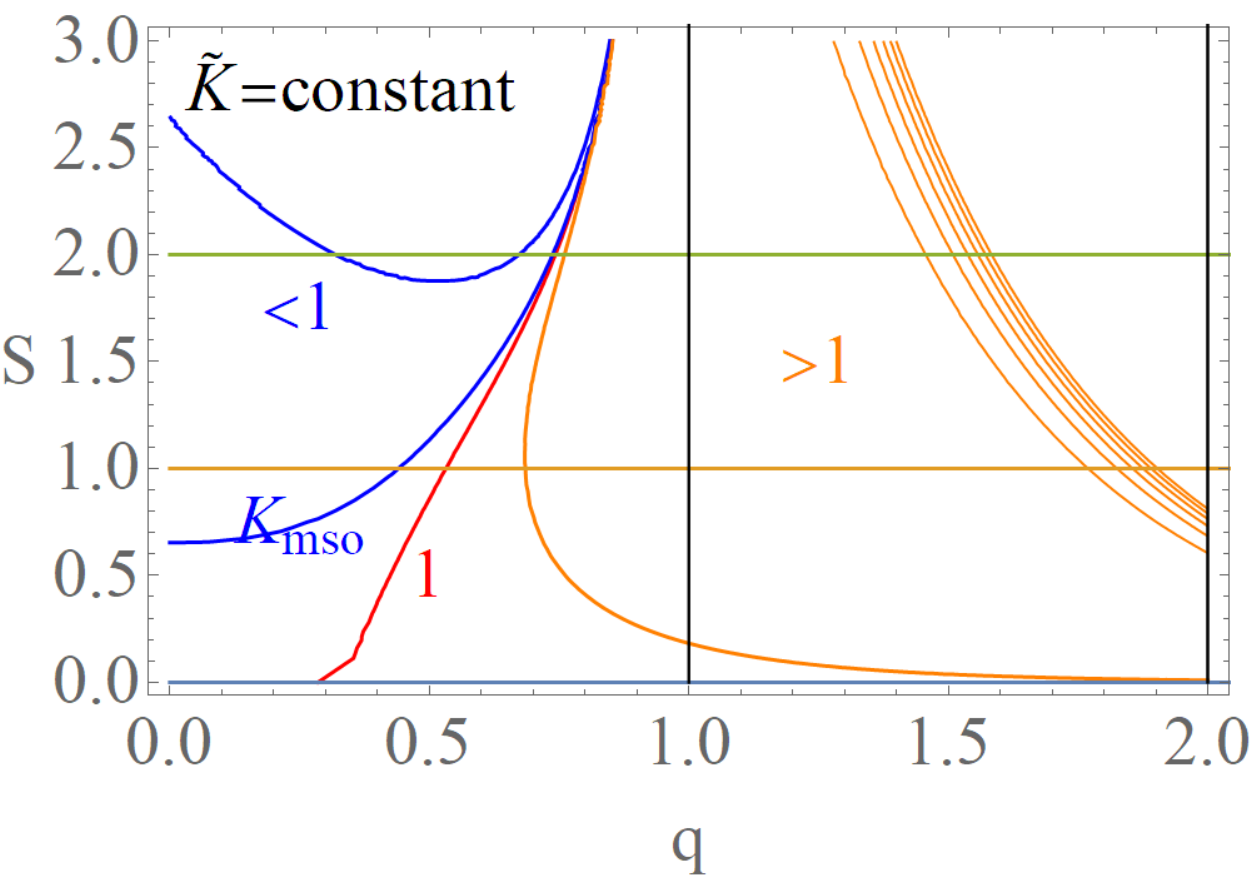}
   \includegraphics[width=5.3cm]{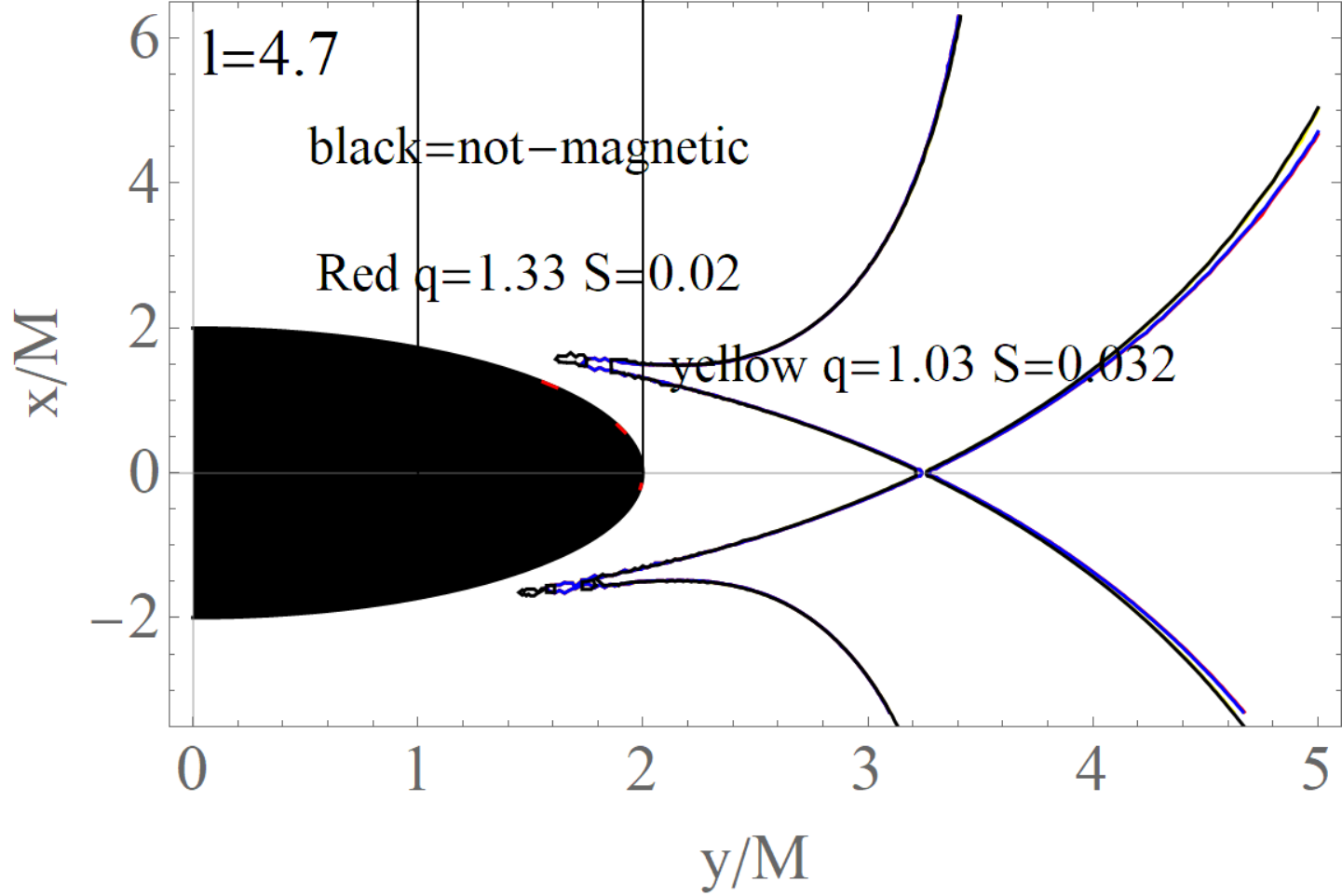}
      \includegraphics[width=5.3cm]{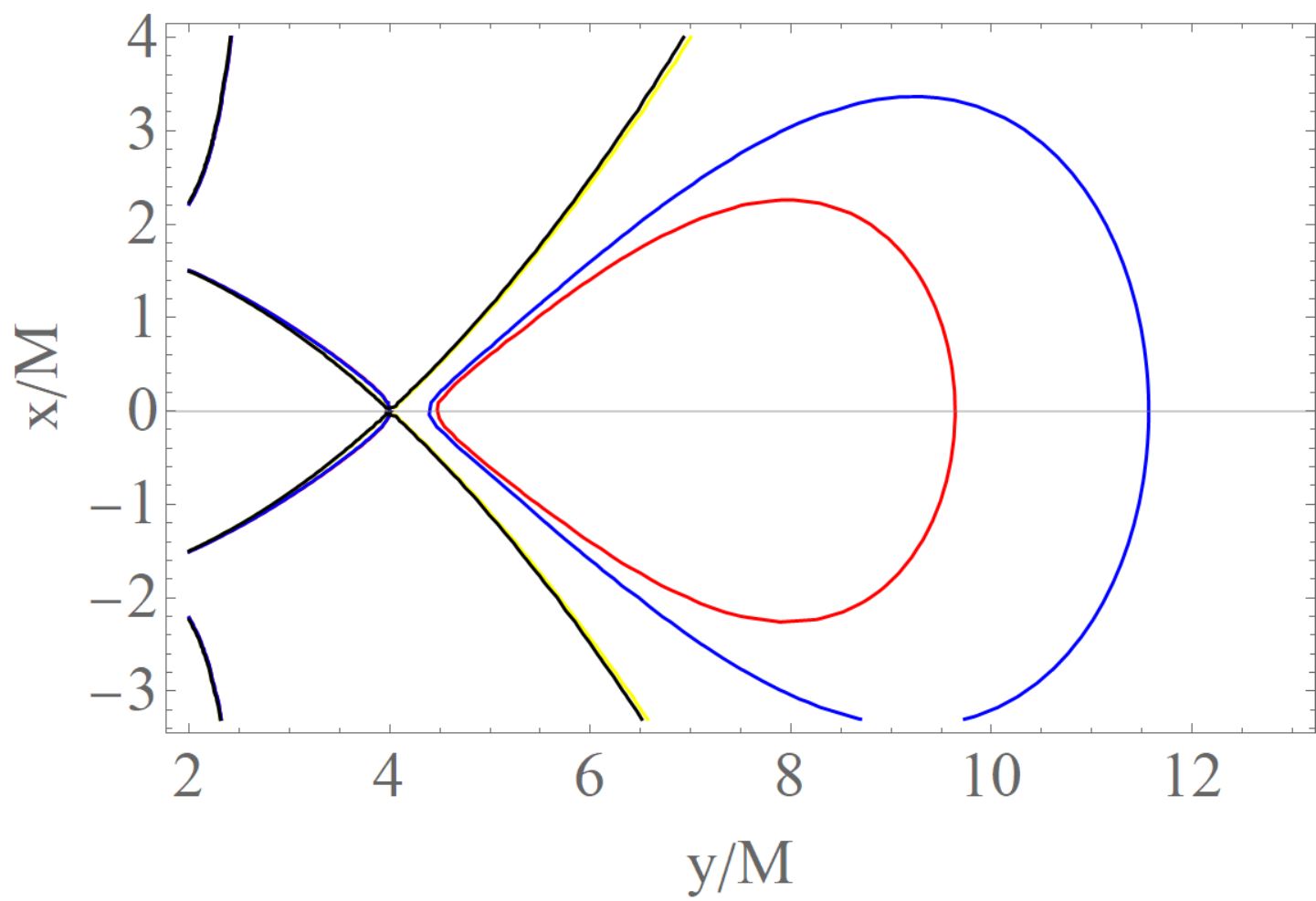}
      \end{center}
  \caption{Analysis of the configurations with a toroidal magnetic field. Left panel:  $\tilde{K}=$constant  (the $K(r)$ function for the magnetized case) in the plane $q$ and $\Sa$ (magnetic parameters).  Range of values $\tilde{K}$ and $\tilde{K}=1$ are shown.  Central panel: proto-jet configuration for fixed fluid specific angular momentum $\ell$ and different values of the magnetic parameters $q$ and $ \Sa$. Right panel: at $\ell=4$ proto-jet, open configurations and closed surfaces, magnetic parameters values  follow  colors  notation reported in  the central panel.   }\label{Fig:Plotnom}
\end{figure*}
As shown in Figs\il\ref{Fig:Plotnom},  although qualitatively an effect appears in the sense of funnels squeezing along the rotational axis, from this first analysis   there is no quantitative relevant collimation effect as shown in \citet{epl} for a  squeezing effect (exploring the disk thickness changing  the physical characteristics of the torus  particularly its magnetic field) on the equatorial plane for the toroidal configurations closed and orbiting a Schwarzschild central \textbf{BH}.

\section{Conclusions}\label{Sec:conclusions}
We  explored   the possibility  of jets collisions   with  misaligned accreting tori  orbiting around a \textbf{SMBH}. We use a model of misaligned  tori orbiting one central static \textbf{SMBH} which may be host for example in
\textbf{AGNs}, since   several  observational evidences   associate such structures to  different periods of accretion of \textbf{SMBHs}.
 We studied  the  geometric limits given by the black hole  geometry on the possibility of   jets, created by tori in vicinity of the central \textbf{SMBH}, to collide with obscuring inclined toroidal structures, which are extended on large distance from the \textbf{SMBH}.
There are several  indications showing a jet emission-accretion disk correlation.
The aggregates of misaligned  multi tori orbiting one central \textbf{SMBH}  with  different matter flows  include proto-jets.  As consequences of the agglomeration internal ringed structure,
 orbiting toroidal jets shell and  jets  may interact with the  surrounding matter, leading to  tori-proto-jets   collision. The  location of  jet ``launching  point"   in proto-jets, which are interpreted as limiting configurations for orbiting funnels of matter in toroidal symmetry,  is associated to the emergence of a  cusp,  that is a minimum of pressure and density in the extended orbiting matter.
The fluid fills any (closed) equipotential surfaces with a purely toroidal flow.

The interaction  with  surrounding matter provides   possible   jets  replenishment mechanism. This mechanism may  follow the impact of accreting material on an inner configuration that can be  associated eventually to the emergence of chaotic processes.
For the analysis of these agglomerations, we  used    \textbf{GRHD} tori  in the \textbf{RAD} frame introduced in   Sec.\il(\ref{Sec:app}).
 Proto-jet emission hypothesis and  possible shells of jets in the \textbf{RAD} frame have been discussed in Sec.\il(\ref{Sec:protp}).
Proto-jet emission   has been constrained by \textbf{GRHD} thick    tori
characteristic   frequencies, as shown in the  analysis of Sec.\il(\ref{Sec:freqs}).
Then constraints on the  \textbf{GRHD} systems are addressed in Sec.\il(\ref{Sec:constr}).
There are  ``darker regions'' along  the fluid rotation axis, bounded by special surfaces of \textbf{GRHD} origins. A torus, quiescent or cusped, can impact a jet at any angle. Thus   the possibility of a torus, quiescent or cusped, filling the proto-jet darker region has to be included. Cusped tori are very much constrained in terms of locations, dimension and velocity component.  For quiescent tori  there is a great degree of indetermination on the  location, velocity components and density. Such toroids  can range from the small quiescent  tori, which are  slower rotating, that is with  $\ell\in]\ell_{mso},\ell_{mbo}[$, and  located close to accreting region,  to the very large tori, which are faster rotating,  $\ell>\ell_{\gamma}$, and  located far away from the central \textbf{BH}.
Some relevant limiting radii have been identified,  with a HD origin and  a GR nature related to the causal structure imposed by the geometric background determined by the central \textbf{SMBH},  for example
the   ``asymptotic radius"  $r_{\infty}$ derived  from  the normalization conditions--see Sec.\il(\ref{Sec:Asymptotic}).
Limiting conditions on frequency and momentum are analyzed in Sec.\il(\ref{Sec:fre-mome-supri}).
In Sec.\il(\ref{Sec:hearLS}) we introduced the stationary observers and light-surfaces determining
replicas  discussed  in Sec.\il(\ref{Sec:replics}), significant for the possible observational evidence of  the jet shells.
Impacting conditions are   discussed in Sec.\il(\ref{Sec:glod-powe}).
Tori characteristics, limiting conditions and pressure gradients in the tori are the focus of Sec.\il(\ref{Sec:polar-gradient}). We proved then that the radial gradient of pressure in fact determines the disk verticality and the center of maximum pressure in the disk  is related to the morphological maximum of the tori surfaces.
In  Section (\ref{Sec:cross-interaction}) we explored the toroidal  surfaces collisions  in different topologies.
Finally, the role of a  toroidal magnetic field in the possible proto-jet collimation  is analyzed  in  section (\ref{Sec:Komi}).  Although from a first analysis,  the  toroidal magnetic field  does not appear to show a significant quantitative effect on the  collimation or shift of the cusp associated with the proto-jets, it cannot be excluded that in conjunction with dragging of frame, in the case of a rotating attractor, a  toroidal  magnetic field may instead be significant.
{(For more discussion on general relativistic exact models of magnetic field around rotating compact sources  and dragging effects see \citet{M17,P16,GP15,HG2006}). }

 Collision  emergence and  the stability properties of the  aggregates  have been  considered at  different inclination angles relative to a fixed distant observer.
Therefore constraints to jet-torus  impact  are provided in terms of misalignment  angles and   tori parameters,  exploring  also    the observational evidences of    \textbf{SMBH} host configurations. (It is  possible to reduce the results  given here, expressed in terms of conditions on the $\ell$ and $K$ parameters, in terms of the   flow velocity  components,   $U^{\phi}$ and $U^t$, the relativistic velocity and the frequencies.). In this respect we investigated the    constraints provided by the  light surfaces, defining  structures  related to recently introduced Killing metric bundles \citep{Pugliese:2020azr,Pugliese:2020ivz,Pugliese:2019rfz,Pugliese:2019efv,remnants}.    These structures pointed out the  existence of orbit-replicas that could host shadowing  effects as replicas  (horizon replicas in  jet shells) of the emissions  in regions close and far from the \textbf{BH}. These orbits are characterized by   equal limiting photon orbital frequency.
The observational relevance of the bundles could be explained  by looking at their exact definition.  For a given function   $\Qa(\omega(r))$ of  the limiting light-like toroidal orbital frequencies, there are two  orbits $(r,r_1)$ and two planes $(\sigma,\sigma_1)$ such that
  there is  $\Qa(\omega(r,\sigma))= \Qa(\omega(r_1,\sigma_1))$ and, as  per definition  of replica,  there is $\omega(r,\sigma)=\omega(r_1,\sigma_1)$.
The regions defined by  $(r,\sigma)$ and $(r_1,\sigma_1)$   can be interpreted as  presenting replicas of the $\Qa$  proprieties,
 where  if  $r$ is a circle very close to the attractor then the second point  $r_1$  is located   far from the attractor  (note that this relation is independent from the azimuthal angle $\phi$ and  it is eventually even reducible to a simplest relation between the couple of radii $(r,r_1)$).

The investigation clarifies   also the role of the pressure gradients of the   orbiting matter and  the essential  role of the  radial gradient of the  pressure in the determination of the disk  verticality.
We investigated the possibility that a toroidal magnetic field could be related to the collimation of proto-jets:
including
a strong toroidal magnetic field, developed in \citet{Komissarov:2006nz,Montero:2007tc}, we addressed  the specific question of jets  collimation.
The assumptions of these simplified models of  thick (stationary) GRHD disks and a central Schwarzschild \textbf{BH}   can be a good approximation of    more refined dynamical models,  providing an estimation of different aspects of    tori   construction as elongation on their symmetry plane and the critical pressure points.
This  analysis   is therefore the  first step to the exploration of
  the wider scenario in which  the \textbf{RAD} orbits a central Kerr attractor is involved, relating directly   jet emission with energy extraction from the  central \textbf{BH}  at the expense in general of rotational energy, considering also
Blandford-- Znajek process,  the   magnetic field lines  torque, magnetic Penrose process\citep{Universe} and the Lense--Thirring precession effects.


\begin{thebibliography}{99}
    \bibitem[Abramowicz\&Fragile(2013)]
{abrafra}
  Abramowicz M. A. \& Fragile P. C.  2013,
   Living Rev. Relativity, {\textbf{16}}, 1


\bibitem[Abramowicz et al.(1978)]
 {AJS78}
 Abramowicz  M.A., Jaroszy{\'n}ski M. J\& Sikora  M. 1978,
\aap, {63}, 221


\bibitem[Adamek\&Stuchlik(2013)] {adamek} Adamek K. \&   Stuchlik Z. 2013,
 Class. Quantum Grav. {30}, 205007

\bibitem[Alig et al.(2013)]{Aligetal(2013)}	
Alig  C.,  Schartmann   M.,   Burkert A., Dolag  K. 2013,
\apj,  771,  2, 119

\bibitem[Aly et al.(2015)]
{Aly:2015vqa}
Aly H.,   Dehnen W.,  Nixon C., \& King A. 2015,
  \mnras,  { 449}, 1,  65
  \bibitem[Balbus\&Hawley(1998)]
{BHawley}
 Balbus S. A.\& Hawley J. F.  1998,
  Rev. Mod. Phys,. {70}, 1
    \bibitem[Blaes(1987)]
{Blaes1987}Blaes   O. M.  1987, \mnras \textbf{227}, 975

\bibitem[Blanchard et al.(2017)]
{Blanchard:2017zfe}
  Blanchard P.~K. 2017,  {\it et al.},
  arXiv:1703.07816 [astro-ph.HE]
\bibitem[Blaschke\&Stuchl{\'{\i}}k(2016)]
  {Blaschke:2016uyo}
 Blaschke M.  \&  Stuchlik Z. 2016,
  \prd , { 94},  8,  086006


\bibitem[Bonnell\&Rice(2008)]
{Bonnell} Bonnell I.~A., \&  Rice W.~K.~M. 2008,\  Science, 321, 1060

 \bibitem[Bonnerot et al.(2016)]
{Bonnerot:2015ara}
 Bonnerot  C., Rossi E.~M., Lodato G.  
 2016,
\mnras,\  {455}, 2,  2253

    \bibitem[Boyer(1965)]
{Boy:1965:PCPS:} Boyer R. H. 1965, MPCPS, \textbf{61}, 527




\bibitem[Carmona-Loaiza et al.(2015)]
{Carmona-Loaiza:2015fqa}
Carmona-Loaiza J.M., Colpi M.,  Dotti M. et al 
2015,
 \mnras,  { 453},  1608


\bibitem[Chatterjee et al.(2020)] {Chatterjee:2020eqc}
Chatterjee K.~, Younsi Z.~, Liska M., Tchekhovskoy A., et al. 2020,
[arXiv:2002.08386 [astro-ph.GA]].


\bibitem[Contopoulos et al.(2012)]{CCpuol}
 Contopoulos I. , Kazanas D.,  Papadopoulos D. B.  2013,
\apj,   \textbf{765},  2, 113


\bibitem[Dexter\&Fragile(2011)]{2011ApJ...730...36D} Dexter  J. \& Fragile P.~C.  2011,  \apj , 730, 36


 \bibitem[Dogan et. al(2015)]
 {Dogan:2015ida}
Dogan  S., Nixon C., King A., et al 
2015,
 \mnras, {449},  2,  1251
   \bibitem[Dyda et al.(2015)]{Dyda:2014pia}
Dyda S,  Lovelace   R.V.E., et al. 2015
  \mnras,  { 446}, 613

\bibitem[Fragile\&Blaes(2008)]{Fragile:2008sv}
Fragile P. and  Blaes O.~M. 2008,
\apj,   \textbf{687}, 757

\bibitem[Franchini et al.(2019)]{franchini}Franchini A., Martin R. G. , Lubow S.H. 2019,
\mnras,  485,  1, 315--325

 \bibitem[Gafton et al.(2015)]
{Gafton:2015jja}
  Gafton   E., Tejeda E. , et al  2015
  \mnras,  { 449}, 1,  771

  \bibitem[Greenhill et al.(2003)]{Greenhill2003}Greenhill L. J. , Kondratko P. T. ,  Lovell J. E. J., Kuiper T. B. H., et al.,2003,  
\apj, 582, L1

{\bibitem[Gutierrez\&Pachon(2015)]{GP15} Gutierrez--Ruiz A.F., Pachon L. A. 2015, \prd,  91, 124047}


\bibitem[Hamersky\&Karas(2013)]{Hamersky:2013cza}
Hamersky J. \&V. Karas 2013,
  \aap,\  { 555}, A32

{\bibitem[Herrera et al.(2006)]{HG2006}Herrera L, Gonzalez G.A., Pachon L.A., Rueda J.A. 2006, Class. Quantum Grav. 23, 2395
}

\bibitem[Herrnstein et al (1996)]{Herrnstein1996}  Herrnstein J. R., Greenhill L. J. , Moran J. M.  1996, \apj, 468,
L17

\bibitem[Horak\&Bursa(2009)]{Horak:2009iz}
  Horak J.\& Bursa M.,
  ``Polarization from the oscillating magnetized accretion torus,'' in
   R. Bellazzini, E. Costa, G. Matt\&G. Tagliaferri
 \emph{X-ray Polarimetry,
} Cambridge University Press
 (2010) arXiv:0906.242


 \bibitem[King\&Nixon(2018)]
{King:2018mgw}
  King A. and Nixon C. 2018,
  \apj,   { 857}, 1,  L7

  \bibitem[Komissarov(2006)]
{Komissarov:2006nz}
  Komissarov S.~S. 2006 ,
  \mnras,\  { 368},  993

\bibitem[Kovar et al.(2010)]
  {Kovar:2010ty}
  Kovar J., O.~Kopacek, Karas V. and Stuchlik Z. 2010,
  Class.\ Quant.\ Grav.\  {\bf 27},  135006


\bibitem[Kovar et al.(2014)]
{Kovar:2014tla}
  Kovar J., Slany P.,  Cremaschini C., et al. 2014 
  \prd,  {\bf 90}, 4,  044029

\bibitem[Kovar et al.(2016)]
{Kovar:2016kqh}
  Kovar J., Slany P., Cremaschini  C., et al. 2016 
  \prd, {\bf 93},  12,  124055


   \bibitem[Kovar et al.(2011)]{Kovar:2011uh}
  Kovar J., Slany P., Stuchlik Z., Karas V., C.~Cremaschini and J.~C.~Miller 2011,
  \prd,  {\bf 84}, 084002

  \bibitem[Kozlowski et al.(1978)]
{KJA78}	
 Kozlowski M., M.  Jaroszynsk\& Abramowicz M. A. 1978,
\aap 63,  1--2, 209--220

\bibitem[Lasota et al.(2016)]
   {Lasota:2015bii}
  Lasota J. P.,  Vieira R. S. S. , Sadowski A., Narayan R.,  \& Abramowicz M. A. 2016,
\aap, {587},  A13
\bibitem[Lei et al.(2008)]
{Lei:2008ui}
 Lei  Q. , Abramowicz M. A. ,  Fragile P. C. , \emph{et al.} 2008 
\aap, {\textbf{498}}, 471
\bibitem[Liska et al. (2019)]{LLL}
Liska M., Tchekhovskoy A. , Ingram A., van der Klis M.  2019, \mnras,  487,  1,   550--56
\bibitem[Lodato\&Pringle(2006)]
{2006MNRAS.368.1196L}   Lodato G. \&  Pringle  J.~E. 2006, \mnras, \textbf{368}, 1196
 \bibitem[Lyutikov(2009)]
  {Lyutikov(2009)}
Lyutikov M. 2009,
\mnras, 396,  3,  1545--1552


\bibitem[Madau(1988)]
{Madau(1988)}
Madau P.  1988,	
\apj, 1, 327,   116-127
\bibitem[Mahlmann et al.(2018)]{Mahlmann:2018ukr}
Mahlmann   J.~F., Cerda-Duran P. and  Aloy M.~A. 2018,
 \mnras, {\bf 477}, 3,  3927


    \bibitem[Miller et al.(2015)]
{natures}
 	Miller J. M.,   Kaastra, J. S., Coleman Miller  M.	C., et al  2015
\nat, {526}, 542--545

\bibitem[Miller-Jones et al.(2019)]{Miller-Jones:2019zla}
Miller-Jones J.~C., Tetarenko A.~J., Sivakoff G.~R., Middleton M.~J., et al. 2019 
Nature, \textbf{569}, 374-377

  {\bibitem[Mirza(2017)]{M17} Mirza, B.M. 201, apj, 847, 73 }

  \bibitem[Montero et al.(2007)]
  {Montero:2007tc}
Montero P.~J. , Zanotti O. ,  Font J. A.,\& Rezzolla  L.  2007,
  \mnras,\  { 378},  1101



\bibitem[Nixon et al.(2013)]
{Nixon:2013qfa}
 Nixon C,  King A, \& Price D. 2013,
  \mnras,  { 434}, 1946

\bibitem[Parker(1955)]{Parker:1955zz}
    Parker E.~N. 1955,
 \apj, {122}, 293

\bibitem[Parker(1970)]{Parker:1970xv}
 Parker  E.~N. 1970,
  \apj,  160, 383
{\bibitem[Petri(2016)]{P16} Petri, J. 2016, \aap,  594, A112}

\bibitem[Pugliese\&Montani(2013)]
{epl}
 Pugliese D. \&Montani G. 2013,
  EPL, { \textbf{101}}, 1, 19001


\bibitem[Pugliese\&Montani(2015)]
{pugtot}
  Pugliese D.\&Montani G. 2015,
  \prd, \textbf{91}, 083011



  \bibitem[Pugliese\&Montani(2018)]
{mnrasB}
  Pugliese D.\&Montani G. 2018,
  \mnras,  {\bf 476}, 4,  4346

  \bibitem[Pugliese\&Montani(2020)]{Pugliese:2020ivz}
Pugliese D.and Montani G. 2020,
Entropy, \textbf{22}, 402


\bibitem[Pugliese\&Quevedo(2018)]{observers}
  Pugliese D. and Quevedo H. 2018,
  Eur.\ Phys.\ J.\ C, {78},  1,  69


\bibitem[Pugliese\&Quevedo(2019a)]{Pugliese:2019rfz}
Pugliese D.and Quevedo H. 2019a,
[arXiv:1910.04996 [gr-qc]]


\bibitem[Pugliese\&Quevedo(2019b)]{Pugliese:2019efv}
Pugliese D and Quevedo H 2019b,
[arXiv:1910.02808 [gr-qc]]


\bibitem[Pugliese\& Quevedo(2019c)]{remnants}
  Pugliese D. and Quevedo H. 2019c,
  Eur.\ Phys.\ J.\ C, 79, 3,  209




\bibitem[Pugliese\& Quevedo(2021)]{Pugliese:2020azr}
Pugliese D.and Quevedo H 2021.,
European Physical Journal C,  81,  258

   \bibitem[Pugliese\&Stuchl\'{\i}k(2015)]
{ringed}
 Pugliese D.\&Stuchl{\'{\i}}k Z. 2015,
\apjs,  { \textbf{221}}, 2,  25


  \bibitem[Pugliese\&Stuchl\'{\i}k(2016)]
  {open}
Pugliese D. \&Stuchl{\'{\i}}k Z. 2016,
 \apjs,  { \textbf{223}}, 2,  27

\bibitem[Pugliese\&Stuchl{\'{\i}}k(2017)]
 {dsystem} Pugliese D.\& Stuchl{\'{\i}}k Z. 2017, \apjs,  \textbf{229}, 2, 40
\bibitem[Pugliese\&Stuchlik(2018a)]{Multy} Pugliese D.\& Stuchlik Z. 2018a,
  Class.\ Quant.\ Grav.,  {\bf 35}, 18,  185008




 \bibitem[Pugliese\&Stuchl{\'{\i}}k(2018a)]
{proto-jet}
  Pugliese D.\&Stuchl{\'{\i}}k Z. 2018a,
  Class.\ Quant.\ Grav.\  {\bf 35}, 10,  105005

  \bibitem[Pugliese\&Stuchl{\'{\i}}k(2018b)]
{long}
  Pugliese D.\& Stuchl\'{\i}k Z. 2018b,
 JHEAp, { \textbf{17}},  1


    \bibitem[Pugliese\&Stuchl{\'{\i}}k(2019)]
{Letter}Pugliese D. \& Stuchl\'{\i}k Z. 2019, 
  Eur.\ Phys.\ J.\ \textbf{C}, {\bf 79}, 4,  288

 \bibitem[Pugliese\&Stuchlik(2020a)]{mnras2}  Pugliese D.\& Stuchlik Z. 2020a, \mnras, 493, 4229


\bibitem[Pugliese\&Stuchlik(2020b)]{cqg2020}
Pugliese D.\&  Stuchlik Z. 2020b,
    Class. Quantum Grav. 37, 195025




\bibitem[Reyes-Ruiz\&Stepinski(1999)] {R-ReS1999}
Reyes-Ruiz  M., Stepinski  T. F.  1999,
 \aap,\  { 342},  892--900


  \bibitem[Sadowski et al.(2016)]
{Sadowski:2015jaa}
 Sadowski A., Lasota J. P.,  Abramowicz M. A.  \&  Narayan R. 2016,
  \mnras,  {456}, 4,  3915
 \bibitem[Shakura(1973)]
 {[S73]}  Shakura N.I.  1973,   Sov. Astronomy, {16}, 756

\bibitem[Shakura\&Sunyaev(1973)]
 {[SS73]}
Shakura  N.I. \& Sunyaev R. A.   1973,\aap, {24}, 337

\bibitem[Sikora(1981)]
  {Sikora(1981)}
Sikora M.  1981, \mnras,  196,  257

  \bibitem[Slany et al.(2013)]
{Slany:2013rml}
  Slany P., Kovar J., Stuchlik Z. and Karas V. 2013,
  \apjs, {\bf 205}, 3







    \bibitem[Stuchlik(1980)]
{1980BAICz..31..129S}  Stuchlik Z. 1980,  BAICz, 31, 129



   \bibitem[Stuchlik(1983)]
{1983BAICz..34..129S}   Stuchlik Z. 1983, BAICz, \textbf{34}, 129
\bibitem[Stuchl{\'{\i}}k(2005)]
{2005MPLA...20..561S}  Stuchlik Z. 2005, Modern Physics Letters \textbf{A}, \textbf{20}, 561

{\bibitem[Stuchlik et al.(2017)]{2017PhRvD..96j4050S}   Stuchlik Z., Blaschke M.,\&  Schee J. 2017, \prd, 96, 104050}

\bibitem[Stuchl{\'{\i}}k \& Hled{\'{\i}}k(1999)]
{1999PhRvD..60d4006S}   Stuchlik Z., \&  Hled{\'{\i}}k S. 1999, \prd, \textbf{60}, 044006
\bibitem[Stuchl{\'{\i}}k et al.(2016)]
{2016PhRvD..94j3513S}  Stuchlik Z.,  Hled{\'{\i}}k S., \& Novotn{\'y} J. 2016, \prd , \textbf{94}, 103513
\bibitem[Stuchl{\'\i}k et al.(2011)]
{2011CQGra..28o5017S}  Stuchlik Z.,  Hled{\'\i}k S., Truparov{\'a} K. 2011, CQGra, 28, 15501






\bibitem[Stuchl{\'\i}k, et al.(2020)]
{Universe}   Stuchlik Z.,  Kolo{\v{s}} M. , Kovar J., et al 
2020, Univ, 6, 26


\bibitem[Stuchl{\'\i}k et al.(2013)]
    {2013A&A...552A..10S}  Stuchlik Z.,  Kotrlov{\'a} A. , T{\"o}r{\"o}k G. 2013, A\&A, 552, A10



\bibitem[Stuchl{\'\i}k \& Kov{\'a}{\v{r}}(2008)]
{2008IJMPD..17.2089S}  Stuchlik Z., \&Kovar J. 2008, Int. Journ. of Modern Physics D, 17, 2089

 \bibitem[Stuchl{\'{\i}}k et al.(2000)]
    {2000A&A...363..425S} Stuchlik Z., Slany P. \&  Hled{\'{\i}}k S. 2000, \aap, \textbf{363}, 425

   \bibitem[Stuchl{\'\i}k et al.(2009)]
    {2009CQGra..26u5013S}   Stuchlik Z.,  Slany P., Kovar J. 2009, CQGra, 26, 215013

 \bibitem[Stuchlik et al.(2005)]{Stuchlik:2004wk}
  Stuchlik Z., Slany P., Torok G.  and Abramowicz M.~A.~ 2005,
  \prd,  {\bf 71}, 024037


\bibitem[Tchekhovskoy et al.(2010)]{TNM}
Tchekhovskoy A., Narayan R., and McKinney J. C.  2010,
\apj,   {711}, 50--63

\bibitem[Teixeira et al.(2014)]{Teixeira:2014una}
Teixeira D.~M., Fragile P.~C., Zhuravlev  V.~V.  et al 
2014,
\apj,   \textbf{796}, 2, 103


\bibitem[Trova et al.(2016)]
{Trova:2016ton}
  Trova A., Karas V., Slany P. et  Kovar J. 2016,
  \apjs,  {\bf 226}, 1,  12

  \bibitem[Trova et al.(2018b)]
{Trova:2018bsf}
  Trova A., Schroven K., Hackmann E., et al,. 2018 
  \prd,  {\bf 97}, 10, 104019


\bibitem[Trova et al.(2018c)]
{Schroven:2018agz}
  Schroven K., Trova A., Hackmann E. et al 
  2018,
  \prd,  {\bf 98}, 2,  023017




\bibitem[Uzdensky(2004)]{Uz}
 Uzdensky D. A. 2004,
\apj,  {\bf 603}, 652--662




\bibitem[Uzdensky(2005)]{Uzdensky:2004qu}
  Uzdensky D.~A. 2005,
  \apj,     {\bf 620}, 889



  \bibitem[Yoshizawa et al.(2003)]
{Y.I.I2003}
  Yoshizawa A., S.Itoh  I., Itoh  K. ,  2003
  \emph{Plasma\&Fluid Turbulence: Theory and Modelling},
  CRC Press

  \bibitem[Zubovas\&King(2008)]{WA}Zubovas K. , King A., 2008
 arXiv:1901.02224 [astro-ph.GA]
\end{thebibliography}
\end{document}